\documentclass[11pt,oneside,preprint,preprintnumbers]{book}

\usepackage{stackengine} 
\usepackage{ragged2e}
\usepackage[T1]{fontenc}
\usepackage[utf8]{inputenc}
\usepackage{bbold}
\usepackage{mathptmx}
\usepackage{fullpage}
\usepackage{amssymb,amsmath}
\usepackage{graphicx}
\usepackage{float}
\usepackage{graphicx, multirow,soul}
\usepackage[citecolor=blue]{hyperref}
\usepackage[all]{hypcap}
\usepackage{url}
\usepackage{minibox}
\usepackage[shortlabels, inline]{enumitem}
\usepackage{color}
\usepackage[T1]{fontenc} 
\usepackage{subfigure}
\usepackage{slashed}
\usepackage{float}
\usepackage{amssymb}
\usepackage{bm}
\usepackage{amsbsy}
\usepackage{enumitem}
\usepackage{textcomp}
\usepackage{ulem}
\usepackage{caption}
\DeclareCaptionFormat{empty}{}
\usepackage[export]{adjustbox}
\usepackage{pdflscape}
\usepackage{array}
\usepackage{gensymb}
\newcolumntype{M}[1]{>{\centering\arraybackslash}m{#1}}
\usepackage{tikz,color}
\usetikzlibrary{shapes.geometric, arrows}
\usetikzlibrary{decorations.markings}
\usetikzlibrary{decorations.shapes}
\usetikzlibrary{decorations.shapes}
\usetikzlibrary{shapes,backgrounds,calc}
\tikzset{decorate sep/.style 2 args=
{decorate,decoration={shape backgrounds,shape=circle,shape size=#1,shape sep=#2}}}
\tikzset{decorate sep/.style 2 args=
{decorate,decoration={shape backgrounds,shape=circle,shape size=#1,shape sep=#2}}}

\tikzstyle{startstop} = [rectangle, rounded corners, minimum width=3cm, minimum height=1cm,text centered, draw=black, fill=blue!20]
\tikzstyle{ss} = [rectangle, rounded corners,trapezium left angle=70, trapezium right angle=110, minimum width=1cm, minimum height=1cm,text centered, draw=black, fill=blue!20]
\tikzstyle{startstop2} = [rectangle, rounded corners, minimum width=3cm, minimum height=1cm,text centered, draw=black, fill=green!20]
\tikzstyle{ss2} = [rectangle, rounded corners, minimum width=1.5cm, minimum height=1cm,text centered, draw=black, fill=green!20]
\tikzstyle{startstop3} = [rectangle, rounded corners, minimum width=3cm, minimum height=1cm,text centered,fill=red!20, draw=black]
\tikzstyle{ss3} = [rectangle,rounded corners, minimum width=1.5cm, minimum height=1cm,text centered,fill=red!20, draw=black]
\tikzstyle{ss4} = [rectangle,rounded corners, minimum width=1.5cm, minimum height=1cm,text centered,fill=green!20, draw=black]
\tikzstyle{ss5} = [rectangle,rounded corners, minimum width=1.5cm, minimum height=1cm,text centered,fill=blue!20, draw=black]
\tikzstyle{io1} = [rectangle split, rectangle split vertical,rectangle split parts=2,  rounded corners, trapezium left angle=70, trapezium right angle=110, minimum width=3cm, minimum height=1cm, text centered,rectangle split part fill={blue!20,green!20}, draw=black]
\tikzstyle{io2} = [rectangle split, rectangle split vertical,rectangle split parts=2,  rounded corners, trapezium left angle=70, trapezium right angle=110, minimum width=3cm, minimum height=1cm, text centered,rectangle split part fill={blue!20,red!20}, draw=black]
\tikzstyle{io3} = [rectangle, rounded corners, minimum width=3cm, minimum height=1cm,text centered, draw=black, fill=blue!20, draw=black]
\tikzstyle{process} = [rectangle, , rounded corners,minimum width=4cm, minimum height=1cm,  draw=black,text width=3.8cm,text centered]
\tikzstyle{decision} = [diamond, minimum width=3cm, minimum height=1cm, text centered, draw=black, fill=green!30]
\tikzstyle{arrow} = [thick,->,>=stealth]
\usepackage{scalefnt}
\usetikzlibrary{shapes.multipart,matrix,positioning}
\newcommand{\fundamental}[2][black!40!green,fill=black!40!green]{\tikz[baseline=-0.75ex]\draw[#1,radius=#2] (0,0) circle ;}
\newcommand{\intermediate}[2][blue,fill=blue]{\tikz[baseline=-0.1em]\draw[#1] (0,0) rectangle (#2,#2);}
\newcommand{\advanced}[2][black]{\begin{tikzpicture}\node[draw,scale=#2,diamond,fill=#1]{};\end{tikzpicture}}
\newcommand{\ket}[1]{|#1\rangle}

\usepackage{setspace}
\usepackage{multicol}
\usepackage{fancyhdr,lastpage}
\usepackage{lastpage}
\usepackage{amsmath}
\usepackage{layout}
\usepackage[left=1in,top=1in,right=1in,bottom=1in,headheight=3ex,headsep=3ex]{geometry}
\usepackage[font={large}]{caption}
\usepackage{tikz}
\usetikzlibrary{shapes}
\usepackage{xcolor}
\usepackage[some]{background}
\definecolor{titlepagecolor}{cmyk}{1,.60,0,.40}
\DeclareFixedFont{\bigsf}{T1}{phv}{b}{n}{1.5cm}
\backgroundsetup{
scale=1,
angle=0,
opacity=1,
contents={\begin{tikzpicture}[remember picture,overlay]
 \path [fill=titlepagecolor] (-0.5\paperwidth,5) rectangle (0.5\paperwidth,10); 
\end{tikzpicture}}
}
\makeatletter                       
\def\printauthor{
    {\large \@author}}              
\makeatother
\author{
  \Large
    Anastasia Perry \\
    \texttt{aperry@imsa.edu}\vspace{20pt} \\
    Ranbel Sun \\
    \texttt{rsun@andover.edu}\vspace{20pt}\\
       Ciaran Hughes \\
    \texttt{chughes@fnal.gov}\vspace{20pt} \\
    Joshua Isaacson \\
    \texttt{isaacson@fnal.gov}\vspace{20pt} \\
     Jessica Turner \\
    \texttt{jturner@fnal.gov}
        }

\usepackage[
    type={CC},
    modifier={by-nc-sa},
    version={4.0},
]{doclicense}
\begin{document}
\large

\begin{titlepage}
\BgThispage
\newgeometry{left=1cm,right=4cm}
\vspace*{2cm}
\noindent
\textcolor{white}{\bigsf Quantum Computing as a  High School Module}
\vspace*{2.5cm}\par
\noindent
\begin{minipage}{0.35\linewidth}
    \begin{flushright}
        \printauthor
    \end{flushright}
\end{minipage} \hspace{15pt}
\begin{minipage}{0.02\linewidth}
  \vspace{15pt}
    \rule{1pt}{310pt}
\end{minipage} \hspace{-10pt}
\begin{minipage}{0.6\linewidth}
  \vspace{15pt}
  \Large 
  Quantum computing is a growing field at the intersection of physics and computer science. This module introduces three of the key principles that govern how quantum computers work: superposition, quantum measurement, and entanglement. The goal of this module is to bridge the gap between popular science  articles and advanced undergraduate texts by making some of the more technical aspects accessible to motivated high school students. Problem sets and simulation-based labs of various levels are included to reinforce the conceptual ideas described in the text. This is intended as a one week course for high school students between the ages of 15-18 years. The course begins by introducing basic concepts in quantum mechanics which are needed to understand quantum computing.
\end{minipage}
\end{titlepage}
\restoregeometry

\newpage
\thispagestyle{empty}
\vspace{-1cm}
\begin{flushright}
{FERMILAB-FN-1077-T}
\end{flushright}

\noindent {\bf{\Large Copyright}}
\doclicenseThis
To view a copy of this license, visit http://creativecommons.org/licenses/by-nc/4.0/. This license allows you to modify  and build upon our work for non-commercial use as long as you credit us and license your new creations under identical terms.

\vspace{0.7cm}
\noindent {\bf{\Large Disclaimer}} \\

\noindent The authors take no responsibility for broken links or if the web interface of referenced material changes. The views and opinions expressed here are those of the authors and do not necessarily reflect the official policy or position of any other agency, organization, employer, or company. The authors are not to be held responsible for misuse, reuse, recycled, cited and/or uncited copies of content.

\frontmatter                            
\tableofcontents                        
\mainmatter                             

\graphicspath{{Chapter0-Overview/}}
\setcounter{chapter}{-1}
\chapter{Course Description}

%
\section{About}
Quantum computing is a growing field at the intersection of physics and computer science. This module introduces three  key principles of quantum computing: superposition, quantum measurement, and entanglement. The goal of this course is to bridge the gap between popular science  articles and advanced undergraduate texts, making some of the more technical aspects accessible to motivated high school students. Problem sets and simulation-based labs of various levels are included to reinforce the concepts described in the text. 

Note that the module is not designed to be a comprehensive introduction to modern physics. Rather, it focuses on topics students may have heard about but are not typically covered in a general course.  Given the usual constraints on teaching time, these materials could be used after the AP exams, in an extracurricular club, or as an independent project resource to give students a taste of what quantum computing is really about.

This is intended as a one-week course for high school students between the ages of 15-18 years. The course begins with the introduction of basic concepts in quantum mechanics needed to understand 
quantum computing.  
%
\section{Prerequisites}

The material assumes knowledge of electricity, magnetism, and waves from high school-level physics. Introductory modern physics (photoelectric effect, wave/particle duality, etc.) is helpful but not required. No computer programming experience is necessary.

Each unit builds up to three different levels of complexity depending on the students' experience with math and abstract reasoning. All problems are labeled according to difficulty. In addition, the intermediate and advanced sections within each chapter are labeled such that one can skip over them if needed. Links to external resources are provided below for those who require a refresher.

\subsection*{\fundamental{5pt} Fundamental}

\begin{itemize}
\item Basic probability -
  \href{https://www.khanacademy.org/math/statistics-probability/probability-library/basic-theoretical-probability/v/basic-probability}{Khan Academy Probability}\footnote{\href{https://www.khanacademy.org/math/statistics-probability/probability-library/basic-theoretical-probability/v/basic-probability}{https://www.khanacademy.org/math/statistics-probability/probability-library/basic-theoretical-probability/v/basic-probability}}

\item Histograms - 
\href{https://www.khanacademy.org/math/ap-statistics/quantitative-data-ap/histograms-stem-leaf/v/histograms-intro}{Khan Academy Histograms}\footnote{\href{https://www.khanacademy.org/math/ap-statistics/quantitative-data-ap/histograms-stem-leaf/v/histograms-intro}{https://www.khanacademy.org/math/ap-statistics/quantitative-data-ap/histograms-stem-leaf/v/histograms-intro}}

\end{itemize}

\subsection*{\intermediate{8pt} Intermediate}

\begin{itemize}
\item Probability multiplication - 
\href{https://www.khanacademy.org/math/statistics-probability/probability-library/multiplication-rule-independent/v/compound-sample-spaces}{Khan Academy Multiplication}\footnote{\href{https://www.khanacademy.org/math/statistics-probability/probability-library/multiplication-rule-independent/v/compound-sample-spaces}{https://www.khanacademy.org/math/statistics-probability/probability-library/multiplication-rule-independent/v/compound-sample-spaces}}

\item Vector decomposition - 
  \href{http://www.physicsclassroom.com/class/vectors/Lesson-1/Vectors-and-Direction}{Physics Classroom}\footnote{\href{http://www.physicsclassroom.com/class/vectors/Lesson-1/Vectors-and-Direction}{http://www.physicsclassroom.com/class/vectors/Lesson-1/Vectors-and-Direction}}

  \end{itemize}

\subsection*{\advanced{0.6pt} Advanced}

\begin{itemize}
\item Matrix multiplication -
\href{https://www.khanacademy.org/math/precalculus/precalc-matrices/multiplying-matrices-by-matrices/v/matrix-multiplication-intro}{Khan Academy Matrix Multiplication}\footnote{\href{https://www.khanacademy.org/math/precalculus/precalc-matrices/multiplying-matrices-by-matrices/v/matrix-multiplication-intro}{https://www.khanacademy.org/math/precalculus/precalc-matrices/multiplying-matrices-by-matrices/v/matrix-multiplication-intro}}

\item Interactive Matrix Multiplication - 
\href{https://www.st-andrews.ac.uk/physics/quvis/simulations\_html5/sims/MatrixMultiplication/MatrixMultiplication.html}{University of St.~Andrews}\footnote{\href{https://www.st-andrews.ac.uk/physics/quvis/simulations\_html5/sims/MatrixMultiplication/MatrixMultiplication.html}{https://www.st-andrews.ac.uk/physics/quvis/simulations\_html5/sims/MatrixMultiplication/MatrixMultiplication.html}}

\item Matrices as transformations - 
\href{https://www.khanacademy.org/math/precalculus/precalc-matrices/matrices-as-transformations/v/transforming-position-vector}{Khan Academy Matrices Transformations}\footnote{\href{https://www.khanacademy.org/math/precalculus/precalc-matrices/matrices-as-transformations/v/transforming-position-vector}{https://www.khanacademy.org/math/precalculus/precalc-matrices/matrices-as-transformations/v/transforming-position-vector}}

\end{itemize}

\section{Learning Objectives}

\begin{enumerate}

\item Introduction to Superposition
  \begin{itemize}
  \item Qualitatively understand what it means for an object to be in a quantum superposition.
  \item Identify the measurement outcome of a system in a classical vs. quantum superposition.
  \end{itemize}
Key Terms: {\it{quantum system, quantum state, quantum superposition}}

\item What is a Qubit?
  \begin{itemize}
    \item Understand the difference between a classical bit and a qubit.
    \item Write a mathematical expression for the superposition of a two-state particle using ``ket'' notation.
    \item Compute the probability of finding the particle in a particular state given a normalized superposition state.
    \item Express a qubits' state as a vector and use matrix multiplication to change the state.
    \end{itemize}    
  Key Terms: {\it{qubit, ket notation, state amplitude, normalization, unitary matrix}}

\item Creating Superposition:  Beam splitter
    \begin{itemize}
      \item Explain how light behaves like a particle in the single-photon beam splitter experiment.
      \item Understand how the beam splitter creates a particle in a superposition state.
      \item Trace the path of light through a Mach-Zehnder interferometer from both a wave interference and particle perspective.
    \end{itemize}
 Key Terms: {\it{photon, beam splitter, phase shift, Mach-Zehnder interferometer}}
    
\item Creating Superposition: Stern-Gerlach
  \begin{itemize}
    \item Explain why electron spin could serve as an example of a qubit.
    \item Understand how the Stern-Gerlach experiment illustrates spin quantization, superposition, and measurement collapse.
    \item Define what is meant by a measurement basis and convert a given spin to a different basis. 
    \item Compute the probability of an electron passing through one or more Stern-Gerlach apparatuses.
  \end{itemize}
Key Terms: {\it{spin, Stern-Gerlach experiment, measurement basis, orthogonal states, no-cloning theorem}}

\item Quantum Cryptography
  \begin{itemize}
    \item Send a message with the one-time pad to understand what is meant by a cryptographic key.
    \item Generate a shared key using the BB84 quantum key distribution protocol.
    \item Show how the principles of superposition and measurement collapse make the protocol secure.
  \end{itemize}
Key Terms: {\it{key, quantum key distribution}}
  
\item Quantum Gates
  \begin{itemize}
  \item Build and test simple quantum circuits on IBM's quantum computer simulator.
  \item Interpret the histograms produced by single qubit gates: the $X$, Hadamard, and $Z$ gates.
  \item Predict the output of multiple gates in a row, including two successive Hadamards.
  \item Use the matrix representation of gates to determine the new state of the system.
  \end{itemize}
Key Terms: {\it{quantum gates, $X$ gate, Hadamard gate, $Z$ gate}}
  
\item Entanglement
  \begin{itemize}
    \item Understand how measurement affects the state of entangled particles.
    \item Write the state of a multi-qubit system in ``ket'' notation.
    \item Identify whether two qubits are entangled   given a particular state.
    \item Predict the output of circuits involving CNOT gates.
    \item Entangle two qubits using gates.
  \end{itemize}
Key Terms: {\it{quantum entanglement, product/separable states, entangled states, CNOT gate}}
  
\item Quantum Teleportation
  \begin{itemize}
  \item Qualitatively understand how the quantum state of a particle could be transmitted from one place to another.
  \item Explain the limitations and paradoxes of quantum teleportation.
  \end{itemize}
Key Terms: {\it{quantum teleportation, no-cloning theorem}}
  
\item Quantum Algorithms
  \begin{itemize}
  \item Understand the benefits and limitations of quantum computers.
  \item Use the Mach-Zehnder interferometer to implement the Deutsch-Jozsa quantum algorithm.
  \item Describe how superposition and interference are leveraged in quantum computing algorithms.
  \end{itemize}
Key Terms: {\it{quantum parallelism, Deutsch-Jozsa algorithm}}

\end{enumerate}

\section*{Alternative Pathways}

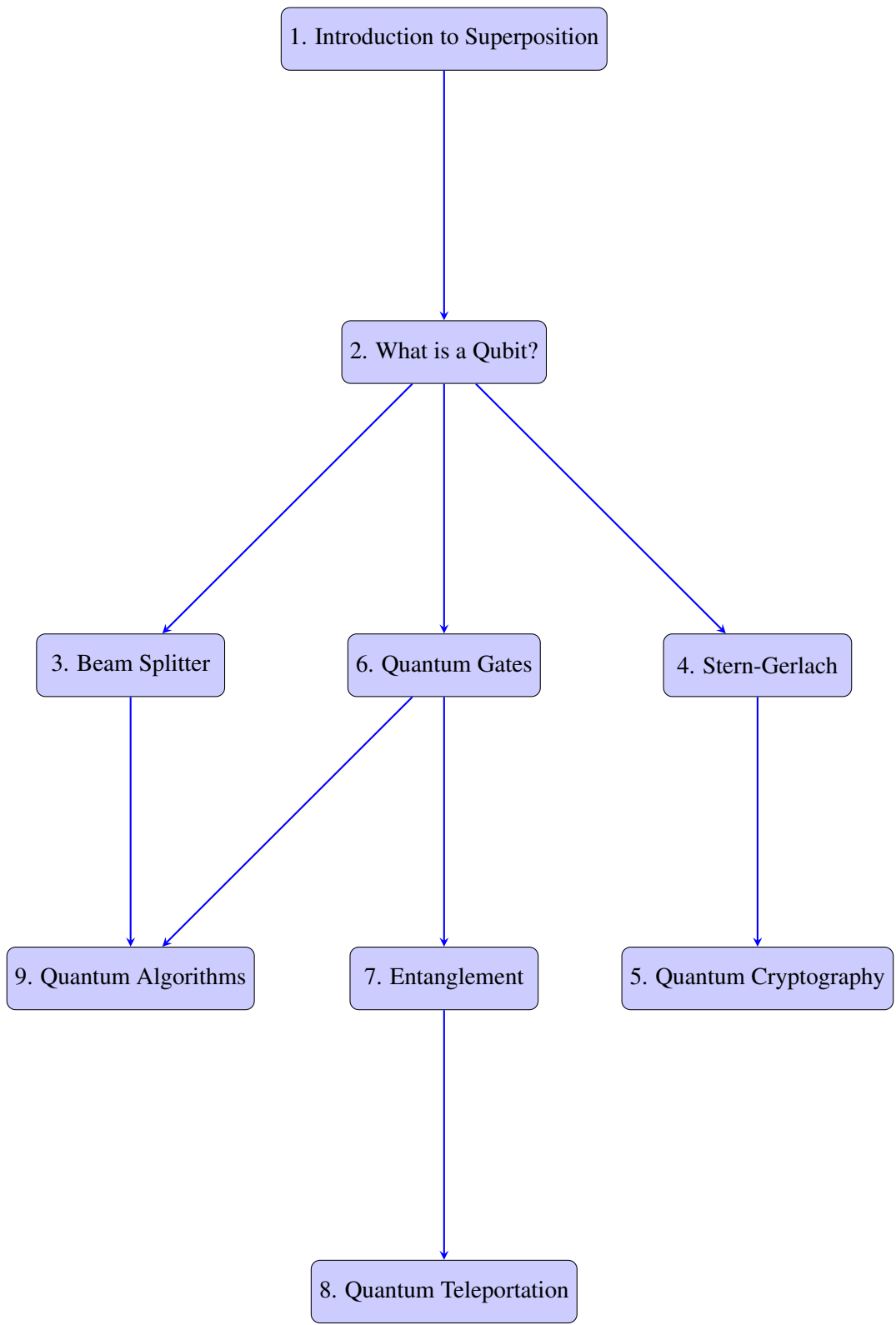
\begin{figure}
\centering
\scalefont{1.0}
\begin{tikzpicture}[node distance=5cm]
\node (in1) [startstop] {1. Introduction to Superposition};
\node (in2) [startstop, below of=in1] {2. What is a Qubit?};
\node (in3) [startstop, below of=in2] {6. Quantum Gates};
\node (in4)  [startstop,left of=in3]{3. Beam Splitter};
\node (in5) [startstop,right of=in3]{4. Stern-Gerlach};
\node (in6) [startstop, below of=in3] {7. Entanglement};
\node (in7)  [startstop,left of=in6]{9. Quantum Algorithms};
\node (in8) [startstop,right of=in6]{5. Quantum Cryptography};
\node (in9) [startstop,below of=in6]{8. Quantum Teleportation};
\draw [blue,arrow,thick] (in1) -- (in2);
\draw [blue,arrow] (in2) -- (in3);
\draw [blue,arrow] (in2) -- (in4);
\draw [blue,arrow] (in2) -- (in5); 
\draw [blue,arrow] (in4) -- (in7);
\draw [blue,arrow] (in3) -- (in7);
\draw [blue,arrow] (in3) -- (in6); 
\draw [blue,arrow] (in5) -- (in8); 
\draw [blue,arrow] (in6) -- (in9); 
\end{tikzpicture}
\caption{Flowchart of learning outcomes.}\label{fig:tikz}
\end{figure}

The units are best studied in numerical order. However, for those with limited time, Figure \ref{fig:tikz} shows the minimum recommended prerequisites for each unit. A few references and examples may have to be skipped over, but the core content should still be understandable.

\graphicspath{{Chapter1-Intro/}}
\chapter{Introduction to Superposition}

In this section, we review the basic concepts of classical and quantum superposition. In Activities Sheet 1, we present the related activities and questions. Before going into specific details on quantum superposition, it is useful to explain how the term ``superposition''  is used in different contexts, i.e., in classical or quantum physics. 
\section{\fundamental{5pt} Classical Superposition}
\begin{figure}[h]
  \centering
  \includegraphics[width=0.75\textwidth]{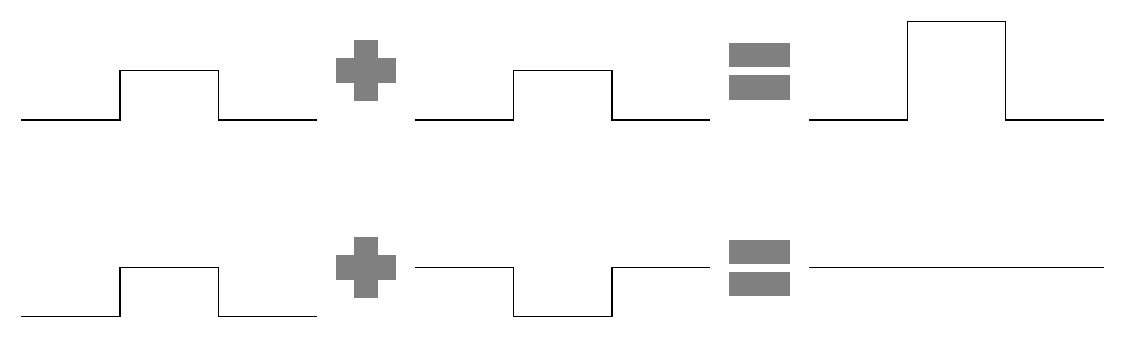}
  \caption{Examples of constructive and destructive interference due to the classical superposition principle.}
  \label{fig:superp}
\end{figure}

In classical physics, the concept of {\bf{superposition}} is used to describe when two physical quantities are added together to make another third physical quantity that is entirely different from the original two. An example of the ``superposition principle''  in classical physics is clear when working with waves. Two pulses on a string which pass through each other will interfere following the principle of superposition as shown Figure~\ref{fig:superp}. Noise-canceling headphones use superposition by
creating sound waves with the same magnitude as the incoming sound wave but with a frequency  completely out of phase,
 thereby canceling the sound wave. This destructive interference is illustrated in the second figure of Figure~\ref{fig:superp}.

\begin{figure}[h]
\centering
  \includegraphics[width=0.35\textwidth]{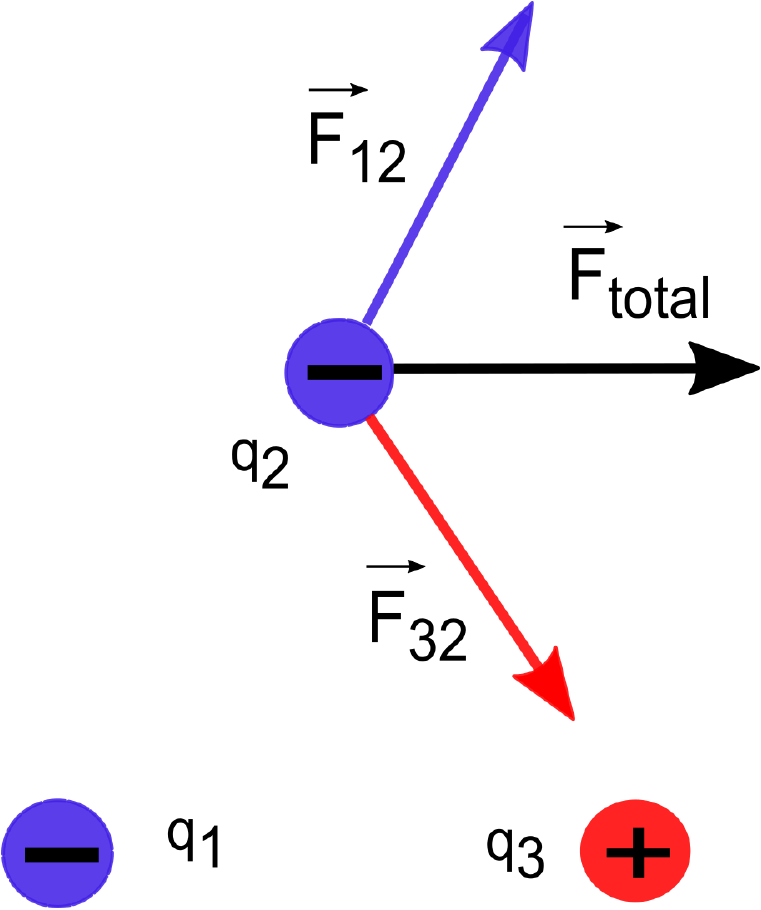}
  \caption{ A classical superposition is used to calculate the total electric force on a charge $q_2$ due to charges $q_1$ and $q_3$.}
  \label{fig:forces}
\end{figure}

Another common application of classical superposition is finding the total magnitude and direction of quantities such as force, electric field, magnetic field, etc. For example, to calculate the total electric force $\vec{F}_{\text{total}}$ on a charge $q_2$ produced by other charges $q_1$ and $q_3$, one would sum the forces produced by each individual charge: $\vec{F}_{\text{total}} = \vec{F}_{12} + \vec{F}_{32}$. The challenge here is that forces are vectors, so vector addition is needed, as shown in Figure~\ref{fig:forces}.

\section{\fundamental{5pt} Quantum Superposition}

\begin{figure}[h]
\centering
  \includegraphics[width=0.75\textwidth]{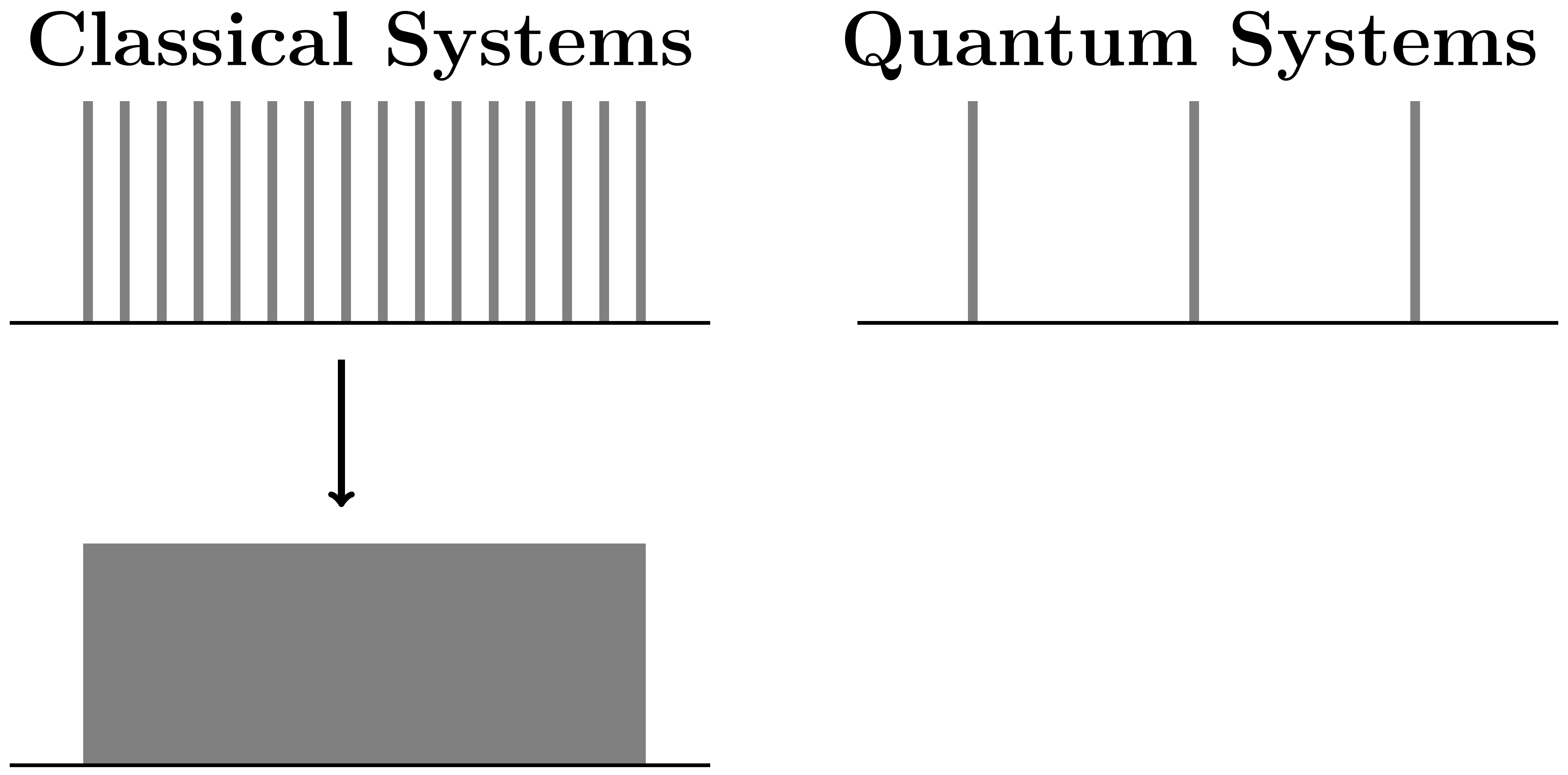}
  \caption{{Quantum effects associated with energy quantization are important at the atomic  and subatomic distances}. In this figure, the grey lines represent allowed energies. In quantum systems, the energies are quantized. As we zoom out of the quantum system to see it through a classical lens (represented by the downward arrow), the energies become more dense and appear continuous. This is the reason quantization is not noticeable in everyday objects.}
  \label{fig:classvquant}
\end{figure}

\begin{figure}[h]
  \centering
  \includegraphics[width=0.5\textwidth]{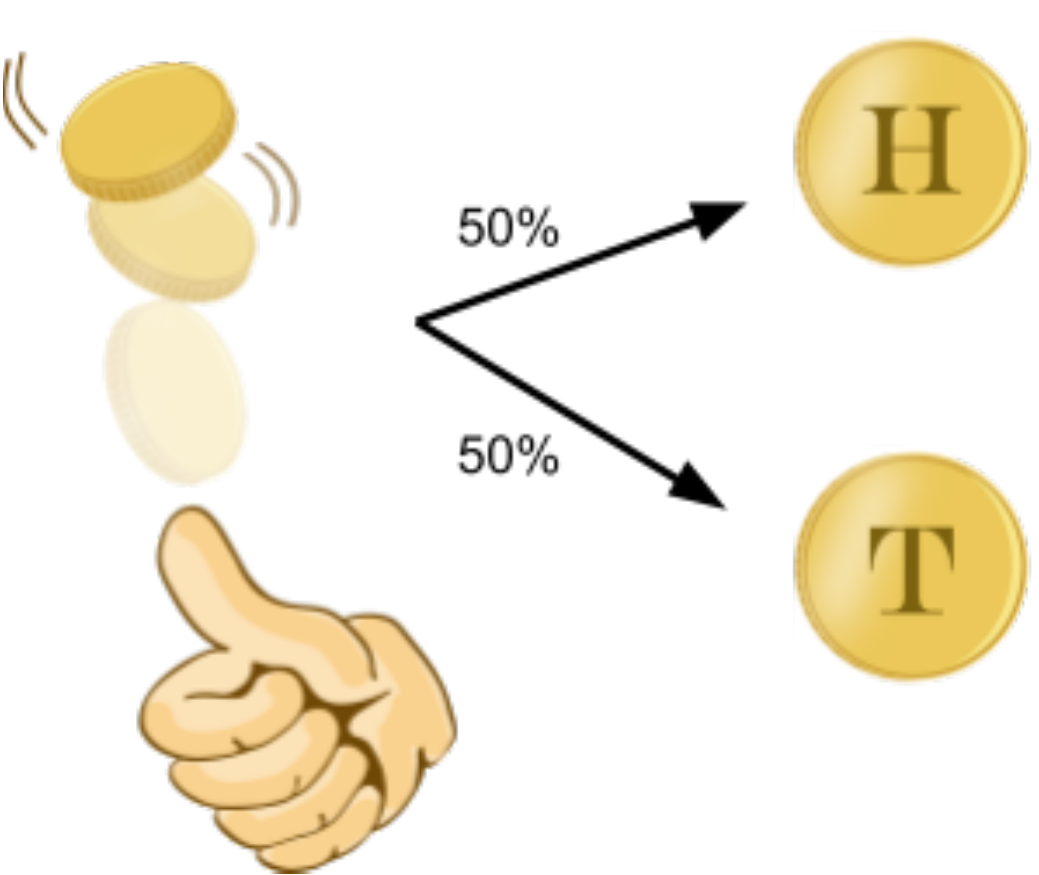}
  \caption{A tossed coin has a $50\%$ chance of landing on heads or tails.}
  \label{fig:HT}
\end{figure}

Quantum superposition is a phenomenon associated with quantum systems, i.e., small objects such as nuclei, electrons, elementary particles, and photons, for which wave-particle duality and other non-classical effects are observed. For example, you would normally expect that an object can have an arbitrary amount of kinetic energy, ranging from 0-$\infty$ Joules. A baseball could be at rest or thrown at any speed. However, according to quantum mechanics, the ball's energy is {\bf{quantized}}, meaning it can only
take on certain values and nothing in between.
A specific example of energy quantization is when energies can only have integer values $E = 0, 1, 2, 3, \ldots$, but not any numbers inbetween.
This is  counterintuitive, as we cannot observe it with our classical eyes. The gaps in energy are too small to be measured on the macroscopic level and as such can be treated as continuous for macroscopic physics. However, the gaps are more pronounced at smaller scales, as shown in Figure~\ref{fig:classvquant}. Bohr successfully modeled the hydrogen atom by quantizing the energy levels of the proton-electron bound state.
One aspect of quantum superposition is easily demonstrated using a coin. A coin has a $50/50$ probability of landing as either heads or tails, as shown in Figure~\ref{fig:HT}. \\

\noindent \textbf{Question 1}: What state is the coin in while it is in the air? Is it heads or tails? 
\\

We can say that the coin is in a superposition of both heads and tails. When it lands, it has a \textbf{definite state}, either heads or tails.
Broadly, the word ``state'' means any particular way that a system can possibly be described. For example, the coin can be either heads, or tails, or a combination of heads or tails while flipped in the air. All of these cases are called states of the coin system.
The measurement destroys the superposition.

At any given time, a system can be described as being in a particular state. The state is related to its quantized values. For example, a tossed coin is either in a heads state or a tails state. An electron orbiting a hydrogen atom could be in the ground state or an excited state. A quantum system is special because it can be in a superposition of these definite states, i.e., both heads and tails simultaneously. It is possible for a quantum object to exist in multiple states at the same time. The outcome of a measurement is to observe some definite state with a given probability. 

In Schr\"odinger's famous thought experiment, Schr\"odinger's cat is placed in a closed box with a single atom that has some probability of emitting deadly radiation at any time. Since radioactive nuclear decay is a spontaneous process, it is impossible to predict for certain when the nucleus decays. Therefore, you do not know whether the cat is alive or dead unless you open and look in the box. (\href{https://www.youtube.com/watch?v=uWMTOrux0LM}{Watch this video}.)\footnote{https://www.youtube.com/watch?v=uWMTOrux0LM} It can be said that the cat is both alive {\bf{AND}} dead with some probability. That is, the cat is in a quantum superposition state until you open the box and measure its state. Upon measurement, the cat is obviously either alive OR dead and the superposition has collapsed to a definite, non-superposition state.

Quantum systems can exist in a superposition state, and measuring the system will collapse the superposition state into one definite classical state. This might be hard to understand from a classical point of view, as we usually do not see quantum superposition in macroscopic objects. Einstein was really bothered by this feature of quantum systems. His friend, Abraham Pais, records: 
``I recall that during one walk, Einstein suddenly stopped, turned to me, and asked whether I really believed that the moon exists only when I look at it.''\footnote{Nielsen, M. A. 1., \& Chuang, I. L. (2000). \emph{Quantum computation and quantum information}. New York: Cambridge University Press, p. 212.}

\section*{\fundamental{5pt} Big Ideas}
\begin{enumerate}
\item A particle in a quantum superposition exists as  a combination of different states at the same time.
\item Measurement destroys the superposition because only one state is seen with   a given probability.
\end{enumerate}

\section{Activities}
\begin{itemize}
\item[\fundamental{5pt}] Quantum Tic-Tac-Toe in Worksheet \ref{sec:WorksheetQTicTacToe}
\end{itemize}

\section{Check Your Understanding}

\begin{figure}[h]
\centering
  \includegraphics[width=0.2\textwidth]{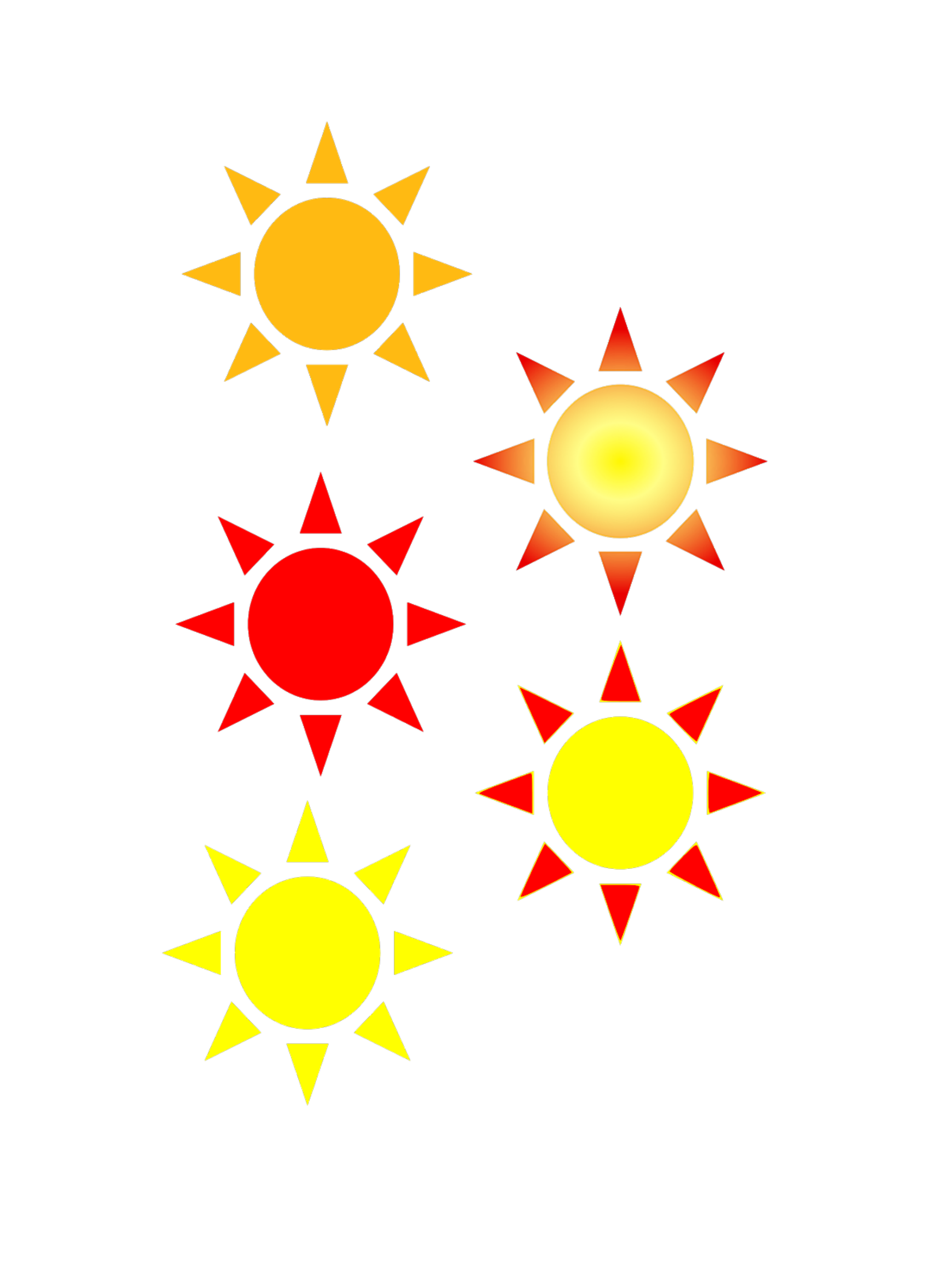}
  \caption{Image of the painted suns.}
  \label{fig:sun}
\end{figure}

\begin{enumerate}
\item \fundamental{5pt} Discuss whether the following quantities are quantized or continuous:
\begin{enumerate}[label=(\alph*)]
\item electric charge
\item  time
\item  length
\item cash
\item paint color
\end{enumerate}

\item \fundamental{5pt} An ink is created by mixing together 50$\%$ red ink and 50$\%$  yellow ink. An artist uses it to stamp a picture of a sun. If the ink behaves like a quantum system in a half-yellow, half-red quantum superposition, what are the different options for what the resulting picture could look like? Some options are shown in Figure~\ref{fig:sun}. 

\item \fundamental{5pt} If \href{https://en.wikipedia.org/wiki/The\_dress}{this controversial picture of a dress}\footnote{https://en.wikipedia.org/wiki/The\_dress} is always seen as blue/black by Student A and always seen as white/gold by Student B, is the dress in a quantum superposition?

\end{enumerate}

%
%
%
%


\graphicspath{{Chapter2-Qubit/}}
\chapter{What is a Qubit?}
\label{chapter:qubit}

In classical computers, information is represented as the binary digits $0$ or $1$. These are called bits. For example, the number $1$ in an $8$-bit binary representation is written as $00000001$. The number $2$ is represented as $00000010$. We place extra zeros in front to write every number with $8$-bits total, which is called one byte. In fact, every classical computer translates these bits into the human readable information on your electronic device. The word document you read or video you watch is encoded in the computer binary language in terms of these $1$'s and $0$'s. Computer hardware understands the $1$-bit as an electrical current flowing through a wire (in a transistor) while the $0$-bit is the absence of an electrical current in a wire. These electrical signals can be thought of as ``on'' (the $1$-bit) or ``off'' (the $0$-bit). Your computer then decodes the classical $1$ or $0$ bits into words or videos, etc. 

\begin{figure}[h]
  \centering
  \includegraphics{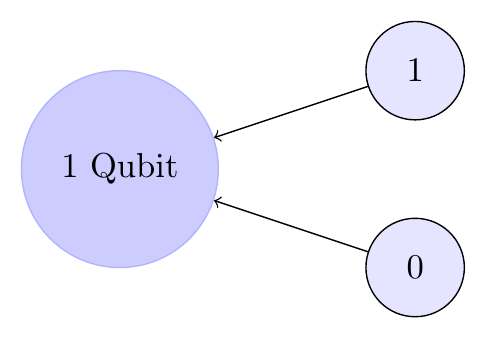}
  \caption{A classical bit can be either $0$ or $1$. A qubit can be in a superposition of both $0$ and $1$.}
  \label{fig:qubit}
\end{figure}

Quantum bits, called {\bf{qubits}}, are similar to bits in that there are two measurable states called the $0$ and $1$ states. However, unlike classical bits, qubits can also be in a superposition state of these $0$ and $1$ states, as shown in Figure~\ref{fig:qubit}. Certain computations that would normally need to be performed on $0$ or $1$ separately on a classical computer could now be completed in a single operation
 using a qubit on a quantum computer. Intuitively, this could make computations much faster. It is important to understand that although a single qubit is in a superposition of two classical bits, when a qubit is measured, the qubit actually only results in one classical bit of information: either $0$ or $1$.

\section{\fundamental{5pt} Mathematical Representation of Qubits}
\subsection*{Dirac bra-ket notation}

In order to work with qubits, it is useful to know how one can express quantum mechanical states with mathematical formulas. 
Dirac or ``bra-ket'' notation is commonly used in quantum mechanics and quantum computing. The state of a qubit is enclosed in the right half of an angled bracket, called the {\bf{``ket''.}} A qubit, $\lvert \Psi \rangle$, could be in a $\lvert 0 \rangle$ or $\lvert 1\rangle$ state which is a superposition of both $\lvert 0\rangle$ and $\lvert 1\rangle$. This is written as 
\begin{equation}
\lvert \Psi \rangle = \alpha \lvert 0 \rangle + \beta \lvert 1 \rangle,
\end{equation}
with $\alpha$ and $\beta$ called the amplitudes of the states. Amplitudes are generally complex numbers (a special type of number used in mathematics and physics). However, to understand the meaning of amplitudes, we can just imagine the amplitudes as
 being ordinary (real) numbers. Amplitudes allow us to mathematically represent all of the possible superpositions.

\begin{figure}[h]
\centering
  \includegraphics[width=0.5\textwidth]{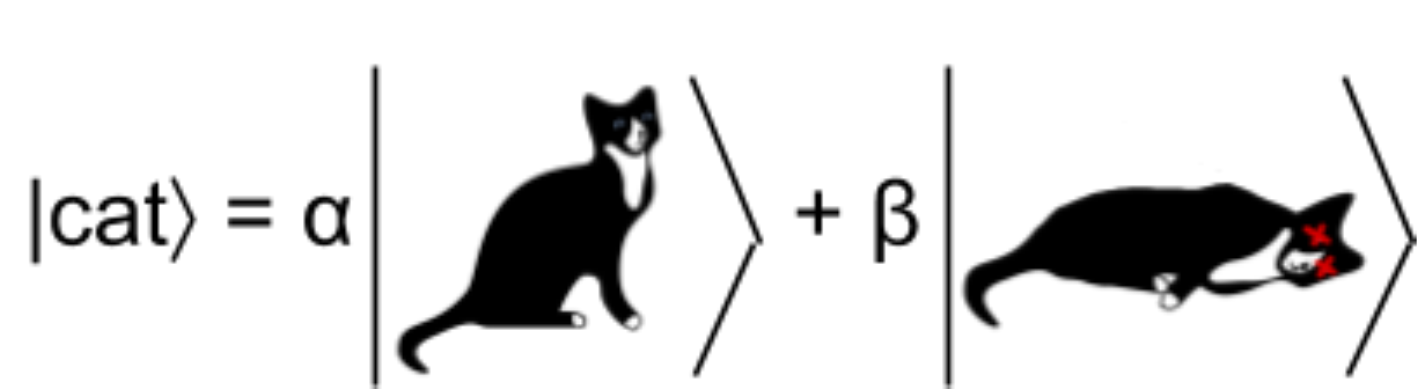}
  \caption{The state of Schr\"{o}dinger's cat expressed in bra-ket notation.}
\end{figure}\label{fig:cat}

{\bf{Amplitudes}} are very important because they tell us the probability of finding the particle in that specific state when performing a measurement.  The probability of measuring the particle in state $\lvert 0 \rangle$ is $\lvert \alpha \rvert^2$, and the probability of measuring the particle in state $\lvert 1 \rangle$ is $\lvert \beta \rvert^2$. Why is it squared? The short answer is that it gives the correct experimental predictions for this choice of representation.\footnote{We know that quantum physics is probabilistic from experiments. The squared coefficients are needed to make a quantity that behaves like a probability distribution, i.e., it is a real number and positive. There cannot be a negative probability by definition.} Squaring $\alpha$ and $\beta$ to find the probability is similar to squaring a waves amplitude to find the energy in the wave. Since the total probability of observing all the states of the quantum system must add up to $100\%$, the amplitudes must follow this rule:
\begin{equation}
\lvert \alpha \rvert^2 + \lvert \beta \rvert^2 = 1.
\end{equation}
This is called a {\bf{normalization}} rule. The coefficients $\alpha$ and $\beta$  can always be rescaled by some factor to normalize the quantum state.

\subsection*{Examples}
\begin{enumerate}

\item The quantum state of a spinning coin can be written as a superposition of heads and tails. Using heads as $\lvert 1\rangle$  and tails as $\lvert 0 \rangle$, the quantum state of the coin is 
\begin{equation}
\lvert \text{coin} \rangle = \frac{1}{\sqrt{2}} \left(  \lvert 1\rangle  + \lvert 0 \rangle  \right). 
\end{equation}
What is the probability of getting heads?

The amplitude of $ \lvert 1\rangle $ is $\beta = 1/\sqrt{2}$, so $\lvert \beta \rvert^2=\left( 1/\sqrt{2}\right)^2=1/2$. So the probability is $0.5$, or $50\%$.

\item A weighted coin has twice the probability of landing on heads vs. tails. What is the state of the coin in ``ket'' notation?
\begin{equation}
\begin{aligned}
P_{\text{heads}} + P_{\text{tails}} &= 1 ~~~(\text{Normalization Condition})\\
P_{\text{heads}} &= 2P_{\text{tails}} ~~~(\text{Statement in Example})\\
\rightarrow P_{\text{tails}} &= \frac{1}{3}= \alpha^2 \\
\rightarrow  P_{\text{heads}} &= \frac{2}{3}= \beta^2\\
\rightarrow \alpha &= \sqrt{ \frac{1}{3}}, ~\beta = \sqrt{\frac{2}{3}}\\
\rightarrow \lvert \text{coin} \rangle  & =\sqrt{ \frac{1}{3}} \lvert 0 \rangle +\sqrt{ \frac{2}{3}} \lvert 1 \rangle.
\end{aligned}
\end{equation}

One common misconception is that the measurement result will be a weighted average of the $|0\rangle$ and $|1\rangle$ states. It is important to note that after you perform the measurement, the particle is no longer in a superposition but takes on a definite state of either $|0\rangle$ or $|1\rangle$.\footnote{When formulating the mathematical representation of quantum mechanics, this is one of four fundamental assumptions that need to be made. The reason for the collapse is still unknown: https://en.wikipedia.org/wiki/Wave\_function\_collapse.} You would not be able to find $\alpha$ or $\beta$ unless you created many particles in the same quantum state and then measured how many collapse into $\lvert0\rangle$ (giving $\alpha$) and how many collapse into $\lvert1\rangle$ (giving $\beta$). You need multiple identical particles to count how many collapse into $\lvert0\rangle$ or $\lvert1\rangle$. 
\end{enumerate}

\section*{\fundamental{5pt} {Big Ideas}}
\begin{enumerate}
\item A particle is in a superposition of states until you measure a property of the particle. When you measure a property of a particle, the particle collapses into one of the observable states. 
\end{enumerate}

\section{\advanced{0.6pt} Matrix Representation} 
When writing one qubit in a superposition $|\psi\rangle = \alpha|0\rangle + \beta|1\rangle$, it is useful to use matrix algebra. In matrix representation, a qubit is written as a two-dimensional vector where the amplitudes are the components of the vector:
\begin{align}
  |\psi\rangle &= 
  \begin{pmatrix}
    \alpha \\
    \beta 
  \end{pmatrix}. 
\end{align}
The states $|0\rangle$ and $|1\rangle$ are usually represented as 
\begin{align}
  |0\rangle &= 
  \begin{pmatrix}
    1 \\
    0 
  \end{pmatrix}, \quad
  |1\rangle = 
  \begin{pmatrix}
    0 \\
    1 
  \end{pmatrix}. 
\end{align}        
A qubit's state can be changed by some physical action such as applying an electromagnetic laser or passing it through an optical device. Mathematically, changing a qubit's state is represented by multiplying the qubit vector $|\psi\rangle$ by some {\bf{unitary matrix}} $U$ so that after the change the state is now $|\psi'\rangle = U|\psi\rangle$. Unitary is a mathematical term which expresses that $U$ can only act on  the qubit in such a way that $|\alpha|^2+|\beta|^2$ does not change.  A matrix $U$ is unitary if the matrix product of $U$ and its conjugate transpose $U^{\dagger}$ (called $U$-dagger) produces the identity matrix: $UU^{\dagger} = U^{\dagger}U=\mathbb{1}$. This is very important because, in all mathematical constructions of quantum mechanics, one fundamental assumption is that each (matrix) operator must be unitary.
 This ensures that after changing the state through some action the total probability to observe all possible states still adds up to $100\%$.
This physical  action of interacting with the state  corresponds mathematically to applying a unitary operator.
 If this did not happen, then we could not interpret the results of quantum mechanics to be probabilities, and the results would disagree with the many experiments we have performed. 

\subsection*{Examples}
\begin{enumerate}

\item What is the conjugate transpose of the following matrix?
\begin{align}
  A  &= 
  \begin{pmatrix}
    1 & i \\
    1 & i 
  \end{pmatrix}. 
\end{align}

The conjugate transpose of a matrix is found by two steps. The first step is to ``conjugate'' all of the complex numbers. The conjugate of a complex number is found by switching the sign of the imaginary part. The complex conjugate of $1$ is just $1$, while the complex conjugate of $+i$ is $-i$. The second step is to transpose the conjugated matrix. Transposing a matrix switches rows with columns, i.e., the first row turns into the first column, second row turns into the second column, etc. Therefore,  
\begin{align}
  A^{\dagger}  &= 
  \begin{pmatrix}
     1 &  1 \\
    -i & -i 
  \end{pmatrix}. 
\end{align}

\item Is the above matrix $A$ unitary?

  \begin{align}
    A A^{\dagger}  &=
    \begin{pmatrix}
      1 & i \\
      1 & i 
    \end{pmatrix}
    \begin{pmatrix}
      1 & 1 \\
     -i &-i 
    \end{pmatrix}  \\
    &= 2
    \begin{pmatrix}
      1 & 1 \\
      1 & 1 
    \end{pmatrix}
    \ne 
    \begin{pmatrix}
      1 & 0 \\
      0 & 1 
    \end{pmatrix}.
  \end{align}
Multiplying $A$ by its conjugate transpose does not produce the identity matrix, so $A$ is not unitary.
  
\item What is the result of applying the unitary operator $X$ onto a $|0\rangle$ state qubit?

  \begin{align}
  X  &= 
  \begin{pmatrix}
    0 & 1 \\
    1 & 0 
  \end{pmatrix}, \qquad
   |0\rangle = 
  \begin{pmatrix}
     1 \\
     0 
  \end{pmatrix}. 
\end{align}
\begin{align}
  X|0\rangle  &= 
  \begin{pmatrix}
    0 & 1 \\
    1 & 0 
  \end{pmatrix}
  \begin{pmatrix}
     1 \\
     0 
  \end{pmatrix}
  = 
  \begin{pmatrix}
     0 \\
     1 
  \end{pmatrix} = |1\rangle. 
\end{align}
The $X$ matrix changes the $|0\rangle$ qubit state to the $|1\rangle$ qubit state.   
\end{enumerate}

\section{\advanced{0.6pt} Bloch Sphere} It is sometimes convenient to visually represent a qubit using a Bloch sphere. The Bloch sphere is an abstract representation with similar geometric properties to the unit circle from trigonometry. However, it only works for a single qubit and cannot be used for two or more qubits. Therefore, we will not go over the Bloch sphere, but you can read further about this  on the \href{https://quantumexperience.ng.bluemix.net/qx/tutorial?sectionId=beginners-guide&page=introduction}{IBM Q website}.\footnote{https://quantumexperience.ng.bluemix.net/qx/tutorial?sectionId=beginners-guide\&page=introduction}

\section*{Physical Realization of Qubits}
In a classical computer, the $0$- and $1$-bit mathematically represent the two allowed voltages across a wire in a classical circuit. Semiconductor devices called transistors are used to control what happens to these voltages. A question frequently posed by new students is ``What is a qubit made out of?''  As quantum computers are based on fundamentally different concepts, they must be built from completely different technology; e.g., it is not possible to have a classical current in a superposition of both flowing and not flowing through a wire. Quantum computers are still in their infancy, and so there are many different candidates for the technology to build them. Some technologies are  based on optics, others use superconductors\footnote{Fermi National Accelerator Laboratory is researching how to make long-lived coherent qubits using their {{superconducting}} radio-frequency cavity expertise, i.e., \href{https://qis.fnal.gov/superconducting-quantum-systems/}{https://qis.fnal.gov/superconducting-quantum-systems/}} or possibly  molecules. It is still unclear if any of these are more beneficial than the others, and it is even more unclear if all future quantum computers will be built from the same technology or if there will be many different types of quantum computers available (in the same way there exists both Xbox and Play station game consoles, but both do the same thing-interactive gaming). We will study two different experiments which illustrate the  properties of the qubits, but the details of building a quantum computer are well beyond the scope of this introduction.

\section{Check Your Understanding}

\begin{enumerate}
  
\item \fundamental{5pt} If a coin is a classical bit of information (heads = $1$ and tails = $0$), how is the number $2$ represented in standard $8$-bit notation using coins? (Hint: Find the $8$-bit representation of the number $2$, then convert to H's and T's.)
  
\item \fundamental{5pt} Using the chart below, can you figure out what this binary message $01000011$ $01000001$ $01010100$ says? (Note: This is actually how your computer and phone decode information from bits to text.)
 \begin{table}[h!] 
\centering
\begin{tabular}{| c | c  | c| c |}
\hline
Character &   Binary Code & Character &   Binary Code   \\
 \hline \hline
A & 01000001 & N& 01001110\\
B & 01000010                & O&  01001111 \\
C & 01000011 			& P&  01010000\\
D & 01000100 			& Q&  01010001\\
E &  01000101			& R& 01010010\\
F &  01000110			& S&  01010011\\
G & 01000111			 & T& 01010100\\
H &  01001000				& U& 01010101\\
I &  01001001				& V& 01010110 \\
J & 	01001010				 & W&01010111 \\
K & 01001011				 & X& 01011000 \\
L &  01001100				& Y&01011001 \\
M &  01001101				& Z& 01011010\\
\hline
\end{tabular}\caption{Table for message.}\label{tab:LFV3b}
\end{table}

\item Assume a flipped coin can be measured as either heads (H) or tails (T). 
  
  \begin{enumerate}[label=(\alph*)]
  \item \fundamental{5pt} If the coin is in a normalized state $\frac{1}{\sqrt{10}} \lvert H \rangle + \frac{3}{\sqrt{10}} \lvert T \rangle$, what is the probability that the coin will be tails?
  \item \intermediate{8pt} During a flip, the coin is in a state $\frac{1}{3} \lvert H \rangle + \frac{2}{3} \lvert T \rangle$. Is this state normalized?
  \item \fundamental{5pt} A machine is built to flip coins and put them into  a state $\frac{1}{2} \lvert H \rangle + \frac{\sqrt{3}}{2} \lvert T \rangle$ when flipped. If $100$ coins are flipped, how many coins should land on tails? 
  \item \intermediate{8pt} A coin starts in the state $\frac{1}{\sqrt{10}} \lvert H \rangle + \frac{3}{\sqrt{10}} \lvert T \rangle$. After a measurement is made on the coin, what could be the state of the coin? 
  \end{enumerate}
  
\item \intermediate{8pt} Multiple qubits are prepared in the same superposition state. By making measurements on these particles, can you write down their initial state?
  
\item \fundamental{5pt} A quantum particle is prepared in an unknown state. It is then measured with the outcome $\lvert 0 \rangle$.
  \begin{enumerate}[label=(\alph*)]
  \item Which of the following could be its initial state before the measurement: $\lvert 0 \rangle$, $\frac{1}{\sqrt{10}}\lvert 0 \rangle + \frac{3}{\sqrt{10}}\lvert 1 \rangle$, $\frac{1}{2}\lvert 0 \rangle + \frac{\sqrt{3}}{2}\lvert 1 \rangle$ or $\frac{1}{\sqrt{2}}\left( \lvert 0 \rangle + \lvert 1 \rangle \right)$?
  \item If you tried to measure the same particle a second time, can you narrow down what the initial state was?
  \item Another particle is prepared in the same unknown state. It is measured in the $\lvert 1 \rangle$ state. What can you say about the initial state now?
  \end{enumerate}
  
\item \advanced{0.6pt} What is the matrix product of the $X$ matrix,
  \begin{align}
    X  &= 
    \begin{pmatrix}
      0 & 1 \\
      1 & 0 
    \end{pmatrix},
  \end{align}
  and $\lvert 0 \rangle $ state qubit?
  
\item \advanced{0.6pt} What is the matrix product of the above $X$ matrix and the $\lvert 1 \rangle $ state qubit?
  
\item \advanced{0.6pt} What is the matrix product of the above $X$ matrix and a qubit in the general state $\lvert \Psi \rangle  = \alpha\lvert 0 \rangle + \beta\lvert 1 \rangle$?
  
\item \advanced{0.6pt} Find the conjugate transpose of the matrix 
  \begin{equation}
  Y=\begin{pmatrix}
  0 & -i\\
  i & 0
  \end{pmatrix}.
  \end{equation}
  
\item \advanced{0.6pt} Show that the matrix 
  \begin{equation}
  U=\frac{1}{\sqrt{2}}\begin{pmatrix}
    1 & 1\\
    1 & -1
  \end{pmatrix}
  \end{equation}
  is unitary.

\item \advanced{0.6pt} Show by example that applying a non-unitary matrix to a qubit results in probabilities that no longer add up to $100\%$. (Hint: Start with any initial state, e.g., $|0\rangle$. Measure the probabilities of finding either $0$ or $1$. Apply a non-unitary matrix to the initial state. Then measure the probabilities of finding either a $0$ or $1$. Do the probabilities add up to $100\%$?)
  
\end{enumerate}

\graphicspath{{Chapter3-Beamsplitter/}}
\chapter{Creating Superposition: The Beam Splitter}
\section{\fundamental{5pt} How is a superposition state created?}

While a flipping coin is a simple model of a qubit, it is not very useful for building a quantum computer because it does not exhibit all of the properties of a true quantum superposition. For example, we cannot manipulate the superposition amplitudes. In this section, we will study some real physical examples of quantum particles in a superposition containing two states. These examples include a photon in a beam splitter, an electron in the double-slit experiment, and an electron in a Stern-Gerlach apparatus.

\section{\fundamental{5pt} Beam Splitter}
\begin{figure}[h]
  \centering
  \includegraphics[width=0.35\textwidth]{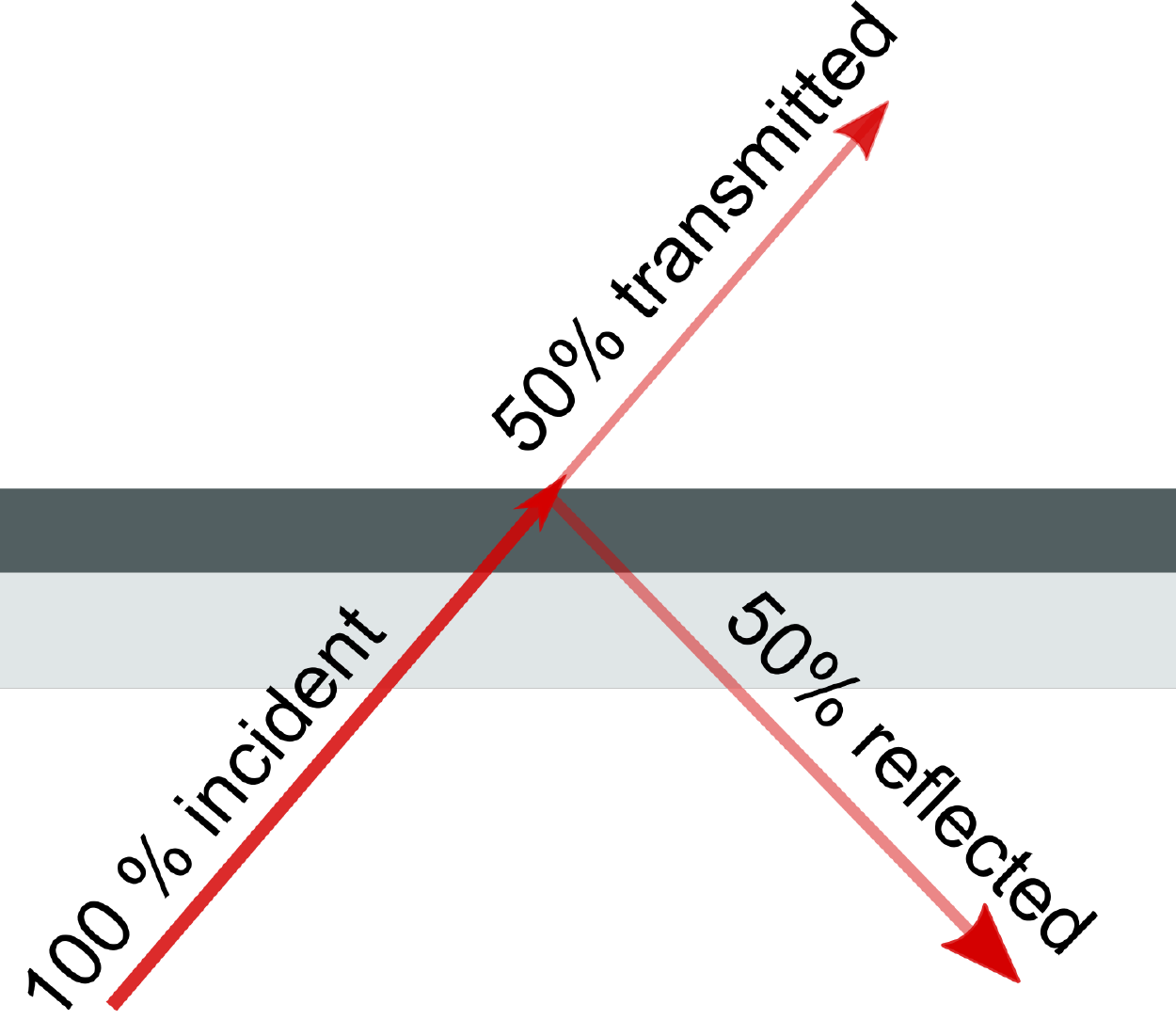}
  \caption{A beam splitter reflects 50$\%$ of the incident light and transmits 50$\%$ of the incident light.}
  \label{fig:reflected}
\end{figure}

In classical optics, a {\bf{beam splitter}} acts like a partially reflective mirror that splits a beam of light into two. In a $50/50$ beam splitter, 50$\%$ of the light intensity is transmitted and 50$\%$  is reflected, as shown in Figure~\ref{fig:reflected}. 

\begin{figure}[h]
  \centering
  \includegraphics[width=0.20\textwidth]{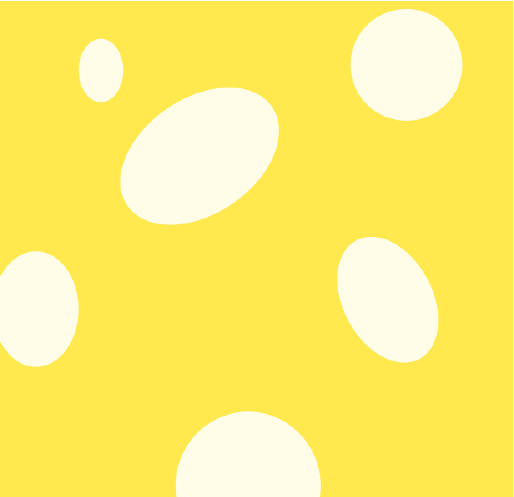}
  \caption{A beam splitter reflects 50$\%$ of the incident light and transmits 50$\%$ of the incident light.}
  \label{fig:cheese}
\end{figure}

One way to visualize the beam splitter is to imagine a barrier with holes randomly cut out like Swiss cheese, as shown in Figure~\ref{fig:cheese}. Imagine this barrier is placed in a pond, and a water wave moves toward the barrier. After the wave hits the barrier, we would observe a smaller wave going through the barrier and another would be reflected off the barrier.

\noindent \textbf{Question 1}: What would happen if a classical particle such as a soccer ball is randomly kicked at the barrier? Assume the ball can fit through the holes.\\

\begin{figure}[h!]
  \centering
  \includegraphics{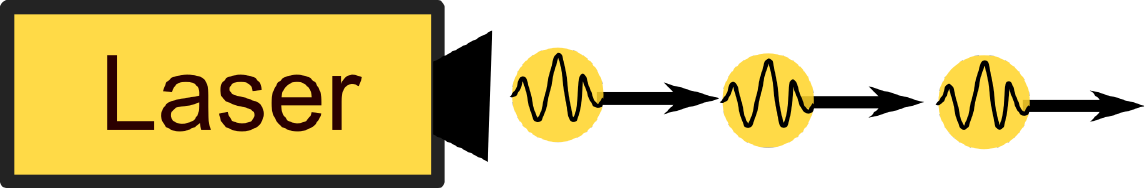}
  \caption{Low-intensity light is a stream of single photons.}
  \label{fig:laser}
\end{figure}

Experiments show that light behaves both like a wave (Young's double-slit experiment) and a particle (photoelectric effect, Compton effect). Classically, light is thought of as a wave consisting of continually oscillating electric  and magnetic fields. However, light can also be thought of as a stream of particles called {\bf{photons}}. Photons have no mass but carry the light's energy from one point to another at the speed of light. A laser beam is comprised of photons. If you turn down the intensity of your laser, you can even send one photon at a time, as shown in Figure~\ref{fig:laser}. In practice, setting up a single photon source and detector requires specialized equipment, so we will instead run a simulator to see what happens.

\noindent \textbf{Question 2}: Open the \href{https://www.st-andrews.ac.uk/physics/quvis/simulations_html5/sims/Mach-Zehnder-Interferometer/Mach_Zehnder_Interferometer.html}{beam splitter simulator}\footnote{https://www.st-andrews.ac.uk/physics/quvis/simulations\_html5/sims/Mach-Zehnder-Interferometer/Mach\_Zehnder\_Interferometer.html}, go to the Controls screen, and fire a single photon. The setup before the photon hits a beam splitter is shown in Figure~\ref{fig:beamexperiment}. Which detectors are triggered when the photon passes through the $50/50$ beam splitter?
\begin{figure}[h!]
\centering
  \includegraphics[width=0.45\textwidth]{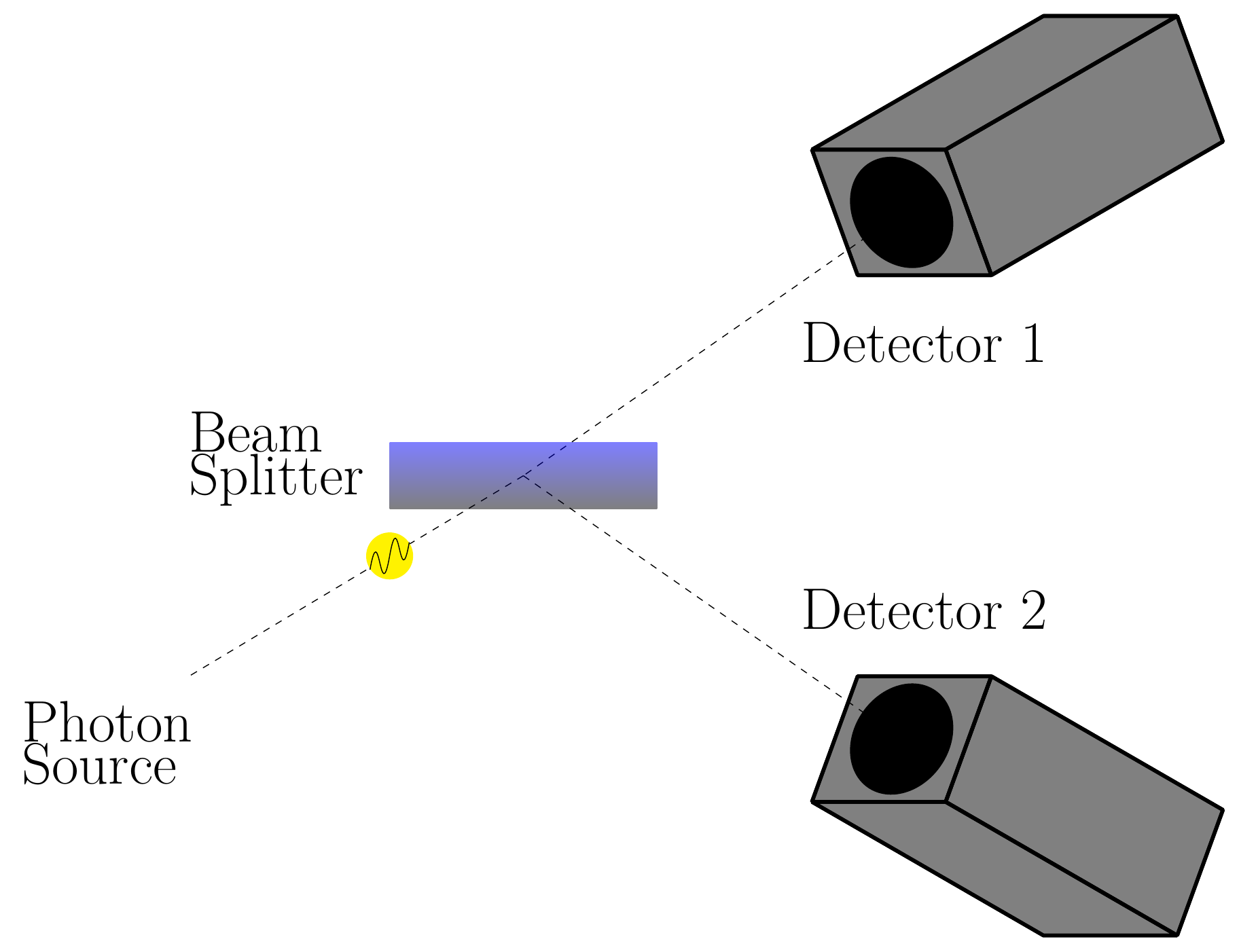}
  \caption{A single photon is sent at a beam splitter and the outcome measured with detectors to see whether it transmits or reflects.}
\end{figure}\label{fig:beamexperiment}
\begin{enumerate}
\item Always detector 1
\item Always detector 2
\item Detector 1 OR detector 2
\item Both detector 1 AND detector 2
\item Neither
\end{enumerate}

\noindent \textbf{Question 3}: Which detector(s) would trigger if a classical \textbf{wave} is sent through the beam splitter? 
\begin{enumerate}
\item Always detector 1
\item Always detector 2
\item Detector 1 OR detector 2
\item Both detector 1 AND detector 2
\item Neither
\end{enumerate}

\noindent \textbf{Question 4}: Which detector(s) would trigger if a classical \textbf{particle} is sent through the beam splitter? 
\begin{enumerate}
\item Always detector 1
\item Always detector 2
\item Detector 1 OR detector 2
\item Both detector 1 AND detector 2
\item Neither
\end{enumerate}

\noindent \textbf{Question 5}:  What does the photon do at the instance it encounters the $50/50$ beam splitter?
\begin{enumerate}
\item Splits in half. Half the photon is transmitted and half is reflected
\item The whole photon goes through with 50$\%$ probability and reflects with 50$\%$ probability
\item The whole photon is both transmitted and reflected, essentially in two places at once
\end{enumerate}

If the photon was split in half, both detectors would be triggered together. As only one detector goes off at a time, the photon could not have split up. In this case, we see that light behaves more like the soccer ball than the water wave.

\begin{figure}[h]
\centering
  \includegraphics[width=0.45\textwidth]{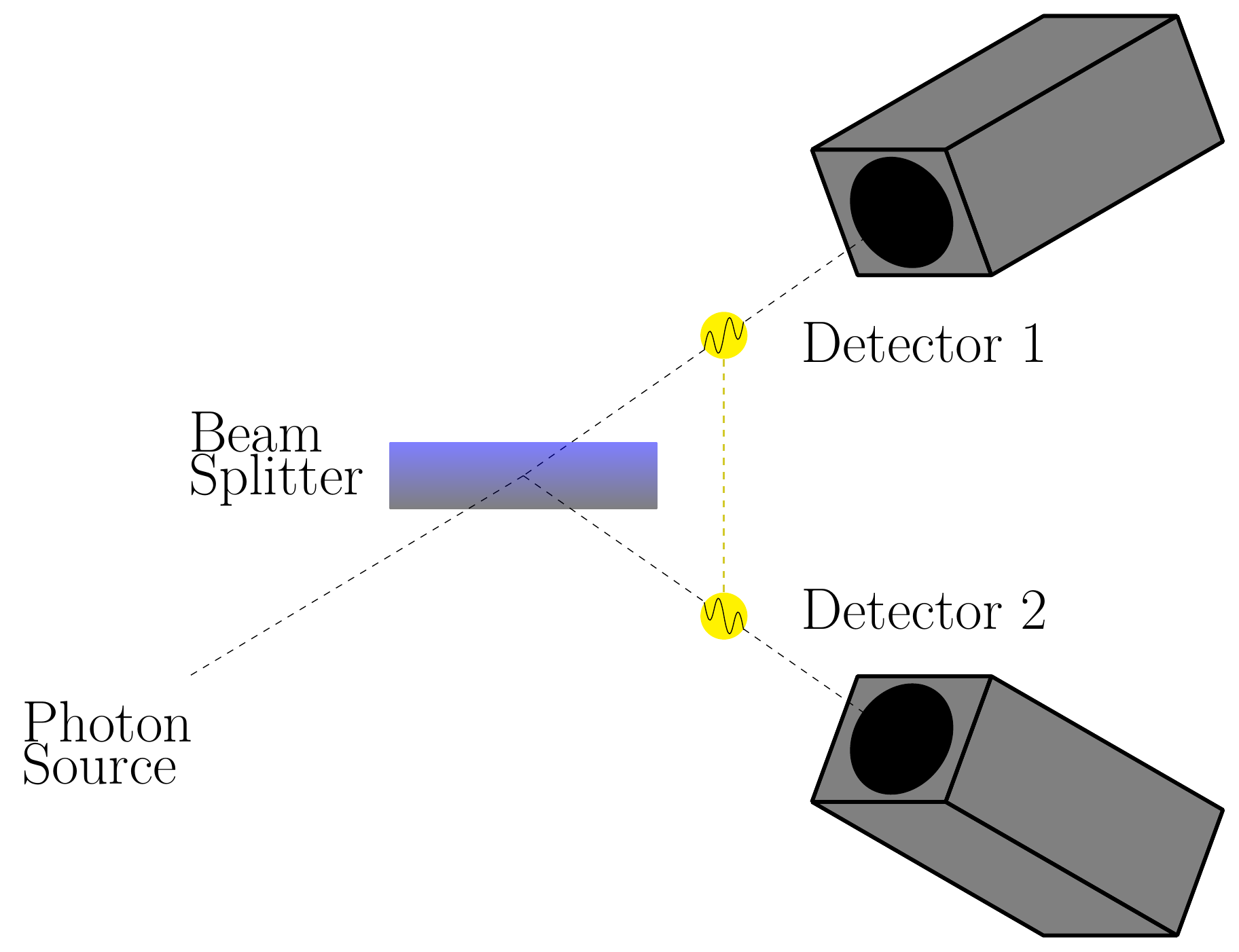}
  \caption{The beam splitter puts the photon into a superposition state.}
  \label{fig:beamsuperposition}
\end{figure}

At this point you may be thinking that the photon was either transmitted or reflected at the beam splitter, and we simply didn't have that information until it hit Detector 1 or 2.  Unfortunately, this would be the incorrect interpretation formed by our classical lizard brain.  This would be like saying the coin was Heads all along, and all we had to do was look at it to determine its state. Just like how a spinning coin will land on heads 50$\%$ of the time and tails 50$\%$ of the time, the single photon is in a superposition of both states all the way until the point when it reaches the detectors. This distinction might seem like a matter of semantics, but {{this  is important as the distinction describes two different ways that the universe operates at the smallest possible distances. Also, it will be important once the system becomes more complicated.}} The experimental setup after the photon hits a beam splitter is shown in Figure~\ref{fig:beamsuperposition}. 

If we let the transmitted path be $\lvert0\rangle$ (detector 1), and the reflected path be $\lvert1\rangle$ (detector 2), then the photon's state after the beam splitter is
\begin{equation}
\lvert \text{photon}\rangle = \frac{1}{\sqrt{2}} \lvert 0\rangle + \frac{1}{\sqrt{2}} \lvert 1\rangle. 
\end{equation}
Upon measurement, will the superposition collapse into either $0$ or $1$? Unfortunately, there is no way to predict which detector will be activated at any given time. Quantum mechanics is inherently probabilistic. 

The phenomenon of superposition  allows quantum computers to perform operations on two bits of information at once with  a single qubit. In fact, it is possible to create a general purpose (also called universal) quantum computer using photons as qubits, beam splitters to create superposition, and pieces of glass that slow down the photons along selected paths (phase shifters).\footnote{ Knill, E.; Laflamme, R.; Milburn, G. J. (2001). "A scheme for efficient quantum computation with linear optics". Nature. Nature Publishing Group. 409 (6816): 46–52.}

\section{\fundamental{5pt} Mach-Zehnder Interferometer}
\begin{figure}[h]
  \centering
  \includegraphics[width=0.9\textwidth]{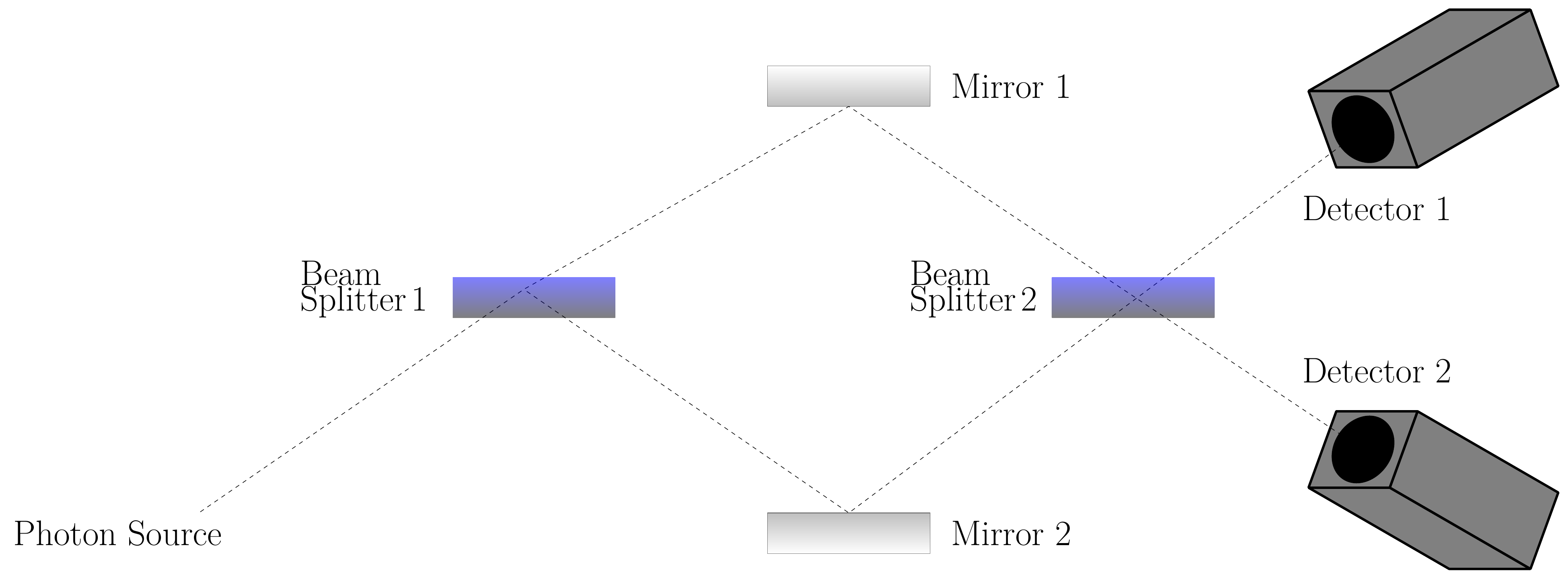}
  \caption{Schematic of the Mach-Zehnder interferometer from the \href{https://www.st-andrews.ac.uk/physics/quvis/simulations_html5/sims/Mach-Zehnder-Interferometer/Mach_Zehnder_Interferometer.html}{beam splitter simulator}.} 
  \label{fig:MZInterferometer}
\end{figure}

To convince ourselves that the photon really did take two paths at once, let's see what happens when a second beam splitter is added. This experimental setup is shown in Figure~\ref{fig:MZInterferometer}. 
The mirrors redirect the photons towards the second beam splitter. This device configuration is known as a {\bf{Mach-Zehnder interferometer}}. The set up is very sensitive to the distances between the mirrors and detectors, which have to be the same or differ by an integer number of the photon's wavelength.\\

\noindent \textbf{Question 6}: If we assume that the photon was reflected by the first beam splitter, which detectors would be triggered?
\begin{enumerate}
\item Always detector 1
\item Always detector 2
\item Detector 1 OR detector 2
\item Both detector 1 AND detector 2
\item Neither
\end{enumerate}

\noindent \textbf{Question 7}: If we assume that the photon was transmitted by the first beam splitter, which detectors would be triggered?
\begin{enumerate}
\item Always detector 1
\item Always detector 2
\item Detector 1 OR detector 2
\item Both detector 1 AND detector 2
\item Neither.
\end{enumerate}

\noindent \textbf{Question 8}: Construct the Mach-Zehnder interferometer in the beam splitter simulator\footnote{https://www.st-andrews.ac.uk/physics/quvis/simulations\_html5/sims/Mach-Zehnder-Interferometer/Mach\_Zehnder\_Interferometer.html} and fire a single photon. Which detectors are triggered?
\begin{enumerate}
\item Always detector 1
\item Always detector 2
\item Detector 1 OR detector 2
\item Both detector 1 AND detector 2
\item Neither
\end{enumerate}

If the photon was either transmitted or reflected by the first beam splitter, it would have a $50/50$ chance of transmission or reflection by the second beam splitter. Thus, both detectors should trigger with equal probability. {{However, strangely the}} experimental results do not agree with this hypothesis, as only one detector is triggered with 100$\%$ probability. {{This weird phenomenon is}} more intuitively understood from the wave perspective of light.

To understand the operation of the interferometer, it is important to note that the beam splitters have a polarity. The beam splitter consists of a piece of glass coated with a dielectric on one side. When light enters the beam splitter from the dielectric side, the reflected light is {\bf{phase shifted}} by $\pi$. Light entering from the glass side will not experience any
 phase shift. The phase shift only occurs when the light travels from a low to high index of refraction ($n_{\text{air}}<n_{\text{dielectric}}<n_{\text{glass}}$). 

\begin{figure}[h]
  \centering
  \includegraphics[width=0.45\textwidth]{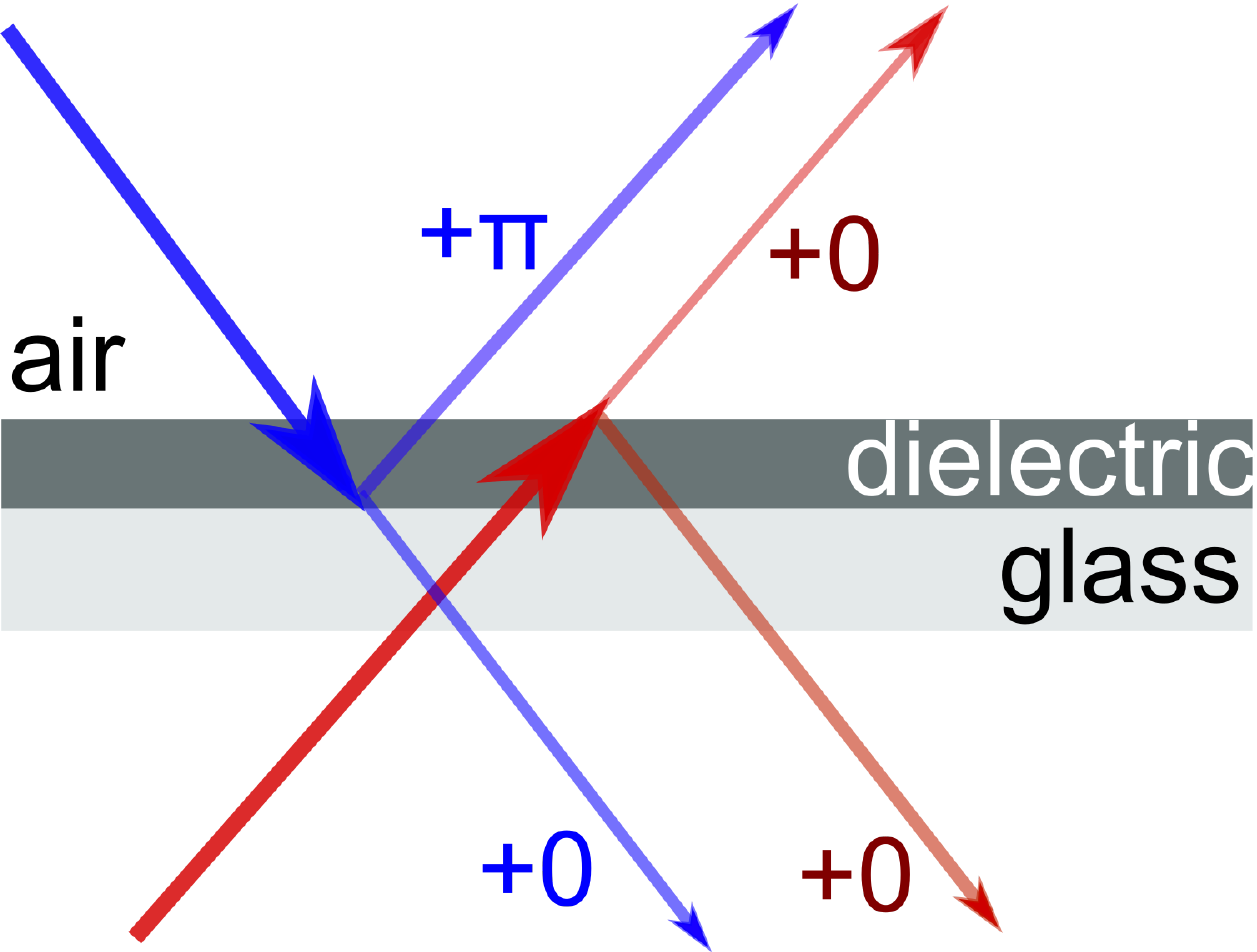}
  \includegraphics[width=0.45\textwidth]{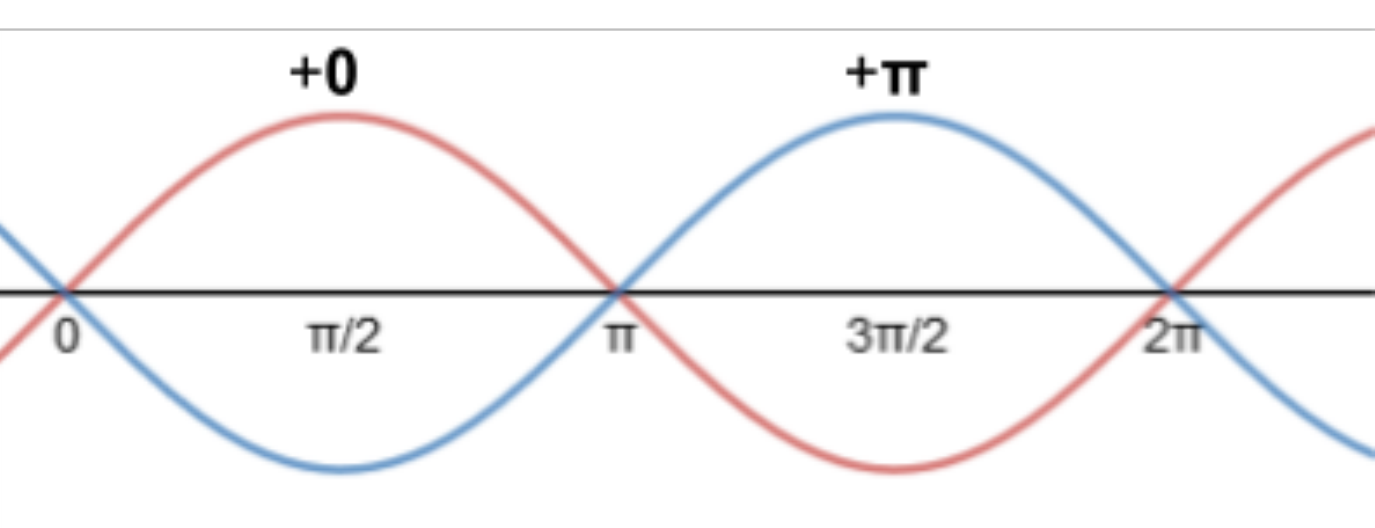}
  \caption{The light through a beam splitter is phase shifted if it is reflected from the dielectric side but not phase shifted from if it is reflected from the glass side.}
  \label{fig:phaseshift}
\end{figure}

What does it mean for a photon to be phase shifted? In this case, it is more intuitive to think about the wave nature of light. The phase shift would invert the electric and magnetic field oscillations relative to the incoming wave. If a $\pi$-shifted wave overlaps with the original wave, destructive interference occurs. This is shown in Figure~\ref{fig:phaseshift}. \\

\textbf{Question 9}: If we assume that light is a classical wave exhibiting interference, can you work out which detectors would be triggered? Note that the first beam splitter has the dielectric side on top, while the second has the dielectric on the bottom, as shown in Figure~\ref{fig:MZInterferometer}.
\begin{enumerate}
\item Always detector 1
\item Always detector 2
\item Detector 1 OR detector 2
\item Both detector 1 AND detector 2
\item Neither
\end{enumerate}

\subsection*{Particle Explanation}
\begin{figure}[h]
\centering
  \includegraphics[width=0.9\textwidth]{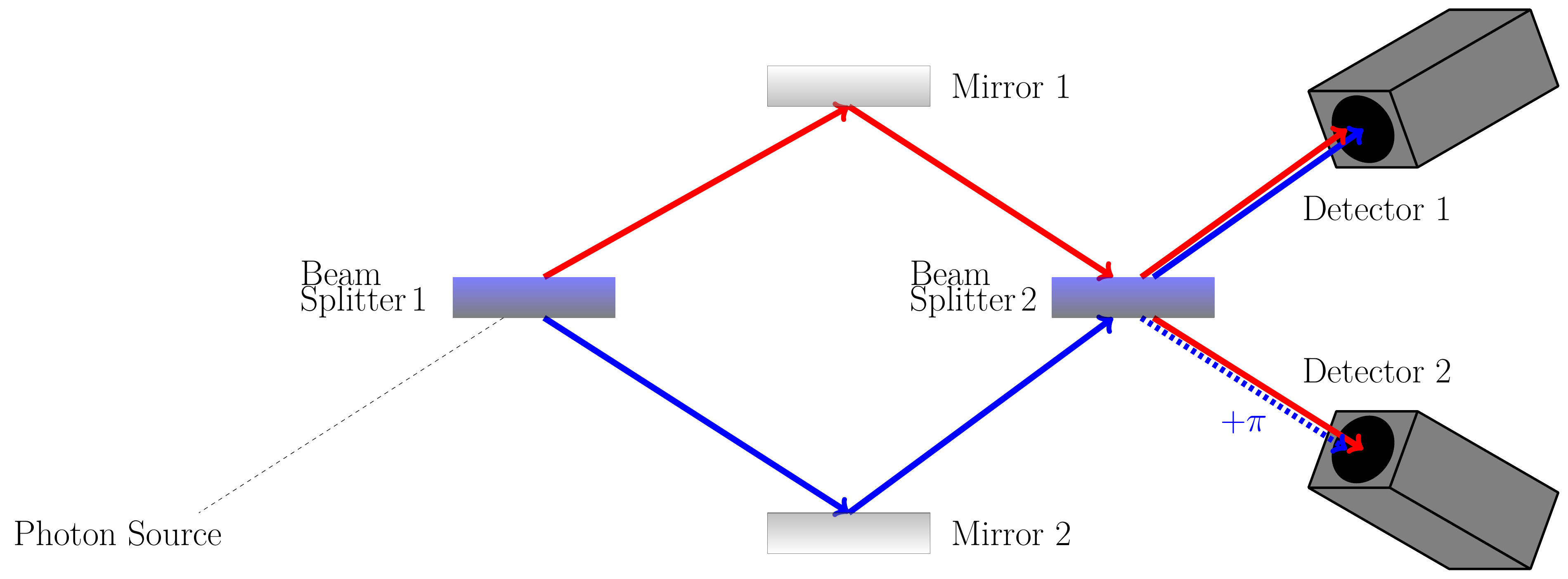}
  \caption{The blue path shows the photon's path if it is reflected by the first beam splitter. The red path shows the path if the photon is transmitted. Red and blue interfere constructively at Detector 1 while destructively at Detector 2.}
  \label{fig:MZTwoBeam}
\end{figure}

The behavior of the interferometer can also be viewed from the particle perspective, though it may be less intuitive. Recall from the single beam splitter experiment that the photon did not split up or clone itself. It was in a superposition state, essentially taking both paths. The second beam splitter treats the photon as if it came in from both
 top and bottom simultaneously. The bottom photon is phase shifted relative to the top photon, resulting in destructive interference at Detector 2. Since there is no phase shift at Detector 1, there is no cancellation and it triggers with 100$\%$ probability, as shown in Figure~\ref{fig:MZTwoBeam}. \\

\noindent \textbf{Question 10}:  If the photon is sent into the Mach-Zehnder interferometer from the upper left instead of the bottom left, which detector(s) would be triggered and with what probability?\\

\begin{figure}[h]
  \centering
  \includegraphics[width=0.45\textwidth]{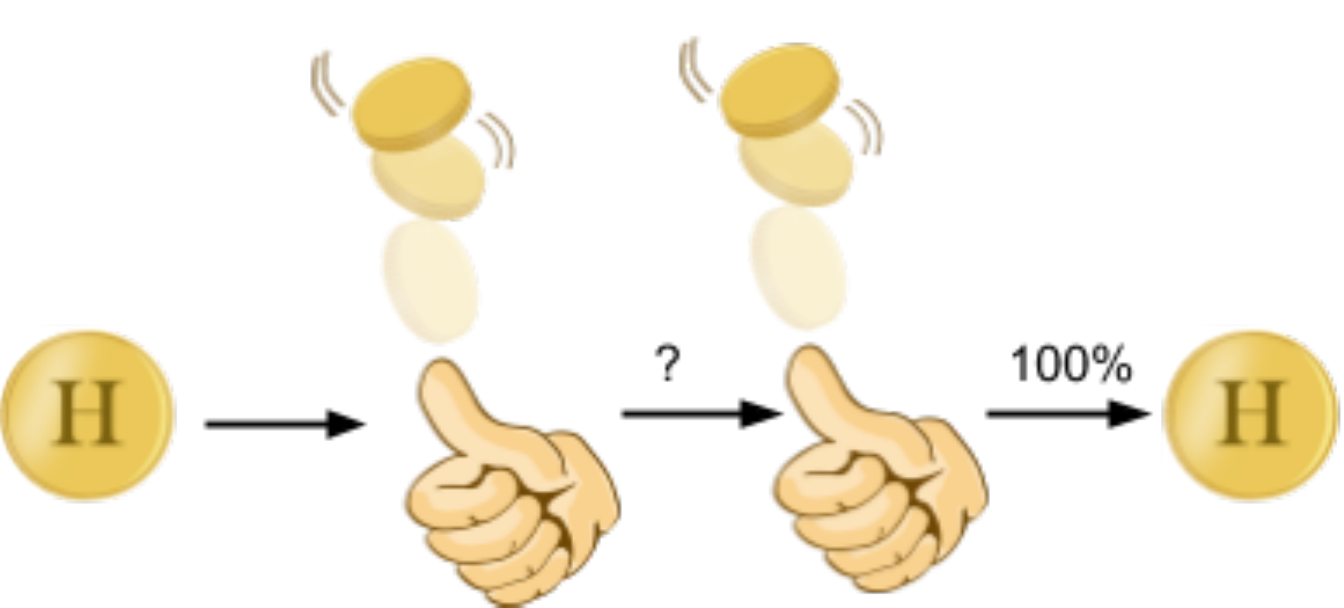}
  \caption{Coin analogy for the interferometer. Sending a photon through one beam splitter puts it in superposition, but adding a second beam splitter undoes the superposition and recovers the original state.}
  \label{fig:coin2}
\end{figure}

Even though the output of the first beam splitter is $50/50$, the second beam splitter can distinguish whether the laser was fired from the top of the bottom. The first beam splitter creates a superposition state, but adding a second one undoes the superposition and recovers the original state. This is a non-classical operation. It would be like starting with the coin heads up, flipping it, flipping it again while it is still in the air, and then always getting heads when it lands! This is highlighted in Figure~\ref{fig:coin2}.

There is hidden information in the superposition state. In the Mach-Zehnder photon qubit, the information is encoded in the form of the phase shift.
In the experiment shown in Figure~\ref{fig:MZTwoBeam}, we choose the phase shift to have a value of $\pi$. However, we could have just as easily chosen the phase shift to have any value between $0$ and $2\pi$ (the angles of a circle). Each separate choice of phase shift would produce a different type of superposition state that would still produce the same measurable $50/50$ outcome.\footnote{A complex amplitude $e^{i\phi}$ with infinite possible phase angles $\phi$ does not affect the probability since $\lvert e^{i\phi}\lvert ^2 =1$.} This phase shift information is present in the amplitudes but not the square of the amplitudes (and hence hidden from us in the Mach-Zehnder experiment-though we could make an other experiment to try to determine this information). Here are two simple examples of distinct states that can be created in {{two different experimental arrangements of}} the Mach-Zehnder experiment which still have the same $50/50$ probability:
\begin{equation}
\frac{1}{\sqrt{2}} \lvert 0\rangle + \frac{1}{\sqrt{2}} \lvert 1\rangle \quad \text{ or} \quad \frac{1}{\sqrt{2}} \lvert 0\rangle - \frac{1}{\sqrt{2}} \lvert 1\rangle.
\end{equation}
In these two states the {{plus or}} minus signs represents two of the many different phase shifts {{that are possible. Each different choice of the phase shift depends on how the experimental arrangement is chosen}}. As you can see, quantum superposition is inextricably linked to wave-particle duality.

Furthermore, in the Mach-Zehnder experiment we created a superposition, performed a phase shift and then observed wave interference. These experimental operations are equivalent to mathematically applying (matrix/gate) operations on a qubit, as we shall see later. As such, the Mach-Zehnder is an example of how we can technologically implement qubits (the photon) and operations (superposition/phase shift, etc) to build a quantum computer.\footnote{It should be noted that the technology has progressed so that most qubits are at present implemented using superconducting transmons and not using a Mach-Zehnder.} In quantum computing, people talk about the superposition of states rather than the wave behavior. Yet, as we have seen, both frameworks lead to the same understanding of the Mach-Zehnder interferometer. Later we will use the interferometer to implement a quantum algorithm.

\section{Check Your Understanding}
\begin{enumerate}
\item \fundamental{5pt} Your friend who is explaining superposition to you says that:

 ``A particle in the state $(1/\sqrt{2})\lvert0\rangle + (1/\sqrt{2})\lvert 1\rangle$ represents a lack of knowledge of the system. Over time, the particle is changing back and forth between the state $\lvert0\rangle$ and $\lvert1\rangle$. The superposition state says that overall, the particle is in each of the two states for half of the time."
 
  What parts of this statement do you agree with and what do you not agree with?
  
\item \intermediate{8pt} Only one detector is triggered if a single photon is sent through the beam splitter experiment shown in Figure \ref{fig:MZInterferometer}. If the laser outputs two photons at the same time, what is the probability that both detectors will be triggered simultaneously? Now how about three photons? Ten photons? Note that this is why a higher power beam of light appears to reach both detectors simultaneously.
  
\item \advanced{0.6pt} In practice, it is difficult to put the detectors the exact same distance from the beam splitter. The difference in distance is measured using the time delay {{$\Delta t$}} between photons. The experiment is shown in Figure~\ref{fig:detlatexp} and the data in Figure~\ref{fig:light}. 
\begin{figure}[h]
\centering
  \includegraphics[width=0.75\textwidth]{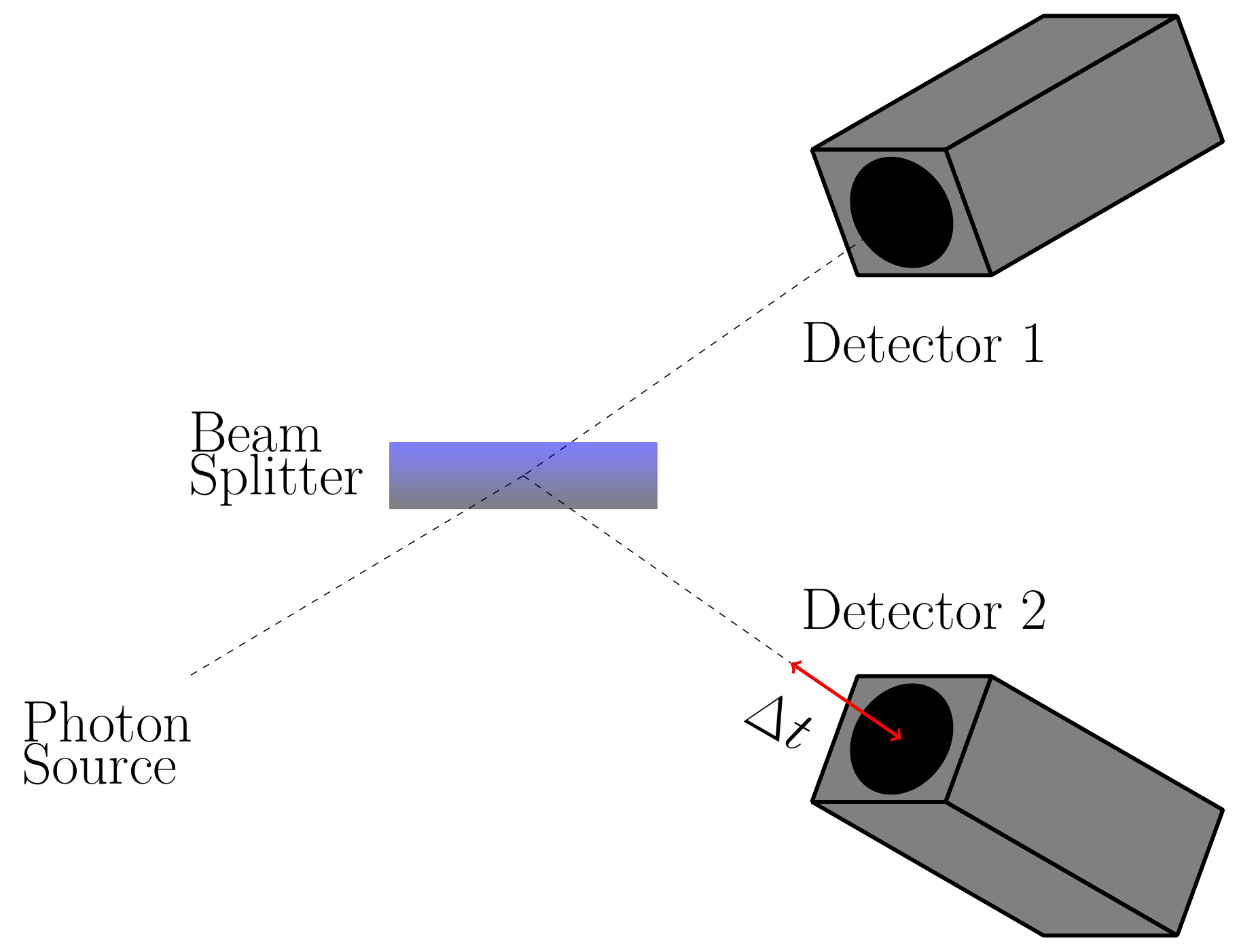}
  \caption{The experiment varies the position of Detector 2 and records the number of coincidences, i.e., the number of times both detectors are triggered simultaneously.}
\label{fig:detlatexp}
\end{figure}
\begin{figure}[h]
\centering
  \includegraphics[width=0.45\textwidth]{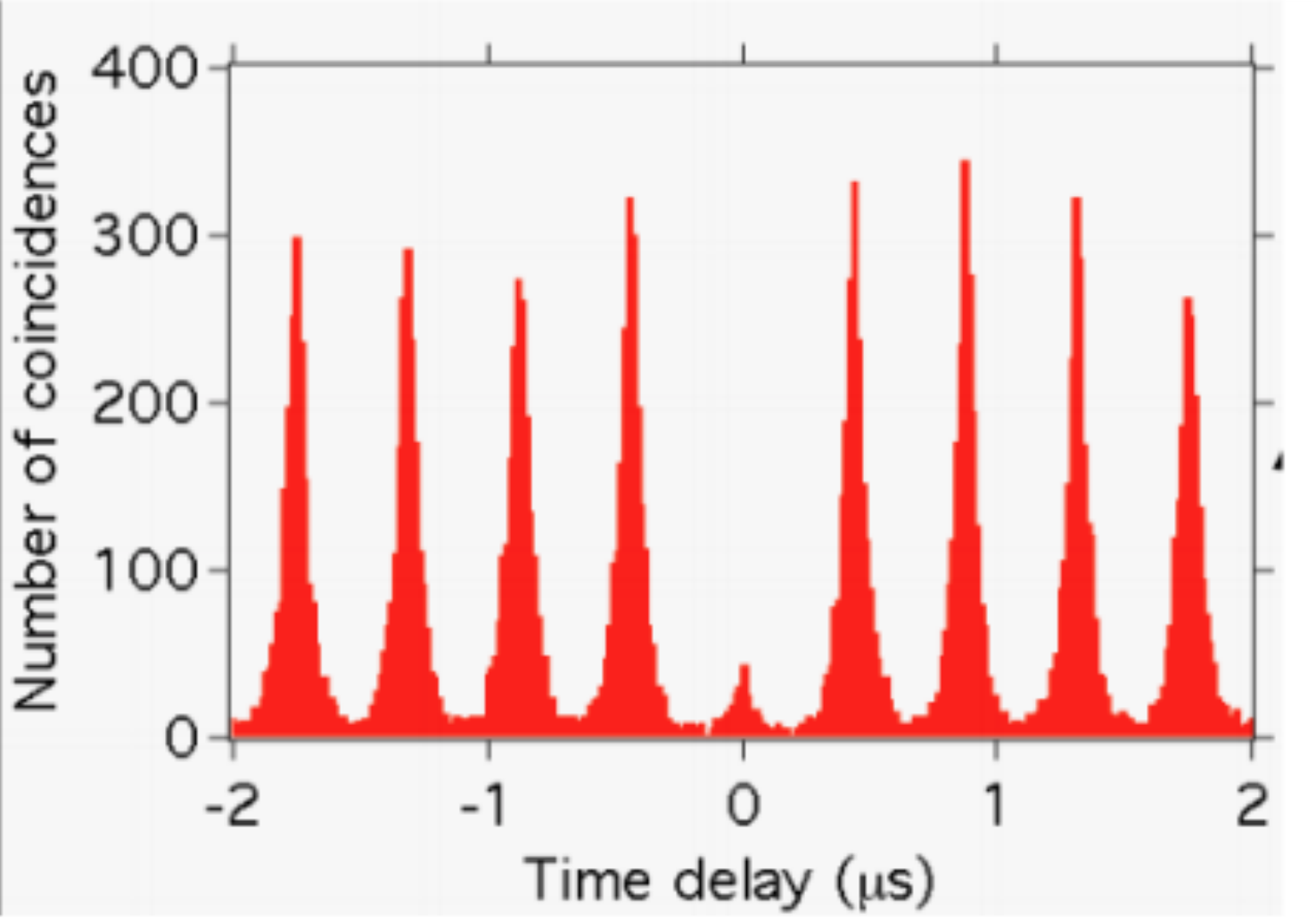}
  \caption{Data is shown below for light bursts sent from the laser every 0.4$\mu$s. Figure reproduced with permission of Martin Laforest and the Communications and Strategic Initiatives Team at the Institute for Quantum Computing, University of Waterloo Outreach department.}
\label{fig:light}
\end{figure}

\begin{enumerate}[label*=\alph*.]
\item Does the data shown in Figure~\ref{fig:light} at {{$\Delta t=0$}} support that light is a particle or a wave?
\item Why are there large coincidence counts when {{$\Delta t\neq0$}}? (Hint: Look at the spacing between the peaks.)
\end{enumerate}

\item \advanced{0.6pt} Using matrices given in Figure~\ref{fig:MZq4}, show how the superposition state is created by applying the beam splitter matrix transformation to the initial photon vector state. 
\begin{figure}[h]
\centering
  \includegraphics[width=0.75\textwidth]{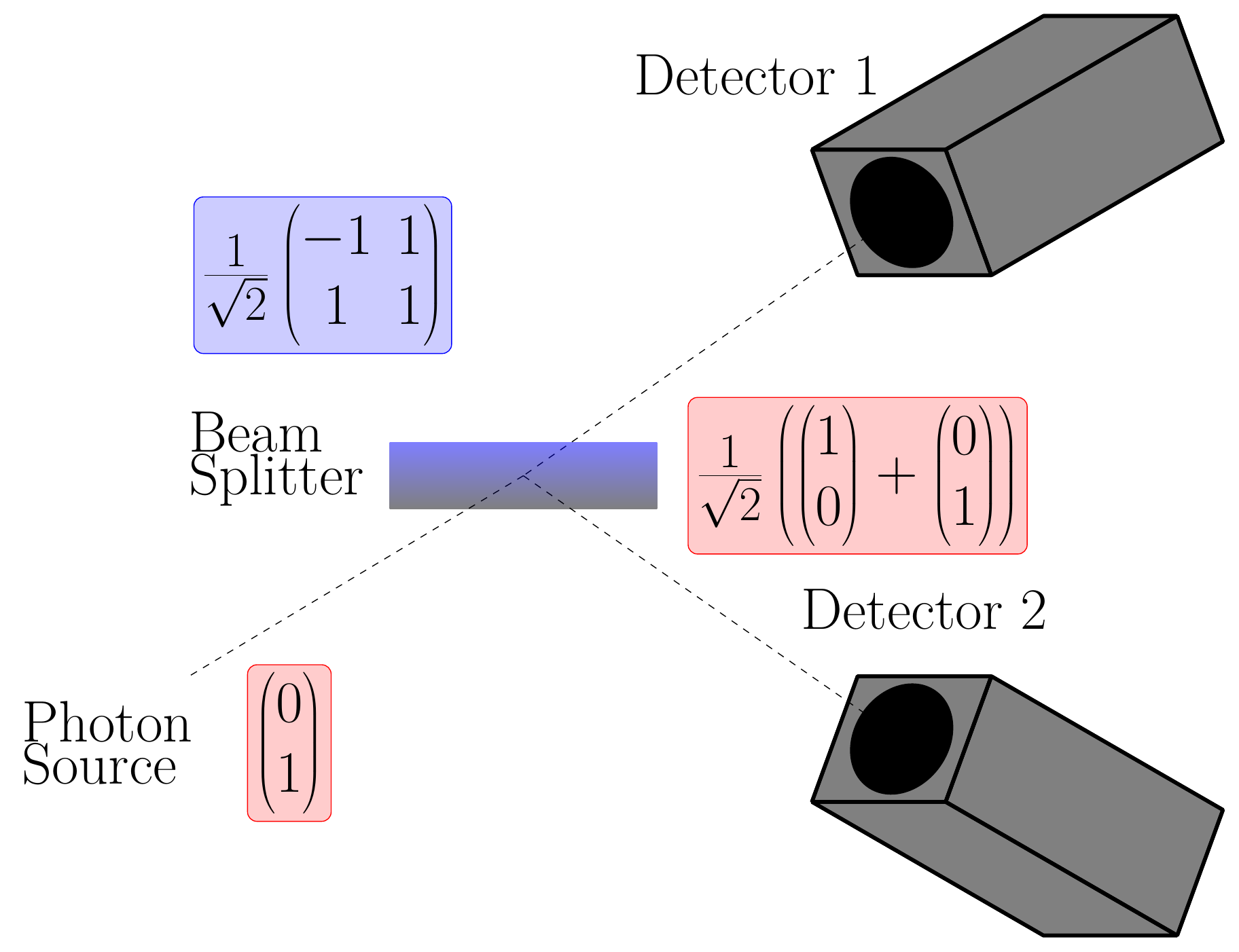}
  \caption{Matrix formulation of the Mach-Zehnder apparatus.}
  \label{fig:MZq4}
  \end{figure}

\item \advanced{0.6pt} Construct the matrix representation for a $30/70$ beam splitter.

\item \intermediate{8pt} Unsettled by the Mach-Zehnder interferometer, you decide to determine once and for all which path the photon takes after the first beam splitter. You place another detector (indicated by the eyeball) on the upper path as shown in Figure~\ref{fig:MZq6}. If the eyeball sees a photon, what would be seen at Detectors 1 and 2? 
\begin{figure}[h]
\centering
  \includegraphics[width=0.75\textwidth]{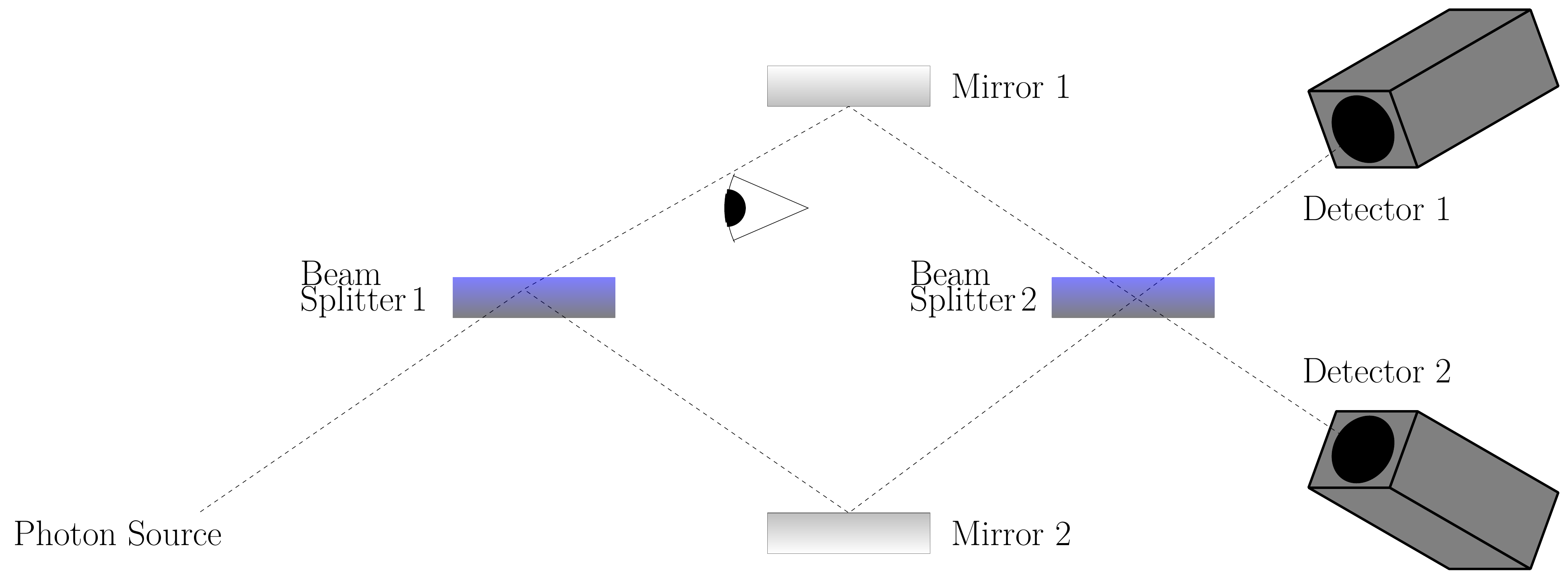}
  \caption{A third detector (your eye) is added to the Mach-Zehnder apparatus.}
  \label{fig:MZq6}
\end{figure}

\end{enumerate}

\graphicspath{{Chapter4-SternGerlach/}}
\chapter{Creating Superposition: Stern-Gerlach}

\section{\fundamental{5pt} Stern-Gerlach Apparatus}

Besides the photon in the interferometer, an electron is another prototype for a qubit.  An electron has many measurable properties such as energy, mass, momentum, etc. Yet, for the purposes of creating a qubit, we want to focus on a property with only two measurable values.  An electron has a two-state property called {\bf{spin}}.

Classically, an electron's spin can be visualized as a rotation about its own axis, like a spinning top or fidget spinner. You learned in high school physics that a moving charge creates a magnetic field according to the right-hand rule. By curling the fingers of your right hand in the direction of the electron's rotation, your thumb points in the direction of the magnetic field created by the charge.  The spinning electron behaves somewhat like a tiny bar magnet.\footnote{This classical picture is just an analogy. In reality, the quantum mechanical property we call ``spin'' is intrinsic to the electron (like its mass or charge) and can be described mathematically just like orbital momentum, but it is not when the electron physically rotates. See \href{https://en.wikipedia.org/wiki/Spin_(physics)}{https://en.wikipedia.org/wiki/Spin\_(physics)}.}

Surprisingly, the {\bf{Stern-Gerlach experiment}} (SGA) showed that the electron spin is quantized into only two values. \href{https://www.youtube.com/watch?v=rg4Fnag4V-E}{This video}\footnote{https://www.youtube.com/watch?v=rg4Fnag4V-E} explains the experimental apparatus used to measure the electron's spin. The key point here is that the vertically oriented apparatus (called the $z$-direction by convention) only measures the spin as either up or down, not randomly oriented at an angle.  Since the spin of an electron has two  measurable states, it can represent a qubit with $\lvert 0 \rangle$ as spin up and $\lvert 1 \rangle$ as spin down.  

\begin{figure}[h!]\centering
  \includegraphics[width=0.5\textwidth]{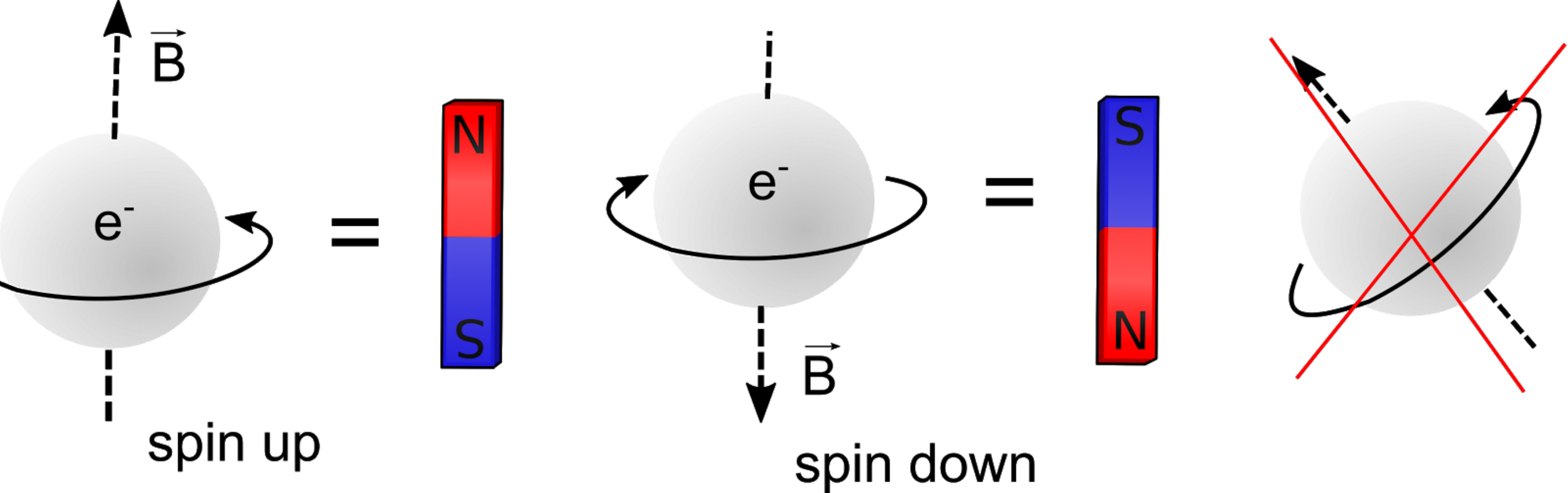}\label{fig:magnet}
    \caption{An electron can spin either up or down and produce a magnetic field.}
\end{figure}

\noindent \textbf{Question 1}: Open up the \href{https://phet.colorado.edu/sims/stern-gerlach/stern-gerlach_en.html}{PhET Stern-Gerlach simulator}\footnote{\href{https://phet.colorado.edu/sims/stern-gerlach/stern-gerlach\_en.html}{https://phet.colorado.edu/sims/stern-gerlach/stern-gerlach\_en.html}} and try sending electrons of various initial spins into the Stern-Gerlach apparatus (SGA). \\

\begin{figure}[h!]\label{fig:SG1}
\centering
  \includegraphics[width=0.5\textwidth]{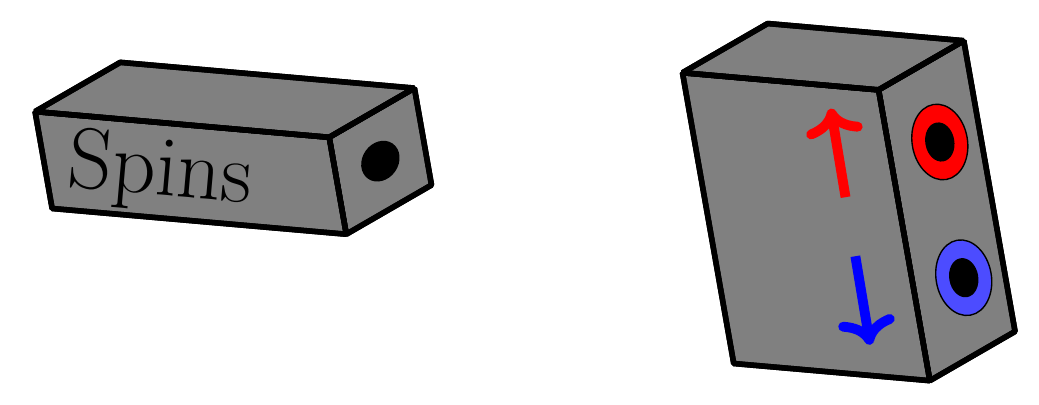}
  \caption{Electron spin produces a magnetic field either in the  up or down direction.
  }
\end{figure}

Are the results what you would expect?
The ``up'' and ``down''  directions are defined by the orientation of the apparatus. There is nothing inherently special about the $z$-direction compared to the $x$- or $y$-direction. An SGA rotated horizontally would measure either spin left or spin right.  An SGA rotated by $45^\circ$ would measure the spin to be either diagonally up or diagonally down.  What is particularly interesting is if we send a single spin up electron into a horizontally oriented SGA. \\

\textbf{Question 2}: Where would you expect a spin up electron to land after passing through a horizontal SGA?\\

\begin{figure}[h!]\label{fig:SG1}
\centering
  \includegraphics[width=0.5\textwidth]{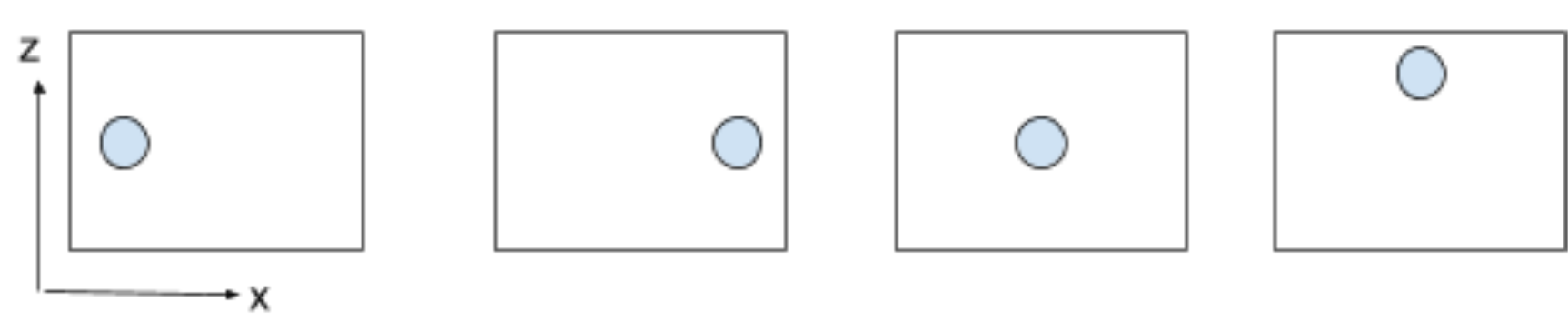}
  \caption{Possible positions.}
\end{figure}
Classically, vertically oriented bar magnets in a horizontal magnetic field would land at the center of the screen. However, recall that the spin can only be measured as left or right and cannot possibly land in the center. The way quantum mechanics solves this problem is to have the electron land either on the left or the right with $50\%$ probability. Sound familiar? Sending a 
spin up electron through a horizontal SGA puts the electron in a superposition state of left and right. \\

Spin in the vertical direction can be represented as a superposition of spins in the horizontal direction. As shown in the simulation, an electron with vertical spin has a $50\%$ chance of being measured as right or left:
\begin{align}
\lvert \uparrow \rangle & = \frac{1}{\sqrt{2}}\lvert \rightarrow \rangle + \frac{1}{\sqrt{2}}\lvert \leftarrow \rangle, \\
\lvert \downarrow \rangle & = \frac{1}{\sqrt{2}}\lvert \leftarrow \rangle - \frac{1}{\sqrt{2}}\lvert \rightarrow \rangle.
\end{align}
In more traditional qubit notation, spin in the $+z$ and $-z$ axis is written as $\lvert 0 \rangle$ and $ \lvert 1\rangle$, while spin in the $+x$ and $-x$ axis is $\lvert + \rangle$ and $\lvert - \rangle$:
\begin{align}
\lvert 0 \rangle & = \frac{1}{\sqrt{2}}\lvert+\rangle + \frac{1}{\sqrt{2}}\lvert-\rangle, \label{eq:1}\\
\lvert 1 \rangle & = \frac{1}{\sqrt{2}}\lvert + \rangle - \frac{1}{\sqrt{2}}\lvert - \rangle. \label{eq:2}
\end{align}
This is non-classical because you {{cannot}} add or subtract horizontal magnetic field vectors to get a vertical magnetic field vector. One analogy might be to think about a person looking at a coin vertically to determine its state. If they see heads or tails, someone looking from the side would see a superposition. If they are forced to make a choice via measurement, they would say heads or tails with 50$\%$ probability. \\
\begin{figure}[h!]\label{fig:weirdeyecoin}
\centering
  \includegraphics[width=0.5\textwidth]{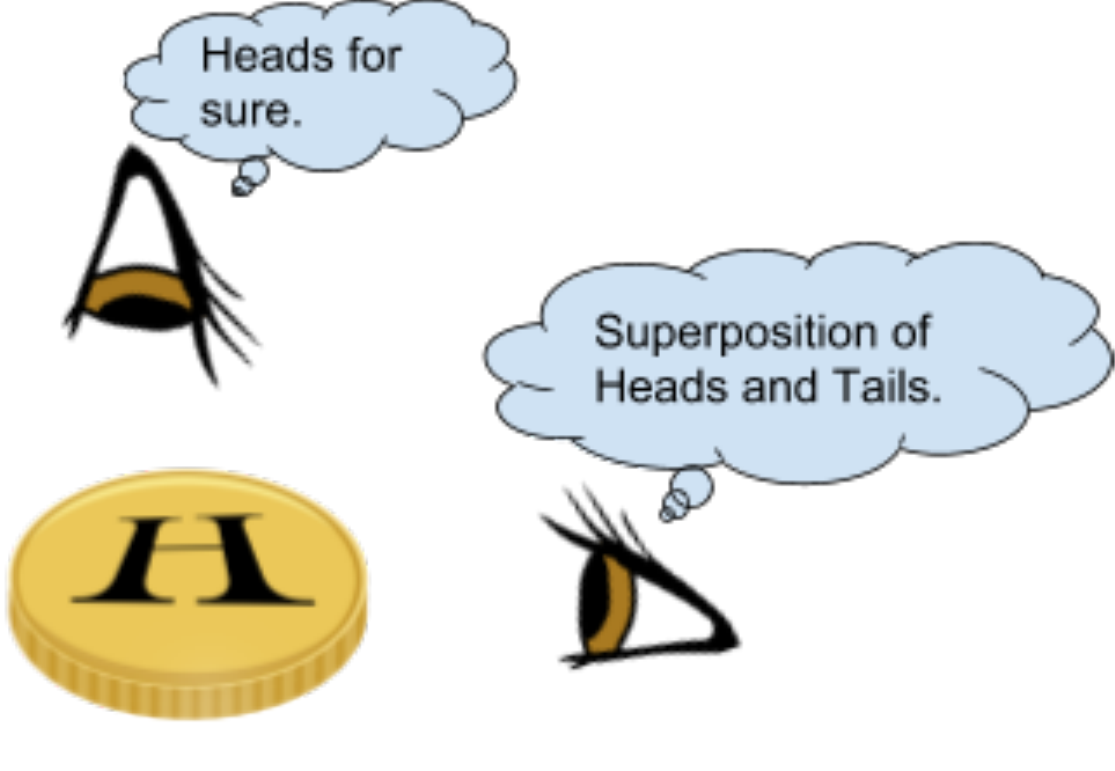}
  \caption{Analogy for how a definite vertical spin is seen as a superposition in the horizontal direction.}
\end{figure}

\textbf{Example:} Write the $\lvert + \rangle$ state in terms of $\lvert 0 \rangle$ and $\lvert 1 \rangle$. \\

\textbf{Solution:} Adding Equations~\ref{eq:1} and \ref{eq:2} we find
\begin{equation}
\lvert 0 \rangle + \lvert 1 \rangle = \frac{2}{\sqrt{2}}\lvert + \rangle.
\end{equation}
Rearranging, we get 
\begin{equation}
\lvert + \rangle  = \frac{1}{\sqrt{2}}\lvert 0 \rangle + \frac{1}{\sqrt{2}}\lvert 1 \rangle.
\end{equation}
Similarly, by subtracting Equations~\eqref{eq:1} and \eqref{eq:2}, we find
\begin{equation}
\lvert - \rangle  = \frac{1}{\sqrt{2}}\lvert 0 \rangle - \frac{1}{\sqrt{2}}\lvert 1 \rangle.
\end{equation}
These equations show that a horizontal spin is a superposition of spin up and spin down.\\

As we saw in the beam splitter example, the minus sign encodes information about the original state of the particle before it is put in superposition. It is possible to choose other complex amplitudes that give the same probability, but the details are mathematically beyond our scope. The conclusion we reached is that spins in one direction can be written as a superposition of spins in another direction. The Stern-Gerlach experiment shows that qubits in superposition are an accurate description of how nature actually works. Therefore, one promising application of quantum computers is simulating systems that occur in nature such as electronic properties of a molecule for use in drug design\footnote{\href{https://analyticsindiamag.com/top-applications-of-quantum-computing-everyone-should-know-about/}{https://analyticsindiamag.com/top-applications-of-quantum-computing-everyone-should-know-about/}}. \\

\section{\fundamental{5pt} Measurement Basis}

 The ``$z$-basis'' is composed of $\lvert 0 \rangle$ and $\lvert 1 \rangle$  while $\lvert + \rangle$ and $\lvert - \rangle$ compose the ``$x$-basis.''  
 A basis is analogous to a coordinate system for quantum states.  Any state can be written in terms of a different choice of basis, similarly to how any vector can be broken down into components along a different choice of axes. \\

In the figure below, a box on a ramp is subject to a force. The vector decomposition of $\vec{F}$ is shown for three different coordinate systems. All three coordinate systems are valid for describing the force, but only the first two are convenient to use in {{actual calculations}}. By choosing $x$-$y$ to be perpendicular, you have made the components mutually exclusive: if a vector is horizontal, you know it's definitely not vertical. The $x$- and $y$- directions can be treated as two independent problems. A more mathematical way of saying the axes are independent is to say they are orthogonal.  \\

\begin{figure}[h!]\label{fig:SG1}
\centering
  \includegraphics[width=0.5\textwidth]{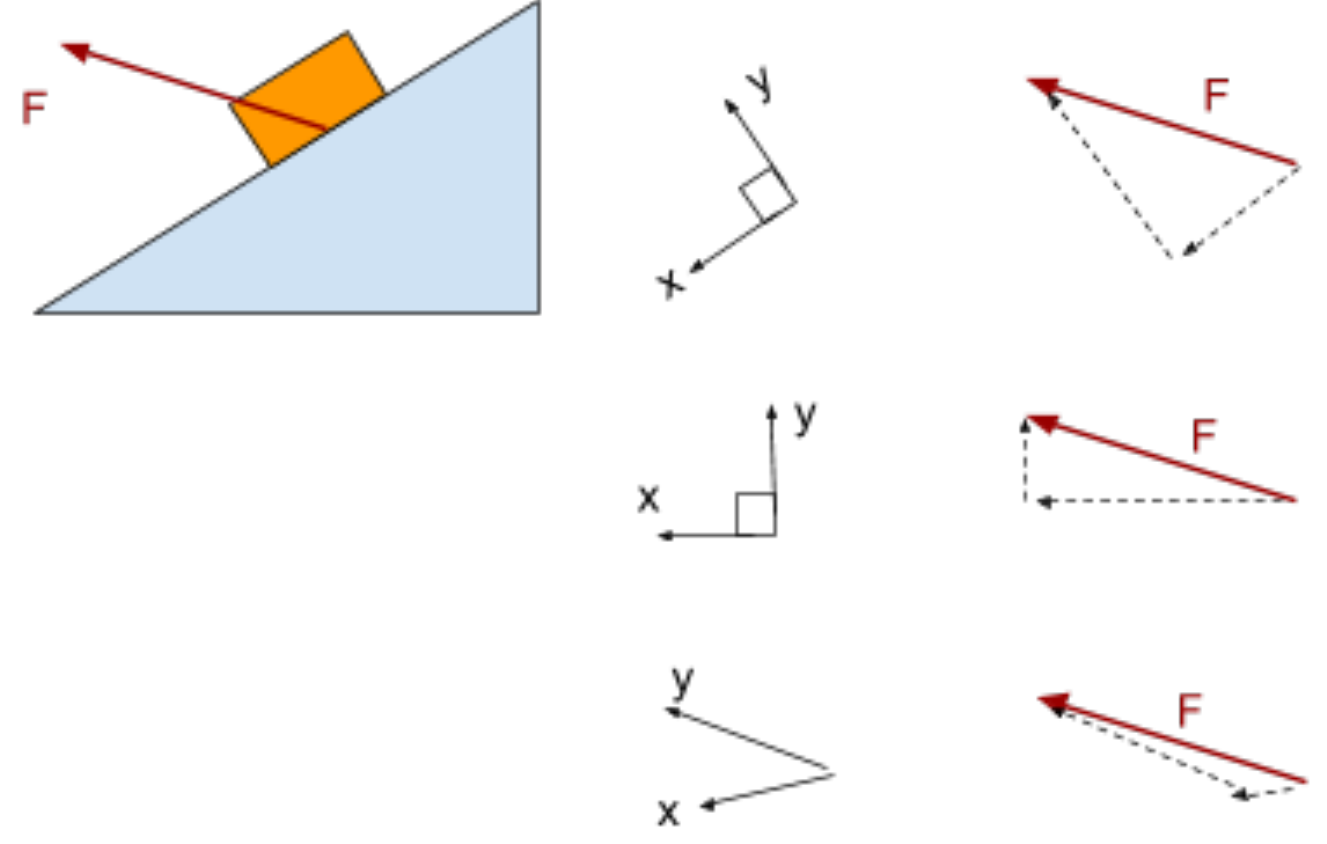}
  \caption{ Rewriting quantum states in terms of a different basis is similar to decomposing a classical vector into a different choice of coordinate system.}
\end{figure}
In quantum mechanics, there are an infinite number of possible choices for a basis. However, the basis should have two properties:
\begin{enumerate}
\item The basis must  describe all possible quantum states for the system.
\item The basis must be orthogonal.
\end{enumerate}

Let us check these conditions for the $z$-basis, which consists of states $\lvert 0 \rangle$ and $\lvert 1 \rangle$: 
\begin{enumerate}
\item Because the Stern-Gerlach experiment shows that an electron is either spin up or spin down, the most general state of the electron would be a superposition of up and down:
\begin{equation}
\lvert \text{electron} \rangle = \alpha \lvert 0 \rangle + \beta \lvert 1 \rangle.
\end{equation}
A linear combination of $\lvert 0 \rangle$ and $\lvert 1 \rangle$ completely describes the electron's state.
\item If you measure the spin as $\lvert 0 \rangle$, it is definitely not $\lvert 1 \rangle$, so $\lvert 0 \rangle$ and $\lvert 1 \rangle$ are orthogonal.
\end{enumerate}
The same argument can be made for the $x$-basis or any other angle of the SGA. 

\section{\intermediate{6pt} Geometric Representation of a Basis}
In this geometric representation of the $z$-basis and $x$-basis, the orthogonal states are drawn perpendicular to one another. If the electron is in a particular state $\lvert 0 \rangle$ in the $z$-basis, the state vector can be decomposed into $1/\sqrt{2}\lvert + \rangle + 1/\sqrt{2}\rangle  \lvert - \rangle$ 
in the $x$-basis. Physically turning the SGA from vertical to horizontal is how one changes from measuring in the $z$ to $x$-basis. Since 
$\lvert 0 \rangle = 1/\sqrt{2}\lvert - \rangle + 1/\sqrt{2}\rangle  \lvert - \rangle$,
the spin up particle became a 50/50 superposition when the measurement device became horizontal. \\

\begin{figure}[h!]
\centering
  \includegraphics[width=0.5\textwidth]{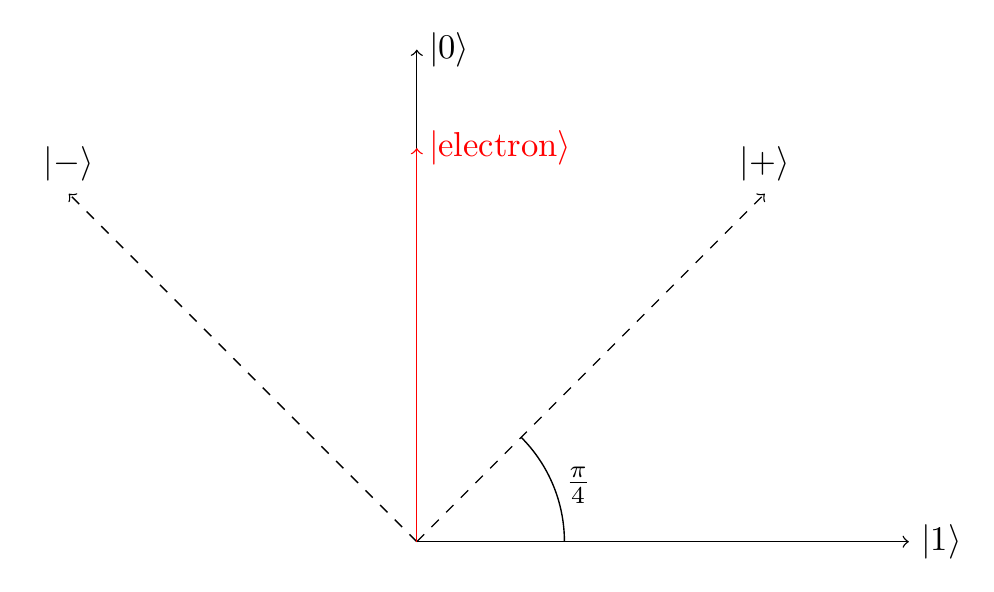}
  \caption{Geometric representation of the $z$-basis and $x$-basis. The state of a spin up electron is shown.}\label{fig:geom}
 \end{figure}

\textbf{Question 3:} Use Figure \ref{fig:geom}  and trigonometry to show that 
$\lvert 1 \rangle = 1/\sqrt{2}\lvert + \rangle - 1/\sqrt{2}\rangle  \lvert -\rangle$.\\

Often, there is hidden information about the state that cannot be measured unless we change to a different basis.  In the $x$-basis, there is no measurable difference between $\lvert 0 \rangle$ and $\lvert 1 \rangle$.  In the $z$-basis, $\lvert 0 \rangle$ would have 100$\%$ probability of going up and 0$\%$ down, while $\lvert 1 \rangle$ would have 0$\%$ probability of going up and 100$\%$ down.

\section{\intermediate{6pt} Effect of Measurement}

You learned that measuring a qubit collapses its superposition state into one of two possibilities. A spinning coin is in a superposition state, but once it lands, it becomes either heads or tails.  The photon is in a superposition state after passing through a beam splitter, but once it reaches the detectors, we know for sure whether it was reflected or transmitted. 
To see the truly strange nature of quantum measurement, let's see what happens when electrons are sent through multiple Stern-Gerlach devices in a row.\\

\textbf{Question 4}: Open the PhET \href{https://phet.colorado.edu/sims/stern-gerlach/stern-gerlach_en.html}{Stern-Gerlach simulator}\footnote{\href{https://phet.colorado.edu/sims/stern-gerlach/stern-gerlach\_en.html}{https://phet.colorado.edu/sims/stern-gerlach/stern-gerlach\_en.html}} and send electrons with randomly oriented spins through a vertical SGA. What is the spin of the electrons that pass through the hole?
\begin{enumerate}[label=(\alph*)]
\item $+z$
\item $-z$
\item Superposition of $+z$ and $-z$
\end{enumerate}

\begin{figure}[h!]
\centering
  \includegraphics[width=0.5\textwidth]{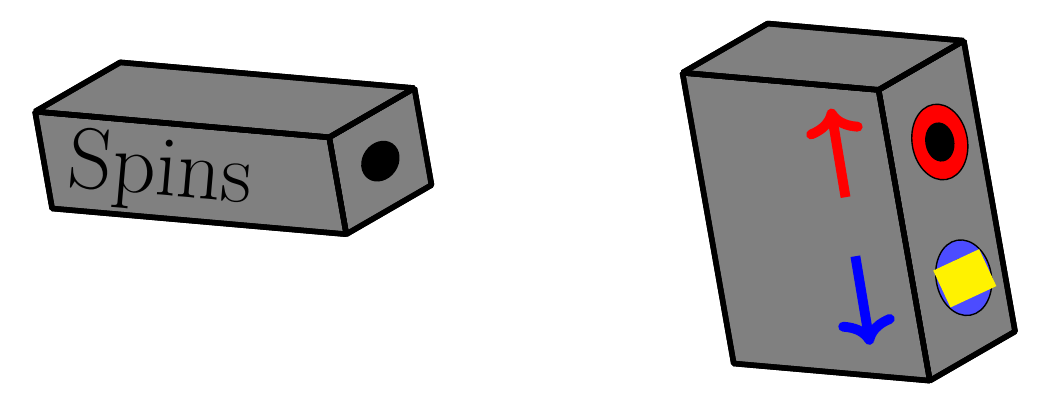}
  \caption{The $z$-axis SGA lets through spin up electrons but blocks spin down electrons.}
\end{figure}

\textbf{Question 5}: Add a second SGA, oriented horizontally. What is the spin of the electrons before entering the second SGA? 
\begin{enumerate}[label=(\alph*)]
\item $+x$
\item $-x$
\item Superposition of $+x$ and $-x$
\end{enumerate}

\begin{figure}[h!]
\centering
  \includegraphics[width=0.5\textwidth]{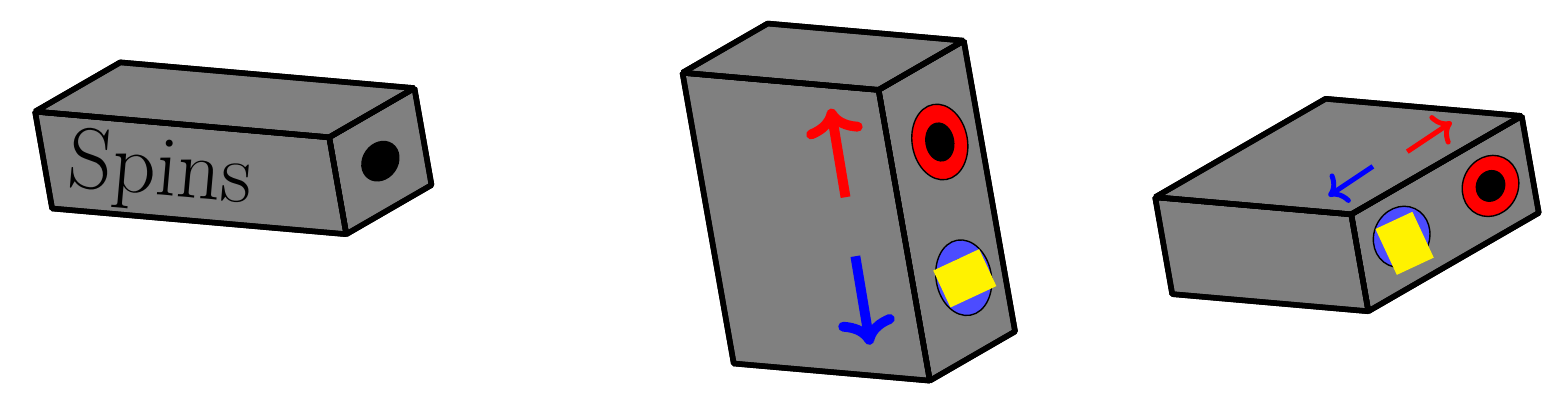}
  \caption{The  $z$ and $x$-axis SGA.}
\end{figure}

\textbf{Question 6}: What is the spin of the electrons after passing through the second SGA?
\begin{enumerate}[label=(\alph*)]
\item $+x$
\item $-x$
\item Superposition of $+x$ and $-x$
\end{enumerate}

\textbf{Question 7}: What is the $z$-spin of the electron coming out of the second SGA? Design an experiment to confirm this in the simulation.
\begin{enumerate}[label=(\alph*)]
\item $+z$
\item $-z$
\item Superposition of $+z$ and $-z$
\end{enumerate}

Given that only spin up electrons passed through the first SGA, one would expect that the electron is still spin up after the second SGA, no matter what is measured in $x$.  However, if you measure the $z$-spin with a third SGA, it has a 50$\%$ chance of being up or down!  

\begin{figure}[h!]
\centering
  \includegraphics[width=0.5\textwidth]{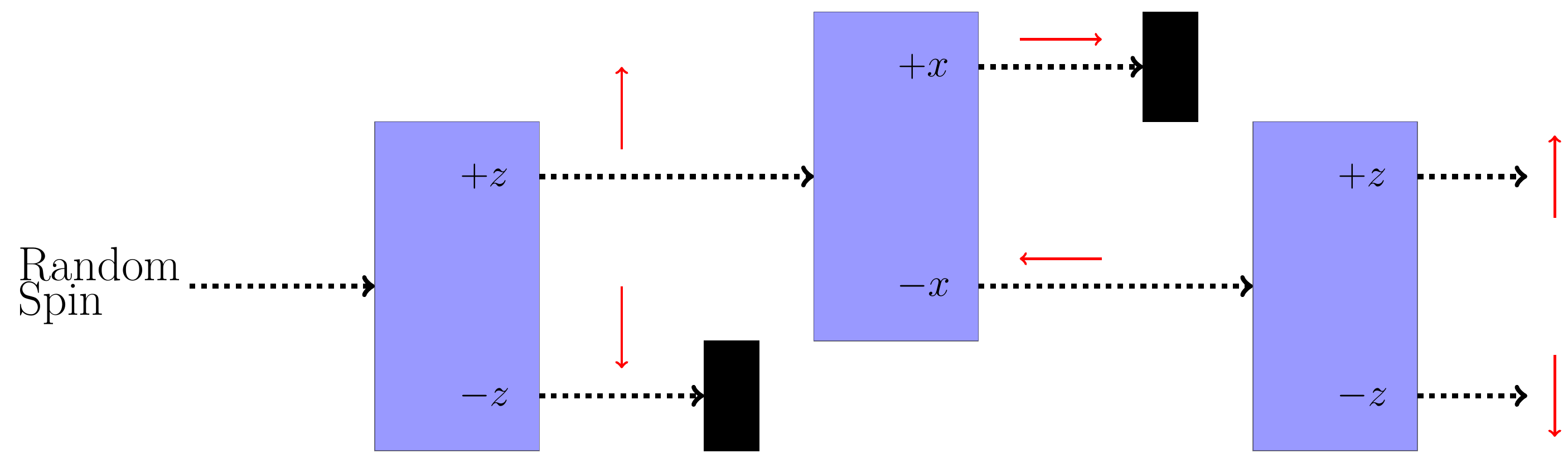}
  \caption{The first SGA selects for $+z$ spin and the second SGA puts it in a superposition of $+x$ and $-x$.  The third SGA shows that measuring the $x$ puts the electron in a superposition of $+z$ and $-z$.}
\end{figure}

By looking at the electron, we fundamentally changed its state. Measuring the $x$-spin of the qubit puts it into a superposition of up and down, even when it started as up to begin with. When you measure the length of an object with a ruler, you don't expect the objects' length to change after you measure the width!\\

Quantum measurement collapse is used in many applications such as cryptography where one could detect if a message has been intercepted. This will be discussed in  further detail in Chapter \ref{chap:crypto}. Moreover, this property of quantum states implies   a qubit in an unknown state cannot be copied.  This property is known as the no-cloning theorem. Classical computers can make a copy of a text and the original stays the same. If you try to copy an unknown qubit you first have to measure it, which collapses its superposition state. Therefore, quantum computers are unlikely to replace your laptop. However, for certain applications, the hidden information in superposition states allows information processing beyond what is possible in a classical computer.

\section{Activities}
\begin{itemize}
\item[\intermediate{8pt}] Polarizer Demo in Worksheet \ref{sec:WorksheetPolDemo}
\item[\intermediate{8pt}] Measurement Basis Lab in Worksheet \ref{sec:WosksheetMeasureLab}
\item[\advanced{0.5pt}] Superposition vs. Mixed States Lab in Worksheet \ref{chapter:WorksheetSuper}  
\end{itemize} 


\section{Check Your Understanding}
\begin{enumerate}

\item \fundamental{5pt} The Stern-Gerlach apparatus is rotated by $90^\circ$ so that the magnetic field is in the $x$-direction as shown in Figure \ref{fig:4.1}.  If electrons from a random source are sent through the apparatus, what pattern would be formed on the screen?

\begin{figure}[h!]
\centering
  \includegraphics[width=0.5\textwidth]{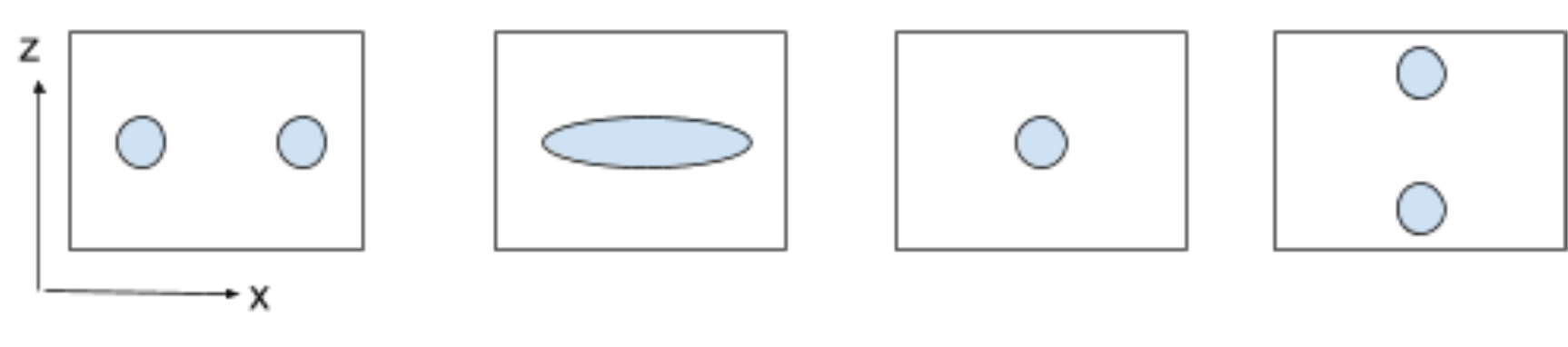}
  \caption{Stern Gerlach apparatus.}
  \label{fig:4.1}
\end{figure}

\item \intermediate{5pt} Would $\lvert 0 \rangle$ and $\lvert + \rangle$ together satisfy the criteria for a valid basis?
\item \intermediate{5pt} An electron is in a superposition state shown in the geometric representation in Figure \ref{fig:4.3}.
\begin{figure}[h!]
\centering
  \includegraphics[width=0.5\textwidth]{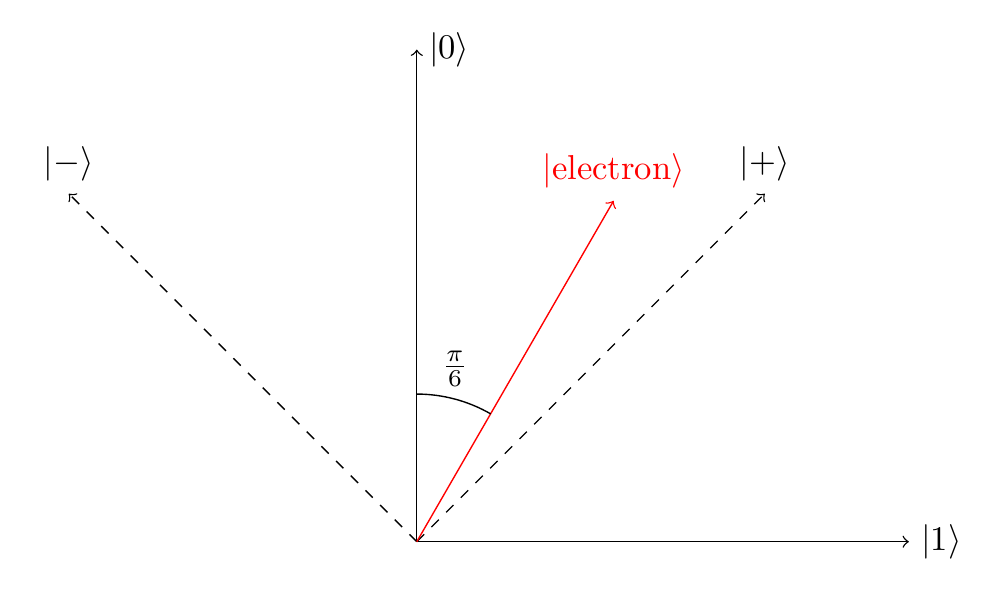}
  \caption{Superposition state of the electron.}
  \label{fig:4.3}
\end{figure}

\begin{enumerate}[label=(\alph*)]
\item What is the state of the electron in the $z$-basis?  i.e. find $\alpha$ and $\beta$ in 
$\lvert \text{electron} \rangle = \alpha\lvert 0 \rangle + \beta\lvert 1 \rangle$
\item  What is the probability of measuring spin up?
\item What is the state of the electron in the $x$-basis?  i.e, find $\alpha$ and $\beta$ in $\lvert \text{electron} \rangle = \alpha\lvert + \rangle + \beta \lvert - \rangle$.
\item What is the probability of measuring the spin in the $ \alpha\lvert + \rangle$ direction?
\end{enumerate}

\item \advanced{0.5pt} To measure the difference between an electron in a spin state $\frac{1}{\sqrt{2}}\lvert 0\rangle + \frac{1}{\sqrt{2}}\lvert 1\rangle$ and one in $\frac{1}{\sqrt{2}}\lvert 0\rangle - \frac{1}{\sqrt{2}}\lvert 1\rangle$, one could use:

\begin{enumerate}[label=\Roman*]
\item A horizontal SGA.
\item A vertical SGA.
\item A $45^\circ$ diagonal SGA.
\end{enumerate}

\begin{enumerate}[label=(\alph*)]
\item  I only
\item  II only
\item  I or III
\item  II or III
\item  I, II, or III
\end{enumerate}

\item \fundamental{5pt} An electron with random spin is sent through two vertical SGAs as shown in Figure \ref{fig:4.5}. What would be the output of the second SGA?
\begin{figure}[h!]
\centering
  \includegraphics[width=0.5\textwidth]{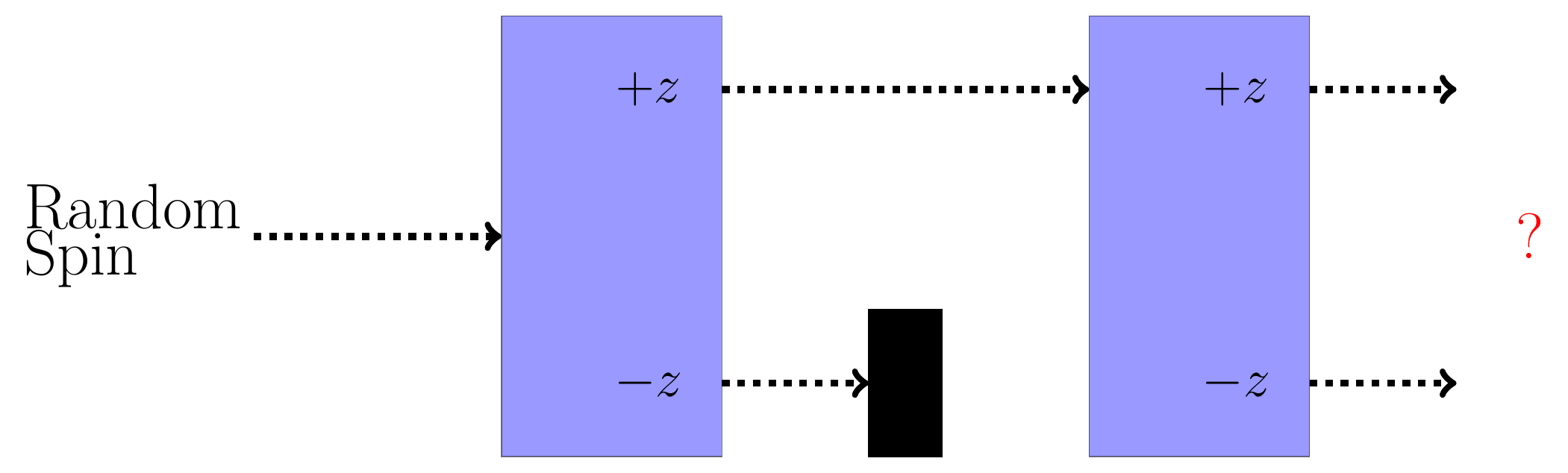}
    \caption{}
    \label{fig:4.5}
\end{figure}

\item \fundamental{5pt} An electron with random spin is sent through two vertical SGAs, where the second SGA is rotated upside down, or $180^\circ$. 
\begin{enumerate}[label=(\alph*)]
\item If the second $+z$ port is blocked as in Figure \ref{fig:4.6a}, what would be the output of the second SGA?
\begin{figure}[h!]
\centering
  \includegraphics[width=0.5\textwidth]{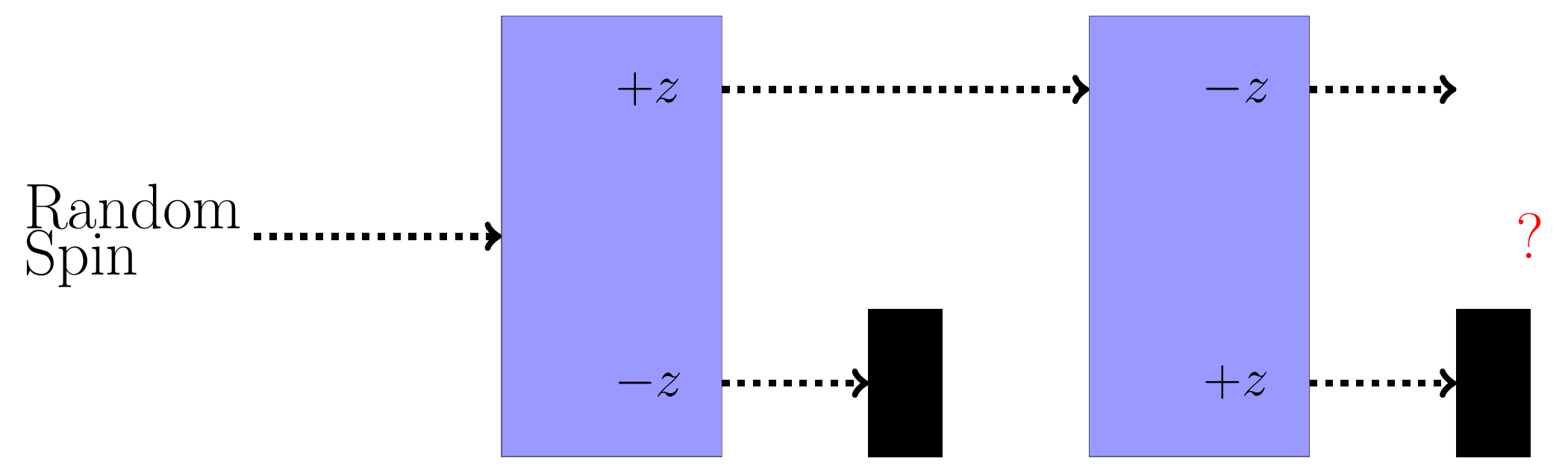}
  \caption{}
  \label{fig:4.6a}
\end{figure}

\item If both ports on the second SGA are open as in Figure \ref{fig:4.6b}, what would you see at the output? 
\begin{figure}[h!]
\centering
  \includegraphics[width=0.5\textwidth]{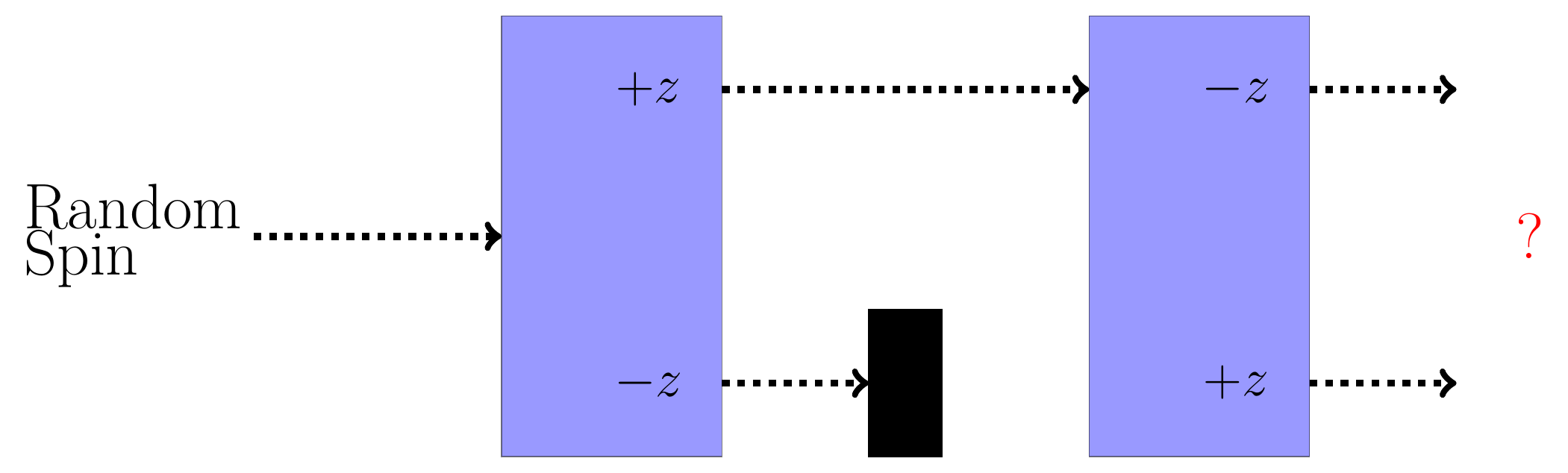}
  \caption{}
  \label{fig:4.6b}
  \end{figure}

\end{enumerate}

\item \fundamental{5pt} An electron with random spin is sent through a horizontal SGA followed by a vertical SGA as in Figure \ref{fig:4.7}. What would be the output of the second SGA?
\begin{figure}[h!]
\centering
  \includegraphics[width=0.5\textwidth]{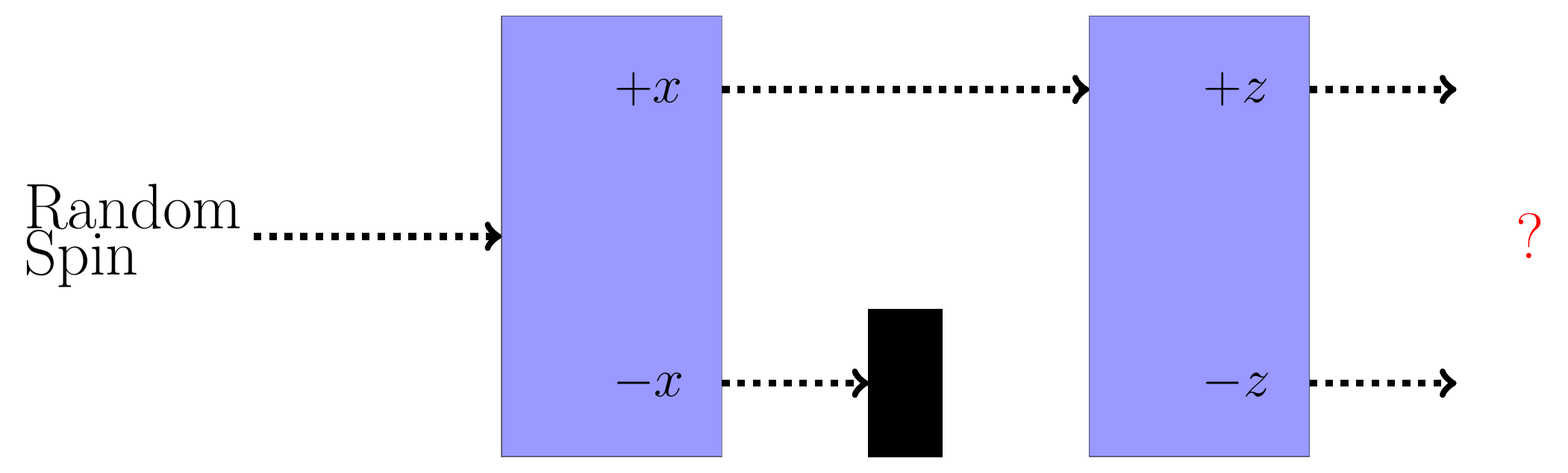}
  \caption{}
  \label{fig:4.7}
  \end{figure}

\item \fundamental{5pt} An electron with random spin is sent through three SGAs as shown in Figure \ref{fig:4.8}. What would be the output of the third SGA?
\begin{figure}[h!]
\centering
\includegraphics[width=0.99\textwidth]{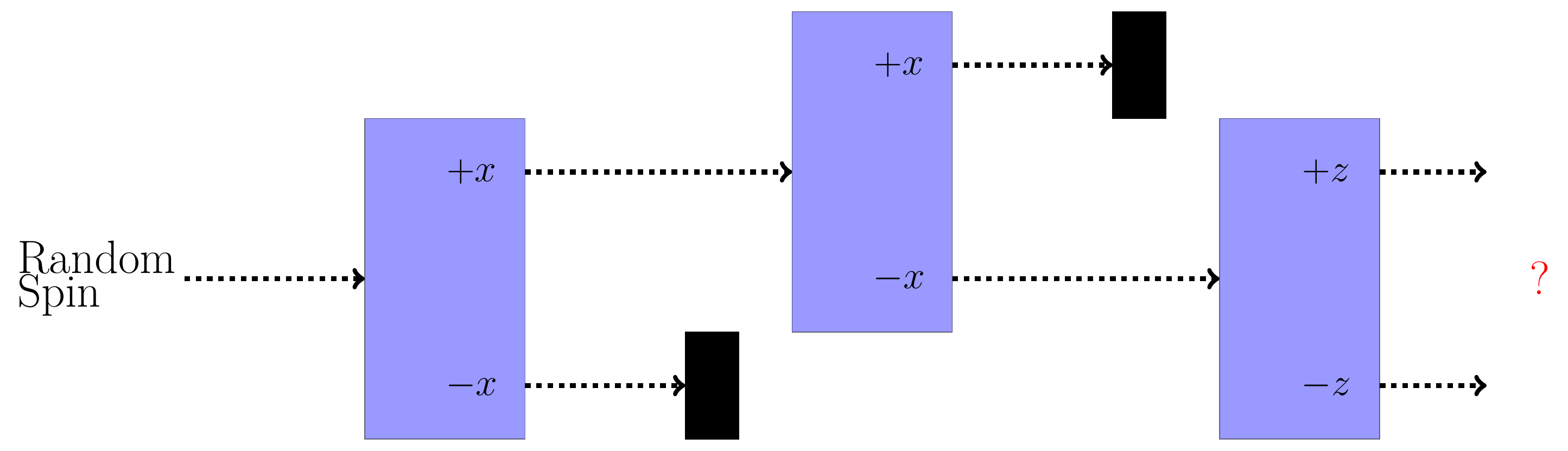}
  \caption{}
  \label{fig:4.8}
  \end{figure}

\item \fundamental{5pt} An electron with random spin is sent through three SGAs as shown in Figure \ref{fig:4.9}. What would be the output of the third SGA?
\begin{figure}[h!]
\centering
  \includegraphics[width=0.99\textwidth]{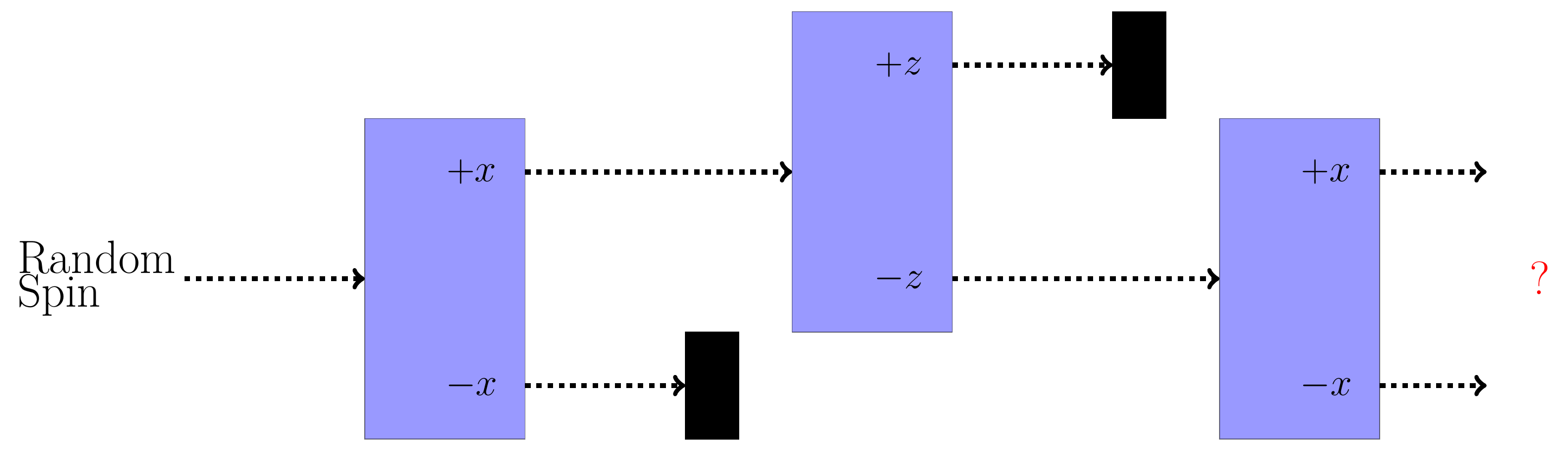}
  \caption{}
  \label{fig:4.9}
  \end{figure}

\item \fundamental{5pt}  An electron with random spin is sent through four SGAs as shown in Figure \ref{fig:4.10}. What would be the output of the fourth SGA?
\begin{figure}[h!]
\centering
  \includegraphics[width=0.99\textwidth]{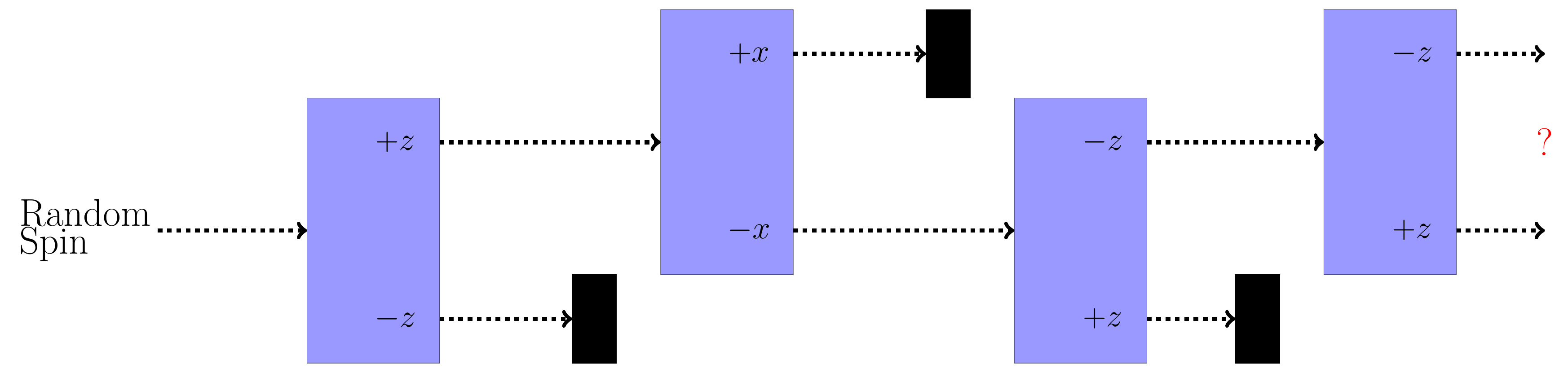}
  \caption{}
  \label{fig:4.10}
  \end{figure}

\end{enumerate}
%
%
%
%
%
%
%
%
%
%
%


\graphicspath{{Chapter5-Crypto/}}
\chapter{Quantum Cryptography}
\label{chap:crypto}

The Internet can be thought of  as a channel of information being sent from you to everyone else connected to the Internet. If you wanted to transmit your sensitive information (such as bank account numbers, military secrets, etc.) over the Internet, then you have to ensure that only the persons you intend  to read your information can  read your sensitive data. Otherwise,  everyone would be able to read your information, e.g.,  access to your bank account details and transfer money out of your account. Therefore, one needs to encrypt any data sent over the Internet. Encryption, in this context, ensures that only the intended sender and receiver can understand any message being sent over an Internet channel. 

Encryption relies on the sender and receiver sharing a secret key (that no one else has) and using that to encrypt and decrypt messages. In this way, since no one else has the secret key, no one else can understand the shared information. Because no one else understands the shared information, they cannot misuse it for their own benefit. The fundamental caveat with encryption is this: you require a secure channel, to share the secret key (if you do not have a secure channel then someone random can just take the secret key and encryption would be pointless), but if you have a secure channel then why do you need to encrypt your data? You need a way around this issue. How do you share a secret key in an insecure channel, where anyone can be listening?

The way around this in the majority of online communications is called public key cryptography.\footnote{https://en.wikipedia.org/wiki/Public-key\_cryptography} A person called Alice makes two keys such that each key knows that only the other key is related to it (think of the keys as siblings). They are called the private and  public key. Alice then gives the public key to everyone in the world but importantly keeps the private key for herself. Anybody else, say Bob, who wants to send a private
message to Alice has to encrypt their message with the public key that Alice generated. There are many different types of encryption protocols that one can use. The special part of public key cryptography is that \textit{only} Alice's private key can decrypt the message that was encrypted using its sibling public key. In this way, only Alice can read the message from Bob. Since no one else has Alice's private key, no one else can read Bob's message. However, if Bob did not use Alice's public key
but used a different public key to encrypt his message, then Alice cannot decrypt that message, as her private key is not a sibling key of the different public key. This whole cryptography scheme relies on the fact that no one can break the encryption protocol. If they could break it, then they could read Alice's message even if they did not have Alice's private key. Note, this is  how your information is protected over the Internet.  Because the encryption protocol is so difficult to break, no one would even attempt to do so. Instead, they may attempt to steal your private key by hacking into your computer.
The one-time pad, also known as the Vernam Cipher, is the only type of encryption protocol known to be perfectly secure.\footnote{Shannon, Claude (1949). ``Communication Theory of Secrecy Systems.'' \textit{Bell System Technical Journal}. 28 (4): 656–715. \href{https://doi.org/10.1002/j.1538-7305.1949.tb00928.x}{doi:10.1002/j.1538-7305.1949.tb00928.x.}} It is assumed that two people exchange a shared key at least as long as the message in a completely secure way. The shared key encrypts the message to
create the cipher, and the cipher is decoded by decrypting with the shared key. The protocol is best understood by trying it out with the associated worksheets. In practice, due to not having a secure channel to share such a complicated key, despite being unbreakable, this method is usually not employed.\footnote{https://en.wikipedia.org/wiki/One-time\_pad} 

The most commonly used modern Internet encryption protocol is called RSA encryption. RSA encryption relies on encrypting messages with keys that are made out of very large integers. To decrypt a message, one would need to factorize this very large integer into its (prime) factors. Factorizing a large integer into its (prime) factors is known to be a problem that classical computers cannot solve in any reasonable amount of time.\footnote{https://en.wikipedia.org/wiki/Integer\_factorization} For example, while it takes just a fraction of a second to multiply two prime numbers together to produce this large integer, finding which  two prime numbers produced the integer would take a classical supercomputer thousands of years. RSA encryption works by encrypting the message with the public key. If an eavesdropper wanted to decrypt this message, they would need to factorize a large integer in the public key, which would take thousands of years. However, the private key related to the public key knows how to check the prime factors of the public key and can decrypt the message easily. 

Alternatively, a bad actor could try to steal the private key, which Internet firewalls protect against. If a private key tries to decrypt a message that was encrypted with a public key not related to it, it has the wrong prime factors associated with the public key and the decryption fails. As such, nearly all Internet encryption relies on a computer not being able to factor large integers in a short amount of time. 

However, in 1995, Peter Shor proposed a quantum computing algorithm, based on superposition and interference, that drastically speeds up the factoring process. A 4000-digit number, which would take a classical computer longer than the lifetime of the universe to factorize, would take less than a day on a quantum computer.  \href{https://quantumexperience.ng.bluemix.net/proxy/tutorial/full-user-guide/004-Quantum\_Algorithms/110-Shor\%27s\_algorithm.html}{Shor's
algorithm}\footnote{https://quantumexperience.ng.bluemix.net/proxy/tutorial/full-user-guide/004-Quantum\_Algorithms/110-Shor\%27s\_algorithm.html} can theoretically break modern encryption schemes, although quantum hardware is not advanced enough yet to make this decryption practical. If it were, all your bank details, military and industrial secrets, water and electric supply, etc., {could} be easily hacked. The details of Shor's algorithm are beyond our scope, so we will instead discuss how the same quantum computer could be used to establish a secure key. Together, the one-time pad and \textbf{quantum key distribution} (QKD) would be a formidable combination.

The \href{https://www.st-andrews.ac.uk/physics/quvis/simulations\_html5/sims/cryptography-bb84/Quantum\_Cryptography.html}{BB84 QKD}\footnote{https://www.st-andrews.ac.uk/physics/quvis/simulations\_html5/sims/cryptography-bb84/Quantum\_Cryptography.html} simulation demonstrates how one could create a shared key using electrons and a Stern-Gerlach apparatus. The BB84 protocol is summarized below. 

\section{\intermediate{5pt} BB84 Protocol}

\subsection*{Before sending the message}
The sender (Alice) and receiver (Bob) publicly agree to the relationship between spins and bit value shown in Table~\ref{tab:Cryptobits}. 
\newcolumntype{L}[1]{>{\raggedright\let\newline\\\arraybackslash\hspace{0pt}}m{#1}}
\newcolumntype{C}[1]{>{\centering\let\newline\\\arraybackslash\hspace{0pt}}m{#1}}
\begin{table}[h!]
\centering
\begin{tabular}{|c|C{1cm}|C{1cm}|C{1cm}|C{1cm}|C{1cm}|}
    \hline
   & \multicolumn{2}{c|}{Alice}& \multicolumn{2}{c|}{Bob} \\ \hline
 Spin &$\uparrow$ & $\leftarrow$ & $\downarrow$ & $\rightarrow$\\ \hline
 Bit value &0&0&1&1\\ \hline
\end{tabular}
\caption{Table for the relationship between Alice and Bob for quantum cryptography.}\label{tab:Cryptobits}
\end{table}

\subsection*{Quantum part}
\begin{enumerate}
    \item Alice randomly chooses either the $x$- or $z$-basis (horizontal or vertical Stern-Gerlach apparatus). 
    \item Alice sends an electron in superposition in the chosen basis through the SGA, measures the spin, and records the corresponding bit value as 0 or 1. The electron is sent to Bob.
    \item Bob randomly chooses either the $x$- or $z$-basis.
    \item Bob measures the spin of the electron and records whether it was 0 or 1.
    \item Repeat steps 1--4 {{as many times as necessary}}.
\end{enumerate}
\subsection*{Example}
\begin{figure}[!h]
    \centering
    \includegraphics[width=0.7\textwidth]{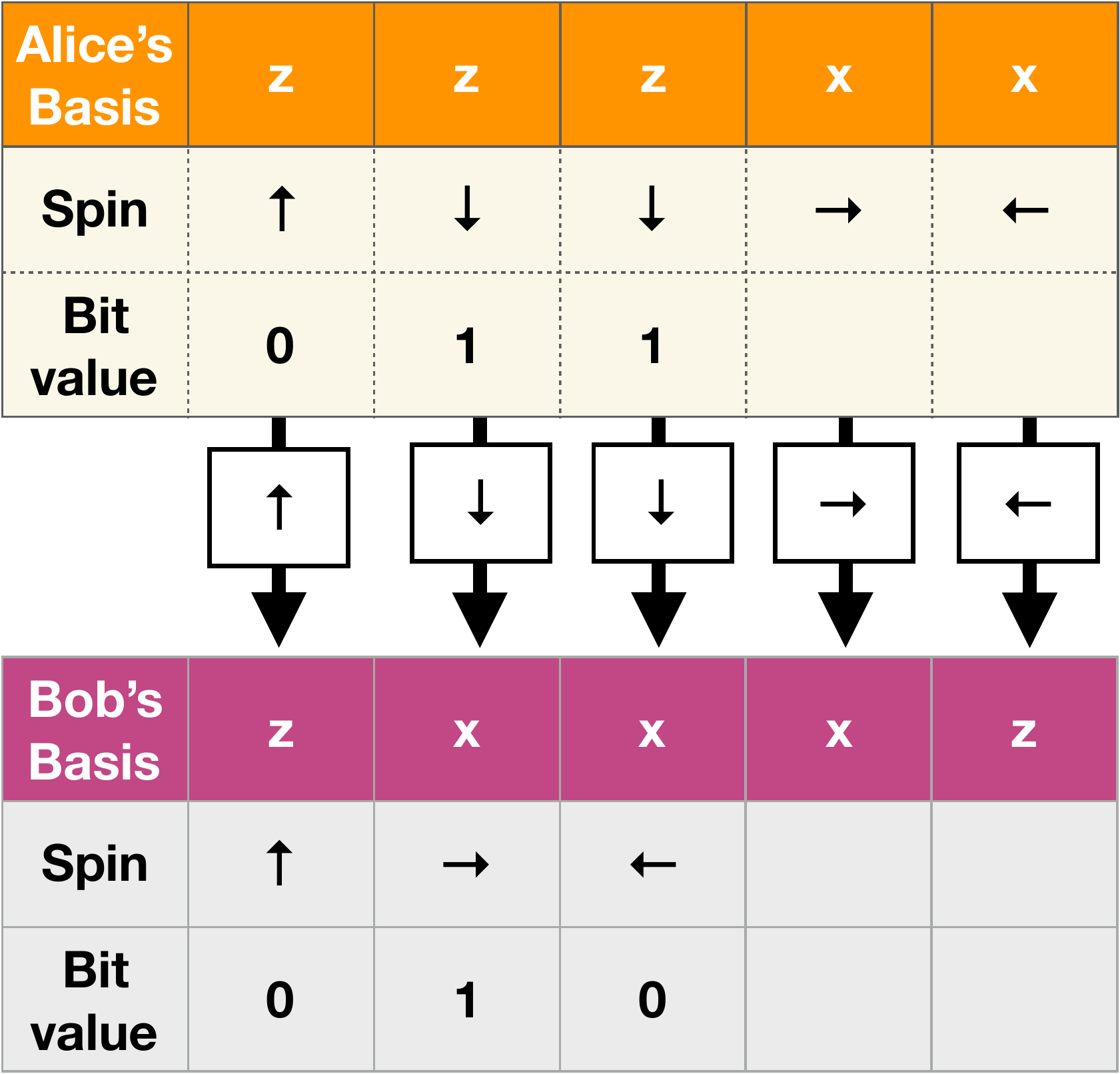}
    \caption{Alice and Bob's measurements of the BB84 protocol.}
    \label{fig:BB84-fig1}
\end{figure}

Alice sends five electrons to Bob. When Alice sends an electron prepared in one basis and Bob measures in the same basis, they measure the same spin.  However, if Bob measures in different basis than Alice, then the electron will be in a superposition state and there will be a 50\% probability of the state collapsing into 0 or 1. Example values for the first three bits of a BB84 experiment are shown in Figure~\ref{fig:BB84-fig1}. Can you fill in the last two bits?

\subsection*{Classical post-processing}
\begin{figure}[!h]
    \centering
    \includegraphics[scale=0.5]{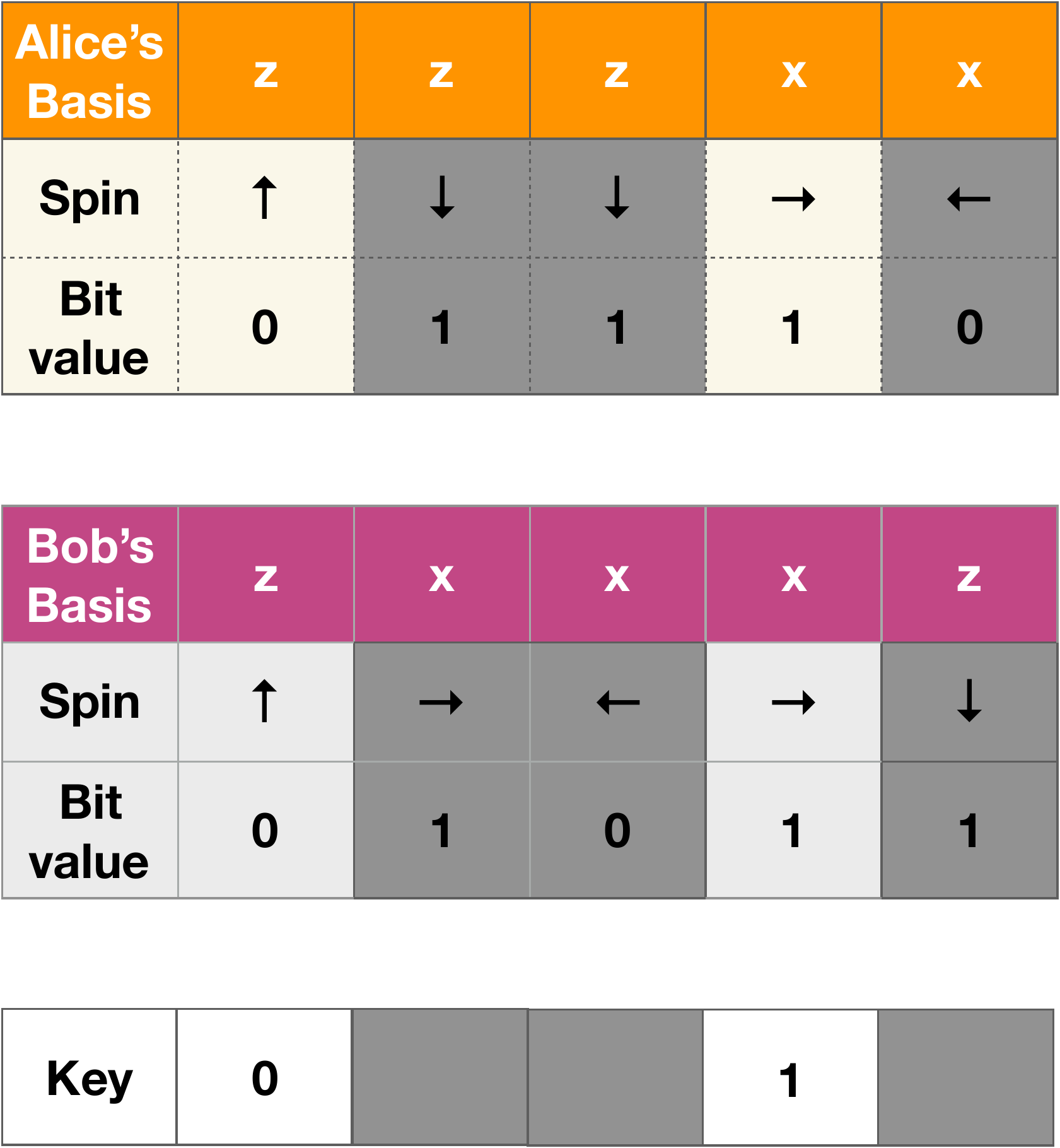}
    \caption{Alice and Bob's measurements of the BB84 protocol completed from Figure~\ref{fig:BB84-fig1}. The discarded bits are grayed out, and the key is 01.}
    \label{fig:BB84-fig2}
\end{figure}

\begin{enumerate}
    \item Alice and Bob publicly share the basis used for each bit measurement \textit{without revealing the actual bit value they measured.}
    \item If they measured in the same basis, they keep that bit. If they measured in a different basis, they discard that bit. This is shown in Figure~\ref{fig:BB84-fig2}. For the measurements performed in the same basis, Alice and Bob are guaranteed to have the same string of bits \textit{unless there was an eavesdropper.}
    \item They publicly compare a subset of the bits, say 20 out of 100 bits. If all 20 are the same, then it is unlikely that there was an eavesdropper. {{These statements will be quantified in the questions in Section \ref{sec:cryptoquestion}.}} The remaining 80 becomes the shared key.
\end{enumerate}

\section{\intermediate{5pt} Detecting an Eavesdropper}
If an eavesdropper (Eve) overhears the post-processing part where Alice and Bob share the basis used for each bit measurement, Eve has no information about whether any bit was either a 0 or 1. The only way for Eve to determine the spin value is to measure it with her own Stern-Gerlach \textit{before} it gets to Bob. However, since the basis is not shared during the transmission, Eve must randomly pick a basis to use. If Alice and Bob randomly choose to not measure in the same basis, they throw
away all the bits, then in this case, it doesn't matter what basis Eve chooses. If Alice and Bob randomly choose to measure in the same basis, however, then there are two outcomes depending on what Eve does: 1) If Eve randomly chooses the same basis as Alice, then she does not alter the state. This is bad, as Eve has successfully eavesdropped information without Alice and Bob knowing. 2) If Eve randomly chooses a different basis than Alice, then she alters the state and puts it into a superposition. Even though Bob is using the same basis as Alice, due to Eve altering the state, Alice and Bob can have a different spin measurement. This is how they can catch an eavesdropper.

\subsection*{Example}
\begin{figure}[h]
    \centering
    \includegraphics[scale=0.6]{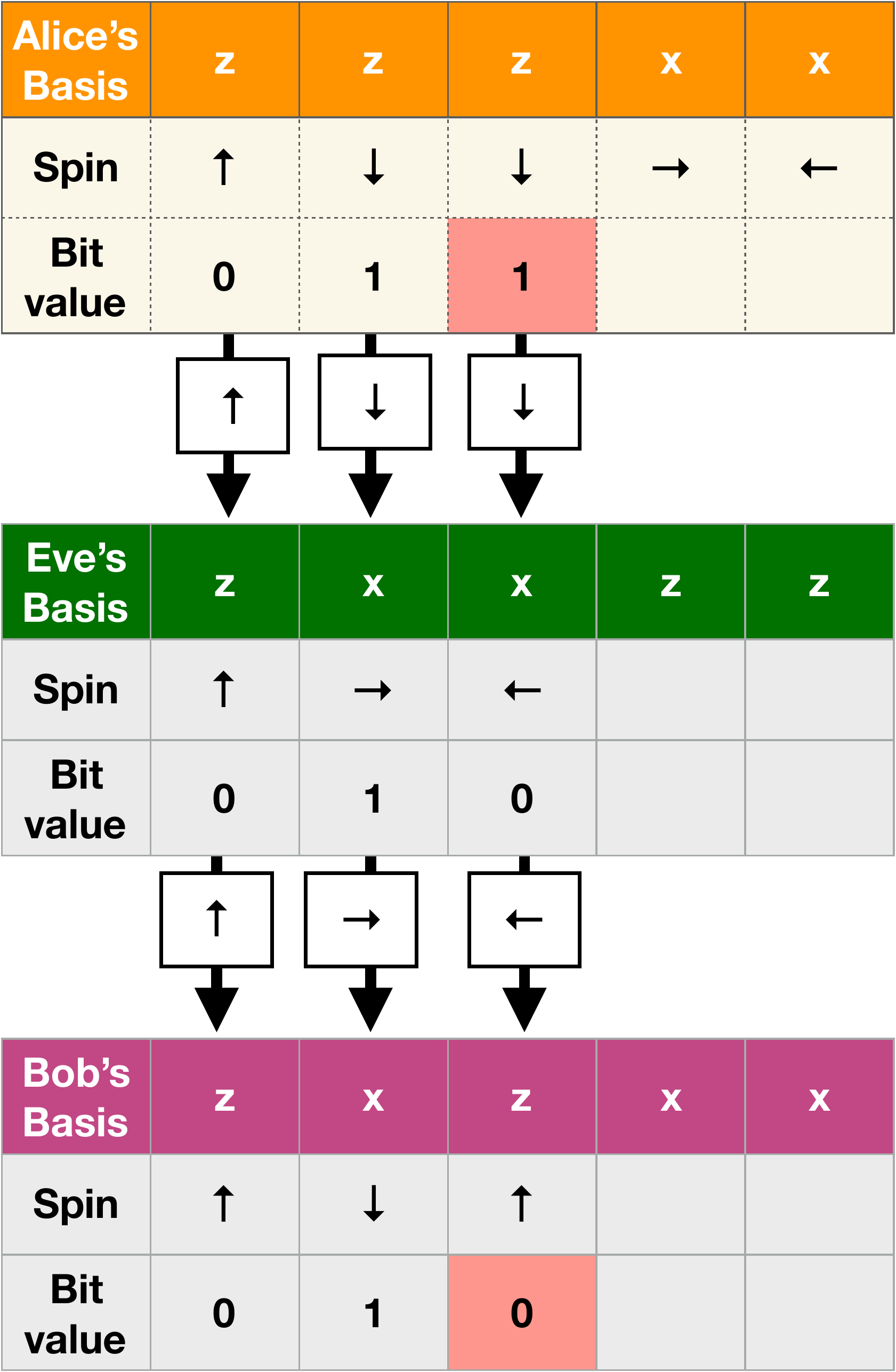}
    \caption{An example of how to catch an eavesdropper using the BB84 protocol. }
    \label{fig:BB84-fig3}
\end{figure}

The eavesdropping situation is shown in Figure~\ref{fig:BB84-fig3}. If Eve chooses the same basis as Alice, the spin is unchanged when it gets to Bob (bit \#1). If Eve chooses a different basis than Alice, the spin will be different when it gets to Bob (bits \#2 and \#3). Eve could get lucky and Bob's bit could agree with Alice (bit \#2). However, Bob is equally likely to measure something different from Alice (bit \#3). Can you fill in what might happen with bits \#4 and \#5?

When Alice and Bob compare a portion of their key bits, a discrepancy would indicate the presence of an eavesdropper. If they compare a sufficient number of key bits and all of them match, they can be reasonably sure that the rest of it is secure. {{This statement will be quantified shortly in the questions.}}

\section{Activities}
\begin{itemize}
\item[\fundamental{5pt}] One-time Pad for Alice/Bob in Worksheet \ref{sec:WorksheetOneTimePad}
\item[\fundamental{5pt}] BB84 Quantum Key Distribution for Alice/Bob/Eve in Worksheet \ref{sec:WorksheetBB84}
\end{itemize} 

\section{Check Your Understanding}
\label{sec:cryptoquestion}
\begin{enumerate}
    \item \fundamental{3pt} If Alice and Bob exchange 1 million bits in order to use the BB84 quantum cryptography protocol, approximately how long will their bit-key string be?  Assume they do not check for eavesdropping.
    \item \fundamental{3pt} Alice and Bob share their lists of measurement basis, but do not share any more information about the bits. What is the probability that Eve will guess the correct bit for a single bit-key?
    \item \intermediate{5pt} Alice and Bob perform 20 bit-key measurement but do not share any information about the bits. What is the probability that Eve will guess the correct 20-bit key?
    \item \fundamental{3pt} If Eve tries all possible key combinations with the one-time pad, can she crack the one-time pad?
    \item \intermediate{5pt} If Eve uses a Stern-Gerlach to measure the spin in between Alice and Bob's measurements, what percentage of the time will she be lucky and get the correct key-bit value without detection?
    \item \intermediate{5pt} If Alice and Bob measure in the same basis and compare 20 bits of their key, what is the probability that Eve could have eavesdropped all 20 bits without being detected?
    \item \fundamental{3pt} Suppose that Eve discovers that the no-cloning theorem is wrong and finds a way to clone the state of each photon. How could she use a cloning machine to learn about the entire key without leaving any trace?
\end{enumerate}

\graphicspath{{Chapter6-Gates/}}

\chapter{Quantum Gates}

\section{\fundamental{5pt} Single Qubit Gates}

As discussed in Chapter \ref{chapter:qubit}, information in classical computers is represented by bits. However, if the bits did not change, then the computer would remain the same forever and would not be very useful! Therefore, it is necessary to change the values of bits depending on what you want the computer to do. For example, if you want a computer to multiply the number $2$ and the number $3$ together to produce the number $6$, then you need to put each of the numbers $2$ and $3$ into an 8-bit binary representation, and then have a computational operation to multiply the two 8-bit values together to produce $6$. The operation of changing bits in a classical computer to do what you want is performed by what are called classical logic gates. 

Classical computers manipulate bits using classical logic gates, such as OR, AND, NOT, NAND, etc. This \href{https://whatis.techtarget.com/definition/logic-gate-AND-OR-XOR-NOT-NAND-NOR-and-XNOR}{link}\footnote{https://whatis.techtarget.com/definition/logic-gate-AND-OR-XOR-NOT-NAND-NOR-and-XNOR} provides a  basic review of classical logic gates. Similarly, quantum computers manipulate qubits using quantum gates. The gates are applied to qubits and the state of the qubits changes depending on which
gate is applied. A quantum algorithm has to be implemented on a quantum computer using the quantum gates. After running a quantum algorithm, the result is retrieved by measuring the qubit's state. The hardware implementation of quantum gates depends on how the qubit and quantum computer has been implemented technologically.\footnote{E.g., topological qubits and superconducting qubits have very different hardware implementations due to their very different nature.} As an example, one could have a qubit based on spin. Then gates could be implemented using an external magnetic field to change the spin (and hence the qubit state). This chapter will focus on gates from the computing perspective rather than the engineering perspective. You will learn about several important gates that act on a single qubit, interpret histograms produced by a quantum computer simulator, and use matrices to describe the operation of these gates.

\section{\intermediate{5pt} $X$ (also called NOT) Gate}

In classical computers, the NOT gate takes one input and reverses its value. For example, it changes the $0$ bit to a $1$ bit, or changes a $1$ bit to a $0$ bit. This is like a light-switch flipping a light from ON to OFF, or from OFF to ON.  A quantum $X$ gate is similar in that a qubit in a definite state $|0\rangle$ will become $|1\rangle$  and vice versa. When the qubit is in a superposition of all basis states, then the superposition also flips, e.g., see Equation (\ref{eqn:XGate}). 

\begin{equation}
  \includegraphics[height = 40pt, valign = c]{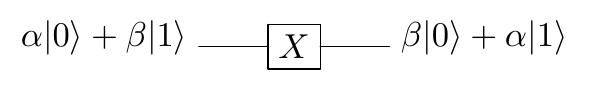}
  \label{eqn:XGate}
\end{equation}

\begin{figure}[h!]
\centering
  \includegraphics[width=0.5\textwidth]{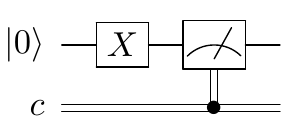}
  \caption{Applying the $X$ gate on the IBM Q simulator and measuring the output.}
  \label{fig:IBMQ-XGate}
\end{figure}

To see how it works, you can try out the \href{https://quantumexperience.ng.bluemix.net/qx/experience}{IBM Q simulator}.\footnote{https://quantumexperience.ng.bluemix.net/qx/experience \\ It can also be run on IBM's real quantum computer, but you get a limited number of trials per day.} Traditionally, all qubits on the IBM Q machine (or any other quantum simulator) start with the incoming qubits in the $|0\rangle$ state. To run this simple gate, drag the $X$ gate onto any qubit. To see the results, add the measurement operation at the end. This is shown in Figure \ref{fig:IBMQ-XGate}. 

\begin{figure}[h!]
\centering
  \includegraphics[width=0.75\textwidth]{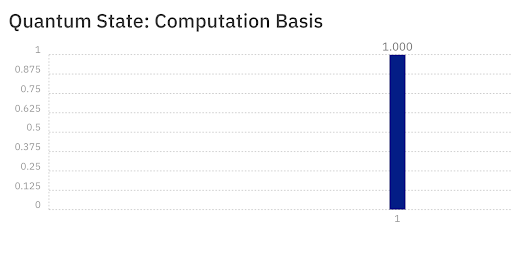}
  \caption{Histogram showing that the qubit is measured in the $|1\rangle$ state with a probability of $1$. Reprint Courtesy of International Business Machines Corporation, \copyright International Business Machines Corporation.}
  \label{fig:IBMQ-XMeasurement}
\end{figure}

After you click the ``Simulate'' button, you should see a histogram showing the measurements of the qubit's final state for $100$ independent trial runs. Since the qubit always starts as the $|0\rangle$ state, applying the $X$ gate produces the $|1\rangle$ state and so the measurement outcome is $|1\rangle$ $100\%$ of the time as  shown in Figure~\ref{fig:IBMQ-XMeasurement}. 

It is worth noting that any computer will have hardware errors. In a classical computer, this could be an electrical short  of the motherboard, degradation of the 
 technology storing memory on a hard drive  {{which}} corrupts the stored classical bits. A real quantum computer will also have hardware errors. The quantum state of a qubit can change accidentally because of these hardware errors. Such errors may arise from the lack of full control of the 
 interference between electromagnetic fields, variations in temperature, or energy dissipation. The accidental and incorrect change of a qubit state gives rise to the wrong answer which is called ``noise''.\footnote{Background noise is an event that causes unwanted or incorrect affects on a signal. } As  quantum computers only measure the state of a qubit, they cannot easily tell if the measurement is correct or incorrect. When we humans look at the measurements to interpret the results, noise can
 cause confusion on which answer is actually correct. Minimizing noise error is the greatest obstacle to building quantum computers.\footnote{Noise can also occur in classical computers. Here, it can be because a wire in the computer which holds the $0$- or $1$-bit breaks, and gives the wrong bit value. However, since classical computation has no probability associated with it, a single classical computation can be rerun twice and should give the exact same result. In practice, your computer
 reruns the same code many times to spot if there has been any errors and chooses the result which occurs most frequently. In this way you do not notice the hardware noise.} For example, noise will cause the histogram in Figure~\ref{fig:IBMQ-XMeasurement} to not have the perfect $100\%$ outcome. Instead, noise will cause the qubit to be in the $|0\rangle$ state incorrectly some of the time, and the measurement histogram will incorrectly be $x\%$ in the $|0\rangle$ state and $(100-x)\%$ in the
 $|1\rangle$ state. If the noise is large, then $x=50\%$ and measurement will be completely random. It should be understood that noise is an effect that occurs in both classical and quantum computers, but because quantum computing technology is in its infancy, the noise is not as well under control.\\

Mathematically, the quantum NOT gate is represented as a matrix $X$ which acts on qubit states using matrix multiplication. The matrix representation is
\begin{equation}
  X=
  \begin{pmatrix}
    0 & 1 \\
    1 & 0
  \end{pmatrix}.
  \label{eqn:XMatrix}
\end{equation}

\section{\intermediate{5pt} Hadamard Gate}

\begin{figure}[h!]
\centering
\includegraphics[width=0.5\textwidth]{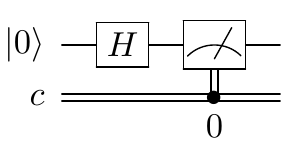}
  \caption{Applying a Hadamard gate and measuring on the IBM Q machine.}
  \label{fig:IBMQ-HGate}
\end{figure}

The Hadamard gate is very important in quantum computing. If the qubit starts in a definite $|0\rangle$ or $|1\rangle$ state, the Hadamard gate puts each into a superposition of $|0\rangle$ and $|1\rangle$ states. Applying a Hadamard gate to the $|0\rangle$ state qubit on the IBM Q simulator and measuring the output is shown in Figure~\ref{fig:IBMQ-HGate}. 

\begin{figure}[h!]
\centering
  \includegraphics[width=0.75\textwidth]{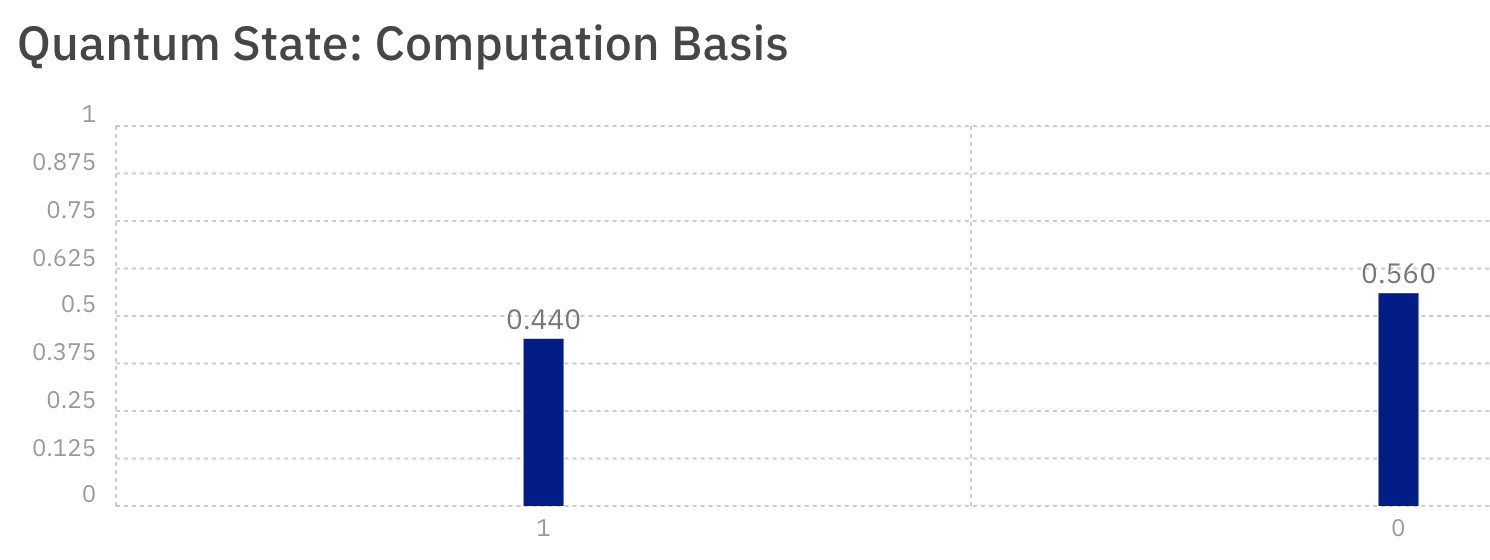}
  \caption{Measurement histogram after applying the Hadamard gate from Figure~\ref{fig:IBMQ-HGate} $100$ times. Reprint Courtesy of International Business Machines Corporation, \copyright International Business Machines Corporation.}
  \label{fig:IBMQ-HMeasurement}
\end{figure}

The result of running the circuit $100$ times is a histogram shown in Figure~\ref{fig:IBMQ-HMeasurement}. Note that each run is independent: before each measurement, the qubit has to be reset to the $|0\rangle$ state and passed through the gate, and then the measurement happens. We repeat this process $100$ times. Each bin in the histogram shows the frequency/probability of measuring $|0\rangle$ or $|1\rangle$. You can clearly see that applying {{the}} Hadamard gate to a single qubit creates a
superposition state of both $|0\rangle$ and $|1\rangle$. The probabilities are not exactly $50/50$ because of statistical error. The more data you collect, the closer the result converges to $50/50$. This is similar to
flipping a coin and counting the number of heads or tails; the greater the number of  flips, the more  likely you are to observe $50/50$ probability of seeing heads/tails. 

Recall that measurement collapses the superposition. Only one classical state can be observed, and all of the other quantum information is lost. Measurement collapse is the reason why a qubit's state cannot be duplicated which is  known as the no-cloning theorem of quantum computing. Once a superposition state is measured, it fundamentally changes into one of the basis states, and hence cannot be duplicated. Still, it is not {known} how or whether measurement collapse
happens.\footnote{https://en.wikipedia.org/wiki/Measurement\_problem}\\

\noindent {\bf{Question 1}}: Create a qubit in the $|1\rangle$ state and pass it through a Hadamard gate. From the measurement histogram, can you tell whether the qubit started as a $|0\rangle$  or $|1\rangle$ initial state?\\

The measurement histogram should look identical whether $|0\rangle$ or $|1\rangle$ were the initial state. Then how can we tell what the initial state was after a Hadamard operation? In the beam splitter, we determined where the photon came from by adding a second beam splitter to create interference. The way to measure and distinguish between them is to add a second Hadamard gate.  \\

\begin{figure}[h!]
\centering
  \includegraphics[width=0.5\textwidth]{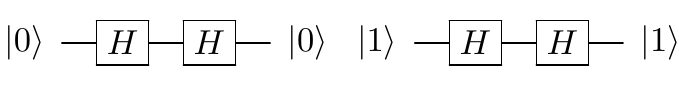}
  \caption{Applying two Hadamard gates to the $|0\rangle$ state or $|1\rangle$ state.}
  \label{fig:2HGate}
\end{figure}

\noindent {\bf{Question 2}}: Build a circuit that applies two Hadamard gates to a qubit in the $|0\rangle$ initial state as shown in Figure \ref{fig:2HGate}. What is the output? Repeat this experiment for the $|1\rangle$ initial state.

\subsection*{Mathematics of the Hadamard Gate} 
The Hadamard gate has the following matrix representation:
\begin{equation}
  H=\frac{1}{\sqrt{2}}
  \begin{pmatrix}
    1 & 1 \\
    1 & -1
  \end{pmatrix}.
  \label{eqn:XMatrix}
\end{equation}

Using matrix multiplication we can show that application of the Hadamard gate to an $|0\rangle$ initial state puts the qubit into the $(1/\sqrt{2})(|0\rangle + |1\rangle)$ state, also called the $|+\rangle$ state which  is shown in Equation (\ref{eqn:0HGate}).
\begin{equation}
    \includegraphics[height = 40pt, valign = c]{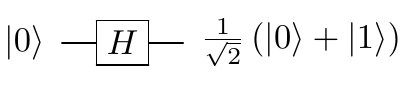}\hspace{-1.2mm}.
  \label{eqn:0HGate}
\end{equation}
 If the initial state is $|1\rangle$,  the Hadamard gate will create the superposition $(1/\sqrt{2})(|0\rangle - |1\rangle)$ state, called the $|-\rangle$ state as shown in Equation (\ref{eqn:1HGate}).
\begin{equation}
    \includegraphics[height = 40pt, valign = c]{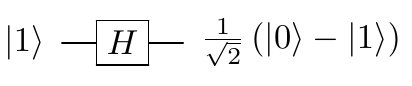}\hspace{-1.2mm}.
  \label{eqn:1HGate}
\end{equation}

In the Stern-Gerlach experiment, you learned that the $|0\rangle$ and $|1\rangle$ states make up the $z$-basis and are associated with spin up and spin down. The $|+\rangle$ and $|-\rangle$ states make up the $x$-basis and are associated with spin right and spin left.  While the Stern-Gerlach could be rotated to measure at any angle, a quantum computer is physically built to only measure in the $z$-basis. Therefore, the spin right $1/\sqrt{2}(|0\rangle + |1\rangle)$ and spin left $1/\sqrt{2}(|0\rangle - |1\rangle)$ look the same when measured by a quantum computer. However, the two states have hidden information that can be recovered by using a second Hadamard gate to change back into the $z$-basis. 

\subsection*{Examples}
\begin{enumerate}
 \item A spin right $1/\sqrt{2}(|0\rangle + |1\rangle)$ is sent through a Hadamard gate, creating a superposition of $|+\rangle$ and $|-\rangle$ given by $1/\sqrt{2}(|+\rangle + |-\rangle)$. By making a basis change substitution, show that this is equivalent to producing a $|0\rangle$ state.

\begin{align}
  \frac{1}{\sqrt{2}}\Big(|+\rangle + |-\rangle\Big) & = \frac{1}{\sqrt{2}}\Big(\frac{1}{\sqrt{2}}|0\rangle + \frac{1}{\sqrt{2}}|1\rangle\Big) + \frac{1}{\sqrt{2}}\Big(\frac{1}{\sqrt{2}}|0\rangle - \frac{1}{\sqrt{2}}|1\rangle\Big), \\
  & = \frac{1}{{2}}|0\rangle + \frac{1}{{2}}|1\rangle + \frac{1}{{2}}|0\rangle - \frac{1}{{2}}|1\rangle, \\
  & = |0\rangle.
\end{align}

\item Use matrix multiplication to show how applying the Hadamard gate twice to a $|0\rangle$ state qubit recovers its original state.

\begin{align}
  H|0\rangle &=\frac{1}{\sqrt{2}}
  \begin{pmatrix}
    1 & 1 \\
    1 & -1
  \end{pmatrix}
  \begin{pmatrix}
    1 \\
    0 
  \end{pmatrix}
  = \frac{1}{\sqrt{2}} 
  \begin{pmatrix}
    1 \\
    1
  \end{pmatrix}, \\
  HH|0\rangle &=\frac{1}{{2}}
  \begin{pmatrix}
    1 & 1 \\
    1 & -1
  \end{pmatrix}
  \begin{pmatrix}
    1 \\
    1 
  \end{pmatrix}
  = 
  \begin{pmatrix}
    1 \\
    0
  \end{pmatrix}. 
  \label{eqn:HH0Matrix} 
\end{align}
In fact, all quantum gates are reversible as a consequence of the unitary matrix condition. Recall that the gates must be unitary so that the probabilities always add up to $1$. Multiplying any unitary matrix by its conjugate transpose will return the identity matrix, i.e., reverses the gate to get the original state by $UU^{\dagger} = U^{\dagger}U = 1$.

\end{enumerate}
\section{\intermediate{5pt} $Z$ Gate}

The $Z$-gate matrix representation is 
\begin{equation}
  Z=
  \begin{pmatrix}
    1 &  0\\
    0 & -1
  \end{pmatrix}.
  \label{eqn:ZMatrix}
\end{equation}

\begin{figure}[h!]
\centering
  \includegraphics[width=0.5\textwidth]{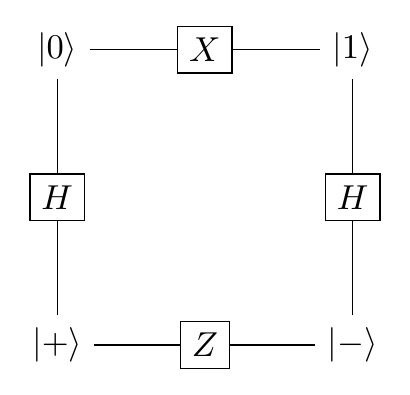}
  \caption{The $X$, $H$, and $Z$ gates change the qubit's state in the $z$- and $x$-basis and are related according to this diagram.}
  \label{fig:XHZGates}
\end{figure}

The $Z$ gate leaves a $|0\rangle$ state unchanged but flips the sign of the $|1\rangle$ state to $-|1\rangle$ by
\begin{equation}
    \includegraphics[height = 40pt, valign = c]{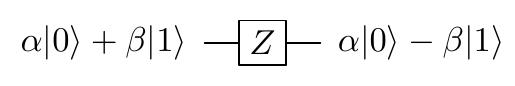}\hspace{-1.2mm}.
  \label{eqn:ZGate}
\end{equation}
This is equivalent to changing the qubit from a $|+\rangle$ state to a $|-\rangle$ state.
The effects of the $X$, $H$, and $Z$ gates are summarized in Figure \ref{fig:XHZGates}.

\section{Check Your Understanding}

\begin{enumerate}
\item \advanced{0.5pt}  Use matrix multiplication to show how applying an $X$ gate flips:

  \begin{enumerate}[label=(\alph*)]
  \item A qubit in the $|0\rangle$ state. 
  \item A qubit in the general $|\psi\rangle = \alpha|0\rangle + \beta | 1\rangle$ state. 
  \end{enumerate}
  
\item \fundamental{5pt} Explain the relationship between a beam splitter and a Hadamard gate.

\item \fundamental{5pt}  A $|0\rangle$ qubit is passed through a Hadamard gate. We measure the qubit state as $|1\rangle$. What is the result if we perform a measurement on the qubit a second time without reinitializing?

  \begin{enumerate}[label=(\alph*)]
  \item $|0\rangle$
  \item $|1\rangle$
  \item $50\%$ chance of $|0\rangle$ or $|1\rangle$
  \end{enumerate}

\item  \fundamental{5pt} Assume a qubit represents a light bulb that can be measured as either ON or OFF.  

  \begin{enumerate}[label=(\alph*)]
  \item The light bulb is originally ON. What gate would you use to turn it OFF?
  \item The light bulb is originally ON and passes through a Hadamard gate.  What do you measure as the output?
  \item The light bulb is originally ON and passed through two Hadamard gates in series. What do you measure as the output?
  \end{enumerate}

\item  \intermediate{5pt} Explain how the Hadamard gate is implemented in the Stern-Gerlach experiment.

\item \intermediate{5pt} Explain the output of the Mach-Zehnder interferometer using what you learned about Hadamard gates.
    
\item \advanced{0.5pt} Use matrix multiplication to demonstrate 
  \begin{enumerate}[label=(\alph*)]
  \item The Hadamard gate applied to a $|1\rangle$ state qubit turns it into a $|-\rangle$.
  \item A second Hadamard gate turns it back into the $|1\rangle$ state.
  \item The output after applying the Hadamard gate twice to a general state $|\psi\rangle = \alpha|0\rangle + \beta | 1\rangle$. 
  \end{enumerate}

\item \fundamental{5pt} Which of the quantum circuits in the Figure~\ref{fig:Q8Gates} would produce the histogram shown in Figure \ref{fig:IBMQ-HMeasurement}?
 
  \begin{figure}[h]
    \centering
    \includegraphics[width=0.75\textwidth]{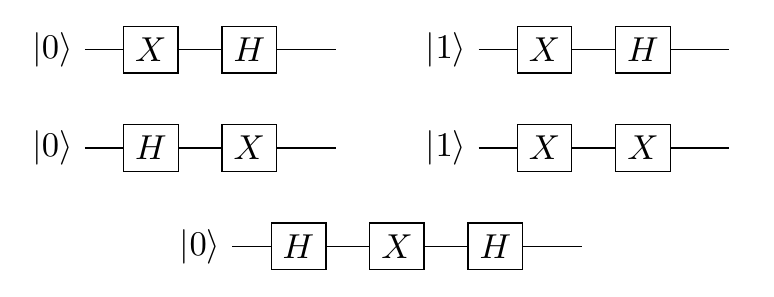}
    \caption{Six quantum circuits.}
    \label{fig:Q8Gates}
  \end{figure}

\item \advanced{0.5pt} Use matrix multiplication to show how applying the $Z$ gate to $|+\rangle$ changes it to $|-\rangle$.

\item \intermediate{5pt} Using only the Hadamard and $Z$ gates, design a quantum circuit that outputs the same result as an $X$ gate.

\item \intermediate{5pt} Using the IBM Q simulator, apply the $Z$ gate to a qubit in the following initial states and interpret the measurement histogram.
  \begin{enumerate}[label=(\alph*)]
  \item $|0\rangle$
  \item $|1\rangle$ (Hint: You need to first flip the $|0\rangle$ state using the $X$ gate.)
  \item $|+\rangle$ (Hint: You need to first create the $|+\rangle$ state using the $H$ gate.)
  \item $|-\rangle$ (Hint: You need to first create the $|-\rangle$ state using the $X$ and $H$ gates.) 
  \end{enumerate}

\item \intermediate{5pt} What is the expected measurement histogram produced by the circuit in Figure \ref{fig:Q12Gates}?
  \begin{figure}[h]
    \centering
    \includegraphics[width=0.7\textwidth]{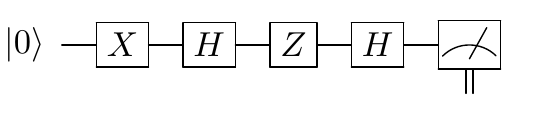}
    \caption{Circuit diagram.}
    \label{fig:Q12Gates}
  \end{figure}

  \item \advanced{0.5pt} Show that the Hadamard gate is unitary and therefore reversible. 

\end{enumerate}

\graphicspath{{Chapter7-Entanglement/}}
\chapter{Entanglement}

So far, we have only discussed the manipulation and measurement of a single qubit. However, \textbf{quantum entanglement} is a physical phenomenon that occurs when multiple qubits are correlated with each other. Entanglement can have strange and useful consequences that could make quantum computers faster than classical computers. Qubits can be ``entangled,'' providing hidden quantum information that does not exist in the classical world. It is this entanglement that is one of the main advantages of the quantum world! 

\begin{figure}
  \centering
  \includegraphics[scale=0.5]{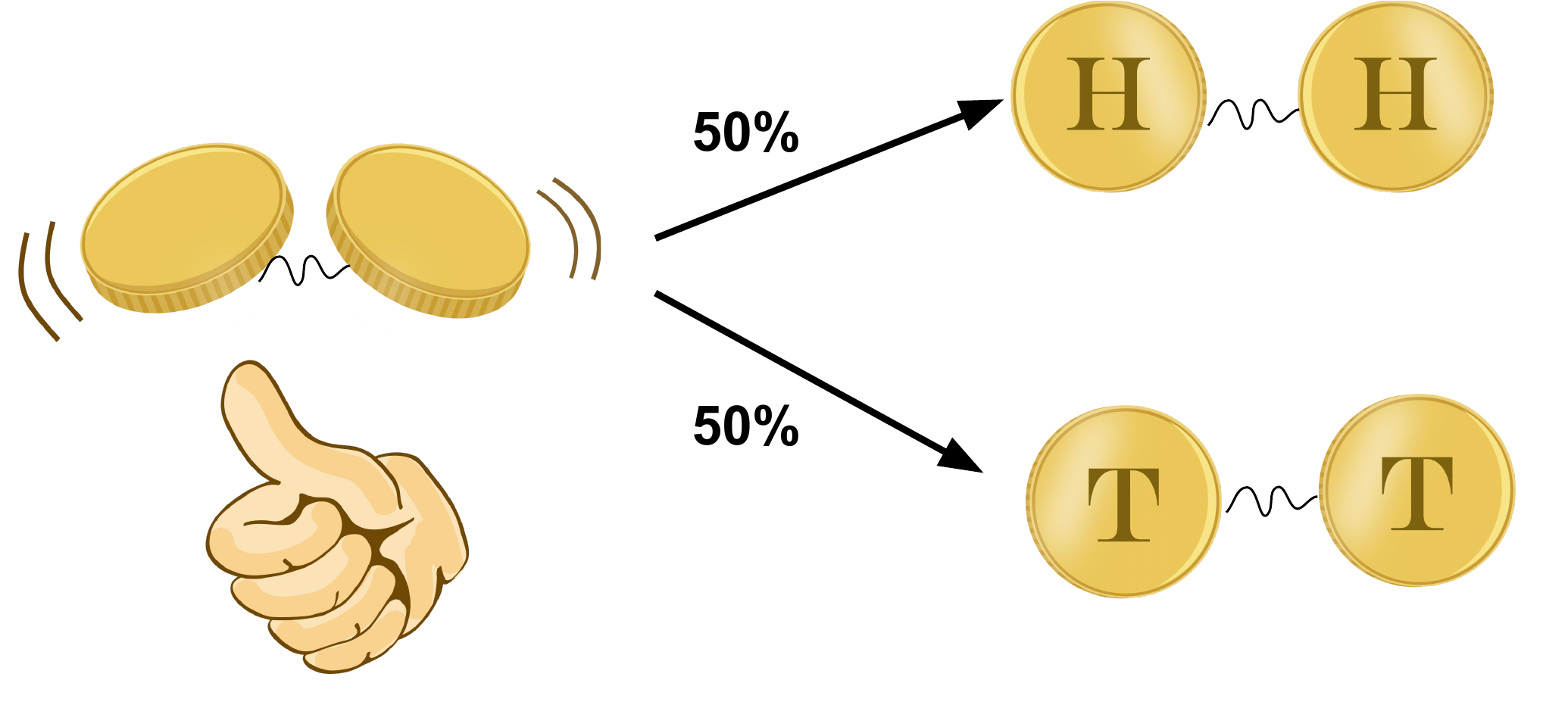}
  \caption{Two coins that are entangled in such a way that they either both land on HH or both land on TT.}
  \label{fig:coinflip}
\end{figure}

To provide one example of the strange behavior of entanglement, suppose we have two fair coins. Classically, if you flipped two fair coins, you would measure the outcomes HH, HT, TH, or TT, each occuring with a 25\% probability. However, by quantum entangling these two fair coins, it is possible to create a state $(1/\sqrt{2})(\lvert HH\rangle + \lvert TT\rangle)$ as  illustrated in Figure \ref{fig:coinflip}. Many other types of entangled coins are possible, but this is one famous example. If you flipped this ``entangled'' pair of coins, they are entangled in such a way that only two measurement outcomes are possible: 1) both coins land on heads; or 2) both coins land on tails; each outcome occurring with 50\% probability. Isn't that weird! 

\begin{figure}[h!]
  \centering
  \includegraphics[scale=0.5]{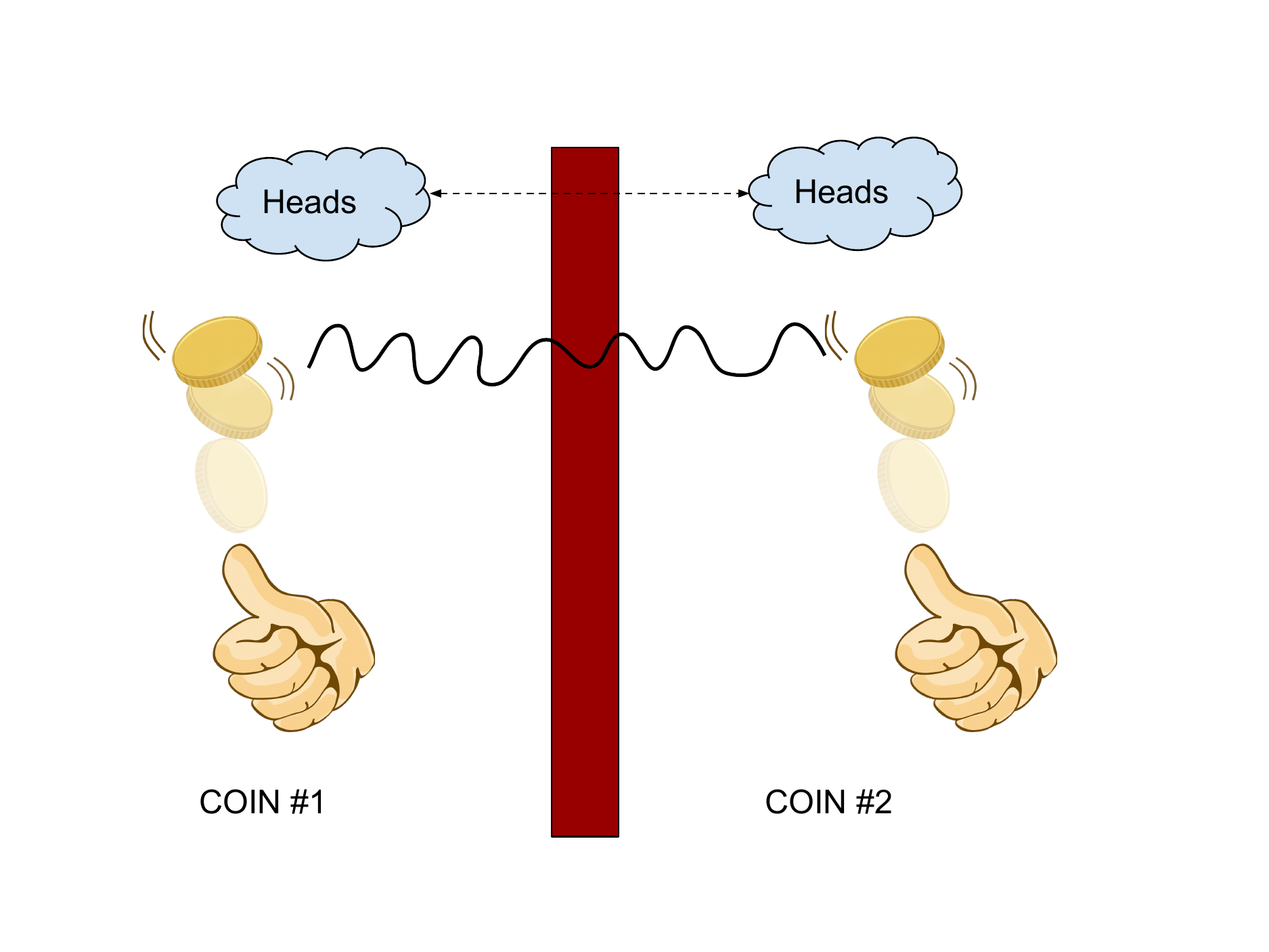}
  \caption{Two coins are separated with no means of  communication between each other. Classically, the flip of the second coin would be unrelated to the first flip. However, entangled coins would still produce correlated results.}
  \label{fig:actiondistance}
\end{figure}

Furthermore, if the two entangled coins are separated by thousands of miles, one coin can be flipped and measured. In this case, if the measured coin produced the outcome heads, then we automatically know that the other coin must also land on heads. If the  measured coin produced the outcome tails, then we automatically know that the other coin must also land of tails! If this isn't strange enough, the two coins could be separated by a distance greater than what light (which travels at the fastest speed in the universe) could travel as shown in Figure \ref{fig:actiondistance}. If the two coins are flipped at the exact same time, somehow the two coins know to land on the same side as the other even though there can be no classical communication between them.\footnote{``Bounding the speed of spooky action at a distance.'' {\it{Physical Review Letters}}. 110: 260407. 2013. \href{https://arxiv.org/abs/1303.0614}{arXiv:1303.0614}.}

How does the other coin instantaneously ``know'' what was measured on the other? Is information somehow being transmitted faster than the speed of light? Einstein called this behavior a ``spooky action at a distance.'' It has since been shown that no information is being transmitted from one place to the other, and so no information is being transmitted faster than the speed of light. Rather, the particles share non-classical information at the time of entanglement, which is then observed in the measurement process. The correlation between entangled qubits is the key that allows quantum computers to perform certain computations much faster than classical computers.

\section{\fundamental{5pt} Hidden Variable Theory}
\begin{figure}[h!]
    \centering
    \includegraphics[scale=0.5]{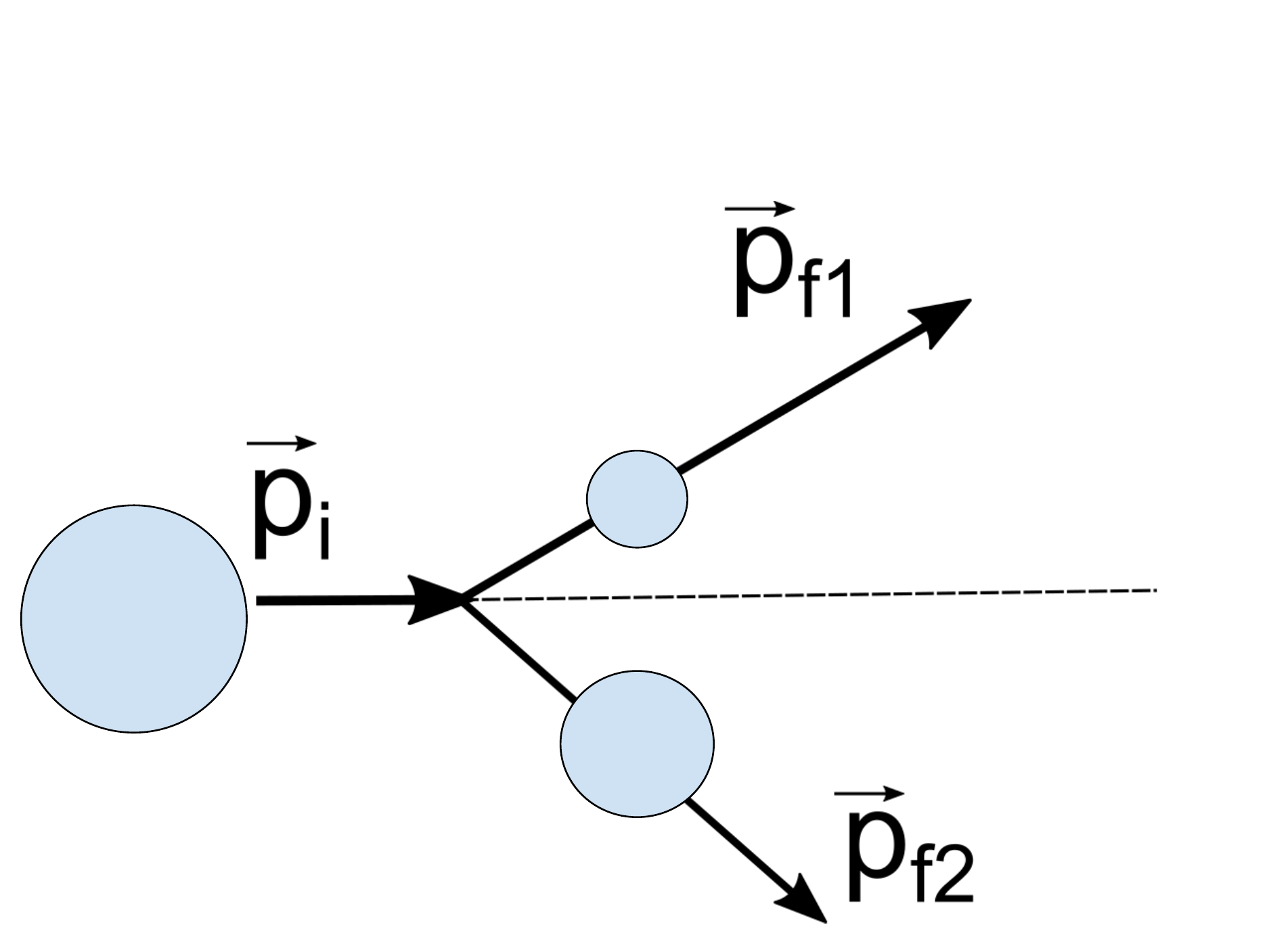}
    \caption{When a particle decays into two smaller particles, the decay products are ``classically entangled'' according to the conservation of momentum.}
    \label{fig:decay}
\end{figure}

It is tempting to think that there may be some classical explanation for entanglement. For example, maybe when causing the coins to interact and entangling them, the same interaction might have changed the coins? Did the entanglement change the fair coins by adding extra mass to the heads side or the tails side, thereby making them unfair? To give a more realistic classical example, if one particle decays into two smaller particles, the momenta of the two particles are related according to the conservation of momentum by $\vec{p}_i = \vec{p}_{f1} + \vec{p}_{f2}$.  Given a known total initial momentum, then by measuring the momentum of one of the smaller particles, we can determine the momentum of the other. By measuring one particle's momentum, we know the other. Momentum is the hidden classical variable that is encoded when the two particles are created. This is shown in Figure \ref{fig:decay}.
{{The natural question then arises if there is a conceptually similar hidden variable in the quantum mechanical situation?}}

However, \href{https://brilliant.org/wiki/bells-theorem/}{Bell's theorem}\footnote{\href{https://brilliant.org/wiki/bells-theorem/}{https://brilliant.org/wiki/bells-theorem/}} showed that the correlation between entangled quantum particles is more than what is possible classically, disproving the idea of a hidden variable. All other potential loopholes have been resolved as of 2016.\footnote{The BIG Bell Test Collaboration (9 May 2018). ``Challenging local realism with human choices.'' {\it{Nature}}. 557: 212--216. \href{https://doi.org/10.1038/s41586-018-0085-3}{doi:10.1038/s41586-018-0085-3}.} As such, entanglement is a purely quantum phenomenon with no classical explanation.

\section{\intermediate{5pt} Multi-Qubit States}
\begin{figure}[h!]
    \centering
    \includegraphics[width=0.5\textwidth]{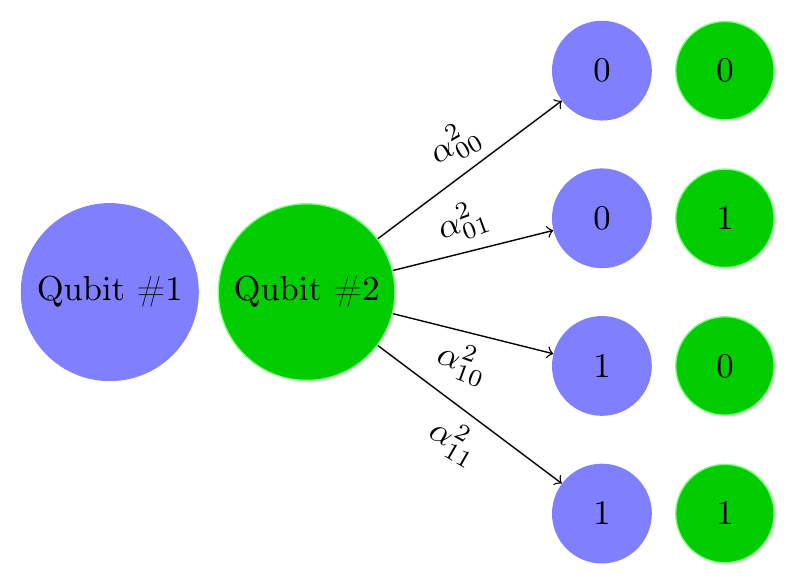}
    \caption{A two-qubit system can collapse into one of four states with probability $\alpha_{ij}^2$.}
    \label{fig:twoqubit}
\end{figure}

Given multiple qubits,  the total state of a system can be written together in a single ket. For example, if coin \#1 is heads and coin \#2 is tails, the two-coin state is expressed as $\lvert HT\rangle$. In general, a system of two qubits which is in a superposition of four classical states may be written as 
\begin{equation*}
    \lvert\psi\rangle = \alpha_{00}\lvert00\rangle + \alpha_{01}\lvert01\rangle + \alpha_{10}\lvert10\rangle + \alpha_{11}\lvert11\rangle. 
\end{equation*}
As we saw for the single qubit states, the coefficients $\alpha_{ij}$ are called the amplitudes and are generally complex numbers.  Measuring the two qubits will collapse the system into one of the four basis states with probability given by $\alpha_{ij}^2$. This is shown in Figure \ref{fig:twoqubit}.

\subsection*{Example}
A system of two qubits is in a superposition state given by $\lvert\psi\rangle = \frac{1}{\sqrt{2}}\lvert00\rangle+\frac{1}{2}\lvert10\rangle-\frac{1}{2}\lvert11\rangle$.
\begin{enumerate}[a)]
    \item What is the probability of measuring both qubits as 1? \\
          $\text{Prob}\left(\lvert11\rangle\right)=\left(\frac{-1}{2}\right)^2=\frac{1}{4}$. 
    \item If we only measure the first qubit and get a value of 1, what is the new state of the system? \\
    
      Since $\lvert00\rangle$ is the only basis state of $\lvert\psi\rangle$ that doesn't have a 1 in the first qubit, we eliminate the state $\lvert00\rangle$ from the possibilities. This results in $\lvert\psi'\rangle=\frac{1}{2}\lvert10\rangle-\frac{1}{2}\lvert11\rangle$. \\
      
      Finally, we renormalize the state so that the probabilities add up to 1. Therefore, the new state is $\lvert\psi'\rangle = \frac{1}{\sqrt{2}}\lvert10\rangle-\frac{1}{\sqrt{2}}\lvert11\rangle$.
\end{enumerate}

\section{\fundamental{5pt} Non-Entangled Systems}
It is possible to have a system of particles that are not entangled with each other. In this case, changing one particle will not cause any change in the other particle. For example, in a classical system, flipping two coins and measuring one coin as heads does not tell you any information about whether or not the other coin will land on heads or tails. These events are said to be independent. If you wanted to calculate the probability of $\ket{HT}$, you would simply multiply the probability of getting H on coin \#1 by the probability of getting T on coin \#2. This is given by
\begin{equation*}
    \text{Prob}\left(\lvert HT\rangle\right) = \left(\frac{1}{2}\right)\left(\frac{1}{2}\right) = \frac{1}{4}. 
\end{equation*}
Non-entangled states are also called product states or separable states because they can be factored into a product of single-qubit states.\footnote{More recently, it has been shown that there can exist quantum correlations in separable states that are not due to entanglement. These are called quantum discord: https://en.wikipedia.org/wiki/Quantum\_discord.} The two single-qubit probabilities multiply to produce the two-qubit probabilities. \\
\subsection*{Example:} One qubit is in a $\alpha_0\lvert0\rangle+\alpha_1\lvert1\rangle$ state, while another is in a $\beta_0\lvert0\rangle+\beta_1\lvert1\rangle$ state.  What is the state of the non-interacting two-qubit system?
\begin{equation*}
    \left(\alpha_0\lvert0\rangle + \alpha_1\lvert1\rangle\right)\left(\beta_0\lvert0\rangle + \beta_1\lvert1\rangle\right) = \alpha_0\beta_0\lvert00\rangle+\alpha_0\beta_1\lvert01\rangle+\alpha_1\beta_0\lvert10\rangle+\alpha_1\beta_1\lvert11\rangle. 
\end{equation*}

\section{\fundamental{5pt} Entangled Systems}
In an entangled system, measuring the value of one qubit changes the probability distribution of the second qubit.\\

\noindent \textbf{Example} Is $\lvert\psi\rangle=\frac{1}{\sqrt{2}}\lvert00\rangle+\frac{1}{\sqrt{2}}\lvert11\rangle$ an entangled state? \\

Yes! To see this, examine qubit \#2. The probabilities for measuring qubit \#2 in the $\lvert0\rangle$ or $\lvert1\rangle$ states are originally $50/50$ respectively. However, if we measured qubit \#1, then the probability for measuring qubit \#2 becomes 100\%. The same argument holds if qubit \#2 is measured first. As such, measuring one of the qubits affects the probability of measuring the other qubit in a certain state, and so they are entangled. Mathematically, an entangled state is a special multi-qubit superposition state that cannot be factored into a product of the individual qubits. \\

\noindent \textbf{Example:} Show that $\lvert\psi\rangle=\frac{1}{\sqrt{2}}\lvert00\rangle+\frac{1}{\sqrt{2}}\lvert11\rangle$ cannot be written as a product of two single qubits. \\

Assume that the state can be written as the product of two states. 
\begin{align}
    \frac{1}{\sqrt{2}}\lvert00\rangle + \frac{1}{\sqrt{2}}\lvert11\rangle &\stackrel{?}{=} \left(\alpha_0\lvert0\rangle+\alpha_1\lvert1\rangle\right)\left(\beta_0\lvert0\rangle+\beta_1\lvert1\rangle\right), \\
 &\stackrel{?}{=}\alpha_0\beta_0\lvert00\rangle+\alpha_0\beta_1\lvert01\rangle+\alpha_1\beta_0\lvert10\rangle+\alpha_1\beta_1\lvert11\rangle. 
\end{align}
Comparing the amplitudes on the left vs. the right, the $\alpha_i$'s and $\beta_j$'s must satisfy:
\begin{equation}
    \alpha_0\beta_0 = \frac{1}{\sqrt{2}}, \quad \alpha_0\beta_1 = 0, \quad \alpha_1\beta_0=0, \quad \alpha_1\beta_1=\frac{1}{\sqrt{2}}. 
\end{equation}
However, this is not possible. For example, take $\alpha_0\beta_1 = 0$. This means that either $\alpha_0 = 0$ or $\beta_1 = 0$. If $\alpha_0 = 0$, then $\alpha_0\beta_0 = 0$, but $\alpha_0\beta_0 =\frac{1}{\sqrt{2}}$ in the above equation. A similar contradiction occurs with $\beta_1=0$. So the initial assumption must be incorrect and this entangled state cannot be written as the product of two separate states. 

\section{\intermediate{5pt} Entangling Particles}
\begin{figure}[h!]
    \centering
    \includegraphics{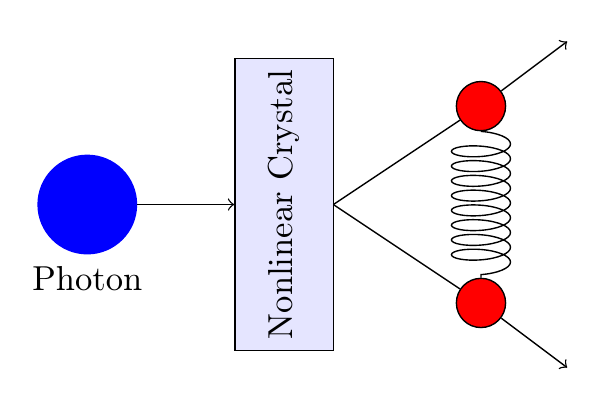}
    \caption{A nonlinear crystal creates two photons with entangled polarizations.}
    \label{fig:crystal}
\end{figure}

As there are many different ways of building a quantum computer, there are many different ways of entangling particles. One method called ``spontaneous parametric down-conversion'' shines a laser at a special nonlinear crystal. The crystal splits the incoming photon into two photons with correlated polarizations. For example, one could produce a pair of photons that always have perpendicular polarizations (see Figure ~\ref{fig:crystal}).

\section{\intermediate{5pt} CNOT Gate}
You have already learned about the $X$, Hadamard, and $Z$ gates. These act on a single qubit. There are also quantum gates that perform a logic operation on {multiple} qubits. The most important multi-qubit gate is the controlled NOT (CNOT) gate.  The CNOT is used to entangle two qubits together and is essential in quantum computing/algorithms. The CNOT takes in two qubits, a control qubit and a target qubit, and outputs two qubits. The control qubit stays the same, while the target obeys the following rule. 
\begin{itemize}
    \item If the control qubit is $\lvert0\rangle$, then leave the target qubit alone.
    \item If the control qubit is $\lvert1\rangle$, then on the target qubit flip $\lvert0\rangle\to\lvert1\rangle$ and $\lvert1\rangle\to\lvert0\rangle$. 
\end{itemize}
The truth table for the CNOT gate is shown in Table \ref{tab:CNOT}.\footnote{\href{https://en.wikipedia.org/wiki/Controlled\_NOT\_gate}{https://en.wikipedia.org/wiki/Controlled\_NOT\_gate}.}
\begin{table}[h!] 
  \centering
  \begin{tabular}{|c|c||c|c|}
    \hline
    \multicolumn{2}{|c||}{Before} &  \multicolumn{2}{|c|}{After} \\
    \hline 
    Control bit & Target bit & Control bit & Target bit  \\
    \hline \hline
    $|0\rangle$ & $|0\rangle$ & $|0\rangle$ & $|0\rangle$ \\
    $|0\rangle$ & $|1\rangle$ & $|0\rangle$ & $|1\rangle$ \\
    $|1\rangle$ & $|0\rangle$ & $|1\rangle$ & $|1\rangle$ \\
    $|1\rangle$ & $|1\rangle$ & $|1\rangle$ & $|0\rangle$ \\
    \hline
  \end{tabular}
  \caption{The truth table for the CNOT gate.}
  \label{tab:CNOT}
\end{table}

From this one can deduce the matrix form of the CNOT gate as 
\begin{align}
  \text{CNOT} &= 
  \begin{pmatrix}
    1 & 0 & 0 & 0 \\
    0 & 1 & 0 & 0 \\
    0 & 0 & 0 & 1 \\
    0 & 0 & 1 & 0 \\
  \end{pmatrix}. 
\end{align}

\begin{figure}[h!]
    \centering
    \includegraphics[width=0.5\textwidth]{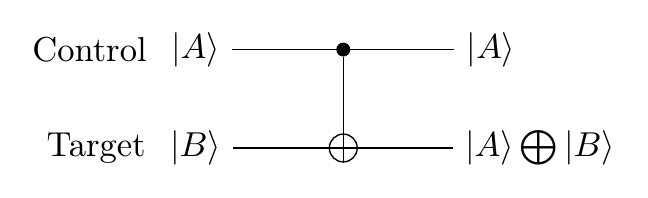}
    \caption{The CNOT gate performs an $X$ gate on the target qubit if the control qubit is $\lvert1\rangle$. }
    \label{fig:CNOT}
\end{figure}

Figure \ref{fig:CNOT} is the circuit for the CNOT gate. 

\subsection*{Examples}
\begin{enumerate}
\item Figure \ref{fig:CNOTIBM} shows the quantum circuit sending $\ket{10}$ through a CNOT gate. What is the output? \\

\begin{figure*}[h]
    \centering
    \includegraphics[width=0.4\textwidth]{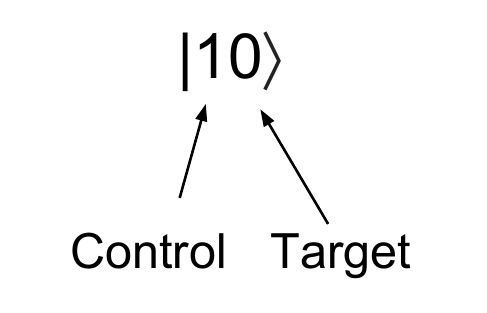}
    \includegraphics[width=0.4\textwidth]{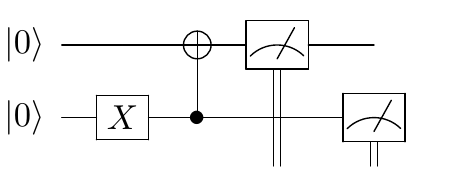}
    \caption{The quantum circuit that sends a control qubit in the $\lvert10\rangle$ state through a CNOT gate.}
    \label{fig:CNOTIBM}
\end{figure*}
The figure shows that the control qubit is q[1] and the target is q[0]. Since the control is in the $\lvert1\rangle$ state, the target qubit is flipped to $\lvert1\rangle$. So measurement will always result in $\lvert11\rangle$.

\item Examine Figure \ref{fig:my_label}. The control qubit is in a superposition of $\lvert0\rangle$ and $\lvert1\rangle$. What is the effect of a CNOT gate?  \\

\begin{figure*}[h]
    \centering
    \includegraphics[width=0.4\textwidth]{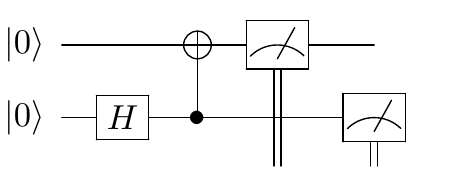}
    \caption{The IBM Q quantum circuit that sends a control qubit in a superposition state through a CNOT gate.}
    \label{fig:my_label}
\end{figure*}

Before the CNOT operation, in ket notation, the control qubit is in the $\frac{1}{\sqrt{2}}\lvert0\rangle+\frac{1}{\sqrt{2}}\lvert1\rangle$ state, while the target qubit is in the $\lvert0\rangle$ state. The two-qubit input state is therefore $\frac{1}{\sqrt{2}}\lvert00\rangle+\frac{1}{\sqrt{2}}\lvert10\rangle$. Applying the rules for the CNOT, the first state $\lvert00\rangle$ does not change as the control qubit is $\lvert0\rangle$. However, for the second state $\lvert01\rangle$, the control qubit is $\lvert1\rangle$ and so the target qubit is flipped from  $\lvert0\rangle$ to $\lvert1\rangle$. The result of the CNOT gate is the state $\frac{1}{\sqrt{2}}\lvert00\rangle+\frac{1}{\sqrt{2}}\lvert11\rangle$. The histogram from measuring this state is shown in Figure \ref{fig:CNOTMeasure}. This is a special state called the Bell state. 

The two qubits are entangled after the CNOT! As illustrated in the previous example, this state cannot be written as the product of two separate qubits. As with the single-qubit gates, the CNOT gate operates on ALL states in the superposition, {{e.g., the CNOT gate acts on the four basis states of a two qubit system simultaneously}}. Quantum algorithms leverages  this parallelism to ensure speed improvements over classical computers. In addition, as with all quantum gates, the CNOT is reversible, meaning the operation can be undone (which can be used to figure out the original qubit states).
\begin{figure*}[h!]
     \centering
    \includegraphics[width=0.7\textwidth]{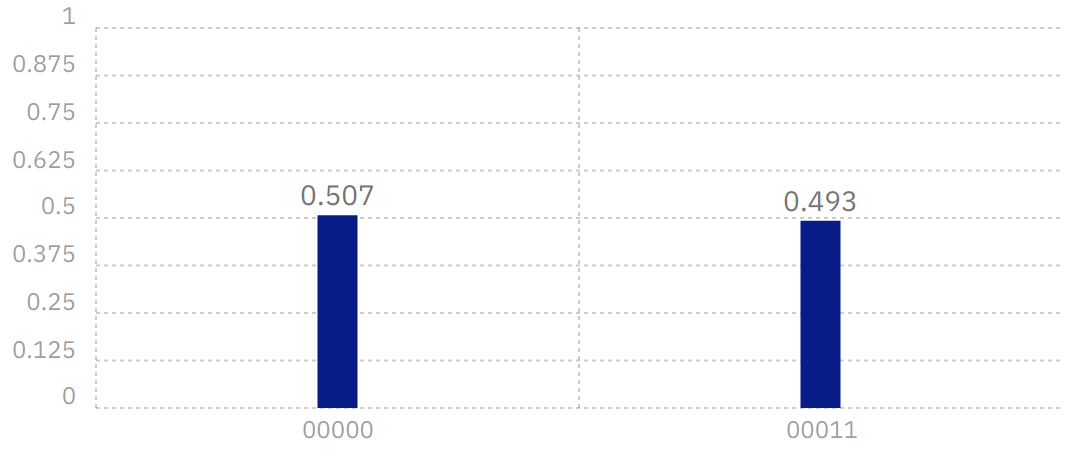}
    \caption{The measurement histogram produced by running the circuit in Figure \ref{fig:my_label}. Reprint Courtesy of International Business Machines Corporation, \copyright International Business Machines Corporation.}       
              \label{fig:CNOTMeasure}
\end{figure*}

\end{enumerate}

\section{Activities}
\begin{itemize}
\item[\intermediate{8pt}] Correlation in Entangled States Lab in Worksheet \ref{chapter:WorksheetCorr}
\item[\fundamental{5pt}] Schr\"{o}dinger's Worm Using Five Qubits in Worksheet \ref{sec:WorksheetWorm}
\end{itemize} 

\section{Check Your Understanding}
\begin{enumerate}
    \item \fundamental{3pt} For each of the questions below, assume that two-qubits start in the state
    \begin{equation}
        \lvert\psi\rangle = \frac{1}{\sqrt{2}}\lvert00\rangle+\frac{1}{2}\lvert10\rangle-\frac{1}{2}\lvert11\rangle.
    \end{equation}
    \begin{enumerate}[a)]
        \item What is the probability of measuring both qubits as 0?
        \item What is the probability of measuring the first qubit as 1?
        \item What is the probability of measuring the second qubit as 0?
        \item What is the new state of the system after measuring the first qubit as 0?
        \item What is the new state of the system after measuring the first qubit as 1?
    \end{enumerate}
    \item \fundamental{3pt} Two fair coins are flipped. What is the state of the two-coin system while the coins are in the air?
    \item \fundamental{3pt} Two six-sided dice are rolled.  What is the total probability of rolling an even number on one die and an old number on the other die?
    \item \intermediate{6pt} Is $\frac{1}{\sqrt{2}}\lvert00\rangle+\frac{1}{\sqrt{2}}\lvert01\rangle$ an entangled state? If so, show that it cannot be written as a product.  If not, what is the individual state of the two qubits?
    \item \intermediate{6pt} Are the following two-qubit states entangled?
    \begin{enumerate}[a)]
        \item $\frac{1}{\sqrt{2}}\ket{01}+\frac{1}{\sqrt{2}}\ket{10}$
        \item $\frac{1}{\sqrt{2}}\ket{01}-\frac{1}{\sqrt{2}}\ket{10}$
        \item $\frac{\sqrt{3}}{2}\ket{00}+\frac{1}{{2}}\ket{11}$
        \item $\frac{1}{\sqrt{2}}\ket{10}+\frac{1}{\sqrt{2}}\ket{11}$
        \item $\frac{1}{2}\ket{00}+\frac{1}{2}\ket{01}+\frac{1}{2}\ket{10}-\frac{1}{2}\ket{11}$
        \item $\frac{1}{\sqrt{2}}\ket{00}+\frac{1}{2}\ket{10}-\frac{1}{2}\ket{11}$
    \end{enumerate}
  \item \fundamental{3pt} Two qubits are passed through a CNOT. The first qubit is the control qubit. What is the output for the following initial states?
    \begin{enumerate}[a)]
        \item $\ket{00}$
        \item $\ket{01}$
        \item $\ket{11}$
        \item $\frac{1}{\sqrt{2}}\ket{01}+\frac{1}{\sqrt{2}}\ket{10}$
        \item $\frac{1}{\sqrt{2}}\ket{00}+\frac{1}{2}\ket{10}-\frac{1}{2}\ket{11}$
    \end{enumerate}
    \item \fundamental{3pt} The output of a CNOT gate is shown in the figure below. What were the inputs?
      \begin{figure*}[h!]
        \centering
        \includegraphics[width=0.5\textwidth]{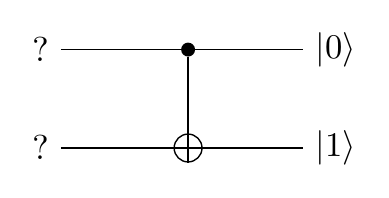}
        \caption{CNOT gate.}
        \label{fig:Q7Etanglement}
      \end{figure*}
    \item \fundamental{3pt} Can you predict the state produced by these quantum circuits?
    \begin{enumerate}[a)]
        \item \includegraphics{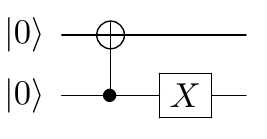}
        \item \includegraphics{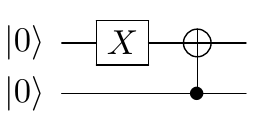}
        \item \includegraphics{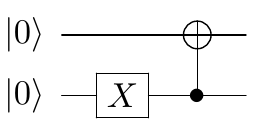}
        \item \includegraphics{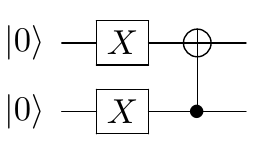}
    \end{enumerate}
    \item \intermediate{6pt} Can you predict which states will be produced by these quantum circuits?
    \begin{enumerate}[a)]
        \item \includegraphics{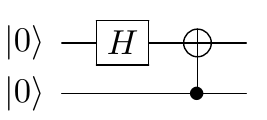}
        \item \includegraphics{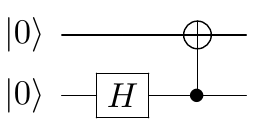}
        \item \includegraphics{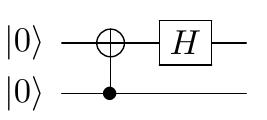}
        \item \includegraphics{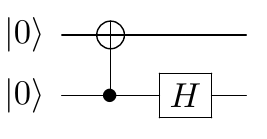}
    \end{enumerate}
    \item \advanced{0.5pt} Can you predict the state produced by these quantum circuits?
    \begin{enumerate}[a)]
        \item \includegraphics{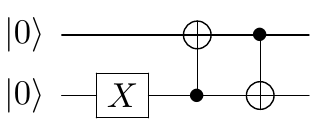}
        \item \includegraphics{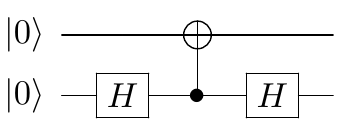}
    \end{enumerate}
    \item \intermediate{6pt} Use the \href{https://quantumexperience.ng.bluemix.net/}{IBM Q} simulator to create the entangled state $\frac{1}{\sqrt{2}}\ket{01}+\frac{1}{\sqrt{2}}\ket{10}$. 
    \item \intermediate{6pt} Suppose Alice has one half of an entangled pair and Bob has the other half.  When Alice makes a measurement on her qubit, Bob’s qubit instantaneously changes its state. Can Alice and Bob use entanglement to transmit information faster than the speed of light? Why or why not?
\end{enumerate}
\graphicspath{{Chapter8-Teleportation/}}
\chapter{Quantum Teleportation}

One interesting application of entanglement is {\bf{quantum teleportation}}, which is a technique for transferring an {\it{unknown}} quantum state from one place to another. In science fiction, teleportation generally involves a machine scanning a person and another machine reassembling the person on the other end.  The original body disintegrates and no longer exists. Similarly, quantum teleportation works by ``scanning'' the original qubit, sending a recipe, and reconstructing the qubit elsewhere. The original qubit is not physically destroyed in the science fiction sense, but it is no longer in the same state. (Otherwise it would violate the previously mentioned {\bf{no-cloning theorem}} $-$ which says that a qubit cannot be exactly copied onto another qubit.\footnote{The no-cloning theorem poses a big problem for correcting errors that happen on quantum computers: \href{https://en.wikipedia.org/wiki/Quantum\_error\_correction}{https://en.wikipedia.org/wiki/Quantum\_error\_correction}.})  The ``scanning'' part poses a problem though. \\

\noindent {\bf{Question 1}}: Create a qubit in the $|1\rangle$ state and pass it through a Hadamard gate. From the measurement histogram, can you tell whether the qubit started as a $|0\rangle$  or $|1\rangle$ initial state?\\

The measurement histogram should look identical if either of the $|0\rangle$ or $|1\rangle$ states is used initially. Then how can we tell what the initial state was after performing a Hadamard operation? In the beam splitter, we determined where the photon came from by adding a second beam splitter to create interference. The way to measure and distinguish between them is to add a second Hadamard gate.  \\

\noindent {\bf{Question 2}}:
 If a qubit is in the unknown state $a|0\rangle + b|1\rangle$, what is the result of a single measurement?
  \begin{enumerate}[label=(\alph*)]
  \item $0$
  \item $1$
  \item $0$ with probability $a^2$ and $1$ with probability $b^2$
  \item A number between $0$ and $1$
  \end{enumerate}

\noindent {\bf{Question 3}}:
What is the result of a second measurement after the first from Question 2? 
  \begin{enumerate}[label=(\alph*)]
  \item $0$ if the first measurement is $0$ or $1$ if the first measurement is $1$
  \item $0$ if the first measurement is $1$ or $1$ if the first measurement is $0$ 
  \item $0$ with probability $a^2$ and $1$ with probability $b^2$
  \item A number between $0$ and $1$
  \end{enumerate}

Given a single qubit, it is not possible to determine how much of a superposition it is in if you only have this single qubit, i.e., you cannot determine the coefficients of $|0\rangle$ and $|1\rangle$ in a general state from one measurement! Note that if the state is known (from measuring many independent qubits that have been prepared identically), then you can just directly send the recipe to prepare this qubit. It is only when the state is unknown and when there is only one qubit that we have to think harder about how to efficiently ``scan'' the particle. 

The way to get around the problem of not being able to measure the qubit (and avoid collapsing the unknown state onto a basis state) is to ``scan'' the qubit indirectly with the help of entangled particles. This \href{https://www.jpl.nasa.gov/news/news.php?feature=4384}{comic}\footnote{https://www.jpl.nasa.gov/news/news.php?feature=4384} illustrates the basic idea. The protocol is as follows:
\begin{enumerate}

\item Alice and Bob meet up and make a qubit each (which we will call qubit \#2 and \#3). At this point, the two qubits are completely independent, i.e., think of the qubits as two different balls that do not contain any information about the other. Then, Alice and Bob decide to entangle their qubits by causing an interaction between the qubits such as application of a two-qubit CNOT gate. Using the previous metaphor, think of entanglement as Alice writing some information on Bob's ball that only she knows how to read, and Bob writing information on Alice's ball that only he knows how to read. For Bob to read Alice's information on his ball, Alice needs to send him a (classical) message with how to understand it, and vice-versa. They do not tell each other how to read the information yet. One possible entangled state (called the Bell-state) that they decide to make is
\begin{equation}
\frac{1}{\sqrt{2}}|00\rangle + \frac{1}{\sqrt{2}}|11\rangle.    
\end{equation}
Alice takes her qubit and walks away, and Bob takes his and walks in a different direction as shown in Figure \ref{fig:QTele1}.
\begin{figure}[h]
  \centering
  \includegraphics[width=0.75\textwidth]{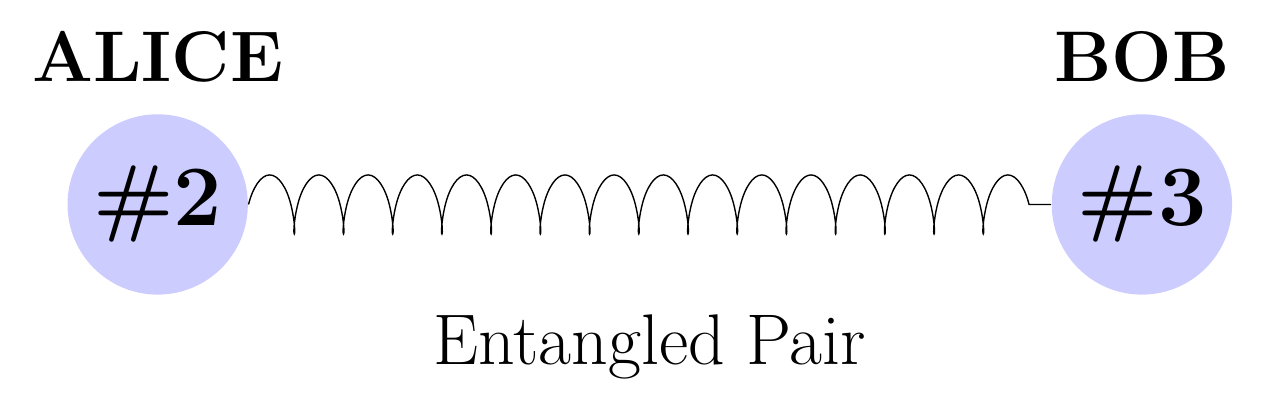}
  \caption{Alice and Bob's qubits are entangled.}
  \label{fig:QTele1}
\end{figure}

\item Now Alice obtains a third different qubit in an unknown state (qubit \#1) that she wants to transfer to Bob. She can only communicate with him classically by email or phone, and it would take too long to physically bring the qubit to Bob. The current situation is shown in Figure \ref{fig:QTele2}. 
\begin{figure}[h]
  \centering
  \includegraphics[width=0.75\textwidth]{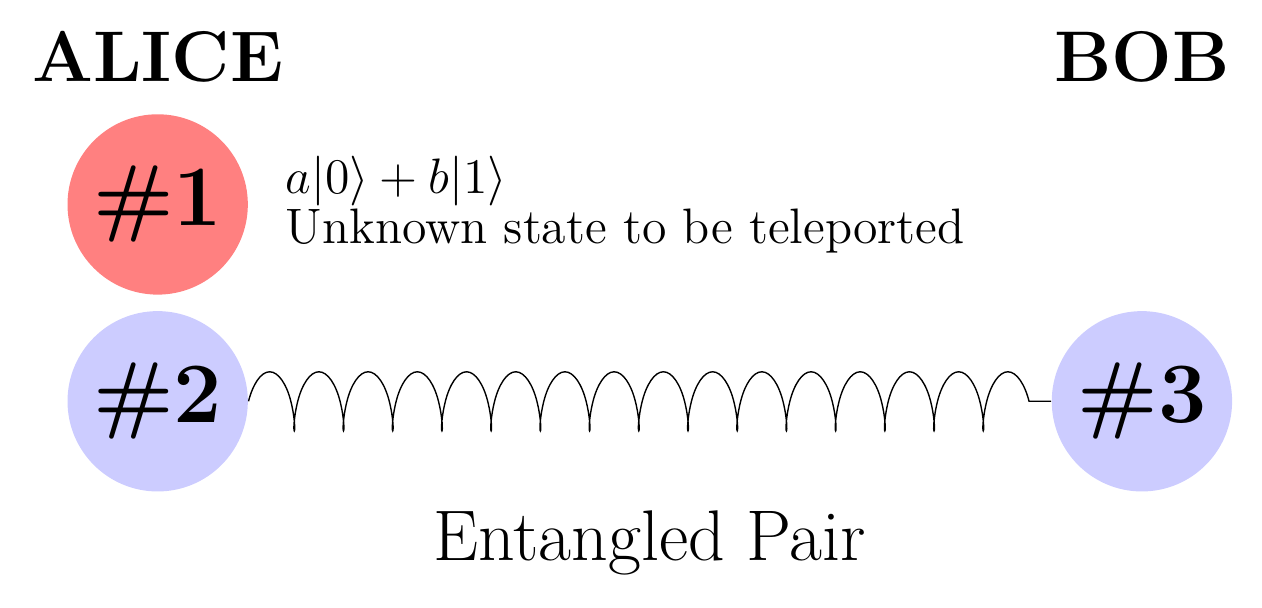}
  \caption{Alice has a qubit ($\#1$) in an unknown state she wants to transfer to Bob.}
  \label{fig:QTele2}
\end{figure}

\item Alice interacts her two qubits using a CNOT gate (qubits \#1 and \#2) and measures the qubit she originally had (qubit \#2). She then sends the unknown qubit to be teleported (qubit \#1) through a Hadamard gate and afterwards measures the output. Recall that the Hadamard gate is used to create a superposition of states. The current situation is shown in Figure \ref{fig:QTele3}. 
\begin{figure}[h]
  \centering
  \includegraphics[width=0.75\textwidth]{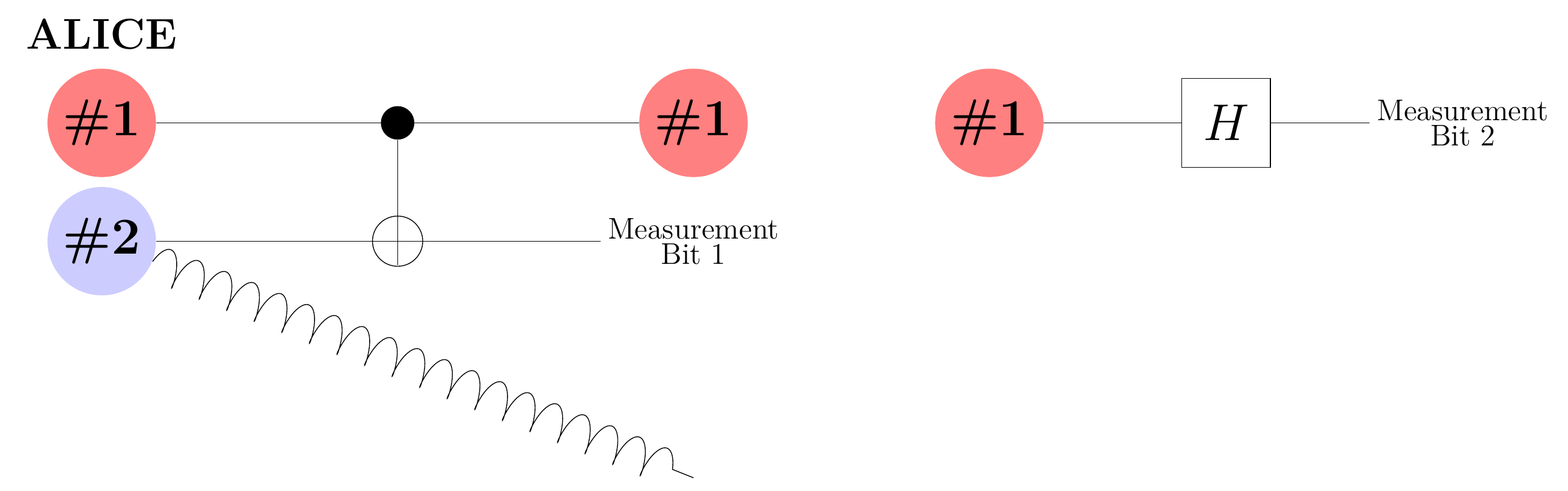}
  \caption{Alice passes her two qubits through a CNOT gate.}
  \label{fig:QTele3}
\end{figure}

Because Alice's original qubit (qubit \#2) was entangled with Bob's, the CNOT interaction with qubit \#1 immediately changes the state of Bob's qubit. By doing the math and drawing the full quantum circuit, one finds that Bob's qubit has changed into one of four possible superposition states. The four possible superposition states that Bob's qubit can be in depends on Alice's original qubit \#2 through the initial entanglement in Step $1$, as well as depending on the unknown qubit \#1 to be teleported from the CNOT gate in Step $3$. The reason we need to measure the state of Alice's qubit \#2 and qubit \#1 is to figure out the way Bob's qubit depends on these two. The current situation is shown in Figure \ref{fig:QTele3-1}. Note that Bob has not done anything with his qubit at this stage. 
\begin{figure}[h]
  \centering
  \includegraphics[width=0.75\textwidth]{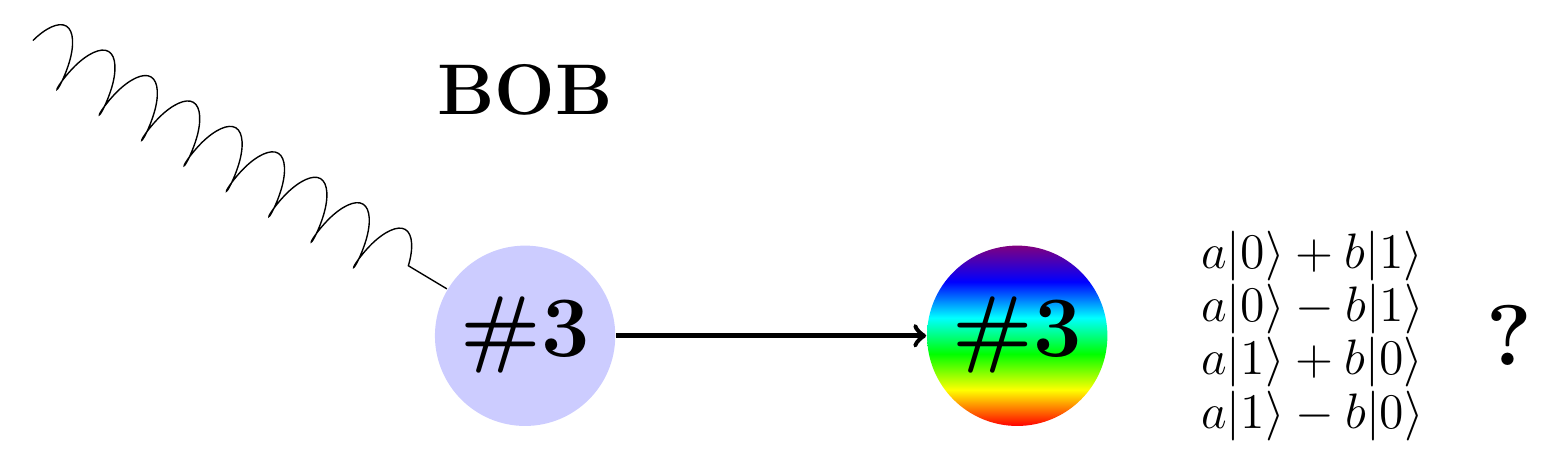}
  \caption{Four possible superposition states of Bob's qubit.}
  \label{fig:QTele3-1}
\end{figure}

\item Alice sends the two classical bits of information from the measurements to Bob by email or phone. 

\item Bob uses the two classical bits as the recipe for turning his qubit (now in an unknown state) into the correct state identical to qubit \#1. Depending on the values of the classical bits, Bob will know which of the four possibilities he has and he can then change it into the correct state using $Z$ and/or $X$ gates. If he has the correct state already, he does nothing. The result of this is shown in Figure \ref{fig:QTele4}.
\begin{figure}[h]
  \centering
  \includegraphics[width=0.75\textwidth]{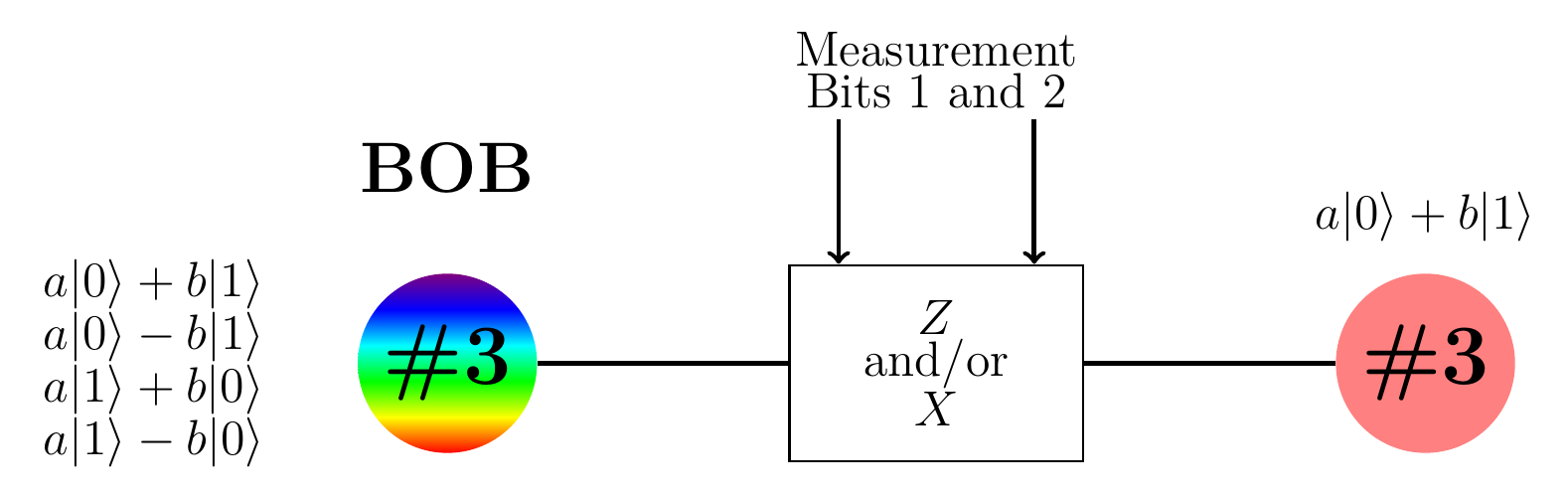}
  \caption{Final situation of entanglement between Bob and Alice.}
  \label{fig:QTele4}
\end{figure}

\end{enumerate}

It is important to understand that neither Alice nor Bob know what qubit \#1's coefficients $a$ or $b$ are at any point in the process.  All they know at the end is that qubit \#1 has been teleported from Alice to Bob. {{The mathematical description of this process is shown in the questions in Section \ref{sec:telequestions}.}}

Why is this protocol interesting? To answer this, imagine Alice and Bob met a long time ago and each took one qubit of the entangled pair. Bob is now traveling around the world and can only communicate with Alice by phone or email. If Alice wanted to transfer quantum data to Bob without quantum teleportation, she would have to meet Bob and physically give Bob her qubit. Quantum teleportation allows Alice to send {\it{quantum}} information using a {\it{classical}} communications channel. All she has to do is make some measurements and email Bob the values. Bob can then apply the correct recipe to his qubit to get the data. Quantum teleportation is a useful way of causing interaction between different parts of a quantum computer (by teleporting a qubit to a different part of the quantum computer you want to interact with) as well as sending information between two people.\footnote{Fermi National Accelerator Laboratory is building a quantum teleportation experiment which will extend over large distances, helping to develop a future quantum internet, e.g., \href{https://qis.fnal.gov/quantum-teleportation-experiment/}{https://qis.fnal.gov/quantum-teleportation-experiment/}}

\section{Check Your Understanding}
\label{sec:telequestions}

\begin{enumerate}

\item \fundamental{5pt} Could quantum teleportation be used to teleport a physical object from one place to another? Why or why not?

\item \fundamental{5pt} What would lead someone to think quantum teleportation can transmit information faster than the speed of light?  Explain why this is not possible.

\item \intermediate{5pt} By the no-cloning theorem, it is not possible to make a copy of an unknown qubit. At what point in the teleportation protocol does the unknown qubit collapse into a definite state? 

\item \intermediate{5pt} In the original protocol, Alice applies the CNOT and then measures Bit $1$ (see Figure \ref{fig:QTele3}). After this, Alice then applies the Hadamard to qubit \#1 and then measures Bit $2$ (see Figure \ref{fig:QTele3}). What happens if she decides to reverse the procedure by measuring Bit $2$ first, before applying the two-qubit CNOT gate?

\item \fundamental{5pt} If Bob knows that his qubit is in the $b|0\rangle + a|1\rangle$ state, which gate(s) would he need to use to change it back into the original needed $a|0\rangle + b|1\rangle$ state?
  \begin{enumerate}[label=(\alph*)]
  \item $X$
  \item $Z$
  \item $X$ then $Z$
  \end{enumerate}

\item \fundamental{5pt}  If Bob knows that his qubit is in the $a|0\rangle - b|1\rangle$ state, which gate(s) would he need to use to change it back into the original needed $a|0\rangle + b|1\rangle$ state?
  \begin{enumerate}[label=(\alph*)]
  \item $X$
  \item $Z$
  \item $X$ then $Z$
  \end{enumerate}

\item \fundamental{5pt}  If Bob knows that his qubit is in the $a|1\rangle - b|0\rangle$ state, which gate(s) would he need to use to change it back into the original needed $a|0\rangle + b|1\rangle$ state?
  \begin{enumerate}[label=(\alph*)]
  \item $X$
  \item $Z$
  \item $X$ then $Z$
  \end{enumerate}

\end{enumerate}

%
%
%
%
%
%
%
%
%
%


\graphicspath{{Chapter9-Algos/}}
\chapter{Quantum Algorithms}

\section{\intermediate{5pt} The Power of Quantum Computing}

\begin{figure}[!h]
    \centering
    \includegraphics{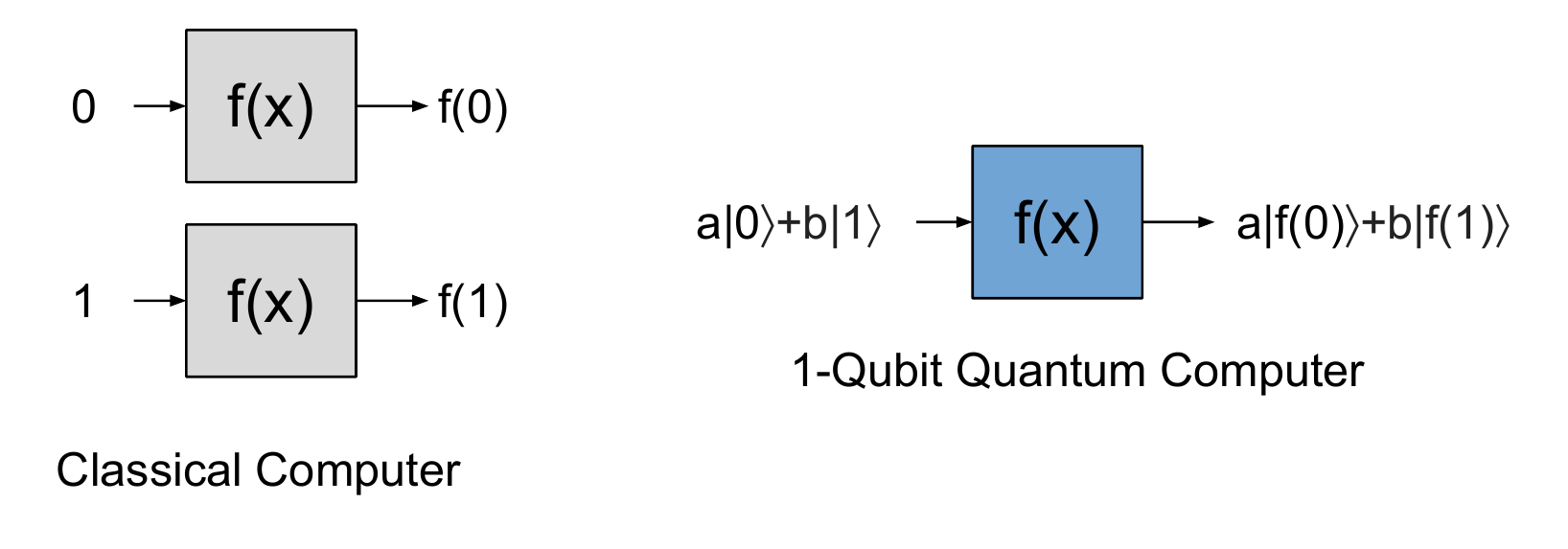}
    \caption{It takes a classical computer two operations to operate on two pieces of information. A quantum computer with one qubit can operate on two classical pieces of information at once.}
    \label{fig:Algorithms1}
\end{figure}

The main advantage that quantum computers have over classical computers is \textbf{parallelism}.  Because qubits can be in a superposition of states, a quantum computer can perform an operation on all of the states simultaneously. Let’s say we want to know the result of applying some function $f(x)$ to some number $x$. Two classical computations are needed to find the result for $x=0$ and for $x=1$, whereas a quantum computer can evaluate both answers in parallel as displayed in Figure \ref{fig:Algorithms1}. 

\begin{figure}[!h]
    \centering
    \hspace*{-1.5cm}
    \includegraphics{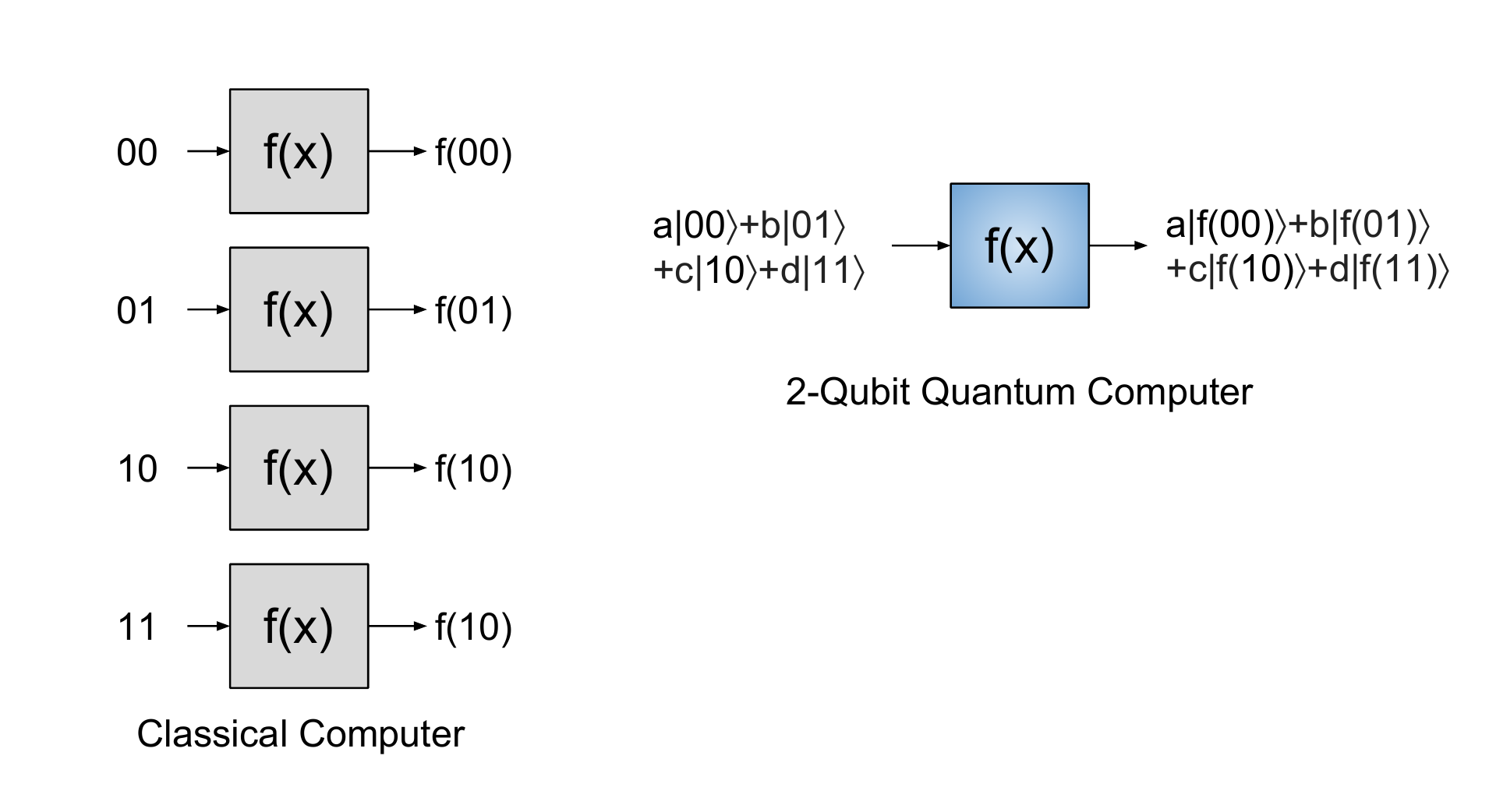}
    \caption{It takes a classical computer four operations to operate on four pieces of information. A quantum computer with two-qubits can operate on four classical pieces of information at once.}
    \label{fig:Algorithms2}
\end{figure}

If we wanted to compute $f(x)$ for $x=2$ (represented as $10$ in binary) and $x=3$ (represented as $11$), we would need to add a second qubit.  The two-qubit quantum computer can then evaluate all four possibilities at once as shown in Figure \ref{fig:Algorithms2}. \\

\noindent \textbf{Question 1}: How many pieces of information can a three-qubit quantum computer process in parallel? Write down all of the states. They are
\begin{equation}
    \ket{000},\ket{001},\ket{010},\ket{100},\ket{011},\ket{110},\ket{101},\ket{111} \rightarrow 8\ \text{pieces of information}. 
\end{equation}
\\
Adding a qubit to a quantum computer doubles its processing power! For a classical computer, you need to double the number of wires in the processor to get double the processing power.\footnote{It is an observation that classical computers double their processing power roughly every $18$ months. This is known as \href{https://en.wikipedia.org/wiki/Moore\%27s_law}{Moore's law}.} However, with a quantum computer, you only need to add a single qubit to double the processing power! Further, an $n$-qubit system can perform certain $2^n$ operations at once!

Separate from the issue of processing power is a concept known as memory. In a classical computer, on a standard $64$-bit laptop, each number can be represented in the $64$-bit binary representation (a simple extension of the $8$-bit binary representation you already learned about). However, if you wanted four numbers on a $64$-bit machine at the same time, then you need to have $4\times 64=256$-bits of memory on your hard drive to store them. On a $64$ bit classical computer, for $M$ different numbers, you need $M\times 64$-bits of memory; i.e., the bits needed for the memory is linear as a function of the number of numbers required. However, on an $n$-qubit quantum computer, there can be $2^n$ different coefficients of the quantum state that could in principle hold the numbers and therefore can be used as memory; i.e., the qubits needed for memory is logarithmic as a function of the number of numbers you want.

Because classical computers are very advanced  and have large processing power and terabytes of memory, classical computers can simulate small quantum computers.  As the  addition of a single qubit would double the memory required, the largest supercomputer in the U.S.\footnote{The \href{https://www.olcf.ornl.gov/olcf-resources/compute-systems/titan/}{Titan} at Oak Ridge Laboratory as of 2018} would only be able to simulate a $46$-qubit quantum computer. As of $2018$, Google has a quantum computer with a quantum chip (called the Bristlecone) which has $72$-qubits. 

\section{\intermediate{5pt} Limitations}

While parallelism sounds amazing in theory, it is not immediately useful on its own. A quantum computation can calculate a superposition of the $2^n$ numbers, however a measurement still needs to be performed in order to extract information from the quantum computer. One measurement will only show one of those answers and afterwards collapse the superposition into a basis state. Think about it as if the $2^n$ numbers are all on a secret scratchpad that we cannot see, and nature shows you one random page at a time, then burns the scratchpad. You would need to run the quantum computer at least $2^n$ times to get all the numbers, therefore negating any advantage over classical computers. As an example of this, the two-qubit quantum computer can calculate the superposition $a\ket{f(00)}+b\ket{f(01)}+c\ket{f(10)}+d\ket{f(11)}$, but measuring this state will result in either $f(00)$, $f(01)$, $f(10)$, {\it{OR}} $f(11)$. If you are unlucky, due to the randomness of quantum physics, you could repeat the computation four times and still not see all of the possibilities.  

Quantum computers are therefore only practical for certain types of problems. Since quantum computers are based on fundamental principles of nature (quantum physics) that includes classical physics, we intuitively expect those types of problems are the ones that can take advantage of more quantumness, e.g., simulating quantum physics directly. Generally, these types of problems look for correlations between different outputs. Due to this, it is generally accepted that quantum computers will not replace classical computers but will be able to perform different calculations that classical computers simply cannot. We will study an example problem which the quantum computer can solve more efficiently than a classical computer.

\section{\advanced{0.5pt} Deutsch-Jozsa Algorithm}

Here we give a proof that quantum computers can be faster than classical computers by explicit construction of a problem. 

\subsection*{The Problem}
Let $f(x)$ be an unknown function that operates on a single qubit. There can only be four different functions that satisfy this requirement, and the four different functions are shown in Table \ref{table:Algorithms3}. 

\begin{table}[h!]
\centering
\begin{tabular}{ c | c | c | c}
$f_{1}$  & $f_{2}$  & $f_{3}$  & $f_{4}$ \\
\hline
$f_{1}\left( 0\right)=0$ & $f_{2}\left( 0\right)=0$ & $f_{3}\left( 0\right)=1$ & $f_{4}\left( 0\right)=1$  \\ 
$f_{1}\left( 1\right)=0$ & $f_{2}\left( 1\right)=1$ & $f_{3}\left( 1\right)=0$ & $f_{4}\left( 1\right)=1$  
\end{tabular}
\caption{There are only four possible single qubit functions.}
\label{table:Algorithms3}
\end{table}

\noindent The question posed to the computer is this:\\

\centerline{"Is $f(x)$ going to output the same two numbers or opposite numbers?"}

A function is called \textbf{constant} if it always outputs the same result for all values of $x$.  A function is called \textbf{balanced} if it outputs $1$ for half of all the possible values of $x$ and $0$ for the other half.\\

\noindent \textbf{Question 2}: Which of the functions in Table \ref{table:Algorithms3} are constant and which are balanced? \\

The functions $f_1$ and $f_4$ are constant, while $f_2$ and $f_3$ are balanced. \\

\noindent \textbf{Question 3}: If you run the classical algorithm and see that $f(0)=1$, could you tell whether the function is constant or balanced? \\

 No, it could either be the balanced function $f_3$ or the constant function $f_4$. A classical computer would have to evaluate both $f(0)$ and $f(1)$ to determine the answer.

\subsection*{Quantum Solution}

\textbf{Procedure}:
\begin{enumerate}
    \item Put a qubit in a superposition of 0 and 1 with a Hadamard gate. 
    \item Operate on the qubit with the unknown function. 
    \item Apply another Hadamard gate. 
    \item Measure the qubit's state. A single measurement will tell you whether the function was constant or balanced.
\end{enumerate}

\noindent We will not go into the math behind the algorithm, but it can be demonstrated using the \href{https://www.st-andrews.ac.uk/physics/quvis/simulations_html5/sims/SinglePhotonLab/SinglePhotonLab.html}{Mach-Zehnder interferometer with phase shifters}\footnote{\href{https://www.st-andrews.ac.uk/physics/quvis/simulations\_html5/sims/SinglePhotonLab/SinglePhotonLab.html}{https://www.st-andrews.ac.uk/physics/quvis/simulations\_html5/sims/SinglePhotonLab/SinglePhotonLab.html}.} as shown in Figure \ref{fig:Algorithms4}.  Recall from the chapter on the beam splitter that the beam splitter will shift the phase of a photon depending on whether the photon hits the glass or dielectric side. The $\pi$ phase shifters are pieces of glass that can be placed along the path to shift the phase an additional 180$^\circ$.  Here is how the algorithm is implemented:

\begin{figure}[!h]
    \centering
    \includegraphics[width=\textwidth]{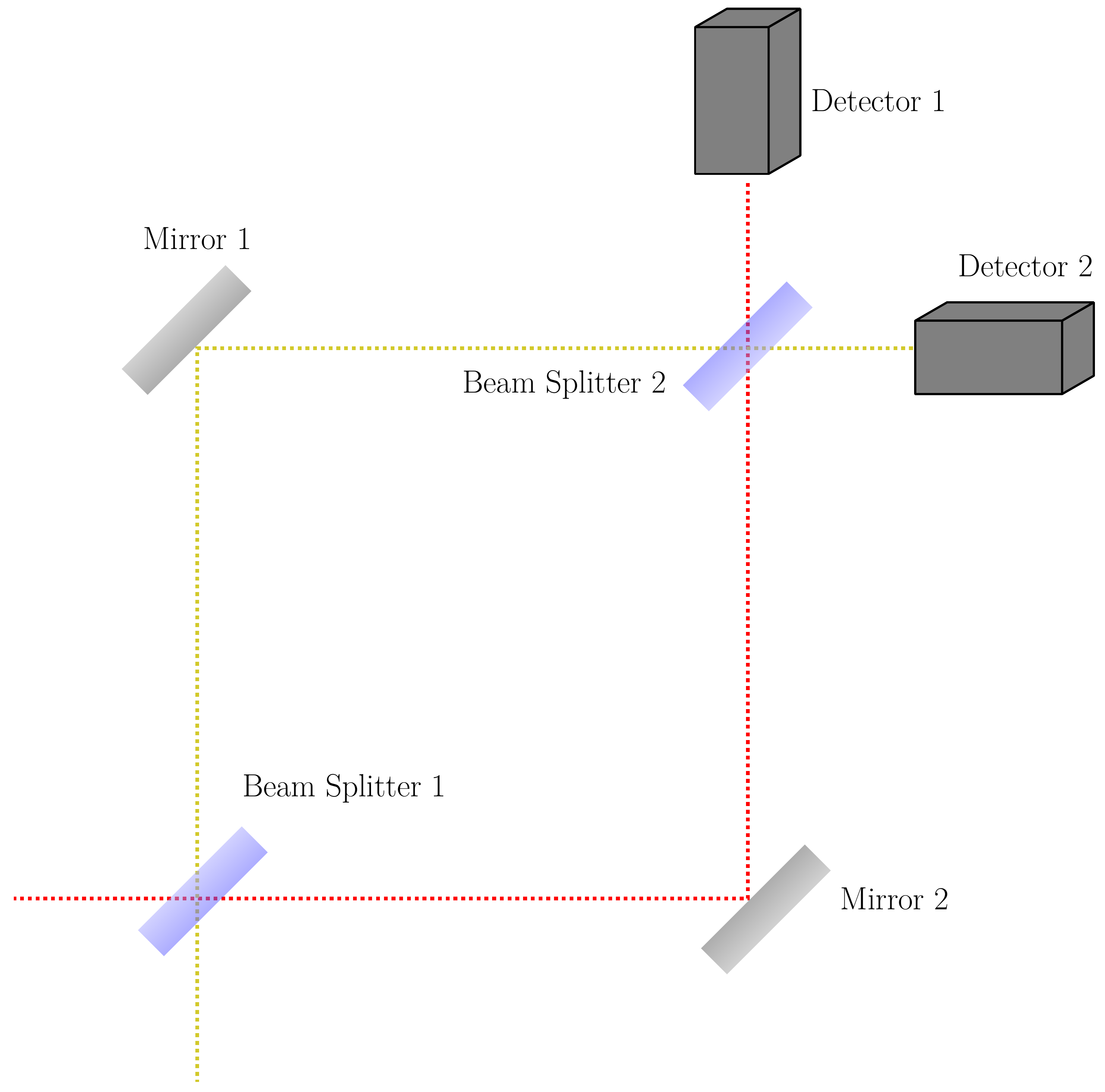}
    \caption{Mach-Zehnder interferometer with optional phase shifters.}
    \label{fig:Algorithms4}
\end{figure}

\begin{enumerate}
\item The two inputs $x=0$ and $x=1$ are represented by the two possible photon paths as shown in Figure \ref{fig:Algorithms5}. A photon taking the yellow path is $x=0$, while a photon taking the red path is $x=1$.  The first beam splitter therefore creates a superposition of 0 and 1 since the photon takes both paths.
    \begin{figure}
        \centering
        \includegraphics{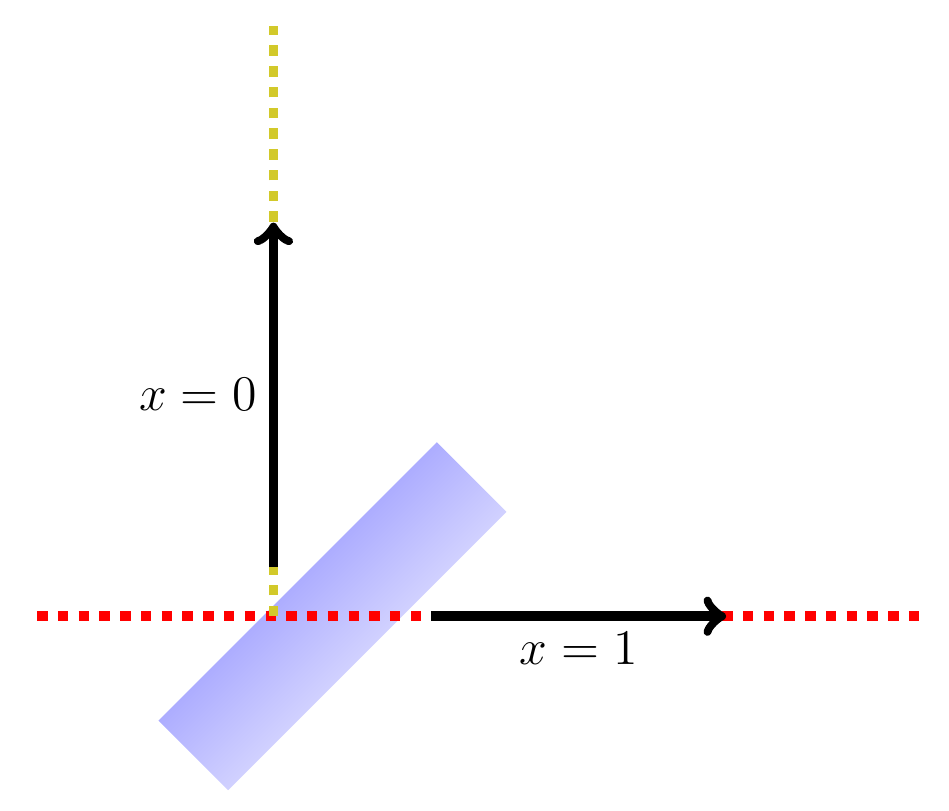}
        \caption{Inputs to the function are photons along two different paths. A photon taking the yellow path is $x=0$, while a photon taking the red path is $x=1$.}
        \label{fig:Algorithms5}
    \end{figure}

  \item The four functions will be modeled by four different phase shifter configurations, as shown in Figure \ref{fig:Algorithms6}.  A phase shifter is placed in the path whenever the function returns a $1$.
    \begin{figure}
        \centering
        \includegraphics[width=\textwidth]{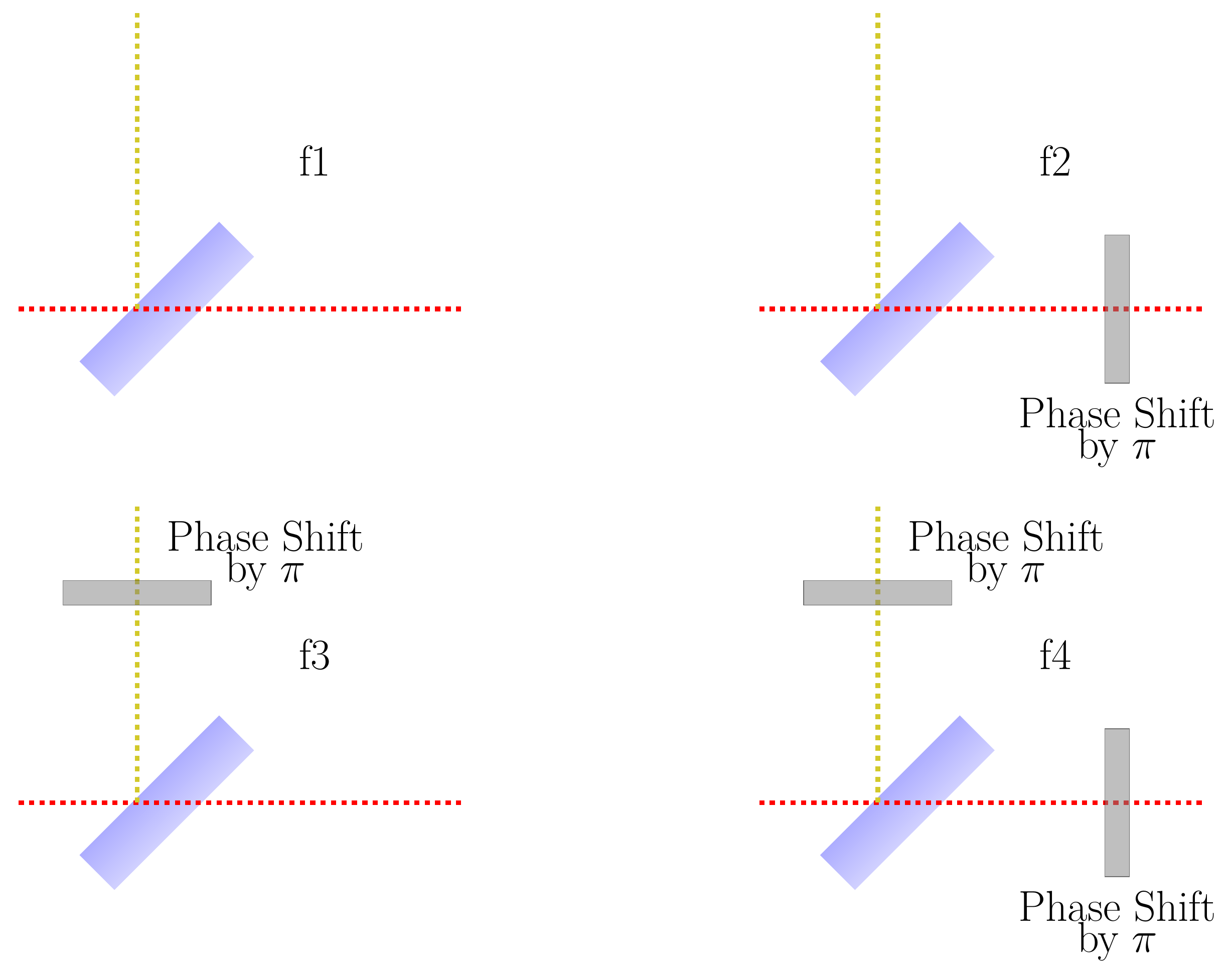}
        \caption{Four different functions modeled by four different phase shifter configurations.}
        \label{fig:Algorithms6}
    \end{figure}
    
    \item The second beam splitter creates the interference necessary to tell whether there was an odd or even number of phase shifters in the way.

    \item Measure which detector is activated. There is only one single measurement made. The single measurement made tells you the answer of the question. 
\end{enumerate}

\noindent \textbf{Question 4}: Which detector would go off for each function?  Can you explain these results by thinking of light as a wave?\\

\noindent \textbf{Question 5}: How many photons would you need to send to determine whether the function was constant or balanced?\\

Thanks to superposition and interference, only one quantum measurement is needed to determine the answer to the Deutsch-Jozsa problem. In fact, the algorithm can be extended to test functions with any number of inputs.  

\section{\intermediate{5pt} Quantum Computers Today}
While the Deutsch problem has no known commercial applications, useful quantum algorithms such as Shor's factoring algorithm rely upon similar concepts. Quantum algorithms are believed to exist that can speed up machine learning algorithms and efficiently simulate the quantum behavior of molecules.  Unfortunately, current quantum computers are still very far from achieving quantum supremacy, i.e., outperforming the best classical computers. As of 2018, companies such as IBM and Google have built different types of quantum computers that contain up to 72 qubits. To give you an idea of where we need quantum computers to be, factoring a 1024-bit modern encryption key using Shor's algorithm would require more than $5,000$ qubits.
{{ In 2019, Google claimed to have performed\footnote{https://www.nature.com/articles/s41586-019-1666-5} the first quantum computation that a classical computer could not do - a milestone known as ``quantum supremacy''. Quantum supremacy  means that a quantum computer can solve a problem that a classical computer cannot. However, the solution of the problem may not be practical useful.
As such, it is important to note that Google have demonstrated quantum supremacy, not the ``quantum usefulness'' milestone. Google perfomed their task on a 53 qubit quantum computer, which took $200$ seconds. They claimed it would take a classical computer $10,000$ years to do the same task. However, shortly after IBM suggested\footnote{https://arxiv.org/abs/1910.09534} that an improved classical supercomputing technique could theoretically perform the task in just $2.5$ days.     }}

There are different technological difficulties when improving a quantum computer. As we have seen one way, a quantum computer can be built using lasers, which are bunches of photons. However, there are also random photons outside of the quantum computer in the environment that may accidentally leak into the quantum computer, and these environmental photons can then cause accidental changes to the quantum state. Such accidental changes are called ``noise''. To reduce the number of these environmental photons, the quantum computer can be cooled down to near absolute zero (around -$450^{\degree}$ Fahrenheit). However, this is difficult. The more qubits you add, the more you need to keep at this low temperature (a technological challenge). Also, the more qubits you add, the more lasers you need to interact the qubits, which can be technologically difficult to keep lots of qubits in one small space but also cause isolated interactions between them. Further, the more qubits you add, the more likely it is that the qubits will interact accidentally with the environment, which will then destroy the system’s quantum properties through a process known as decoherence.  However, given how classical computers went from being the size of a room in the 1960s to an iPhone within a few decades, governments and industries are investing billions of dollars towards making quantum computers realistic. Ultimately, quantum computers are destined to complement classical computers, not replace them, so don't expect to have a quantum phone in your pocket anytime soon!\footnote{Theoretical physicists and computational scientists at Fermi National Accelerator Laboratory are working on improving algorithms and the foundations of quantum science in order to expand the problems that (near term) quantum devices can solve, e.g., \href{https://qis.fnal.gov/quantum-computing-for-hep/}{https://qis.fnal.gov/quantum-computing-for-hep/}}

\section{Check Your Understanding}
\begin{enumerate}
    \item \fundamental{3pt}  \vspace{-0.74cm}
    \begin{enumerate}[a)]
        \item How many different classical pieces of information can be represented by eight classical bits (1 byte)?
        \item What about a quantum computer with eight-qubits?
        \item What advantage does the quantum computer have over the classical computer?
    \end{enumerate}
    
    \item \intermediate{5pt} Explain how superposition and interference allows the Deutsch-Jozsa algorithm to beat the classical algorithm.

    \item \intermediate{5pt} Figure \ref{fig:Algorithms7} shows the gate implementation for testing a three-qubit function $f(x)$. A constant function will always result in $\ket{000}$ {{ or $\ket{111}$}}.
      \begin{figure}
        \centering
        \includegraphics[width=\textwidth]{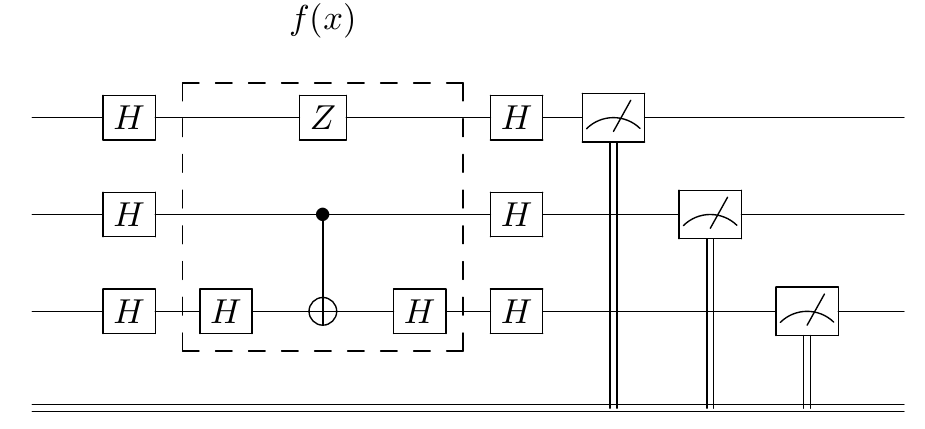}
        \caption{The gate implementation for testing the different possible three-qubit functions.}
        \label{fig:Algorithms7}
      \end{figure}
    \begin{enumerate}[a)]
        \item How many evaluations would be needed on a classical computer to tell whether this function is constant or balanced?
        \item By running this algorithm on IBM Q, can you determine whether this function is constant or balanced?
    \end{enumerate}
\end{enumerate}

\graphicspath{{Chapter10-Worksheets/}}
\chapter{Worksheets}

\section{\intermediate{8pt} Correlation in Entangled States Lab}
\label{chapter:WorksheetCorr}
\textbf{Objectives:}
\begin{itemize}
    \item Experimentally determine the difference between two particles in a product state vs. an entangled state using the \href{https://www.st-andrews.ac.uk/physics/quvis/simulations_html5/sims/entanglement/entanglement.html}{entanglement simulator}.\footnote{\href{https://www.st-andrews.ac.uk/physics/quvis/simulations\_html5/sims/entanglement/entanglement.html}{https://www.st-andrews.ac.uk/physics/quvis/simulations\_html5/sims/entanglement/entanglement.html}} 
    \item Apply the idea of basis changing to explain the correlation that is observed.
\end{itemize}

\section*{Questions}
Alice and Bob each measure one of two qubits with a Stern-Gerlach apparatus. Start with both SGAs along the z-axis

\begin{figure}[!h]
    \centering
    \includegraphics[width=0.9\textwidth]{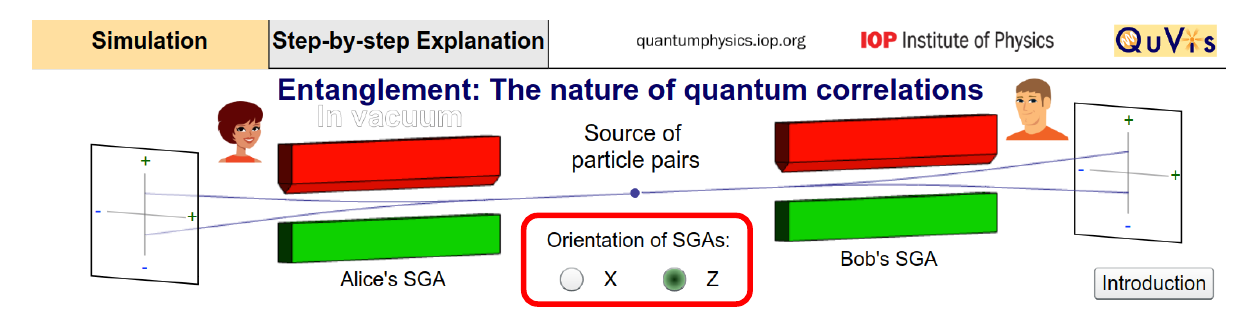}
    \caption{Figure reproduced from the \href{https://www.st-andrews.ac.uk/physics/quvis/}{QuVis website}, licensed under creative commons CC-BY-NC-SA.}
\end{figure}

\begin{enumerate}
    \item Try sending pairs of particles in a product state $\ket{\uparrow_A}\ket{\downarrow_B}$. What do Alice and Bob measure individually?
    \item Try sending pairs of particles in an entangled state: $\frac{1}{\sqrt{2}}\left(\ket{\uparrow_A}\ket{\downarrow_B}-\ket{\downarrow_A}\ket{\uparrow_B}\right)$. What do Alice and Bob measure individually?
    \item If Alice measures her spin, would you be able to predict Bob's result:
    \begin{enumerate}[a)]
        \item In the product state?
        \item In the entangled state?
    \end{enumerate}
\end{enumerate}

Now rotate both SGAs along the x-axis.
\begin{figure}[!h]
    \centering
    \includegraphics[width=0.9\textwidth]{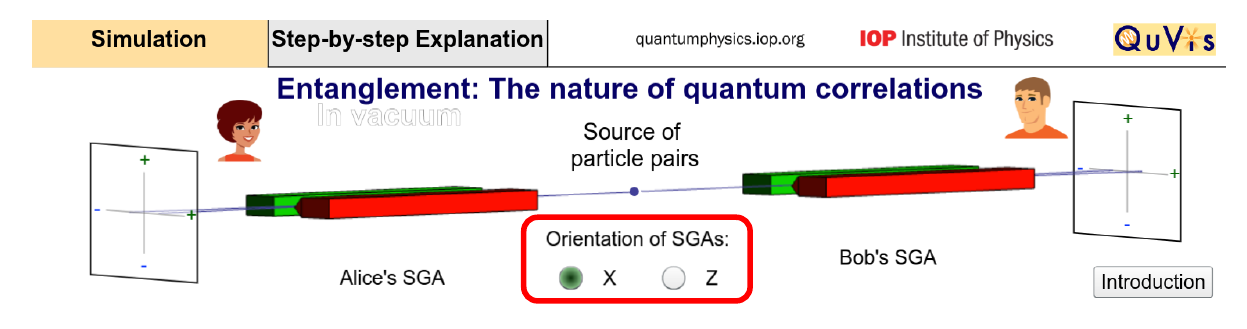}
    \caption{Figure reproduced from the \href{https://www.st-andrews.ac.uk/physics/quvis/}{QuVis website}, licensed under creative commons CC-BY-NC-SA.}
\end{figure}

\begin{enumerate}[resume]
    \item Try sending pairs of particles in a product state $\ket{\uparrow_A}\ket{\downarrow_B}$. What do Alice and Bob measure individually?
    \item Try sending pairs of particles in an entangled state $\frac{1}{\sqrt{2}}\left(\ket{\uparrow_A}\ket{\downarrow_B}-\ket{\downarrow_A}\ket{\uparrow_B}\right)$. What do Alice and Bob measure individually?
    \item If Alice measures her spin, would you be able to predict Bob's result:
    \begin{enumerate}[a)]
        \item In the product state?
        \item In the entangled state?
    \end{enumerate}
    \item Convert the product state $\ket{\uparrow_A}\ket{\downarrow_B}$ into the $x$-basis and use it to explain the observations in the $x$-basis. Recall that $\ket{\uparrow}=\frac{1}{\sqrt{2}}\left(\ket{+}+\ket{-}\right)$ and $\ket{\downarrow}=\frac{1}{\sqrt{2}}\left(\ket{+}-\ket{-}\right)$.
    \item Convert the entangled state $\frac{1}{\sqrt{2}}\left(\ket{\uparrow_A}\ket{\downarrow_B}-\ket{\downarrow_A}\ket{\uparrow_B}\right)$ into the $x$-basis and use it to explain the measurements in the $x$-basis.
    \item Suppose that there are two possible sources of particles. Source \#1 randomly emits two particles in either the state $\ket{\uparrow_A}\ket{\downarrow_B}$ or $\ket{\downarrow_A}\ket{\uparrow_B}$ with equal probability. Source \#2 emits two particles in the entangled state $\frac{1}{\sqrt{2}}\left(\ket{\uparrow_A}\ket{\downarrow_B}-\ket{\downarrow_A}\ket{\uparrow_B}\right)$. How can Alice and Bob tell whether the source is \#1 or \#2?
\end{enumerate}

\newpage

\section*{Answers}

\begin{enumerate}
    \item Alice always measures up; Bob always measures down.
    \item They each see up and down 50\% of the time.
    \item \begin{enumerate}[a)]
        \item Yes.
        \item Yes. Every time Alice measures up, Bob measures down and vice versa.
    \end{enumerate}
    \item They each see up and down 50\% of the time
    \item They each see up and down 50\% of the time
    \item \begin{enumerate}
        \item No, the results are random.
        \item Yes. Every time Alice measures $+$, Bob measures $-$ and vice versa. The entangled state is still correlated in the $x$-basis.
    \end{enumerate}
    \item \begin{equation*}
        \ket{\uparrow\downarrow}=\frac{1}{\sqrt{2}}\left(\ket{+}+\ket{-}\right)\times\frac{1}{\sqrt{2}}\left(\ket{+}-\ket{-}\right)=\frac{1}{2}\ket{++}-\frac{1}{2}\ket{+-}+\frac{1}{2}\ket{-+}-\frac{1}{2}\ket{--}.
    \end{equation*}
    All four possible states are observed. The middle two terms do not cancel out because they are different states: Alice measures $+$ and Bob measures $-$, or Alice measures $-$ and Bob measures $+$.
    \item \begin{equation*}
        \frac{1}{\sqrt{2}}\left(\ket{\uparrow}\ket{\downarrow}-\ket{\downarrow}\ket{\uparrow}\right)=\frac{1}{\sqrt{2}}\left(-\ket{+-}+\ket{-+}\right).
    \end{equation*}
    Only two states are observed where Alice and Bob always get opposite results.
    \item They cannot tell them apart in the $z$-basis, but they could measure the particles in the $x$-basis. If Alice and Bob always get opposite results, the source emits entangled particles. If there is no correlation, the particles are not entangled.
\end{enumerate}

\newpage
\section{\intermediate{8pt} Polarizer Demo}
\label{sec:WorksheetPolDemo}

For students who have learned about polarization, the creation of superposition states can be demonstrated using three \href{https://www.arborsci.com/polarizing-filters.html}{polarizing filters}. When unpolarized light is sent through a vertical filter, only vertically polarized light is able to pass through.  Sending vertically polarized light through a horizontal filter results in no light passing through, since the vertical and horizontal polarizations are mutually exclusive. Surprisingly, adding a diagonal filter in between recovers the light!  The diagonal polarizer introduced a horizontally polarized component, similar to how passing a spin-up electron through a horizontal SGA created a horizontal superposition.

\begin{figure}[!h]
    \centering
    \includegraphics[width=0.8\textwidth]{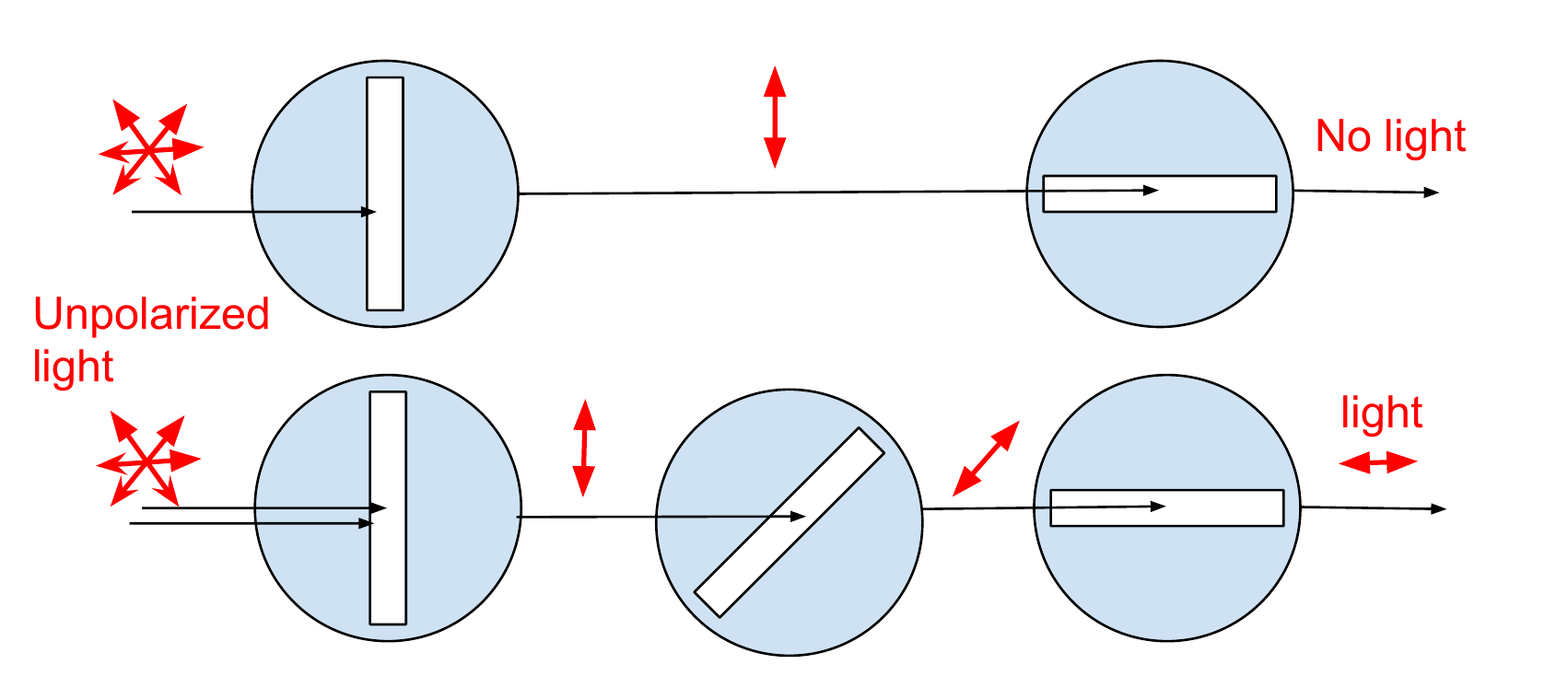}
    \caption{}
\end{figure}

\textbf{Question}: Relate the behavior of the polarizers to what you saw in the SGAs.

\newpage

\section{\fundamental{5pt} Quantum Tic-Tac-Toe}
\label{sec:WorksheetQTicTacToe}

Quantum Tic-Tac-Toe was developed by Alan Goff in $2004$ as a metaphor to teach quantum concepts such as superposition, entanglement, and measurement collapse. It has been found to be a helpful strategy in teaching quantum mechanics to undergraduate students at Purdue, especially for students who struggle with grasping the concepts.\footnote{Hoehn R.et al (2014). ``Using Quantum Games to teacher quantum mechanics, Part 1.'' \emph{Journal of Chemical Education} 91 (3), 417-422. Retrieved from \href{https://pubs.acs.org/doi/ipdf/10.1021/ed400385k}{https://pubs.acs.org/doi/ipdf/10.1021/ed400385k}}

Quantum Tic-Tac-Toe resembles the classical Tic-Tac-Toe game in its setup and objective of completing three in a row. However, the game uses characteristics of quantum systems, so instead of using one marker $X$ or $O$, the players use pairs of $X$s and $O$s, which are traditionally called ``spooky,'' after Einstein's reference to entanglement as ``spooky action at a distance''.\footnote{Einstein, Podolsky, and  Rosen (1935) ``Can quantum-mechanical description of physical reality be considered complete?'' \emph{Physical Review, 47} : 777-780. Retrieved from \href{https://journals.aps.org/pr/pdf/10.1103/PhysRev.47.777}{https://journals.aps.org/pr/pdf/10.1103/PhysRev.47.777}}
Using indices for each marker's move is important when determining the winner of the game. Additionally, we use a color code for each player and connect the spooky markers to help students better visualize the game process. We also number the squares for future reference.

\begin{figure}[!h]
    \centering
    \includegraphics[width=0.35\textwidth]{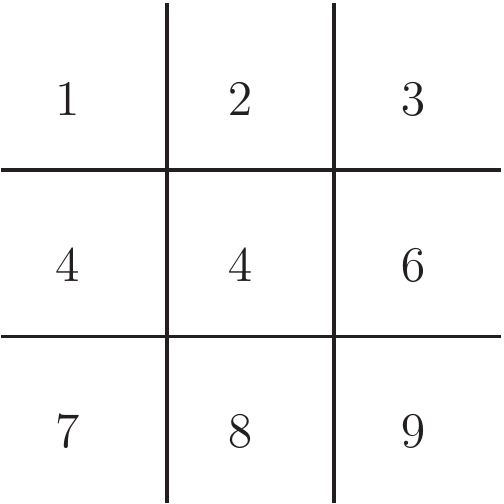}
     \includegraphics[width=0.35\textwidth]{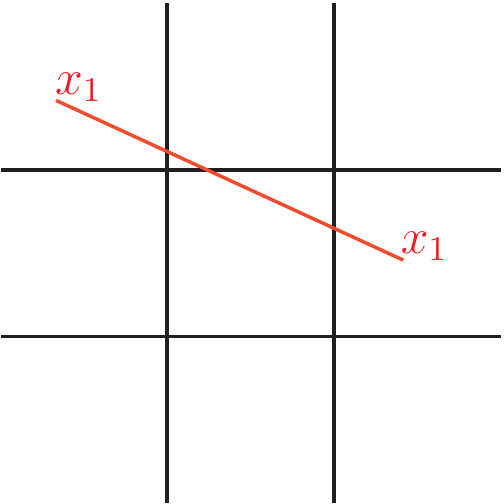}
    \caption{The Quantum Tic-Tac-Toe layout with numbered squares (left): one player's move with spooky markers $x_{1}$ (right).}\label{fig:tic1}
\end{figure}

\section*{The Rules}
\begin{enumerate}
\item The X player goes first. We note that keeping indices helps to track the game. The markers can be placed in any of the two spaces on the game board (Figure~\ref{fig:tic1}). 

\item The O player goes next. The markers can be placed in any two squares, even ones that are already occupied by other X or O markers. Notice in Figure 2 that the index for the O player also starts with 1, representing its first move placing markers in squares 1 and 6.

\item Player X goes again and can place their spooky markers at any two squares, even ones occupied by other Xs or Os. The game goes on until the players create a ``cyclic loop" as seen in the following example: 

\begin{figure}[!h]
    \centering
    \includegraphics[width=0.35\textwidth]{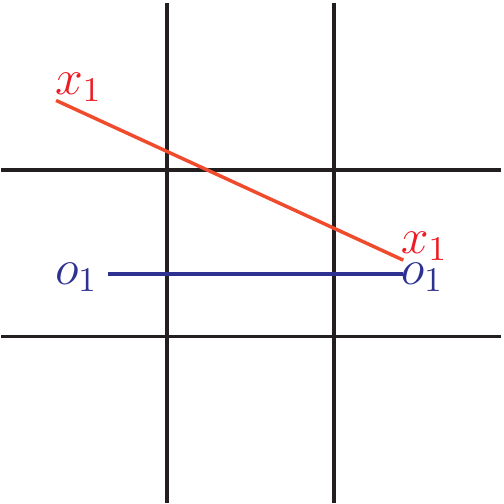}
    \caption{Example of the second player's move. }
\end{figure}

\begin{figure}[!h]
    \centering
    \includegraphics[width=0.35\textwidth]{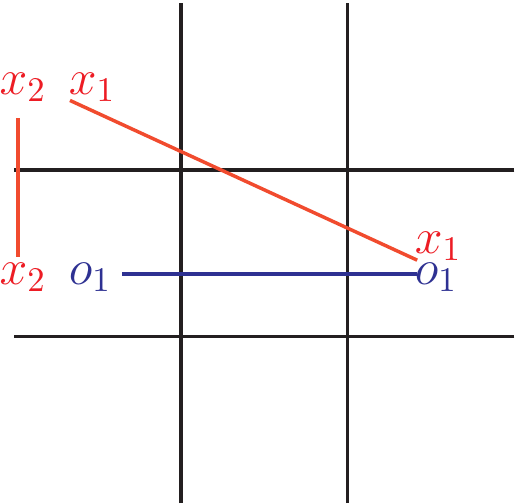}
    \caption{The cyclic loop is created by the player X. Using lines between the spooky markers helps in identifying the loop. }
\end{figure}

\item \textbf{Collapsing the quantum state}. When a loop is created, the players have to collapse their state. The are three options for who makes the decision on how the markers will be collapsed. The fair choice would be by the player who did not create the cycle (in this case, player O). When the markers are forced to collapse, only one of the two squares for each move can be chosen, so player O can choose either square 4 or 6. Depending on their choice, the outcome would be different (Figure 4). Once the states are collapsed, the ``spooky markers''  change into classical markers and they fully occupy the state of one particular square. 

\begin{figure}[!h]
    \centering
    \includegraphics[width=0.35\textwidth]{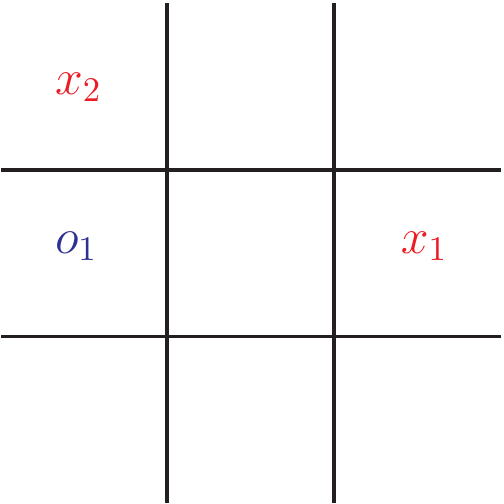}
    \hspace{1.5cm}
        \includegraphics[width=0.35\textwidth]{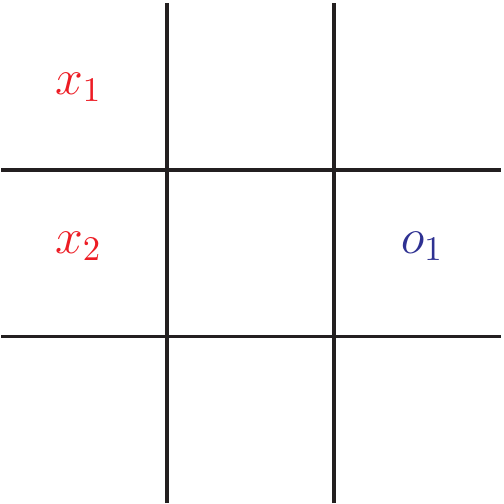}
    \caption{The two collapse outcomes due to player O's decision.}
\end{figure}

\item The next player can place his spooky markers in any two squares except the ones that are occupied by the collapsed markers. The game goes on until another cycle is created and the players are forced to collapse the state. 

\item \textbf{Winning the game}. In some cases both players will create three in a row after collapsing their spooky markers. In this case, the player with the smallest sum of indexes wins.  For example, in Figure 5 player X wins because their has the smaller sum.

\begin{figure}[!h]
    \centering
    \includegraphics[width=0.35\textwidth]{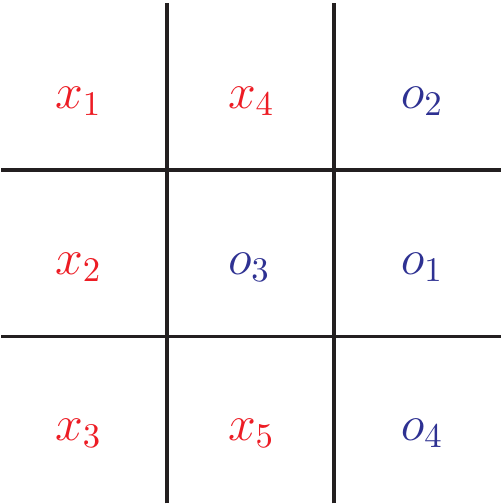}
    \caption{Player X wins, because the sum of their indexes is $1 + 2 + 3 = 6$. Player O got three in a row, but the sum of their indexes is $2 + 1 + 4 = 7$.}
\end{figure}

\end{enumerate}

\textbf{Some other rules can be added or modified. }
One of the requirements could be that players cannot place both markers in the same square like the one shown in Figure~\ref{fig:ttt8}.
\begin{figure}[!h]
    \centering
    \includegraphics[width=0.35\textwidth]{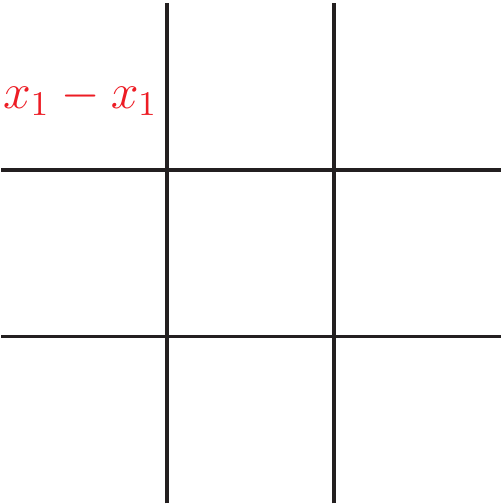}
    \caption{Player X wins because the sum of its indexes is $1 + 2 + 3 = 6$. Player O got three in a row, but the sum of the indexes is $2 + 1 + 4 = 7$}
\end{figure}\label{fig:ttt8}
Another way to make the collapse more quantum (or more random) is using a coin flip to decide which player chooses the collapse.

Other modifications may include assigning different point values for three in a row, such as the winner with lowest sum of the indexes gets 1 point, while the other player gets $1/2$ point. 

One of the main challenges of playing the game is to observe when a cycle has been created so the state of the spooky markers can be collapsed at the right time. A computer-simulated game will automatically keep track of this and will force students to collapse their markers, such as  this \href{http://qttt.rohanp.xyz/}{game simulator}.\footnote{\href{http://qttt.rohanp.xyz/}{http://qttt.rohanp.xyz/}}

We found that using color codes and connecting lines helps visually track loops. Another way is to create a model of the game where students can see the connections and collapse the states using physical pieces. It would be interesting to see students' responses as to which medium helps them understand the game principle better.

\section*{Connection to quantum physics}

How are the game rules and principles connected to the real applications of quantum mechanics? There are three major themes that can be drawn from the game: superposition, the effect of measurement, and entanglement. 

\subsection*{Superposition}
In classical physics all objects have defined states. However, quantum systems can exist in a superposition of several classical states at the same time. The example could be electron with a spin that is in superposition of up and down, or a photon in a superposition of vertical and horizontal polarization. QTTT spooky markers exist in two separate locations on the game board, representing their state as a superposition state of two classical TTT markers. 

\subsection*{Measurement}
When measuring the state of a quantum system, the quantum state of a system collapses and  only one classical state is observed with some probability. In QTTT, the rule of creating the loop forces players to collapse their markers (measure their quantum state). In this case the player decides how to collapse the markers, which corresponds to the scientist choosing the way of measuring quantum system, such as axis orientation. The rule of forcing the measurement when the loop is created does not have exact corresponding physical meaning.  Quantum systems can exist in a superposition state for an extended time, and the measurement is not forced, but chosen by the observer.

\subsection*{Entanglement}
Entanglement is the quantum phenomenon of creating two or more particles, whose states cannot be described separately, but have some correlation even when they are separated by a significant distance. When measuring the state of one of the entangled particles, the state of the other particle can be known even without measurement. Einstein called it ``spooky action at a distance." When the players collapse their states after creating a loop in QTTT, they know for sure in which state each marker would collapse into.  

\newpage
\section{\fundamental{5pt} Schr\"{o}dinger's Worm Using Five Qubits}
\label{sec:WorksheetWorm}

\section*{Getting Started}
\begin{itemize}

\item \textbf{Objectives:} Design, build, and test quantum circuits that model systems in superposition and entanglement. 
\item  \textbf{Setup:} Open the \href{https://quantumexperience.ng.bluemix.net/qx}{IBM Q simulator}\footnote{\href{https://quantumexperience.ng.bluemix.net/qx}{https://quantumexperience.ng.bluemix.net/qx}} and start a new experiment.
\end{itemize}

\begin{figure}[h!]
\centering
\includegraphics[width=0.75\textwidth]{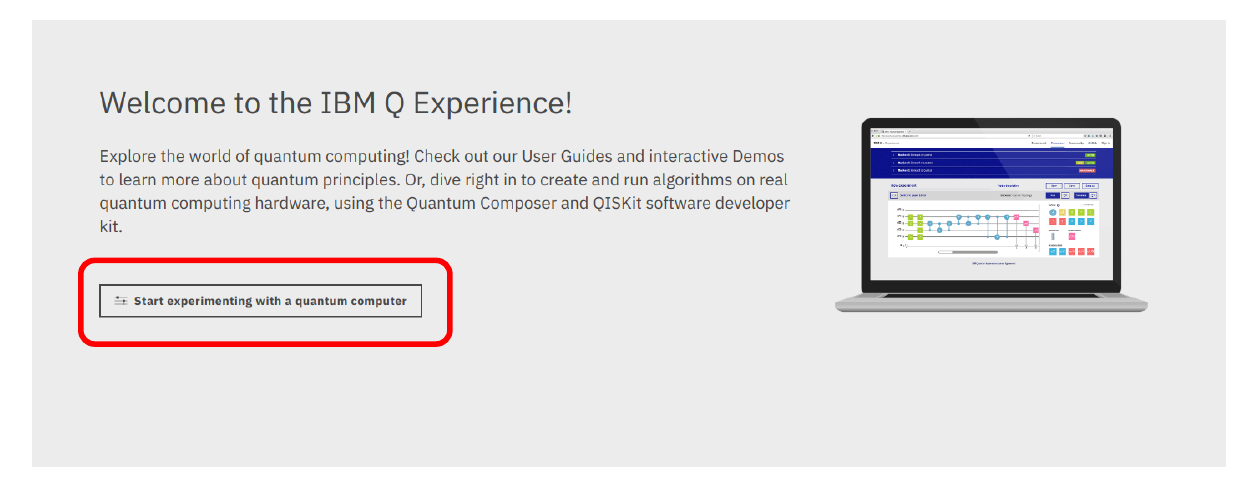}
\caption{Select ``Start experimenting with a quantum computer.'' Reprint Courtesy of International Business Machines Corporation, \copyright International Business Machines Corporation.}
\end{figure}

\begin{figure}[h!]
\centering
\includegraphics[width=0.75\textwidth]{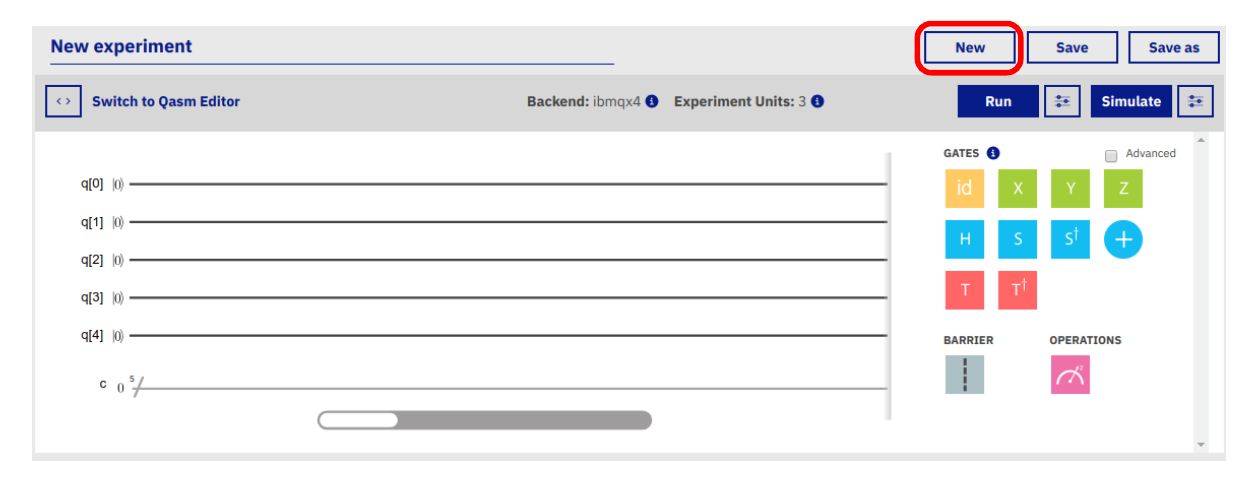}
\caption{Select ``New" option at top right. Reprint Courtesy of International Business Machines Corporation, \copyright International Business Machines Corporation.}
\end{figure}

Choose the ``Custom Topology" backend with the default 5-qubit setting to enable unrestricted gate placements. 
\begin{figure}[h!]
\centering
\includegraphics[width=0.75\textwidth]{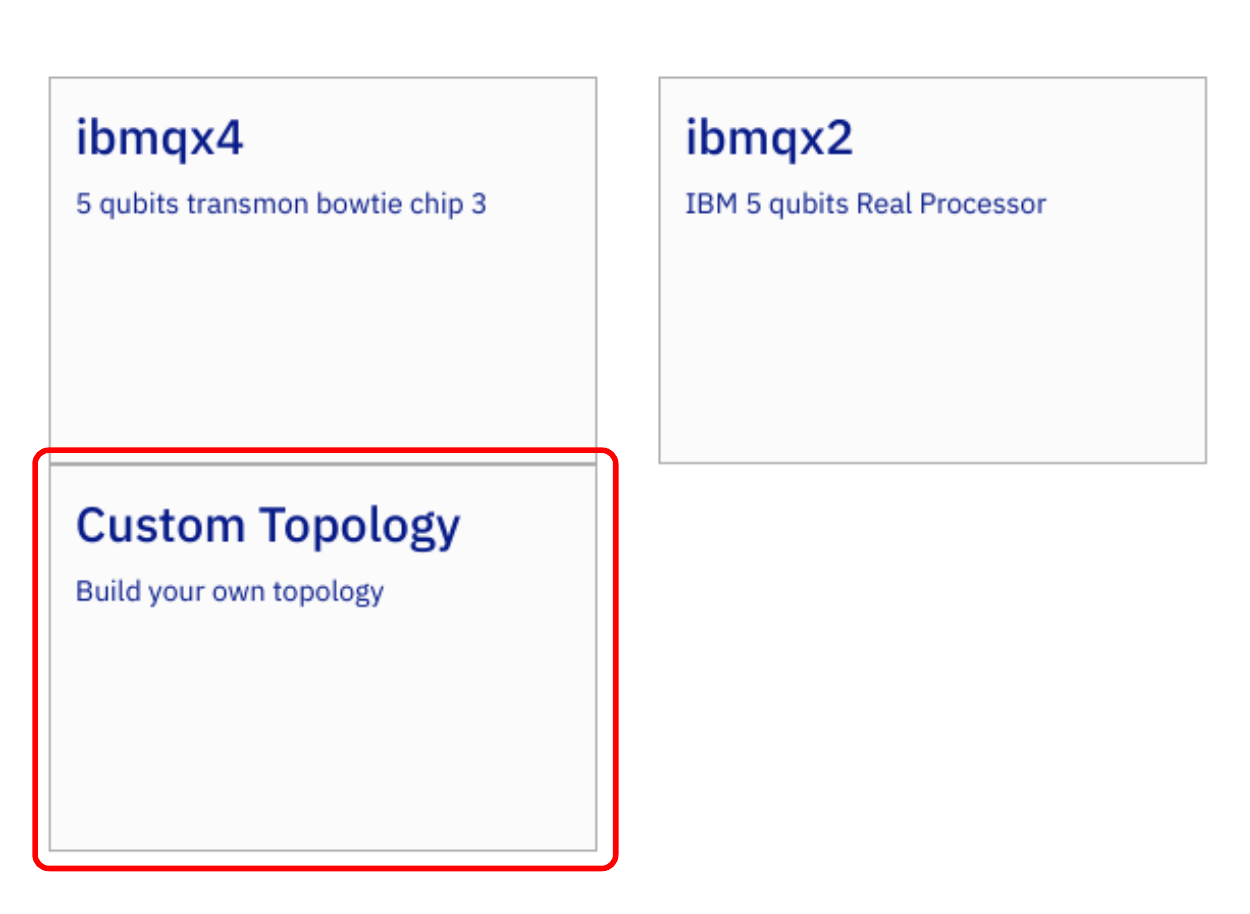}
\caption{Select ``Custom Topology"  option at the bottom left. Reprint Courtesy of International Business Machines Corporation, \copyright International Business Machines Corporation.}
\end{figure}

\newpage
\section*{Part I:  Superposition}
The worm is alive when all five squares are black and dead when only four are black.  Use a 0 to represent a white square and 1 to represent a black square. 
\begin{figure}[h!]
\centering
\includegraphics[width=0.3\textwidth]{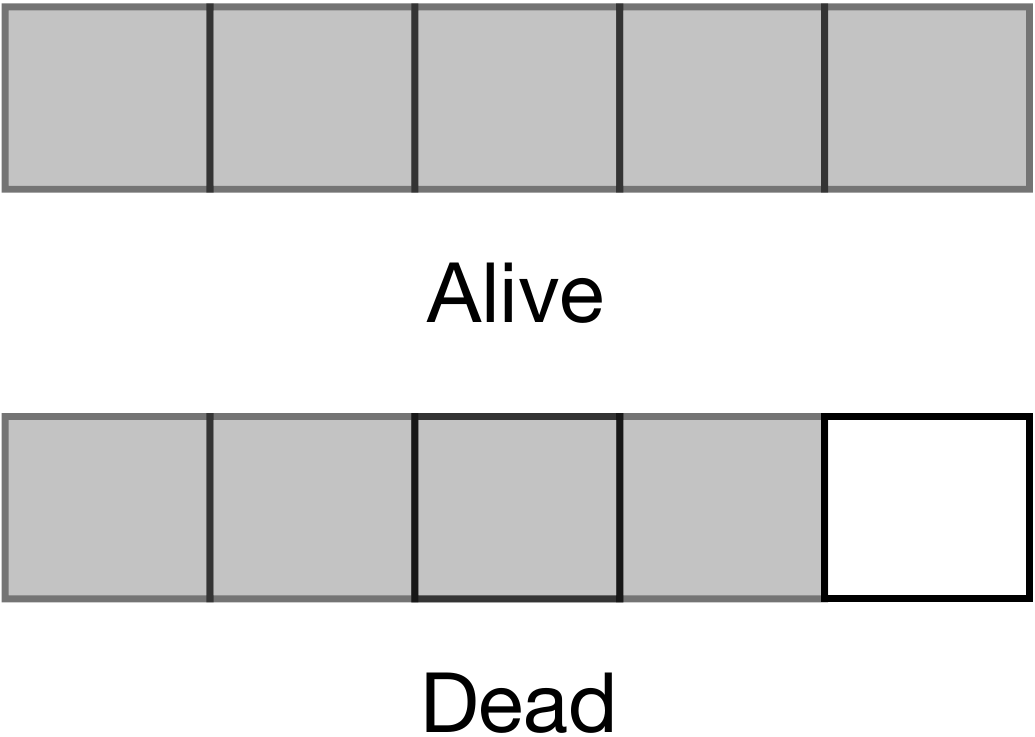}
\caption{Dead or alive worms.}
\end{figure}

\begin{enumerate}
\item What is the classical state of the live 5-bit worm?
\item What is the classical state of the dead 5-bit worm?
\item Use IBM Q to create a worm in a superposition state of alive and dead. Let q[0] correspond to the bit on the far right.
\item Run the simulation and interpret the histogram.
\item How can you modify the circuit so that the worm is first put in a superposition state and then brought to life?  
\item How can you modify the circuit so that the worm in a superposition state becomes definitely dead?
\end{enumerate}

\section*{Part II:  Entanglement}
\begin{figure}[h!]
\centering
\includegraphics[width=0.3\textwidth]{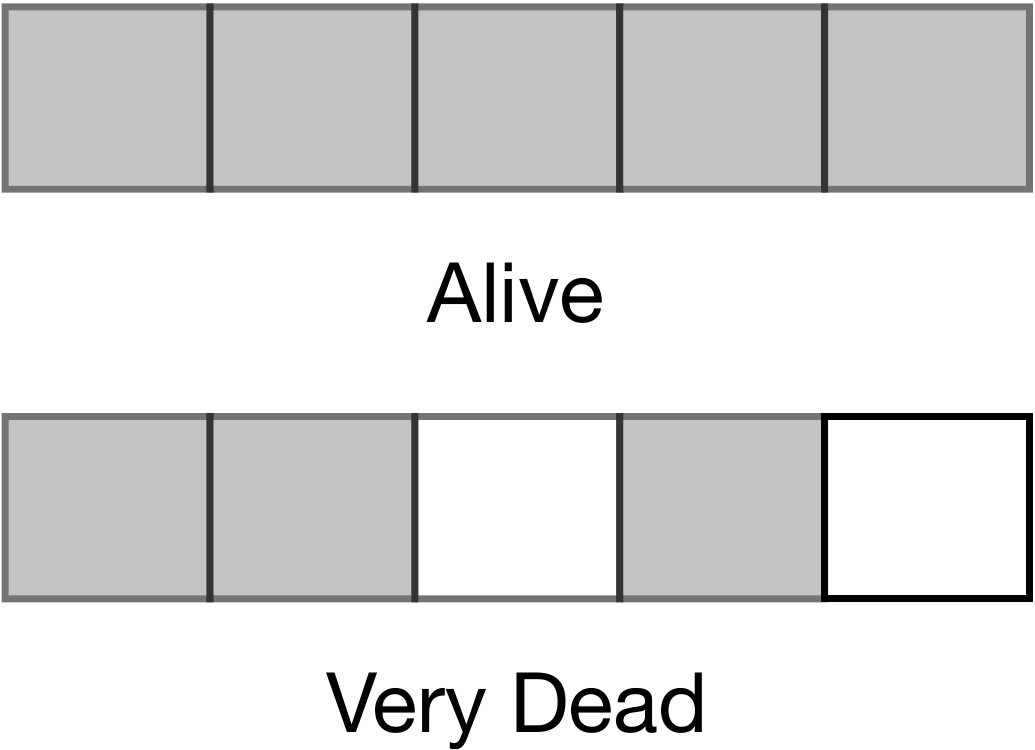}
\caption{Very dead or alive worms.}
\end{figure}

The worm is next to a hungry bird such that it is either alive or chomped to pieces.

\begin{enumerate}
\item What is the classical state of the very dead worm?
\item Create a circuit that produces a worm in a superposition state of alive and very dead. (Hint: Two of the qubits are entangled.)
\item Run the simulation and interpret the histogram.
\item How can you modify the circuit so that the worm in a superposition state becomes either definitely dead or definitely alive?
\end{enumerate}

\newpage

\section*{Answers}

\subsection*{Part I:  Superposition}

\begin{enumerate}
\item $\lvert 11111\rangle$
\item $\lvert 11110\rangle$
\item     \includegraphics[width=0.5\textwidth]{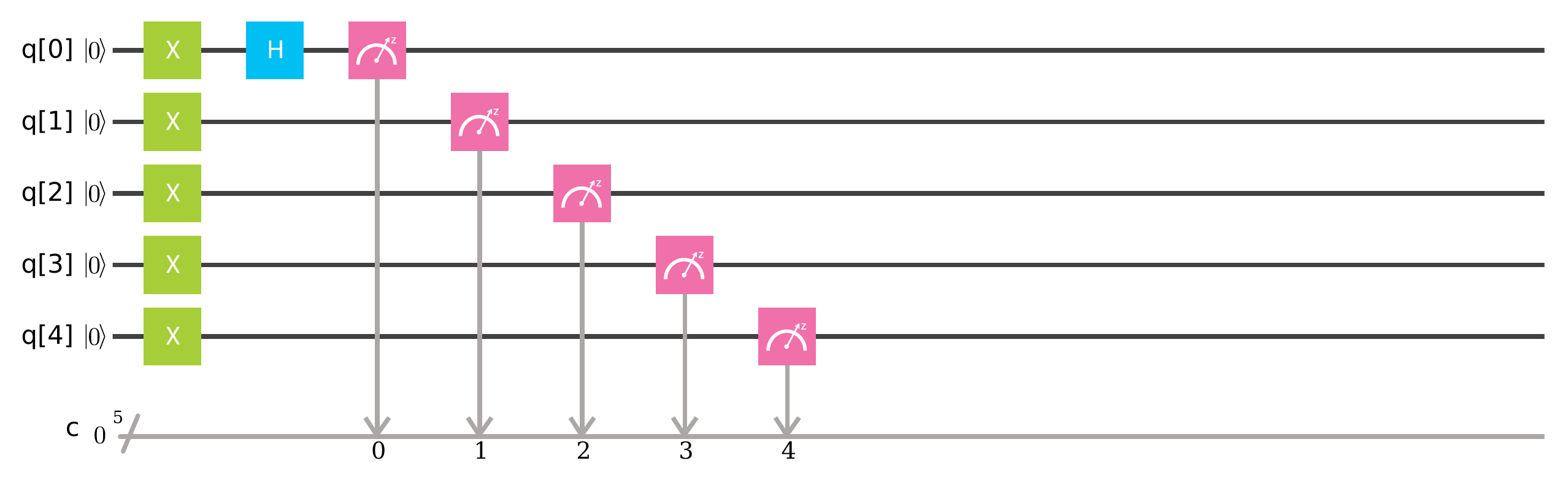} \\
 Reprint Courtesy of International Business Machines Corporation, \copyright~ International Business Machines Corporation. \\
 \includegraphics[width=0.5\textwidth]{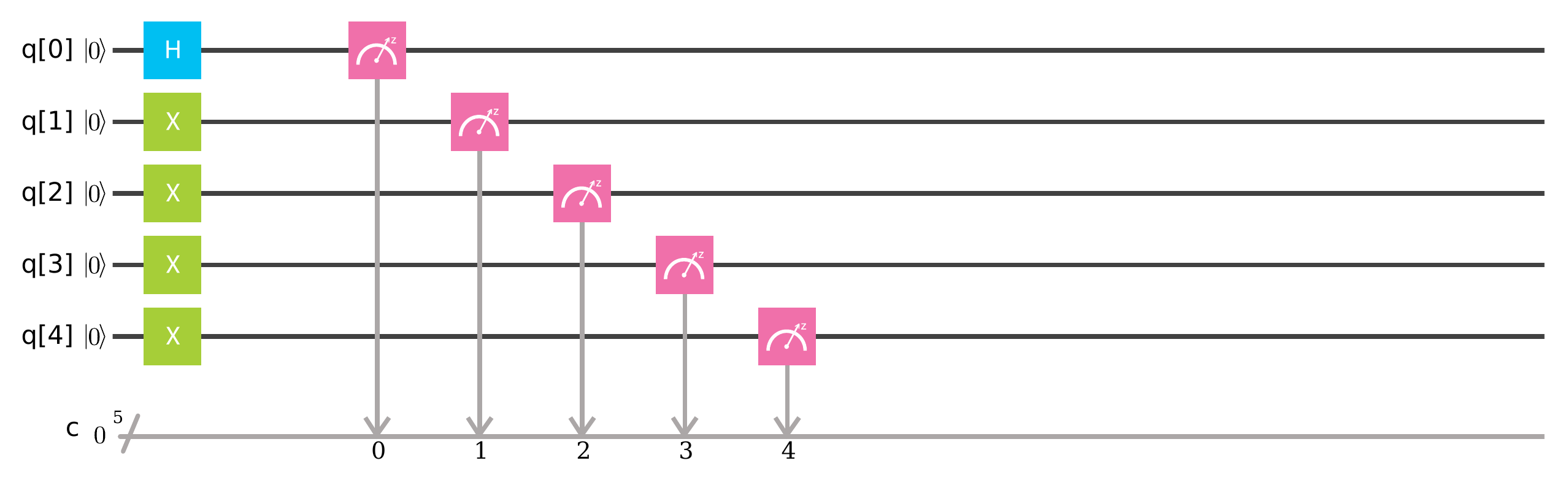} \\
  Reprint Courtesy of International Business Machines Corporation, \copyright~ International Business Machines Corporation. \\
\item The histogram shows about $50/50\%$ chances of both measurements: $11111$ and $11110$. The worm is in a superposition of the dead and alive states.\\
  \includegraphics[width=0.5\textwidth]{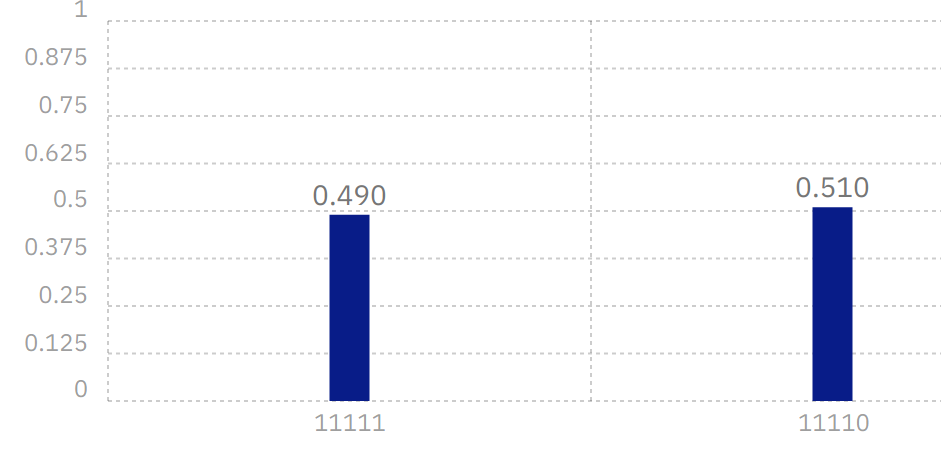} \\
    Reprint Courtesy of International Business Machines Corporation, \copyright~ International Business Machines Corporation. \\
\item Adding a second Hadamard gate undoes the superposition.
\\
\includegraphics[width=0.5\textwidth]{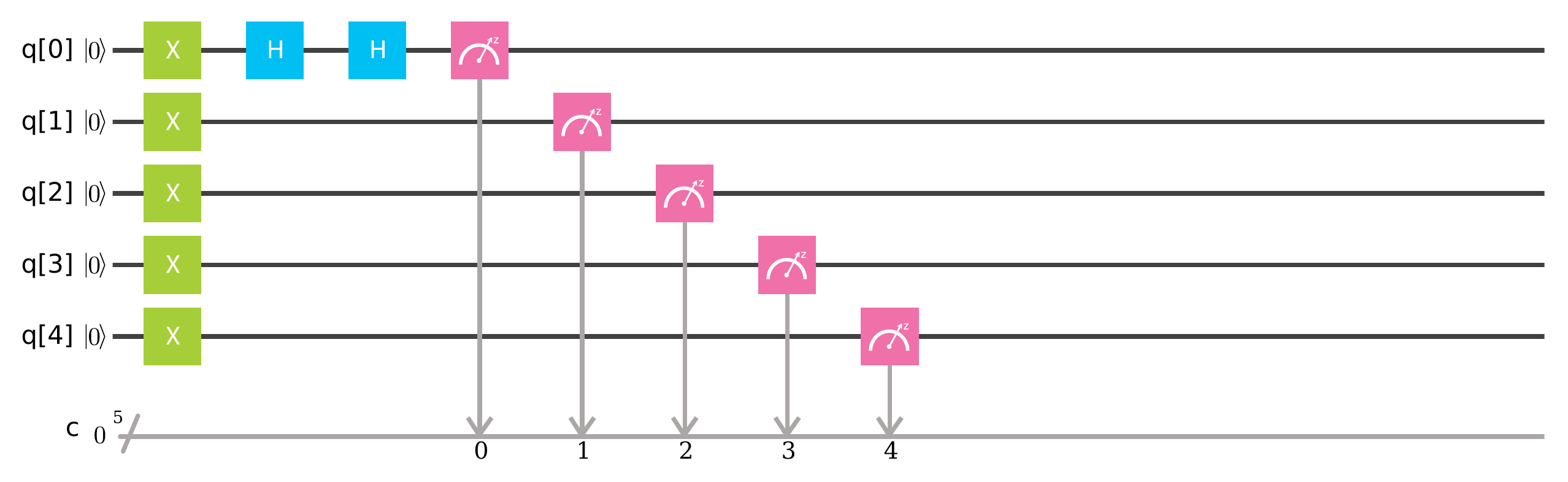} \\
    Reprint Courtesy of International Business Machines Corporation, \copyright~ International Business Machines Corporation. \\
    \includegraphics[width=0.35\textwidth]{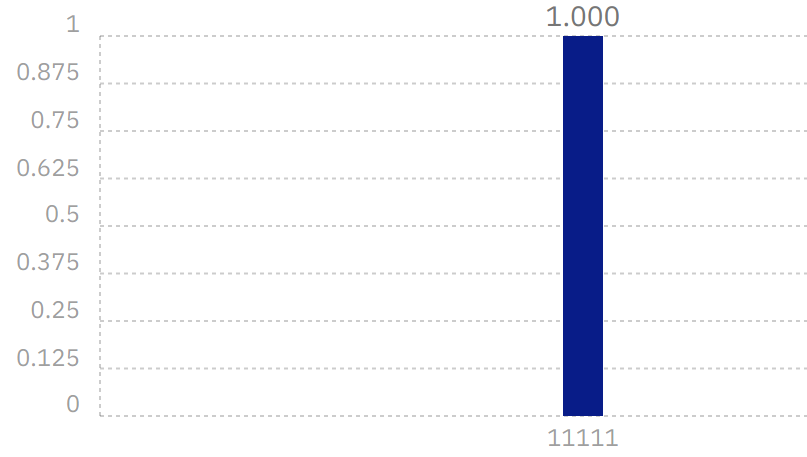} \\
    Reprint Courtesy of International Business Machines Corporation, \copyright~ International Business Machines Corporation. \\
\item Adding an $X$ gate flips the bit.
\\
\includegraphics[width=0.35\textwidth]{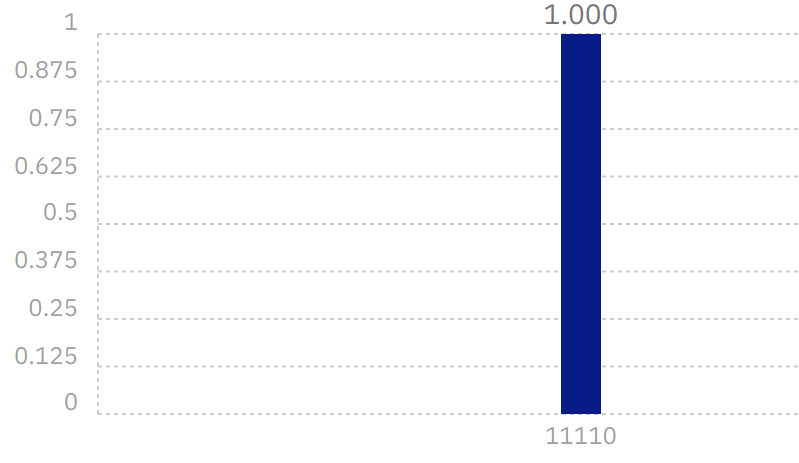} \\
    Reprint Courtesy of International Business Machines Corporation, \copyright~ International Business Machines Corporation. \\
    \includegraphics[width=0.5\textwidth]{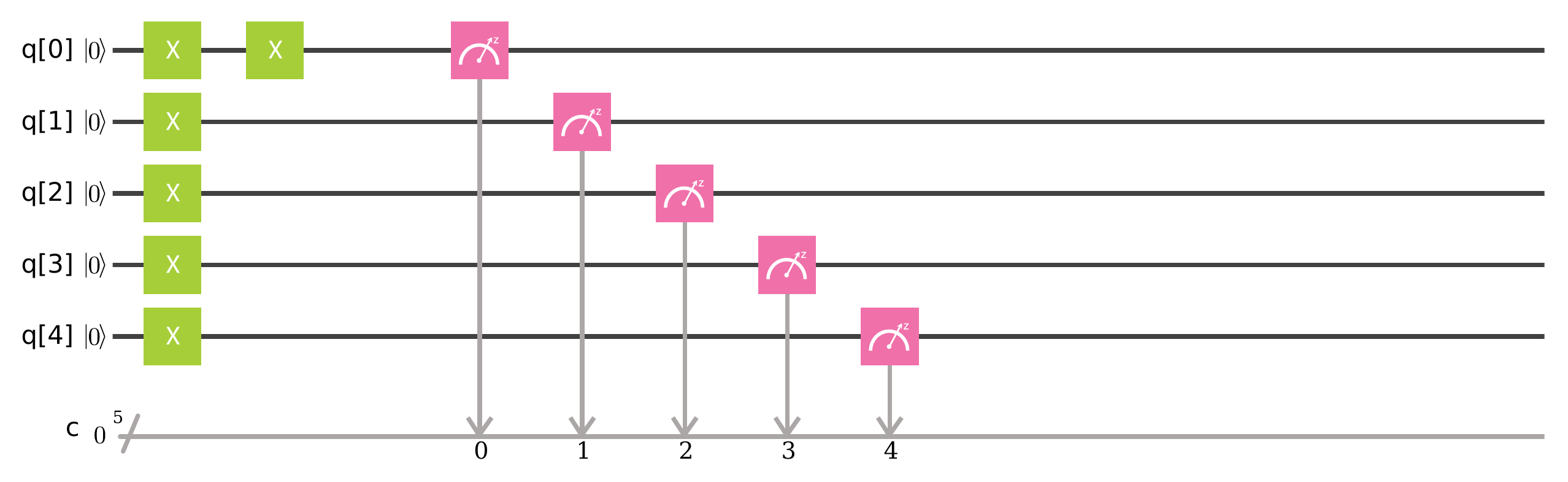} \\
    Reprint Courtesy of International Business Machines Corporation, \copyright~ International Business Machines Corporation. \\
\end{enumerate}
\newpage
\subsection*{Part II:  Entanglement}

\begin{enumerate}
\item $\lvert 11110\rangle$
\item  A CNOT gate is used to entangle the two qubits such that they are either both white or both black.
\\
\includegraphics[width=0.5\textwidth]{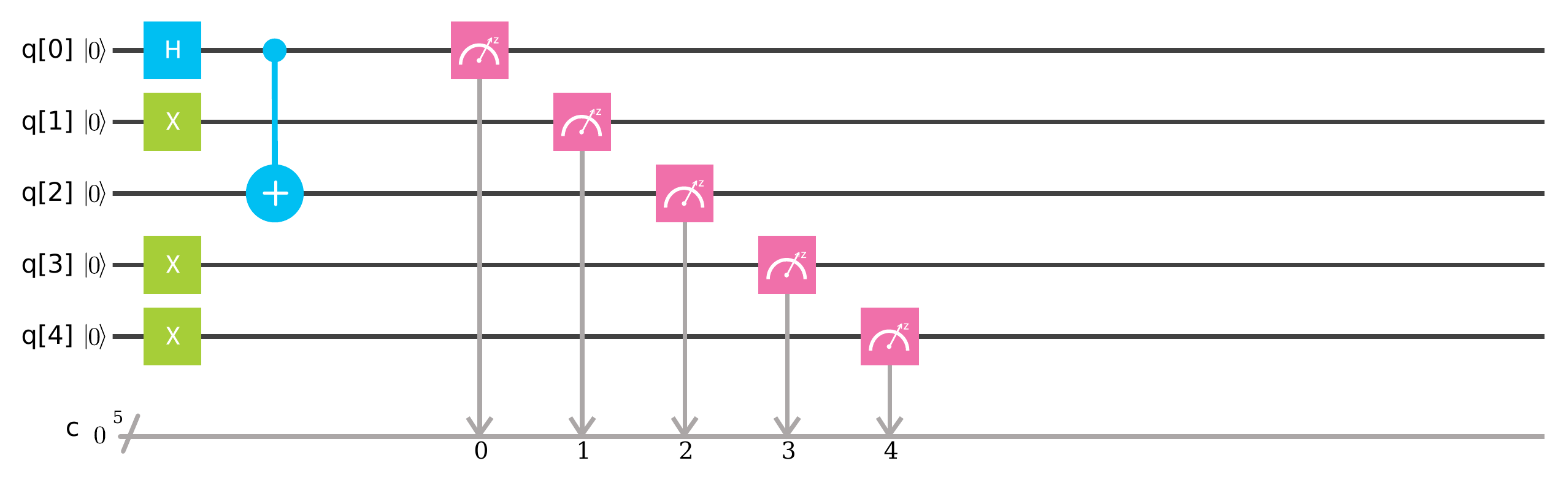} \\
    Reprint Courtesy of International Business Machines Corporation, \copyright~ International Business Machines Corporation. \\
\item $50\%$ chance of alive and $50\%$ chance of very dead
\\
\includegraphics[width=0.5\textwidth]{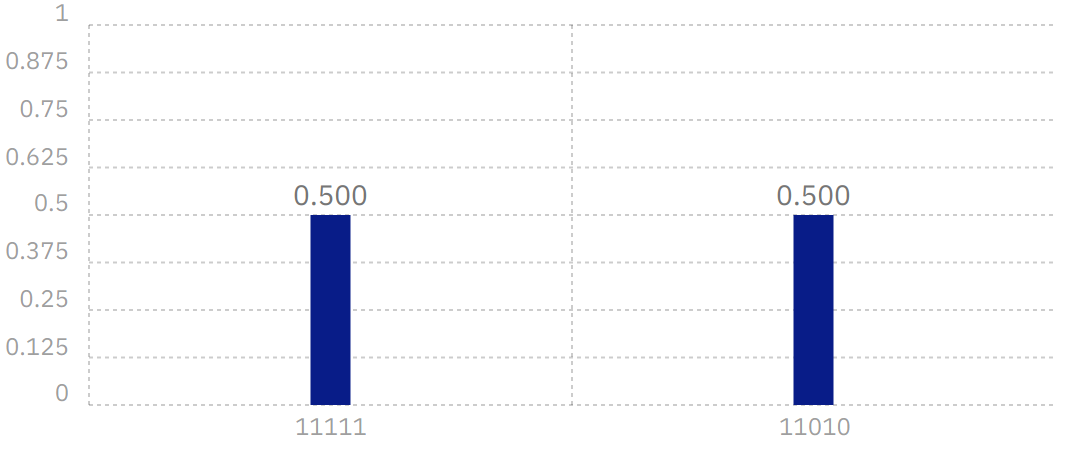} \\
Reprint Courtesy of International Business Machines Corporation, \copyright~ International Business Machines Corporation. \\
\item Adding a second CNOT undoes the entanglement.
\\
\includegraphics[width=0.35\textwidth]{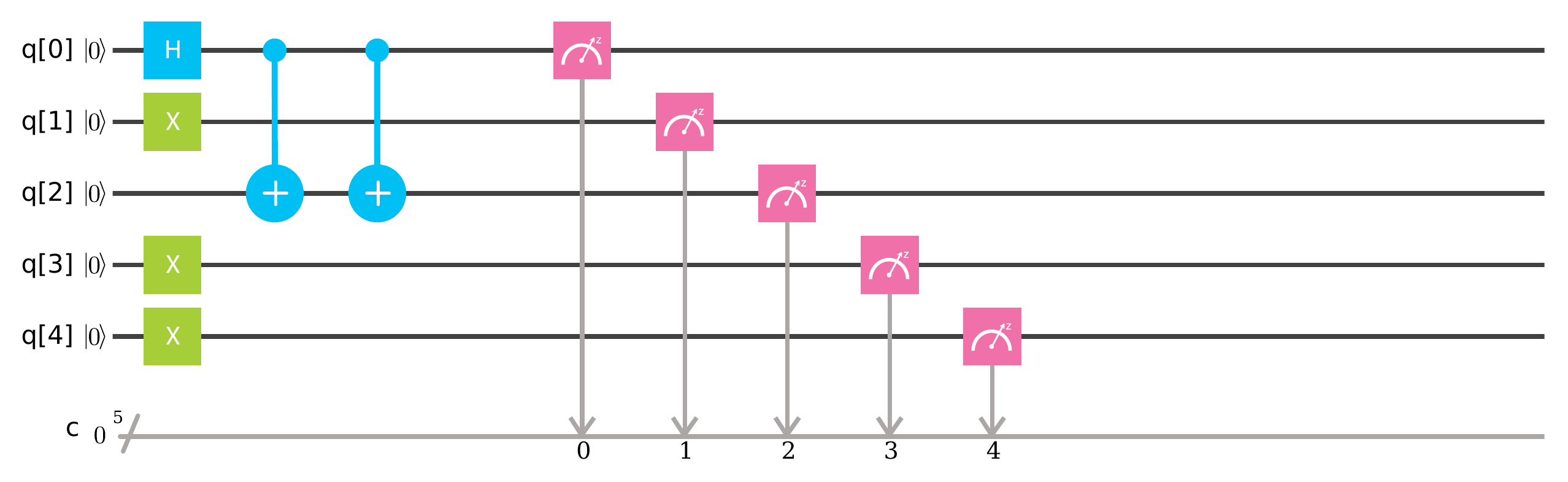} \\
    Reprint Courtesy of International Business Machines Corporation, \copyright~ International Business Machines Corporation. \\
    \includegraphics[width=0.35\textwidth]{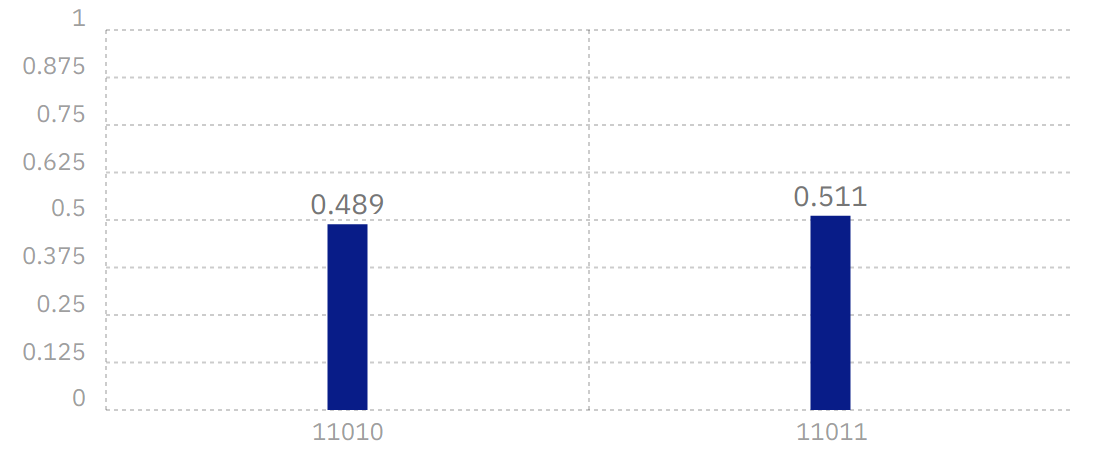} \\
Reprint Courtesy of International Business Machines Corporation, \copyright~ International Business Machines Corporation. \\
Adding a second Hadamard undoes the superposition.
\\
\includegraphics[width=0.35\textwidth]{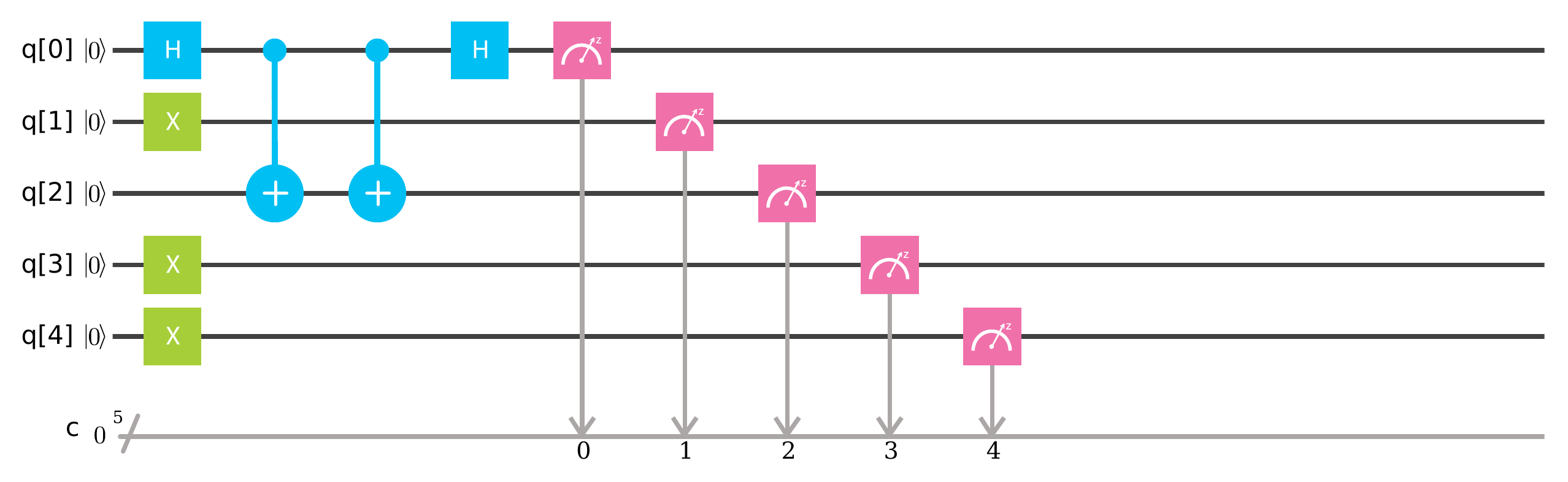} \\
Reprint Courtesy of International Business Machines Corporation, \copyright~ International Business Machines Corporation. \\
\includegraphics[width=0.25\textwidth]{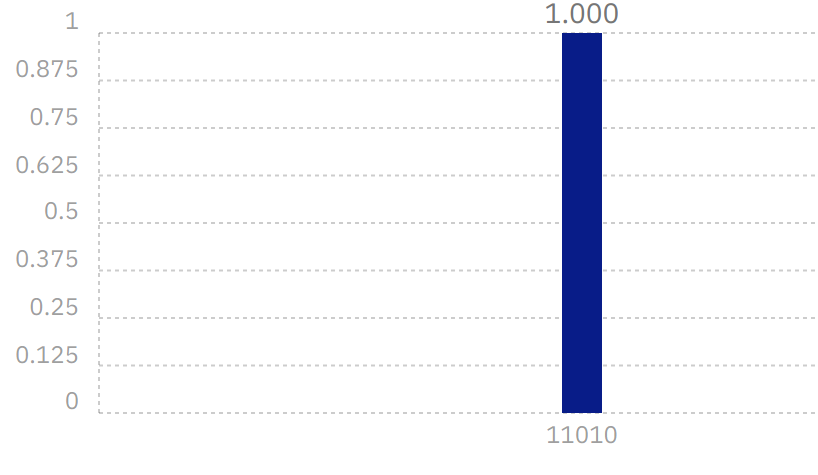} \\
Reprint Courtesy of International Business Machines Corporation, \copyright~ International Business Machines Corporation. \\
\\
\end{enumerate}

\newpage

\section{\advanced{0.5pt} Superposition vs. Mixed States Lab}
\label{chapter:WorksheetSuper}
\section*{Objectives:}
\begin{itemize}
    \item Experimentally determine the difference between particles in a \textbf{superposition state} and a \textbf{mixed state} using the \href{https://www.st-andrews.ac.uk/physics/quvis/simulations\_html5/sims/superposition/superposition-mixed-states.html}{superposition states and mixed states simulator}.\footnote{\href{https://www.st-andrews.ac.uk/physics/quvis/simulations\_html5/sims/superposition/superposition-mixed-states.html}{https://www.st-andrews.ac.uk/physics/quvis/simulations\_html5/sims/superposition/superposition-mixed-states.html}}
    \item Apply the idea of basis changing to explain the experimental results.
    \item Compute the probability amplitudes given measurement results.
\end{itemize}

\section*{Questions:}
\begin{enumerate}
    \item We send 100 electrons of unknown spin into a Stern-Gerlach apparatus.  We measure that 50 are spin up and 50 are spin down.  We can conclude that:
    \begin{enumerate}[a)]
        \item 100 electrons were in a 50/50 superposition state of up and down (superposition state).
        \item The electrons were a {{classical}} mixture of 50 electrons spin up and 50 spin down (mixed state).
        \item Not enough information
    \end{enumerate}
    \item Use the simulator to compare the measurement outcomes of the mixed particles vs. the superposition particles. What are the similarities and differences? \\
    \includegraphics[width=0.9\textwidth]{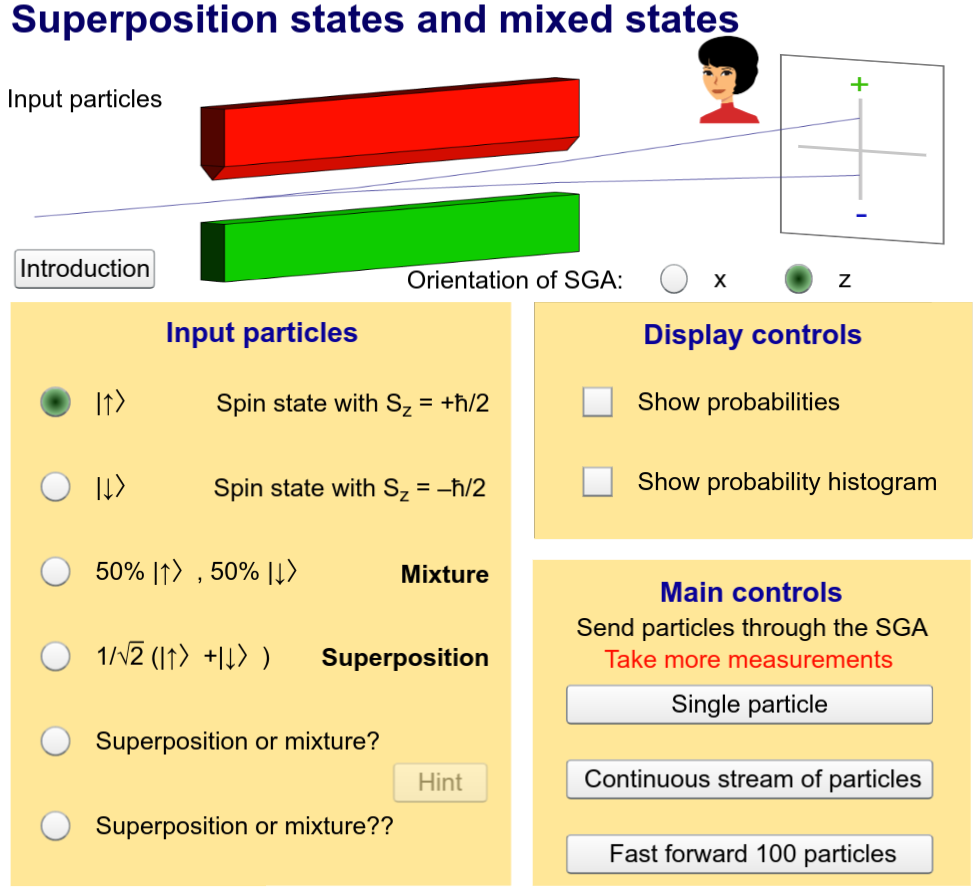} \\ Figure reproduced from the \href{https://www.st-andrews.ac.uk/physics/quvis/}{QuVis website}, licensed under creative commons CC-BY-NC-SA. \\
    \item By making a basis change with $\ket{0}=\frac{1}{\sqrt{2}}\ket{+}+\frac{1}{\sqrt{2}}\ket{-}$ and $\ket{1}=\frac{1}{\sqrt{2}}\ket{+}-\frac{1}{\sqrt{2}}\ket{-}$, can you explain the similarities and differences mathematically?
    \item Which of the two inputs labelled ``Superposition or mixture?'' and ``Superposition or mixture??'' is a random mixture and which is a superposition?
    \item The mixture consists of a fraction $A$ of spin up particles and a fraction $B$ of spin down particles. Find these fractions, $A$ and $B$.
    \item The superposition state can be written as $\alpha\ket{0}+\beta\ket{1}$. Find the amplitudes $\alpha$ and $\beta$ assuming they are real and positive.
    \item Use a basis change to show that the amplitudes $\alpha$ and $\beta$ give the correct probabilities in both the $x$- and $z$- basis.
\end{enumerate}

\newpage

\section*{Answers}
\begin{enumerate}
    \item C. The output is indistinguishable in the $z$-basis
    \item Superposition particles always have the same measurement outcome in the $x$-basis, while mixed particles have random spins in the $x$-basis.
    \item Changing the superposition state from the $z$- to $x$-basis:
    \begin{equation*}
        \frac{1}{\sqrt{2}}\left(\ket{0}+\ket{1}\right) = \frac{1}{\sqrt{2}}\left(\frac{1}{\sqrt{2}}\ket{+}+\frac{1}{\sqrt{2}}\ket{-}+\frac{1}{\sqrt{2}}\ket{+}-\frac{1}{\sqrt{2}}\ket{-}\right)=\ket{+},
    \end{equation*}
    so only +$x$ will be measured. Whereas in a mix of $\ket{0}=\frac{1}{\sqrt{2}}\ket{+}+\frac{1}{\sqrt{2}}\ket{-}$ and $\ket{1}=\frac{1}{\sqrt{2}}\ket{+}-\frac{1}{\sqrt{2}}\ket{-}$, both +$x$ and -$x$ will be measured with 50/50 probability.
    \item ? is the mixture since the $x$-basis measurements are 50/50, showing no correlation.
    \item In the $z$-basis, we measure about 20\% spin up and 80\% spin down. Thus, $A=\frac{1}{5}$ and $B=\frac{4}{5}$.
    \item In the $z$-basis, we measure about 20\% spin up and 80\% spin down. The probabilities are $\frac{1}{5}$ and $\frac{4}{5}$, but the amplitudes are the square root of the probability: $\alpha=\frac{1}{\sqrt{5}}$ and $\beta=\frac{2}{\sqrt{5}}$.
    \item Changing the superposition state from the $z$ to $x$-basis:
    \begin{equation*}
        \alpha\ket{0}+\beta\ket{1}=\frac{1}{\sqrt{5}}\left(\frac{1}{\sqrt{2}}\ket{-}\right)+\frac{2}{\sqrt{5}}\left(\frac{1}{\sqrt{2}}\ket{+}-\frac{1}{\sqrt{2}}\ket{-}\right)=\frac{3}{\sqrt{10}}\ket{+}-\frac{1}{\sqrt{10}}\ket{-},
    \end{equation*}
    By squaring the amplitudes, we find 90\% probability of +$x$ and 10\% probability of -$x$.
\end{enumerate}

\newpage

\section{\intermediate{8pt} Measurement Basis Lab}
\label{sec:WosksheetMeasureLab}

\section*{Objectives}
\begin{itemize}
    \item Use the \href{https://phet.colorado.edu/sims/stern-gerlach/stern-gerlach_en.html}{PHET Stern-Gerlach Simulator}\footnote{\href{https://phet.colorado.edu/sims/stern-gerlach/stern-gerlach\_en.html}{https://phet.colorado.edu/sims/stern-gerlach/stern-gerlach\_en.html}} to see how changing the orientation of the Stern-Gerlach Apparatus (SGA) affects the spin measurement. 
    \item Perform calculations to write the spin in a different measurement basis.
\end{itemize}

\begin{figure}[h!]
    \centering
    \includegraphics[width=0.75\textwidth]{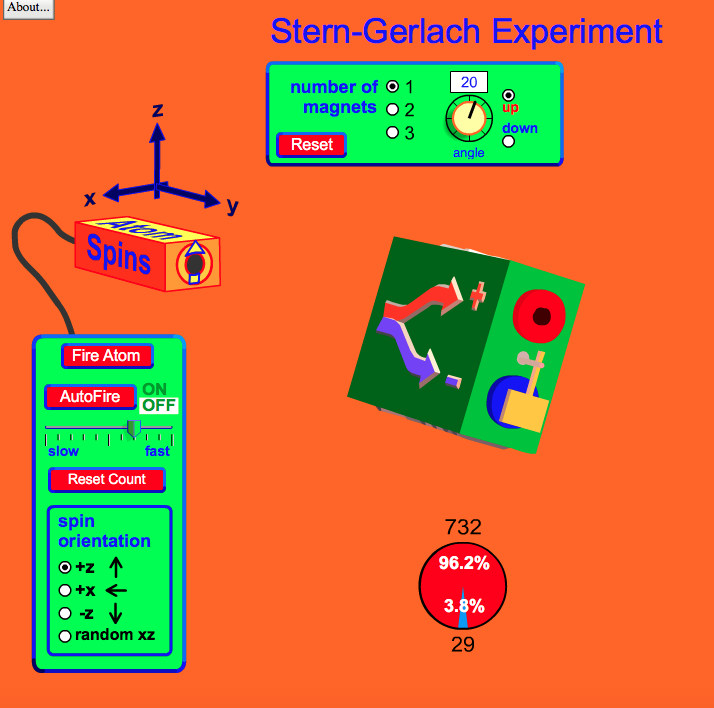}
 \caption{Figure reproduced from the \href{https://phet.colorado.edu/sims/stern-gerlach/stern-gerlach_en.html}{PHET Stern-Gerlach Simulator website}, licensed under creative commons CC-BY.}
\end{figure}

\begin{center}
\begin{tabular}{| c | p{3cm} | p{3cm} |}
    \hline
    Angle of SGA ($\theta_{SGA}$) & Probability of going through & Probability of being blocked \\
    \hline
    $0^\circ$ & & \\[0.5cm]
    \hline
    $15^\circ$ & & \\[0.5cm]
    \hline
    $30^\circ$ & & \\[0.5cm]
    \hline
    $45^\circ$ & & \\[0.5cm]
    \hline
    $60^\circ$ & & \\[0.5cm]
    \hline
    $75^\circ$ & & \\[0.5cm]
    \hline
    $90^\circ$ & & \\[0.5cm]
    \hline
    $105^\circ$ & & \\[0.5cm]
    \hline
    $120^\circ$ & & \\[0.5cm]
    \hline
    $135^\circ$ & & \\[0.5cm]
    \hline
    $150^\circ$ & & \\[0.5cm]
    \hline
    $165^\circ$ & & \\[0.5cm]
    \hline
    $180^\circ$ & & \\[0.5cm]
    \hline
\end{tabular}
\end{center}

\section*{Questions}
\begin{enumerate}
    \item Send spin up electrons through a single SGA and record the measurement probabilities for different SGA angles.
    \item Generate a scatter plot of the data.
    \item What function describes the shape of the graph?
    \item Write the state of the spin up electron as a superposition for an arbitrary SGA angle ($\theta_{SGA}$). In other words, find $\alpha$ and $\beta$ in $\ket{\text{electron}} = \alpha \ket{\text{goes through}} + \beta\ket{\text{blocked}}$. The diagram below may help, but note that $\theta \neq \theta_{SGA}$. \\
        \begin{center}
        \includegraphics[width=0.75\textwidth]{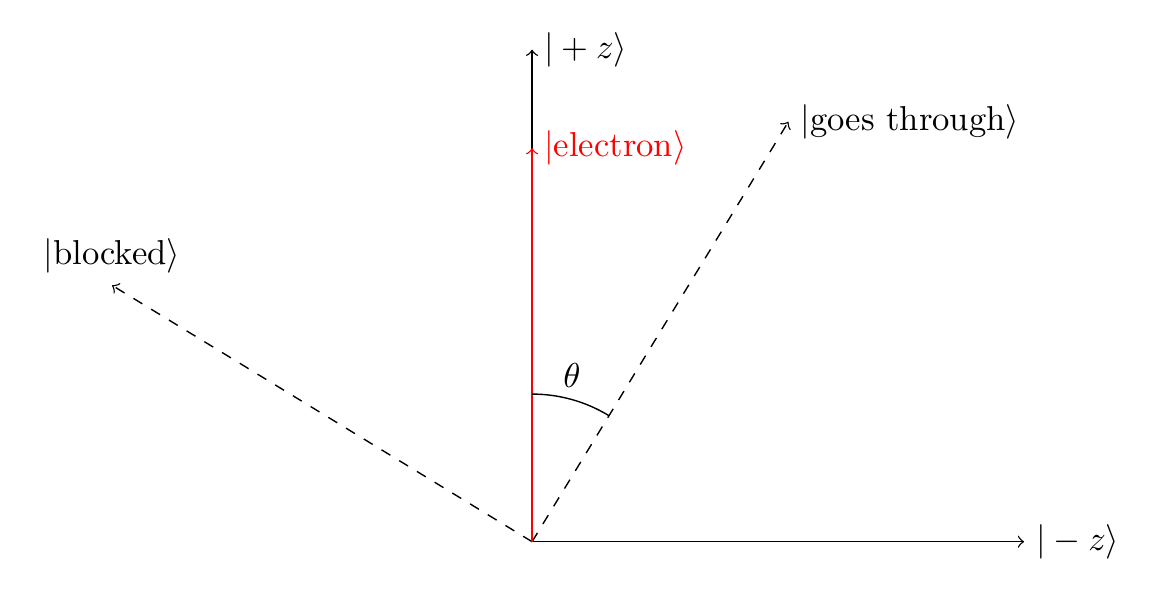}
        \end{center}
    \item Do the theoretical probabilities match the simulated data?
\end{enumerate}

\begin{enumerate}[resume]
    \item What would your scatter plot look like if you sent electrons through with the random $xz$ spin option?
    \item What is the theoretical probability of spin down electrons passing through a SGA angled at $45^\circ$?
    \item What is the theoretical probability of spin $+x$ electrons passing through a SGA angled at $45^\circ$?
\end{enumerate}

\newpage

\section*{Answers}
\begin{enumerate}
    \setcounter{enumi}{1}
    \item Sample Data: \\
        \begin{center}
        \includegraphics[width=0.75\textwidth]{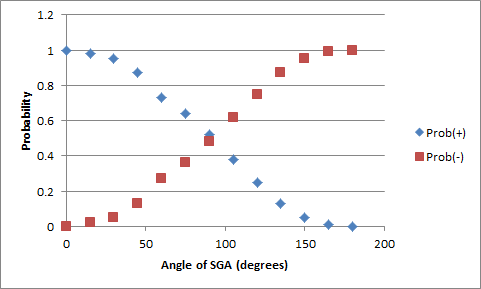}
        \end{center}
    \item Cosine squared function
    \item $\alpha = \cos \theta = \cos\frac{\theta_{SGA}}{2},\quad \beta=\sin\frac{\theta_{SGA}}{2}$
    \setcounter{enumi}{5}
    \item 50/50 probability independent of angle. The graph would be constant at Probability=0.5.
    \item Opposite of the probabaility for the spin up electron, so $1-\cos^2( 22.5^\circ) = 0.146$.
    \item $\sin^2 (22.5^\circ) = 0.146$
\end{enumerate}

\newpage

\section{\fundamental{5pt} One-Time Pad }
\label{sec:WorksheetOneTimePad}
\subsection*{ One-Time Pad - Alice}
\begin{table}[h!]\label{table:ABTab}
\centering
\begin{tabular}{| c | c | }
\hline
Character & Binary Code \\
\hline
$A$ & $01000001$ \\
\hline
$B$ & $01000010$ \\
\hline
$C$ & $01000011$ \\
\hline
$D$ & $01000100$ \\
\hline
$E$ & $01000101$ \\
\hline
$F$ & $01000110$ \\
\hline
$G$ & $01000111$ \\
\hline
$H$ & $01001000$ \\
\hline
$I$ & $01001001$ \\
\hline
$J$ & $01001010$ \\
\hline
$K$ & $01001011$ \\
\hline
$L$ & $01001100$ \\
\hline
$M$ & $01001101$ \\
\hline
$N$ & $01001110$ \\
\hline
$O$ & $01001111$ \\
\hline
$P$ & $01010000$ \\
\hline
$Q$ & $01010001$ \\
\hline
$R$ & $01010010$ \\
\hline
$S$ & $01010011$ \\
\hline
$T$ & $01010100$ \\
\hline
$U$ & $01010101$ \\
\hline
$V$ & $01010110$ \\
\hline
$W$ & $01010111$ \\
\hline
$X$ & $01011000$ \\
\hline
$Y$ & $01011001$ \\
\hline
$Z$ & $01011010$\\
\hline
\end{tabular}
\caption{One-time pad (Alice).}
\end{table}
\newpage 
Before parting ways, you and Bob agree on a key.   Using a coin with heads $=0$ and tails $=1$, randomly generate a key of the same length as the message. Make sure that you and Bob have the same key.\\

Shared Key: \\

\begin{tabular}{| p{1.75cm} | p{1.75cm} | p{1.75cm} | p{1.75cm} | p{1.75cm} | p{1.75cm} | p{1.75cm} | p{1.75cm} |}
    \hline
     & & & & & & & \\[0.75cm]
    \hline
     & & & & & & & \\[0.75cm]
    \hline
\end{tabular}

\vspace{1cm}

\underline{Encoding}:\\
\begin{enumerate}
\item Choose a secret letter to send to Bob in binary.
Message:\\
\hspace*{-1cm}
\begin{tabular}{| p{1.75cm} | p{1.75cm} | p{1.75cm} | p{1.75cm} | p{1.75cm} | p{1.75cm} | p{1.75cm} | p{1.75cm} |}
    \hline
     & & & & & & & \\[0.75cm]
    \hline
     & & & & & & & \\[0.75cm]
    \hline
\end{tabular}

\vspace{1cm}
\item  Add the key to your message, bit by bit, to encode the message. In binary, $0+0=0$, $0+1=1+0=1$, and $1+1=0$. For example, if the key $=0110$ and the message $=1101$, then the cipher text $=1011$, as $0110 + 1101= 1011$.\\
\begin{center}
Cipher Text:\\
\end{center}
\hspace*{-1cm}
\begin{tabular}{| p{1.75cm} | p{1.75cm} | p{1.75cm} | p{1.75cm} | p{1.75cm} | p{1.75cm} | p{1.75cm} | p{1.75cm} |}
    \hline
     & & & & & & & \\[0.75cm]
    \hline
     & & & & & & & \\[0.75cm]
    \hline
\end{tabular}

\item Send the cipher text to Bob.
\end{enumerate}
\underline{Decoding}
\begin{enumerate}
\item Write down the cipher received from Bob. \\
    \hspace*{-1cm}\begin{tabular}{| M{2cm} | p{1.5cm} | p{1.5cm} | p{1.5cm} | p{1.5cm} | p{1.5cm} | p{1.5cm} | p{1.5cm} | p{1.5cm} |}
    \hline
    Cipher from Bob & & & & & & & & \\[0.75cm]
    \hline
    Shared Key & & & & & & & & \\[0.75cm]
    \hline
\end{tabular}
\item  Add the key to Bob's message, bit by bit, to decode the message. \\

    \hspace*{-1cm}\begin{tabular}{| M{2cm} | p{1.5cm} | p{1.5cm} | p{1.5cm} | p{1.5cm} | p{1.5cm} | p{1.5cm} | p{1.5cm} | p{1.5cm} |}
    \hline
    Decoded Message & & & & & & & & \\[0.75cm]
    \hline
\end{tabular}
\item What was the message?
\end{enumerate}
\underline{Eavesdropping}
\begin{enumerate}
\item Swap cipher texts with another group.   How could you recover the original message?
\item How many different keys would you need to try?
\item If the original message had five letters instead of one letter, how many different keys would you need to try?
\item You intercept a five letter message and, by chance, find a key that decrypts it to read HELLO. What other words could it possibly be?
\end{enumerate}
\underline{Questions}
\begin{enumerate}
\item Why does adding the key to the cipher recover the original message?
\item Why is the one-time pad theoretically unbreakable?
\item What is the practical security flaw in the one-time pad?
\end{enumerate}

\newpage

\subsection*{\fundamental{5pt} One-Time Pad (Bob)}
\begin{table}[h!]\label{table:ABCTab}
\centering
\begin{tabular}{| c | c | }
\hline
Character & Binary Code \\
\hline
$A$ & $01000001$ \\
\hline
$B$ & $01000010$ \\
\hline
$C$ & $01000011$ \\
\hline
$D$ & $01000100$ \\
\hline
$E$ & $01000101$ \\
\hline
$F$ & $01000110$ \\
\hline
$G$ & $01000111$ \\
\hline
$H$ & $01001000$ \\
\hline
$I$ & $01001001$ \\
\hline
$J$ & $01001010$ \\
\hline
$K$ & $01001011$ \\
\hline
$L$ & $01001100$ \\
\hline
$M$ & $01001101$ \\
\hline
$N$ & $01001110$ \\
\hline
$O$ & $01001111$ \\
\hline
$P$ & $01010000$ \\
\hline
$Q$ & $01010001$ \\
\hline
$R$ & $01010010$ \\
\hline
$S$ & $01010011$ \\
\hline
$T$ & $01010100$ \\
\hline
$U$ & $01010101$ \\
\hline
$V$ & $01010110$ \\
\hline
$W$ & $01010111$ \\
\hline
$X$ & $01011000$ \\
\hline
$Y$ & $01011001$ \\
\hline
$Z$ & $01011010$\\
\hline
\end{tabular}
\caption{One-time pad (Bob).}
\end{table}

\newpage

Before parting ways, you and Alice agree on a key.   Using a coin with heads $=0$ and tails $=1$, randomly generate a key of the same length as the message. Make sure that you and Alice have the same key. \\

Shared Key: \\

\begin{tabular}{| p{1.75cm} | p{1.75cm} | p{1.75cm} | p{1.75cm} | p{1.75cm} | p{1.75cm} | p{1.75cm} | p{1.75cm} |}
    \hline
     & & & & & & & \\[0.75cm]
    \hline
     & & & & & & & \\[0.75cm]
    \hline
\end{tabular}

\vspace{1cm}

\underline{Encoding}:\\
\begin{enumerate}
\item Choose a secret letter to send to Alice in binary.
Message:\\
\hspace*{-1cm}
\begin{tabular}{| p{1.75cm} | p{1.75cm} | p{1.75cm} | p{1.75cm} | p{1.75cm} | p{1.75cm} | p{1.75cm} | p{1.75cm} |}
    \hline
     & & & & & & & \\[0.75cm]
    \hline
     & & & & & & & \\[0.75cm]
    \hline
\end{tabular}

\vspace{1cm}

\item Add the key to your message, bit by bit, to encode the message. In binary, $0+0=0$, $0+1=1+0=1$, and $1+1=0$. For example, if the key $=0110$ and the message $=1101$, then the cipher text $=1011$. $0110+ 1101 =1011$.\\

\begin{center}
Cipher Text:\\
\end{center}
\hspace*{-1cm}
\begin{tabular}{| p{1.75cm} | p{1.75cm} | p{1.75cm} | p{1.75cm} | p{1.75cm} | p{1.75cm} | p{1.75cm} | p{1.75cm} |}
    \hline
     & & & & & & & \\[0.75cm]
    \hline
     & & & & & & & \\[0.75cm]
    \hline
\end{tabular}

\item Send the cipher text to Alice.
\end{enumerate}

\underline{Decoding}
\begin{enumerate}
\item Write down the cipher received from Alice. \\
    \hspace*{-1cm}\begin{tabular}{| M{2cm} | p{1.5cm} | p{1.5cm} | p{1.5cm} | p{1.5cm} | p{1.5cm} | p{1.5cm} | p{1.5cm} | p{1.5cm} |}
    \hline
    Cipher from Alice & & & & & & & & \\[0.75cm]
    \hline
    Shared Key & & & & & & & & \\[0.75cm]
    \hline
\end{tabular}
\item  Add the key to Alice's message, bit by bit, to decode the message. \\
    \hspace*{-1cm}\begin{tabular}{| M{2cm} | p{1.5cm} | p{1.5cm} | p{1.5cm} | p{1.5cm} | p{1.5cm} | p{1.5cm} | p{1.5cm} | p{1.5cm} |}
    \hline
    Decoded Message & & & & & & & & \\[0.75cm]
    \hline
\end{tabular}
\item What was the message?
\end{enumerate}
\underline{Eavesdropping}
\begin{enumerate}
\item Swap cipher texts with another group.   How could you recover the original message?
\item How many different keys would you need to try?
\item If the original message had five letters instead of one letter, how many different keys would you need to try?
\item You intercept a five-letter message and, by chance, find a key that decrypts it to read HELLO. What other words could it possibly be?
\end{enumerate}
\underline{Questions}
\begin{enumerate}
\item Why does adding the key to the cipher recover the original message?
\item Why is the one-time pad theoretically unbreakable?
\item What is the practical security flaw in the one-time pad?
\end{enumerate}

\clearpage
\pagebreak

\section{\fundamental{5pt} BB84 Quantum Key Distribution}
\label{sec:WorksheetBB84}
\subsection*{BB84 Quantum Key Distribution - Alice}

\underline{No Eavesdropper}
\begin{enumerate}
    \item Randomly choose to prepare the electron in either the $x$- or $z$-basis.
    \item The electron that's sent through your Stern-Gerlach apparatus will either be in a 0 or 1 state. You can randomize this by flipping a coin.
    \item Pass the correct spin card to Bob face down. \\
        \includegraphics[width=0.5\textwidth]{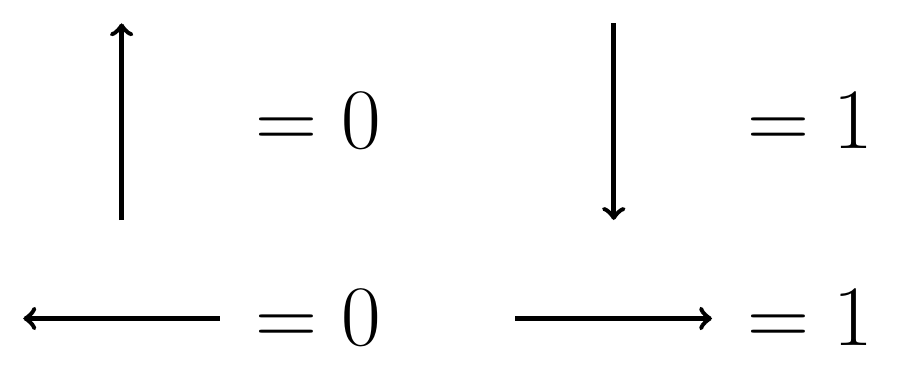}\hspace{2cm}\includegraphics[width=0.3\textwidth]{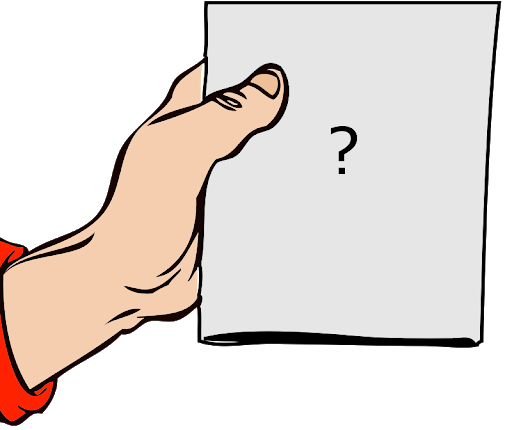}
    \item Once you have filled up the chart, tell Bob the basis used for each bit. If Bob tells you to "discard" the bit, cross it out on your chart.
    \item Check to see that you and Bob end up with the same sifted key.
\end{enumerate}

\hspace{-1cm}
\begin{tabular}{| M{1.5cm} | p{1.5cm} | p{1.5cm} | p{1.5cm} | p{1.5cm} | p{1.5cm} | p{1.5cm} | p{1.5cm} | p{1.5cm} |}
    \hline
    Basis: $x$ or $z$ & & & & & & & & \\[0.75cm]
    \hline
    Bit Value: 0 or 1 & & & & & & & & \\[0.75cm]
    \hline
\end{tabular}

\vspace{1cm}
\hspace{2cm} SIFTED KEY:$\ \rule{6cm}{0.15mm}$

\clearpage
\pagebreak

\underline{With Eavesdropper}
\begin{enumerate}
    \item Repeat the procedure, but instead of passing the spin card directly to Bob, it will first pass through Eve.
    \item Compare the sifted key bits one at a time. How can you tell if Eve intercepted the message?
\end{enumerate}
\vspace{1cm}
\hspace{2cm} SIFTED KEY:$\ \rule{6cm}{0.15mm}$
\vspace{1cm}

\begin{tabular}{| M{1.5cm} | p{1.5cm} | p{1.5cm} | p{1.5cm} | p{1.5cm} | p{1.5cm} | p{1.5cm} | p{1.5cm} | p{1.5cm} |}
    \hline
    Basis: $x$ or $z$ & & & & & & & & \\[0.75cm]
    \hline
    Bit Value: 0 or 1 & & & & & & & & \\[0.75cm]
    \hline
\end{tabular}

\clearpage
\newpage

\subsection*{BB84 Quantum Key Distribution - Bob}

\underline{No Eavesdropper}
\begin{enumerate}
    \item Randomly choose between the $x$- or $z$-basis.
    \item Receive the spin card from Alice and flip it over.
    \begin{itemize}
        \item If your basis is the same as the card's, record the bit value.
        \item If your basis is different, the output of your Stern-Gerlach apparatus will be random. Randomly pick 0 or 1.
    \end{itemize}
        \hspace{2cm}\includegraphics[width=0.5\textwidth]{WorksheetFigs/Bits.pdf}\hspace{2cm}\includegraphics[width=0.3\textwidth]{WorksheetFigs/WorksheetBB84Alice.png}
    \item Once you have filled up the chart, Alice will tell you the basis used for each bit. If you measured in a different basis, tell Alice to "discard" the bit and cross it out on your chart.
    \item Check to see that you and Alice end up with the same sifted key.
\end{enumerate}

\begin{tabular}{| M{1.5cm} | p{1.5cm} | p{1.5cm} | p{1.5cm} | p{1.5cm} | p{1.5cm} | p{1.5cm} | p{1.5cm} | p{1.5cm} |}
    \hline
    Basis: $x$ or $z$ & & & & & & & & \\[0.75cm]
    \hline
    Bit Value: 0 or 1 & & & & & & & & \\[0.75cm]
    \hline
\end{tabular}

\vspace{1cm}
\hspace{2cm} SIFTED KEY:$\ \rule{6cm}{0.15mm}$

\clearpage
\pagebreak

\underline{With Eavesdropper}
\begin{enumerate}
    \item Repeat the procedure, but instead of getting the spin card directly from Alice, it will first pass through Eve.
    \item Compare the sifted key bits one at a time. How can you tell if Eve intercepted the message?
\end{enumerate}
\vspace{1cm}
\hspace{2cm} SIFTED KEY:$\ \rule{6cm}{0.15mm}$
\vspace{1cm}

\begin{tabular}{| M{1.5cm} | p{1.5cm} | p{1.5cm} | p{1.5cm} | p{1.5cm} | p{1.5cm} | p{1.5cm} | p{1.5cm} | p{1.5cm} |}
    \hline
    Basis: $x$ or $z$ & & & & & & & & \\[0.75cm]
    \hline
    Bit Value: 0 or 1 & & & & & & & & \\[0.75cm]
    \hline
\end{tabular}

\clearpage
\newpage

\subsection*{BB84 Quantum Key Distribution - Eve}

\underline{With Eavesdropper (You!)}
\begin{enumerate}
    \item Randomly choose between the $x$- or $z$-basis.
    \item Receive the spin card from Alice and flip it over.
    \begin{itemize}
        \item If your basis is the same as the card's, record the bit value and pass it along to Bob.
        \item If your basis is different, the output of your Stern-Gerlach apparatus will be random. Randomly pick 0 or 1 for your bit value, and pass the card along to Bob.
    \end{itemize}
    \item Listen in as Alice and Bob compare their basis. If Bob says to "discard" the bit, cross it out on your chart.
    \item Compare your sifted key to Alice and Bob's key. Was your eavesdropping successful?
\end{enumerate}
\vspace{1cm}
\hspace{2cm} SIFTED KEY:$\ \rule{6cm}{0.15mm}$
\vspace{1cm}

\begin{tabular}{| M{1.5cm} | p{1.5cm} | p{1.5cm} | p{1.5cm} | p{1.5cm} | p{1.5cm} | p{1.5cm} | p{1.5cm} | p{1.5cm} |}
    \hline
    Basis: $x$ or $z$ & & & & & & & & \\[0.75cm]
    \hline
    Bit Value: 0 or 1 & & & & & & & & \\[0.75cm]
    \hline
\end{tabular}

\graphicspath{{Acknowledgements/}}
\backmatter
\chapter{Acknowledgments}

It is a pleasure to thank Marge Bardeen, Harry Cheung, and Spencer Pasero for helpful discussions on various aspects of this project, from inception to completion. We are grateful to Daniel Carney, William Jay, Yin Lin, Jim Simone,  Julia Stadler, Liner de Souza Santos  and  Anders Ellers Thomsen for  reading and providing feedback on the draft document. It is also a pleasure to thank LaMargo Gill for her remarkably thorough proofreading of this document. We thank Heath O'Connell and Aaron Sauers for their useful advice regarding information content. We thank Olivia Vizcarra and the Fermilab theory group for facilitating this project. 

This work would not be possible without funding from the Robert Noyce Teacher Scholarship and the Fermilab Teacher Research Associates (TRAC) program. 
This work was supported by Fermi Research Alliance, LLC, under Contract No. DE-AC02-07CH11359 with the U.S. Department of Energy, Office of Science, Office of High Energy Physics
 and  partial support was received  by an HEP-QIS QuantISED award titled ``Quantum Information Science for Applied Quantum Field Theory.''

Various sources were used as inspiration for building this course. We  acknowledge IBM Q experience\footnote{https://quantumexperience.ng.bluemix.net/qx} for their useful web interface, and note that specific figures (as indicated in their captions) are owned by IBM as per their \href{https://quantumexperience.ng.bluemix.net/qx/terms}{end-user license agreement}.\footnote{https://quantumexperience.ng.bluemix.net/qx/terms} We urge the reader to review this end-user license agreement before using the IBM Q web interface. Additionally, we would like to attribute the PhET Interactive Simulations for their useful interactive videos, and as per their license also acknowledge the University of Colorado Boulder and \href{https://phet.colorado.edu}{https://phet.colorado.edu}. Furthermore, we credit the \href{https://www.st-andrews.ac.uk/physics/quvis/}{Quantum Mechanics Visualization Project} (QuVis)\footnote{https://www.st-andrews.ac.uk/physics/quvis/}, hosted by the University of St.~Andrews, for useful interactive simulations. Finally, we thank Martin Laforest and the Communications and Strategic Initiatives Team at the Institute for Quantum Computing, University of Waterloo's \href{https://uwaterloo.ca/institute-for-quantum-computing/outreach}{outreach department}\footnote{https://uwaterloo.ca/institute-for-quantum-computing/outreach} for supplying material which formed the inspiration for  Chapters 3, 5 and 9 of this module.

\graphicspath{{Chapter12-Answers/}}
\chapter{Answers}

\section*{Chapter 1 Solutions}
\begin{enumerate}
\item \,
\begin{enumerate}[label=(\alph*)]
\item quantized to the charge of the electron: $e=1.6\times 10^{-19}$ C.
\item time is continuous
\item space is continuous
\item quantized to $\$0.01$
\item continuous because the frequency of light (which causes color) is continuous 
\end{enumerate}

\item It would either look all yellow or all red. 
\item No.  If we showed 100 copies of the picture to Student A, they would always see blue/black. In a 50/50 quantum superposition, they would see around 50 pictures as blue/black and the rest as white/gold. The two states must be an intrinsic property of the dress rather than something that depends on the observer. 
\end{enumerate}

\section*{Chapter 2 Solutions}

\begin{enumerate}
\item With the $8$-bit representation, this would require eight coins arranged as TTTTTTHT.

\item The decoded message says ``CAT''

\item 
  \begin{enumerate}[label=(\alph*)]
  \item 9/10 or 90$\%$. The probability is the square of the amplitude.
  \item No, $1/9+4/9\neq 1$. Normalization means the total probabilities add up to 1.
  \item 75 coins.
  \item Measurement collapses the superposition onto either $|H\rangle$ or $|T\rangle$. 
  \end{enumerate}
    
\item Yes, by measuring how many of the qubits collapse to the $|0\rangle$ state and how many collapse to the $|1\rangle$ state, we can determine what the square of the amplitudes are, and therefore what the initial state of all the particles was. Note, for the interested reader, by measuring the state in a different basis, it may be possible to gain information about the complex phase of the state. 
  
\item  
  \begin{enumerate}[label=(\alph*)]
  \item All of them are possible as they all can produce $|0\rangle$ after measurement. 
  \item After measurement the state in the question is in $|0\rangle$, since the superposition collapsed. Since it is in the $|0\rangle$ state after measurement, if you try to measure the same state again you will always measure $\lvert 0 \rangle$. No new information is provided about the state after the collapse.
  \item If $|0\rangle$ is measured from the unknown state, and a second identical state is prepared and is measured in the $|1\rangle$ state, then you know the unknown state contains some nonzero superposition of both $|0\rangle$ and $|1\rangle$, e.g., it is $|\psi\rangle = \alpha | 0\rangle + \beta|1\rangle$ with $\alpha\ne 0$ and $\beta\ne 0$. So you can rule out $|\psi\rangle = \lvert 0 \rangle$ as the initial state, but all of the other three states given in (a) are still possible. Measurements of many more particles are needed to determine the numerical values of $\alpha$ and $\beta$ in order to find the exact state. 
  \end{enumerate}
  
\item
  \begin{align}
    |1\rangle  &= 
  \begin{pmatrix}
    0  \\
    1  
  \end{pmatrix}.
  \end{align}

\item 
  \begin{align}
    |0\rangle  &= 
  \begin{pmatrix}
    1 \\
    0  
  \end{pmatrix}.
  \end{align}

\item   \begin{align} \lvert \Psi \rangle  = \beta\lvert 0 \rangle + \alpha\lvert 1 \rangle. 
\end{align}

\item
  \begin{equation}
Y^\dagger=\begin{pmatrix}
0 & -i\\
i & 0
\end{pmatrix}.
\end{equation}

\item 
  \begin{equation}
U^{\dagger}U=\frac{1}{2}\begin{pmatrix}
2 & 0\\
0& 2
\end{pmatrix}.
\end{equation}

\end{enumerate}

\section*{Chapter 3 Solutions}

\begin{enumerate}

\item All parts of the statement are false. The particle is in both states the entire time before measurement and has a 50$\%$ chance of being measured as 0 or 1.

\item A single photon can either be in a $0$ or $1$ state. So the possible states that can make up a superposition of two photons, labeled photon A and photon B (and hence the different states that can be measured), are  $\lvert0_A0_B\rangle$, $\lvert1_A0_B\rangle$, $\lvert0_A1_B\rangle$, and $\lvert1_A1_B\rangle$. Both detectors are activated at the same time when we have one of the two photons in the $0$ detector and one in $1$ detector. This is either the $0_A1_B$ state or the $1_A0_B$ state. There are four total possible states, and two of them trigger both detectors so the probability is $2/4=50\%$. When there are three photons, the possible states are $000$, $010$, $011$, $101$, $110$, $001$, $100$ and $111$. Both detectors are activated for any state that has both a $0$ and a $1$. So both detectors will not be triggered by only the $000$ and $111$ states. Since there are six states that trigger both detectors, and eight states total, the probability of both detectors being triggered is $6/8 = 75\%$. Ten photons have $2^{10}$ possible outcomes, where only two outcomes do not trigger both detectors (one state is all zeros, and the other is all ones). So the probability is $(2^{10}-2)/(2^{10})=99.8\%$.

\item 
\begin{enumerate}[label*=\alph*.]

\item The coincidence counts are low when the detectors are an equal distance from the beam splitter. This points to light behaving like a particle entering only one detector at a time.
  
\item  Photons from different 0.4$\mu$s bursts can arrive at the detectors simultaneously.
\end{enumerate}

\item \begin{equation}
\frac{1}{\sqrt{2}}\begin{pmatrix}
-1 & 1\\
1& 1
\end{pmatrix}
\begin{pmatrix}
0\\1
\end{pmatrix}=
\begin{pmatrix}
1\\
1
\end{pmatrix}.
\end{equation}

\begin{equation}
\frac{1}{\sqrt{2}}\begin{pmatrix}
-1 & 1\\
1& 1
\end{pmatrix}
\begin{pmatrix}
0\\1
\end{pmatrix}=
\begin{pmatrix}
1\\
1
\end{pmatrix}.
\end{equation}

\item A $30/70$ superposition state would take the form: 
\begin{equation}
\sqrt{\frac{3}{10}}\lvert 0\rangle + \sqrt{\frac{7}{10}}\lvert 1\rangle.
\end{equation}
The desired beam splitter matrix $M$ should perform the operation:
\begin{equation}
M \begin{pmatrix}
0\\1
\end{pmatrix}=
\begin{pmatrix}
\sqrt{\frac{3}{10}}\\
\sqrt{\frac{7}{10}}
\end{pmatrix}
\implies 
M=\begin{pmatrix}
1 & \sqrt{\frac{3}{10}}\\
1& \sqrt{\frac{7}{10}}
\end{pmatrix}
\end{equation}
would give the correct probabilities, but it is not unitary. And all quantum matrices must be unitary. Using the Hadamard matrix as a reference, the unitary $30/70$ beam splitter matrix is 
\begin{equation}
M=\begin{pmatrix}
 \sqrt{\frac{7}{10}} & \sqrt{\frac{3}{10}}\\
-\sqrt{\frac{3}{10}}& \sqrt{\frac{7}{10}}
\end{pmatrix}.
\end{equation}

\item Congratulations you have figured out the path of the photon! However, by seeing the location of the photon after the first beam splitter, you have collapsed its superposition state.  Therefore, it goes into the second beam splitter from the top, where it exits in a superposition state with $50\%$ probability of triggering Detector 1 or 2.

\end{enumerate}

\section*{Chapter 4 Solutions}

\begin{enumerate}
\item 

\begin{minipage}{\linewidth}
\centering
  \includegraphics[width=0.25\textwidth]{xzanswe.png}
\end{minipage}

\item No. $\lvert + \rangle =1/\sqrt{2}\lvert 0 \rangle + 1/\sqrt{2}\lvert 1 \rangle$ 
are not independent of each other. $\lvert 0 \rangle$ and $\lvert + \rangle$ satisfy condition $\#$1 but not condition $\#$2. 

\item 
\begin{enumerate}[label=(\alph*)]
\item $\cos(\pi/6)\lvert 0 \rangle + \sin(\pi/6)\lvert 1 \rangle$
\item $ \alpha^2=\cos^2(\pi/6)=0.75$
\item  $\cos(\pi/6 + \pi/4)\lvert 0 \rangle + \cos(\pi/6 + \pi/4)\lvert 1 \rangle$
\item $\cos^2(\pi/6 + \pi/4)\approx 0.067$
\end{enumerate}

\item D. The two states are $\lvert + \rangle$ and $\lvert - \rangle$.  The horizontal SGA would distinguish them with 100$\%$ certainty.  They would also produce different probabilities in the diagonal SGA.
\item $100\%$ $+z$

\item
\begin{enumerate}[label=(\alph*)]
\item  Nothing comes out since the $+z$ entering the second SGA is blocked.
\item $100\%$ $+z$.
\end{enumerate}

\item $50\%$ up, $50\%$ down
\item Nothing:  the second SGA blocks the electron.
\item $50\% +x$, 50$\% -x$
\item The $-z$ selected by the 50$\%$ $+x$, 50$\%$ $-x$

\end{enumerate}

\section*{Chapter 5 Solutions}
\begin{enumerate}
    \item Their basis will match half of the time, so their key will be 500,000 bits long.
    \item The probability is $50\%$. Knowing that it’s the $z$-basis could mean either 0 or 1 with equal probability. Similarly in the $x$-basis. 
    \item $\left({1}/{2}\right)^{20} \approx 10^{-6}$. 
    \item Not with certainty. As mentioned in the one-time pad exercise, different keys could give a meaningful message.  You couldn’t tell which one was the correct one.
    \item If Alice and Bob both use the $z$-basis, the different cases are:
    \begin{enumerate}
        \item Alice sends +$z$, Eve measures in $z$, Bob measures +$z$. \checkmark
        \item Alice sends -$z$, Eve measures in $z$, Bob measures -$z$. \checkmark
        \item Alice sends +$z$, Eve measures in $x$, Bob measures +$z$ (will happen with 50\% probability). \checkmark
        \item Alice sends +$z$, Eve measures in $x$, Bob measures -$z$ (will happen with 50\% probability). 
        \item Alice sends -$z$, Eve measures in $x$, Bob measures +$z$ (will happen with 50\% probability). 
        \item Alice sends -$z$, Eve measures in $x$, Bob measures -$z$ (will happen with 50\% probability). \checkmark
    \end{enumerate}
    Therefore there is a $4/6$ probability that Eve has not been detected. 
    \item $\left({4}/{6}\right)^{20} \approx 0.0003$. 
    \item Copy the state of each electron, passing the originals along to Bob. Once the correct basis is revealed, pass those cloned electrons through SGAs oriented in the correct basis and get the key.
\end{enumerate}

\section*{Chapter 6 Solutions}

\begin{enumerate}

\item 

  \begin{enumerate}[label=(\alph*)]
  \item
    \begin{equation}
    X|0\rangle =
    \begin{pmatrix}
      0 & 1 \\
      1 & 0
    \end{pmatrix}
    \begin{pmatrix}
      1 \\
      0 
    \end{pmatrix}
    = 
    \begin{pmatrix}
      0 \\
      1
    \end{pmatrix}
    = |1\rangle .
  \end{equation}
    
  \item 
    \begin{equation}
    X|\psi\rangle =
    \begin{pmatrix}
      0 & 1 \\
      1 & 0
    \end{pmatrix}
    \begin{pmatrix}
      \alpha \\
      \beta 
    \end{pmatrix}
    = 
    \begin{pmatrix}
      \beta \\
      \alpha
    \end{pmatrix}
    = \beta|0\rangle + \alpha|1\rangle .
  \end{equation}
  \end{enumerate}
  
\item The $50/50$ beam splitter is an example of a Hadamard gate because it puts the photon into a $50/50$ superposition.

\item A; the superposition has collapsed.

\item 
  \begin{enumerate}[label=(\alph*)]
  \item An $X$ gate reverses the state of the qubit.
  \item It could either be ON or OFF with equal probability.
  \item The light bulb will be ON with $100\%$ probability. Applying a second Hadamard undoes the first Hadamard. This is a non-classical result because the first Hadamard creates the $50/50$ superposition no matter whether its input is originally ON or OFF. 
  \end{enumerate}

\item Sending spin up electrons into a horizontal SGA is identical to applying a Hadamard gate to a $|0\rangle$ qubit. 

\item The Mach-Zehnder experiment is essentially two Hadamard gates in a row. The second gate undoes the superposition and returns a definite state.

\item
  \begin{enumerate}[label=(\alph*)]
    \item 
      \begin{equation}
        H|1\rangle =\frac{1}{\sqrt{2}}
        \begin{pmatrix}
          1 & 1 \\
          1 & -1
        \end{pmatrix}
        \begin{pmatrix}
          0 \\
          1 
        \end{pmatrix}
        = \frac{1}{\sqrt{2}} 
        \begin{pmatrix}
          1 \\
          -1
        \end{pmatrix} = |-\rangle.
      \end{equation}
    \item
      \begin{equation}
        HH|1\rangle =\frac{1}{{2}}
        \begin{pmatrix}
          1 & 1 \\
          1 & -1
        \end{pmatrix}
        \begin{pmatrix}
          1 \\
          -1 
        \end{pmatrix}
        = 
        \begin{pmatrix}
          0 \\
          1
        \end{pmatrix} = |1\rangle. 
      \end{equation}
    \item
      \begin{equation}
        HH|1\rangle =\frac{1}{{2}}
        \begin{pmatrix}
          1 & 1 \\
          1 & -1
        \end{pmatrix}
        =
        \begin{pmatrix}
          1 & 1 \\
          1 & -1
        \end{pmatrix}
        \begin{pmatrix}
          \alpha \\
          \beta
        \end{pmatrix}
        = 
        \begin{pmatrix}
          \alpha \\
          \beta
      \end{pmatrix}\hspace{-1mm}. 
      \end{equation}      
  \end{enumerate}

\item $|0\rangle \rightarrow X \rightarrow X \rightarrow |1\rangle$ and  $|0\rangle \rightarrow H \rightarrow X \rightarrow H \rightarrow |1\rangle$.

  \item
  \begin{equation}
    Z|+\rangle =\frac{1}{\sqrt{2}}
    \begin{pmatrix}
      1 & 0 \\
      0 & -1
    \end{pmatrix}
    \begin{pmatrix}
      1 \\
      1
    \end{pmatrix}
    =
    \begin{pmatrix}
      \alpha \\
      \beta
    \end{pmatrix}
    = \frac{1}{\sqrt{2}}
    \begin{pmatrix}
      1 \\
      -1
    \end{pmatrix}=|-\rangle. 
  \end{equation}      
  
\item $X=HZH$.

\item 
  \begin{enumerate}[label=(\alph*)]
  \item The $Z$ gate does not affect the $|0\rangle$ state.
  \item The sign on the $|1\rangle$ state is changed, but this does not affect probabilities and so cannot be seen in the histogram.
  \item The $|+\rangle$ is changed to a $|-\rangle$, which shows up as $50\%$ $|0\rangle$ and $|1\rangle$.
  \item The $|-\rangle$ is changed to a $|+\rangle$, which shows up as $50\%$ $|0\rangle$ and $|1\rangle$.
  \end{enumerate}

\item $100\%\ |0\rangle$ as shown in Figure \ref{fig:XHZGates}.

\item 
  \begin{equation}
    H^{\dagger}H = \frac{1}{{2}}
    \begin{pmatrix}
      1 & 1 \\
      1 & -1
    \end{pmatrix}
    \begin{pmatrix}
      1 & 1 \\
      1 & -1
    \end{pmatrix}
    = \frac{1}{{2}}
    \begin{pmatrix}
      2 & 0 \\
      0 & 2
    \end{pmatrix}
    = 
    \begin{pmatrix}
      1 & 0 \\
      0 & 1
    \end{pmatrix}.
  \end{equation}      
  Performing the Hadamard operation twice is the same as multiplying by the identity matrix.  Thus, the qubit is unchanged. 
\end{enumerate}

\section*{Chapter 7 Solutions}
\begin{enumerate}
\item
  \begin{enumerate} [a)]
  \item The probability of measuring $\lvert00\rangle$ is 
    \begin{equation}
      \mathcal{P}(\lvert00\rangle) = \lvert\langle 00\lvert \psi\rangle\lvert^2 = \left(\frac{1}{\sqrt{2}}\right)^2. 
    \end{equation}
    We get this by taking the coefficient of the $\lvert00\rangle$ term and then squaring it. 
  \item The probability of measuring the first qubit as 1, $\mathcal{P}(\text{first qubit }\lvert1\rangle)$, is the sum of all outcomes which have the first qubit in the $\lvert1\rangle$ state. In this example, this is $\mathcal{P}(\text{first qubit }\lvert1\rangle) = \mathcal{P}(\lvert10\rangle) + \mathcal{P}(\lvert11\rangle)$, which is equal to
    \begin{equation}
      \left(\frac{1}{2}\right)^2+\left(\frac{-1}{2}\right)^2=\frac{1}{2}. 
    \end{equation}
  \item The probability of measuring the second qubit as 0, $\mathcal{P}(\text{second qubit }\lvert0\rangle)$, is the sum of all outcomes which have the second qubit in the $\lvert0\rangle$ state. In this example, this is $\mathcal{P}(\text{second qubit }\lvert0\rangle) = \mathcal{P}(\lvert00\rangle) + \mathcal{P}(\lvert10\rangle)$, which is equal to
    \begin{equation}
      \frac{1}{2}+\frac{1}{4}=\frac{3}{4}.
    \end{equation}
  \item After measuring the first qubit as $0$, then we know that the only part of $\lvert\psi \rangle$ that has the first qubit as $0$ is $(1/\sqrt{2})\ket{00}$. However, we need to renormalize the state to make sure it has a probability of one. So the new state of the system after the measuring the first qubit as $0$ is $\lvert\psi' \rangle = \ket{00}$. 
  \item After measuring the first qubit as $1$, then we know that the only parts of $\lvert\psi \rangle$ that have the first qubit as $1$ are $(1/{2})\ket{10} - (1/{2})\ket{11}$. However, we need to renormalize the state to make sure it has a probability of one. So the new state of the system after the measuring the first qubit as $1$ is $\lvert\psi' \rangle = \frac{1}{\sqrt{2}}\ket{10}-\frac{1}{\sqrt{2}}\ket{11}$.
  \end{enumerate}

\item The two-coin system can have four possible states: $\ket{HH}$, $\ket{HT}$, $\ket{TH}$, or $\ket{TT}$. The equal superposition of these states while the coin is in the air is $\lvert\psi\rangle = \frac{1}{2}\ket{HH}+\frac{1}{2}\ket{HT}+\frac{1}{2}\ket{TH}+\frac{1}{2}\ket{TT}$. 

\item $2\times\text{Prob}\left(even\right)\times\text{Prob}\left(odd\right)=2\times\frac{3}{6}\times\frac{3}{6}=\frac{1}{2}$.

\item Not entangled.  Knowing that the first qubit is 0 does not narrow down whether the second qubit is 0 or 1. Qubit 1 is $\ket{0}$ and Qubit 2 is $\frac{1}{\sqrt{2}}\left(\ket{0}+\ket{1}\right)$. The tensor product gives the state in the question. 

\item
  \begin{enumerate} [a)]
  \item Yes
  \item Yes
  \item Yes
  \item No
  \item No
  \item Yes
  \end{enumerate}

\item
  \begin{enumerate} [a)]
  \item $\ket{00}$
  \item $\ket{01}$
  \item $\ket{10}$
  \item $\frac{1}{\sqrt{2}}\ket{01}+\frac{1}{\sqrt{2}}\ket{11}$
  \item $\frac{1}{\sqrt{2}}\ket{00}+\frac{1}{2}\ket{11}-\frac{1}{2}\ket{10}$
  \end{enumerate}

\item The control qubit stays the same as 0. Since the control is 0, the target was unaffected.  Therefore, the input was $\lvert01\rangle$.

\item Note which qubit is the control qubit. Also note that the IBM Quantum Experience website writes qubits in order of the bottom one first, e.g., \stackanchor{q[0] $\ket{0}$}{q[1] $\ket{1}$} $=\ket{10}$. Whereas in this course we use the opposite convention where qubits are in order of the top one first, e.g., \stackanchor{q[0] $\ket{0}$}{q[1] $\ket{1}$} $=\ket{01}$. This applies to questions 8, 9, and 10. 

  \begin{enumerate}[a)] 
  \item $\ket{01}$
  \item $\ket{10}$
  \item $\ket{11}$
  \item $\ket{01}$ 
  \end{enumerate}

\item Note which qubit is the control qubit. 
  \begin{enumerate}[a)] 
  \item The states change from the start to the end after every gate as: $\ket{00}\to \frac{1}{\sqrt{2}}(\ket{0}+\ket{1})\ket{0} \to \frac{1}{\sqrt{2}}(\ket{00}+\ket{10})$. Note in the last step the second qubit is the control qubit, and because it is $\ket{0}$ there is no change in the first qubit. There is no entanglement in case as the CNOT does not actually change anything. 
  \item The states change from the start to the end after every gate as: $\ket{00}\to \ket{0}(\frac{1}{\sqrt{2}}\ket{0}+\frac{1}{\sqrt{2}}\ket{1}) \to \frac{1}{\sqrt{2}}(\ket{00}+\ket{11})$. Note in the last step the second qubit is the control qubit, and when the control qubit is $\ket{0}$ it does not change the target qubit, whereas when the control qubit is $\ket{1}$ it changes target qubit.  There is entanglement in case as the CNOT actually changes the state.
  \item $\ket{00}$ and $\ket{10}$
  \item $\ket{00}$ and $\ket{01}$
  \end{enumerate}
  
\item 
  \begin{enumerate}[a)]
  \item The states change from the start to the end after every gate as: $\ket{00}\to\ket{01}\to\ket{11}\to\ket{10}$.
  \item $ \ket{00} \to \ket{0}(\frac{1}{\sqrt{2}}\ket{0}+\frac{1}{\sqrt{2}}\ket{1}) \to \frac{1}{\sqrt{2}}(\ket{00}+\ket{11}) \to \frac{1}{\sqrt{2}} \ket{0}(\frac{1}{\sqrt{2}}\ket{0}+\frac{1}{\sqrt{2}}\ket{1}) + \frac{1}{\sqrt{2}} \ket{1}(\frac{1}{\sqrt{2}}\ket{0}-\frac{1}{\sqrt{2}}\ket{1}) = \frac{1}{2} (\ket{00} + \ket{01} + \ket{10} -\ket{11}) $. 
  \end{enumerate}

\item \includegraphics{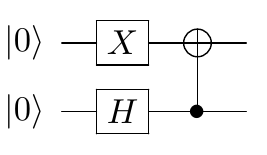}

\item No. Alice measures a random value. This automatically changes Bob's state. However, Alice would need to send a classical message to Bob to find out which state Bob measured; therefore information is still bounded by the classical speed of light. This scenario is known as Bell's theorem. 

\end{enumerate}

\section*{Chapter 8 Solutions}

\begin{enumerate}

\item No, the qubits stay in place. Teleportation only changes the state of an existing qubit.

\item If one entangled particle is measured, the state of the other changes instantaneously no matter how far apart. This was originally why people thought entanglement could transfer data faster than the speed of light. However, no information is communicated between these two particles. You don't know the state of the other entangled qubit unless information about the measurement is transmitted classically. This avoids Einstein's initial reservations about quantum mechanics which he called ``spooky action at a distance.''

\item Once it is measured after the Hadamard gate. 

\item The qubit would collapse into a definite state, and there would be no information about the coefficients $a$ and $b$ when applying the CNOT gate.

\item The $X$ gate flips the $|0\rangle$ into $|1\rangle$ and $|1\rangle$ into $|0\rangle$.

\item The $Z$ gate flips the sign on the $|1\rangle$.

\item The $X$ gate changes the state into $a|0\rangle - b|1\rangle$ and the $Z$ gate flips the sign on the $|1\rangle$ to produce the desired state. 

\end{enumerate}

\section*{Chapter 9 Solutions}

\begin{enumerate}
\item
  \begin{enumerate}[a)]
  \item Both the classical and quantum computer can represent $2^8=256$ classical pieces of information.
  \item Both the classical and quantum computer can represent $2^8=256$ classical pieces of information.
  \item The quantum computer can create a superposition of up to 256 possibilities and do a computation on all of them. However, the output will be only one classical value.
  \end{enumerate}

    \item The photon is put in a \textbf{superposition} such that the function can evaluate both $x=0$ and $x=1$ simultaneously. If the second beam splitter was not there, there would be a 50/50 chance of the photon being in either path and the detectors would not provide a definite answer. The second beam splitter creates the \textbf{interference} necessary to create the 100\% probability of being in the right situation.
    \item \begin{enumerate}[a)]
        \item The most naive classical algorithm is one where you evaluate every function value and make a list of the results. In this case there are $2^3=8$ different function values that would need to be evaluated separately. If each element in the list is identical then the function is constant. 
        \item \includegraphics[width=0.75\textwidth]{Answer3.png} \\ Reprint Courtesy of International Business Machines Corporation, \copyright~ International Business Machines Corporation.\\  Since there are results other than $\ket{000}$, the function is not constant. Also, since the function has four non-zero outcomes (each with 25\% probability) out of the eight possible outcomes, the function is balanced. 
    \end{enumerate}
\end{enumerate}



\end{document}